\PassOptionsToPackage{unicode}{hyperref}
\PassOptionsToPackage{hyphens}{url}
\PassOptionsToPackage{dvipsnames,svgnames,x11names}{xcolor}
\documentclass[
  12pt]{article}

\usepackage{amssymb}
\usepackage{mathrsfs}
\usepackage{iftex}

\usepackage{dsfont} 
\usepackage{graphicx,psfrag,epsf}
\usepackage{enumerate}
\usepackage{array}
\usepackage{url} 
\usepackage{amsmath}
\usepackage{tcolorbox}
\usepackage{subcaption}
\usepackage{bm}
\usepackage{titletoc}
\usepackage{hyperref}
\usepackage{arydshln}
\usepackage{algorithm,algorithmic}
\usepackage{enumitem}
\usepackage{caption}
\usepackage{xr-hyper}

\newtheorem{assumption}{Assumption}

\let\hat\widehat
\let\tilde\widetilde

\newcommand{\ba}{\bm{a}}
\newcommand{\bb}{\bm{b}}

\newcommand{\be}{\bm{e}}
\newcommand{\bbf}{\bm{f}}
\newcommand{\bg}{\bm{g}}

\newcommand{\bv}{\bm{v}}

\newcommand{\bx}{\bm{x}}
\newcommand{\by}{\bm{y}}
\newcommand{\bz}{\bm{z}}

\newcommand{\Ab}{\mathbf{A}}
\newcommand{\Bb}{\mathbf{B}}

\newcommand{\Db}{\mathbf{D}}

\newcommand{\Fb}{\mathbf{F}}
\newcommand{\Gb}{\mathbf{G}}
\newcommand{\Hb}{\mathbf{H}}
\newcommand{\Ib}{\mathbf{I}}

\newcommand{\Lb}{\mathbf{L}}
\newcommand{\Mb}{\mathbf{M}}

\newcommand{\Pb}{\mathbf{P}}

\newcommand{\Sbb}{\mathbf{S}}

\newcommand{\Ub}{\mathbf{U}}
\newcommand{\Vb}{\mathbf{V}}
\newcommand{\Wb}{\mathbf{W}}
\newcommand{\Xb}{\mathbf{X}}
\newcommand{\Yb}{\mathbf{Y}}

\newcommand{\bG}{\bm{G}}

\newcommand{\bS}{\bm{S}}

\newcommand{\bZ}{\bm{Z}}

\newcommand{\cA}{\mathcal{A}}
\newcommand{\cB}{\mathcal{B}}

\newcommand{\cD}{\mathcal{D}}
\newcommand{\cE}{\mathcal{E}}

\newcommand{\cG}{\mathcal{G}}

\newcommand{\cN}{\mathcal{N}}

\newcommand{\cP}{\mathcal{P}}

\newcommand{\cS}{{\mathcal{S}}}

\newcommand{\cU}{\mathcal{U}}

\newcommand{\EE}{\mathbb{E}}

\newcommand{\PP}{\mathbb{P}}

\newcommand{\RR}{\mathbb{R}}

\newcommand{\bbeta}{\bm{\beta}}

\newcommand{\bdelta}{\bm{\delta}}

\newcommand{\btheta}{\bm{\theta}}

\newcommand{\blambda}{\bm{\lambda}}
\newcommand{\bmu}{\bm{\mu}}

\newcommand{\bDelta}{\bm{\Delta}}
\newcommand{\bTheta}{\bm{\Theta}}
\newcommand{\bLambda}{\bm{\Lambda}}

\newcommand{\bSigma}{\bm{\Sigma}}
\newcommand{\bUpsilon}{\bm{\Upsilon}}
\newcommand{\bPhi}{\bm{\Phi}}
\newcommand{\bPsi}{\bm{\Psi}}
\newcommand{\bOmega}{\bm{\Omega}}

\newcommand*{\zero}{{\bm 0}}
\newcommand*{\one}{{\bm 1}}

\newcommand{\diag}{{\rm diag}}
\def\T{{ \intercal }}

\ifx\BlackBox\undefined
\newcommand{\BlackBox}{\rule{1.5ex}{1.5ex}}  
\fi

\ifx\QED\undefined
\def\QED{~\rule[-1pt]{5pt}{5pt}\par\medskip}
\fi

\ifx\proof\undefined
\newenvironment{proof}{\par\noindent{\bf Proof\ }}{\hfill\BlackBox\\[2mm]}
\fi

\ifx\theorem\undefined
\newtheorem{theorem}{Theorem}
\fi
\ifx\example\undefined
\newtheorem{example}{Example}
\fi
\ifx\property\undefined
\newtheorem{property}{Property}
\fi
\ifx\lemma\undefined
\newtheorem{lemma}{Lemma}
\fi
\ifx\proposition\undefined
\newtheorem{proposition}{Proposition}
\fi
\ifx\remark\undefined
\newtheorem{remark}{Remark}
\fi
\ifx\corollary\undefined
\newtheorem{corollary}{Corollary}
\fi
\ifx\definition\undefined
\newtheorem{definition}{Definition}
\fi
\ifx\conjecture\undefined
\newtheorem{conjecture}{Conjecture}
\fi
\ifx\fact\undefined
\newtheorem{fact}{Fact}
\fi
\ifx\claim\undefined
\newtheorem{claim}{Claim}
\fi
\ifx\cond\undefined
\newtheorem{cond}{Condition}
\fi

\ifPDFTeX
  \usepackage[T1]{fontenc}
  \usepackage[utf8]{inputenc}
  \usepackage{textcomp} 
\else 
  \usepackage{unicode-math}
  \defaultfontfeatures{Scale=MatchLowercase}
  \defaultfontfeatures[\rmfamily]{Ligatures=TeX,Scale=1}
\fi
\usepackage{lmodern}
\ifPDFTeX\else  
\fi
\IfFileExists{upquote.sty}{\usepackage{upquote}}{}
\IfFileExists{microtype.sty}{
  \usepackage[]{microtype}
  \UseMicrotypeSet[protrusion]{basicmath} 
}{}
\makeatletter
\@ifundefined{KOMAClassName}{
  \IfFileExists{parskip.sty}{%
    \usepackage{parskip}
  }{
    \setlength{\parindent}{2pt}
    \setlength{\parskip}{6pt plus 2pt minus 1pt}}
}{
  \KOMAoptions{parskip=half}}
\makeatother
\usepackage{xcolor}
\makeatletter
\ifx\paragraph\undefined\else
  \let\oldparagraph\paragraph
  \renewcommand{\paragraph}{
    \@ifstar
      \xxxParagraphStar
      \xxxParagraphNoStar
  }
  \newcommand{\xxxParagraphStar}[1]{\oldparagraph*{#1}\mbox{}}
  \newcommand{\xxxParagraphNoStar}[1]{\oldparagraph{#1}\mbox{}}
\fi
\ifx\subparagraph\undefined\else
  \let\oldsubparagraph\subparagraph
  \renewcommand{\subparagraph}{
    \@ifstar
      \xxxSubParagraphStar
      \xxxSubParagraphNoStar
  }
  \newcommand{\xxxSubParagraphStar}[1]{\oldsubparagraph*{#1}\mbox{}}
  \newcommand{\xxxSubParagraphNoStar}[1]{\oldsubparagraph{#1}\mbox{}}
\fi
\makeatother

\usepackage{longtable,booktabs,array}
\usepackage{calc} 
\usepackage{etoolbox}
\makeatletter
\patchcmd\longtable{\par}{\if@noskipsec\mbox{}\fi\par}{}{}
\makeatother
\IfFileExists{footnotehyper.sty}{\usepackage{footnotehyper}}{\usepackage{footnote}}
\makesavenoteenv{longtable}
\usepackage{graphicx}
\makeatletter
\def\maxwidth{\ifdim\Gin@nat@width>\linewidth\linewidth\else\Gin@nat@width\fi}
\def\maxheight{\ifdim\Gin@nat@height>\textheight\textheight\else\Gin@nat@height\fi}
\makeatother
\setkeys{Gin}{width=\maxwidth,height=\maxheight,keepaspectratio}
\makeatletter
\def\fps@figure{htbp}
\makeatother

\makeatletter
\@ifpackageloaded{caption}{}{\usepackage{caption}}
\AtBeginDocument{%
\ifdefined\contentsname
  \renewcommand*\contentsname{Table of contents}
\else
  \newcommand\contentsname{Table of contents}
\fi
\ifdefined\listfigurename
  \renewcommand*\listfigurename{List of Figures}
\else
  \newcommand\listfigurename{List of Figures}
\fi
\ifdefined\listtablename
  \renewcommand*\listtablename{List of Tables}
\else
  \newcommand\listtablename{List of Tables}
\fi
\ifdefined\figurename
  \renewcommand*\figurename{Figure}
\else
  \newcommand\figurename{Figure}
\fi
\ifdefined\tablename
  \renewcommand*\tablename{Table}
\else
  \newcommand\tablename{Table}
\fi
}
\@ifpackageloaded{float}{}{\usepackage{float}}
\floatstyle{ruled}
\@ifundefined{c@chapter}{\newfloat{codelisting}{h}{lop}}{\newfloat{codelisting}{h}{lop}[chapter]}
\floatname{codelisting}{Listing}

\makeatother
\makeatletter
\@ifpackageloaded{caption}{}{\usepackage{caption}}
\@ifpackageloaded{subcaption}{}{\usepackage{subcaption}}
\makeatother

\ifLuaTeX
  \usepackage{selnolig}  
\fi
\usepackage[]{natbib}
\usepackage{bookmark}

\IfFileExists{xurl.sty}{\usepackage{xurl}}{} 
\hypersetup{
  pdftitle={Title},
  pdfauthor={Author 1; Author 2},
  pdfkeywords={3 to 6 keywords, that do not appear in the title},
  colorlinks=true,
  linkcolor={blue},
  filecolor={Maroon},
  citecolor={Blue},
  urlcolor={Blue},
  pdfcreator={LaTeX via pandoc}}

\newcommand{\anon}{1}

\begin{document}

\def\spacingset#1{\renewcommand{\baselinestretch}%
{#1}\small\normalsize} \spacingset{1}


\if1\anon
{
  \title{\bf  Identifiability and  Inference for Generalized \\Latent Factor Models}
  \author{Chengyu Cui and Gongjun Xu\hspace{.2cm}\\
    Department of Statistics, University of Michigan}
  \maketitle
} \fi

\if0\anon
{
  \bigskip
  \bigskip
  \bigskip
  \begin{center}
    {\LARGE\bf  Identifiability and  Inference for Generalized \\Latent Factor Models}
\end{center}
  \medskip
} \fi

\bigskip
\begin{abstract}Generalized latent factor analysis not only provides a useful latent embedding approach in statistics and machine learning, but also serves as a widely used tool across various scientific fields, such as psychometrics, econometrics, and social sciences.  Ensuring the identifiability of latent factors and the loading matrix is essential for the model's estimability and interpretability, and various identifiability conditions have been employed by practitioners. However, fundamental statistical inference issues for latent factors and factor loadings under commonly used identifiability conditions remain largely unaddressed, especially for correlated factors and/or non-orthogonal loading matrix. In this work, we focus on the maximum likelihood estimation for generalized latent factor models and establish statistical inference properties under popularly used identifiability conditions. The developed theory is further illustrated through numerical simulations and an application to a personality assessment dataset.
\end{abstract}

\noindent%



\spacingset{1.19}
\captionsetup[figure]{width=1\linewidth}
\section{Introduction}
\label{sec:intro}
Generalized latent factor analysis provides a popular statistical tool to analyze heterogeneous large-scale data with discrete and/or continuous responses, with extensive applications in various fields, including psychology \citep{maccallum1999sample,reckase2009}, economics \citep{ross2013arbitrage,stock2002forecasting}, and genetics \citep{preacher2002exploratory}. These models capture the underlying relationships between multiple responses by constructing low-dimensional latent factors that model the generative processes of the observed data. Latent factors along with their corresponding factor loadings often receive significant attention for scientific interpretation in various domain-specific applications. For instance, in psychological assessments, responses are typically recorded as Likert scale ratings indicating the level of agreement with various items, with latent factors representing respondents' psychological traits and factor loadings representing items' characteristics.

The identifiability issue poses a significant challenge in latent factor modeling, which is a prerequisite for model estimability and interpretability. Factor models are not identifiable without additional restrictions.  In the case of unidimensional latent factors, the identification problem is limited to the indeterminacy in the scale of the factor loadings and latent factors. A common approach in this case is to normalize the latent factors to resolve this indeterminacy. In the multidimensional case, identification becomes considerably more complicated. In particular, altering the latent structure through any invertible linear transformation leaves the distribution of responses unchanged, provided the loading structure is adjusted accordingly. To address the identifiability issue, practitioners usually impose certain constraints on the factors and factor loadings \citep[e.g.,][]{andersonintroduction,ullman2012structural}. 
However, much of the research predominantly examines the theoretical aspects of factor models assuming orthogonal factors and loadings \citep{bai2003inferential,bai2006confidence,fan2017sufficient,Wang2018MaximumLE}. Such assumption, though mathematically convenient, 
may not fully align with the complexities encountered in practical scenarios.
For instance,  different dimensions of intelligence (e.g., verbal, mathematical) or different types of mental anxiety (e.g., social, general) are considered to be related rather than independent, as many psychological theories posit \citep{brown2015confirmatory,kline2023principles}.
In his pioneering work on psychological assessment, \cite{thurstone1947multiple} says ``{\it it seems just as unnecessary to require that mental traits shall be uncorrelated in the general population as to require that height and weight be uncorrelated in the general population}.'' 
Correlated factors are thus more commonly used than orthogonal ones in scientific applications, particularly in fields such as psychology, education, and social sciences. 

Practical approaches to address the identifiability issue of latent factor models often involve design information that specifies relationships between specific item responses and the latent factors \citep{andersonintroduction}. By prescribing certain elements in the loading matrix, some factors are designed to influence responses in a specific manner. The patterns of factor loadings reflect scientific hypotheses on the factor structure of observed responses, for which reason such framework is also referred to as confirmatory factor analysis \citep{joreskog1969general,thompson2004exploratory}.
Confirmatory factor analysis has been popularly used in econometrics, psychometrics, and social sciences \citep{benitez2020perform,henseler2020using}.
Despite enhanced interpretability and widespread application, the theoretical properties of these models, particularly in modern large-scale settings with both a large sample size and high-dimensional manifest variables, remain largely unexplored. 

There are several challenges in deriving statistical inference theory under practical identifiability conditions.
The first one arises from the nonlinearity in the generalized latent factor model. Most existing works focus on the statistical properties of linear factor models~\citep{bai2003inferential,bai2012statistical,bing2020adaptive,Rohe2020VintageFA}. 
Analyses under linear factor models typically rely on the covariance structure of continuous responses, where the linear form allows the noise component to be cleanly separated from the signal. In the nonlinear case, however, these techniques are no longer applicable. Another challenge lies in accommodating various practical identifiability conditions and establishing valid statistical inference under each specification.
Although recent works have investigated nonlinear latent factor models~\citep{chen2019joint,chen2021nonlinear,Wang2018MaximumLE,chen2024note}, they predominantly assume orthogonal latent factors and factor loadings. This assumption is often difficult to satisfy in practice, where non-orthogonal designs are far more common. Despite their widespread use, establishing sufficient and necessary conditions for identifiability—and the corresponding inference results—remains an open problem. 
Statistical inferences for non-orthogonal cases are significantly more challenging because, different from the orthogonal setting, the joint log-likelihood function under the non-orthogonal identifiability constraints is not strictly concave. In particular, even after imposing these constraints, the high-dimensional Hessian matrix of the constrained joint log-likelihood function can be ill-conditioned, with its smallest eigenvalue converging to zero. This hinders further derivation of consistency results and requires different theoretical techniques from the existing studies. 


\subsection{Overview of Contribution}
In this article, we address these open problems by studying the sufficient and necessary identifiability conditions and developing statistical inference theory for the maximum likelihood estimators under these conditions.
Our main contributions are as follows: 
\begin{enumerate}[label=(\arabic*), leftmargin=0.65cm]
\item We study the minimal identifiability conditions for generalized latent factor models that cover a broad class of practically relevant settings. Specifically, we establish in Theorem~\ref{thm_iff_ic1} and Theorem~\ref{thm_iff_ic4} the necessary and sufficient identifiability conditions of prescribing zeros in the loading matrix for orthogonal and non-orthogonal factor designs, respectively.
Our findings are of theoretical and practical significance, as these identifiability conditions are highly adaptable and easy to satisfy in practice {and also provide a foundation for developing valid inference procedures across practical settings.}
\item We establish non-asymptotic $l_2$ and $l_{\infty}$ error bounds for the maximum likelihood estimation of generalized latent factor models under these identifiability conditions. To achieve these results, we establish sharp non-asymptotic concentration bounds for the Hessian matrix of the log-likelihood function. Notably, to address the ill-conditioning of the Hessian matrix under non-orthogonal identifiability conditions, we develop novel techniques to establish estimation consistency under those conditions. Furthermore, we demonstrate that our error bounds attain the oracle rates across a broad range of asymptotic regimes,  {where the oracle rate for the loading estimator is defined as the rate achieved when the latent factors are assumed to be known, and the oracle rate for the latent factor estimator is defined analogously.}


\item We derive the asymptotic normality of maximum likelihood estimators under all identifiability conditions, under the double asymptotic regime where both sample size and the number of manifest variables are large. Specifically, we establish asymptotic normality for estimators across different identifiability conditions, enabling conducting statistical inference in these settings. Notably, the asymptotic distribution under non-orthogonal designs is derived despite the presence of an ill-conditioned Hessian matrix. {Interestingly}, the asymptotic covariance matrices of the limiting distributions differ under various identifiability conditions, suggesting some commonly used inference procedures in practice may be problematic (see Section~\ref{sec:mainres} for more details). This finding is further substantiated through a simulation study given in Section~\ref{sec:simu}.

\end{enumerate} 

\subsection{Related Work}\label{sec_related_work}
{
Early psychometrics work studied identifiability of sparse factor loadings through rotation methods~\citep{kaiser1958varimax,jennrich1966rotation,browne2001overview,jennrich2006rotation}. Unlike these studies, our analysis establishes minimal conditions for identifiability that do not depend on any particular rotation criterion. Moreover, whereas prior work focused mainly on the identifiability issue only or on the empirical performance of the estimators~\citep{dunn1973anote,joreskog1969general,jennrich2006rotation}, we develop rigorous statistical guarantees, including consistency and asymptotic distribution results, for estimation in a generalized latent factor model under various identifiability conditions.}

{A notable related work is \cite{Rohe2020VintageFA}, which studies statistical properties of varimax \citep{kaiser1958varimax} under the orthogonal factor setting. They made a substantial contribution by showing that, under mild distributional assumptions, varimax yields identifiable loadings that can be consistently estimated. Building on this, \citet{bing2025optimal} proposed a computationally efficient varimax-based procedure and established minimax optimal rates for the estimation of the loading matrix. Our work differs in two respects. First, we allow a broader class of identifiability conditions that accommodates both uncorrelated and correlated factors. Second, we develop distributional theory for the resulting estimators, enabling further statistical inference under various identifiability conditions.}

{Our work is also related to the low-rank matrix factorization literature, in which the product of latent factor and loading matrix forms a low-rank matrix that is the primary object of interest. Many existing studies focus on the linear model setting and establish statistical guarantees for recovering this matrix and uncertainty quantification for its individual entries~\citep{chen2019inference,chen2020noisy,choi2024matrix,wang2025robust,cui2026convexity}. For the nonlinear case, \cite{chen2024note} established the entry-wise consistency of the recovered matrix under a generalized latent factor model framework similar to ours,  but neither addresses the identifiability issue of the factorization nor develops any distributional theory.} 

The rest of the article is organized as follows. Section~\ref{sec:setup} introduces the generalized latent factor model and the sufficient and necessary identifiability conditions.
Section~\ref{sec:mainres} provides statistical theory of the maximum likelihood estimators under these identifiability conditions, including non-asymptotic error bounds and asymptotic distributions. Section~\ref{sec:simu} reports simulation studies.  We apply the proposed method to analyze a personality assessment dataset in Section~\ref{sec:data}. The article concludes with final remarks in Section~\ref{sec:end}, and all proofs and technical details are included in the Supplementary Material.

{\textbf{Notation}.} For $a,b\in\RR$, let $a\vee b = \max(a,b)$ and $a\wedge b = \min(a,b)$. We use $O_p(\cdot)$ and $o_p(\cdot)$ as the probability versions of $O(\cdot)$ and $o(\cdot)$ for convergence in probability. 
For any integer $N$, let $[N] = \{1, \dots, N\}$. For any set $\cS$, let $|\cS|$ denote its cardinality. 
For a matrix $\Ab=\{a_{ij}\}\in\RR^{n\times m}$, let $a_{ij}$ and $\ba_i$ denote its $(i,j)$th entry and $i$th row. For a vector $\bx\in\RR^n$, define $\|\bx\|=(\sum_{i=1}^n x_i^2)^{1/2}$ and $\|\bx\|_\infty=\max_i |x_i|$. For a matrix $\Ab\in\RR^{n\times m}$, define $\|\Ab\|_F=(\sum_{i=1}^n\sum_{j=1}^m a_{ij}^2)^{1/2}$ and $\|\Ab\|_{\max}=\max_{i,j}|a_{ij}|$.
For subsets $\cS_1\subseteq[n]$ and $\cS_2\subseteq[m]$, let $\Ab_{[\cS_1,\cS_2]}$ denote the corresponding submatrix. We write $\Ab_{[,\cS_2]}$ and $\Ab_{[\cS_1,]}$ when all rows or all columns are retained, respectively, and $\Ab_{[-\cS_1,-\cS_2]}$ for the submatrix obtained by removing rows in $\cS_1$ and columns in $\cS_2$. For a square matrix $\Ab$, let $\lambda_{\min}(\Ab)$ and $\lambda_{\max}(\Ab)$ denote its smallest and largest eigenvalues, and let $\diag(\Ab)$ denote the diagonal matrix formed from the diagonal entries of $\Ab$.
Let $\Ib_n$ be the $n \times n$ identity matrix, $\zero_n$ be an all-zero vector of dimension $n$, and $\zero_{n\times m}$ be an all-zero matrix of dimension $n\times m$. 
{A matrix $\Pb\in\RR^{K\times K}$ is called a signed permutation matrix if $\Pb^\T\Pb=\Ib_K$ and each row and column has exactly one nonzero entry, which is equal to $1$ or $-1$. Let $\cP_K$ denote the set of all $K\times K$ signed permutation matrices.}

\section{Model Setup and Identifiability Conditions}\label{sec:setup}
\subsection{Model Setup}\label{sec_model_setup}
Let $N$ and $J$ denote the number of individuals and manifest variables (items), respectively. For $i\in [N]$ and $j\in [J]$, let $Y_{ij}$ denote the response of individual $i$ to item $j$. 
Suppose there are $K$ latent dimensions, where $K$ is known.  We assume each $i$ has a $K$-dimensional latent factor $\bbf_i = (f_{i1},\cdots,f_{iK})^\T$, and each manifest variable is associated with a $K$-dimensional factor loading $\blambda_j = (\lambda_{j1},\cdots,\lambda_{jK})^\T$. Consider the Big Five personality measurement as an example \citep{goldberg1999broad}, which includes designed items to assess $K=5$ personality factors: ``Agreeableness'', ``Conscientiousness'', ``Extraversion'', ``Neuroticism'', and ``Openness''. In this application, each component of $\bbf_i$ represents a latent personality trait for the $i$th individual, and each loading entry of $\blambda_j$ characterizes the item information by measuring the significance of how item $j$ loads on each of the latent traits. Note that such latent personality factors are typically correlated in practice, and it is of scientific interest to make statistical inferences for both the latent factors and factor loadings. 

We consider a general distribution of the response $Y_{ij}$ given $\blambda_j$ and $\bbf_i$ with the probability density/mass function as:
\begin{equation}
    g_{ij}(Y_{ij} = y_{ij}\mid\eta_{ij})\text{, where }\;\eta_{ij}~=~\blambda_j^\T\bbf_i,\label{eq_factor_m}
\end{equation}
with $y_{ij}$ denoting a realization of $Y_{ij}$.  Here $\eta$ is the unknown parameter, and given $\eta$, the distribution function $g_{ij}(\cdot|\cdot)$ is pre-specified and may vary across $i$ and $j$. Additional technical conditions on $g_{ij}(\cdot|\eta)$ are introduced later in Assumption~\ref{assumption:smoothness} in Section~\ref{sec:mainres}.

Our model setup is flexible and allows for a variety of density (or mass) functions to model different types of response data across a wide range of applications. { One important class of such link functions arises from the exponential family, where the density of each response is parameterized by treating $\eta_{ij}$ as the natural parameter~\citep{skrondal2004generalized}:
\begin{equation*}
    g_{ij}(Y_{ij} \;=\; y_{ij}\mid\eta_{ij}) = h(y_{ij})\exp\big\{y_{ij}\eta_{ij} - b(\eta_{ij})\big\}.
\end{equation*}Common examples under this setting include linear factor models, widely used in macroeconomic forecasting~\citep{stock2002forecasting,bai2006confidence}, where
      $\log g_{ij}(y\mid \eta_{ij})\propto -(y-\eta_{ij})^2$; logistic factor models commonly used in educational and psychological measurement \citep{reckase2009} with
      $g_{ij}(y \mid \eta_{ij}) = {\exp(\eta_{ij})^y}/{(1+\exp(\eta_{ij}))}$; Poisson factor model often used in genetics analysis~\citep{wang2025data} with $g_{ij}(y\mid\eta_{ij}) \propto \exp(-e^{\eta_{ij}}+\eta_{ij}y)$. Beyond the exponential family, our framework also accommodates other link functions such as the probit and tobit models. For the probit model,  $
      g_{ij}(y \mid \eta_{ij}) = \{\Phi(\eta_{ij})\}^y\{1-\Phi(\eta_{ij})\}^{1-y}$ where $\Phi(\cdot)$ stands for the cumulative distribution function of standard normal distribution. For tobit models, $
      g_{ij}(y \mid \eta_{ij})=\phi(y-\eta_{ij})1_{(y>0)} + \Phi(-\eta_{ij})1_{(y=0)} $ where $\phi(\cdot)$ is the probability density function of standard normal distribution.}

Denote by $\Yb=\{Y_{ij}\}_{N\times J}\in\RR^{N\times J}$ the matrix that stacks all the responses and let $\bLambda = (\blambda_1,\cdots,\blambda_J)^\T$ and $\Fb=(\bbf_1,\cdots,\bbf_N)^\T$ be the matrices that stack all the factor loadings and latent factors, respectively. Without loss of generality, assume that $\bLambda$ and $\Fb$ each have full column-rank. We further assume that the responses are conditionally independent given all parameters $\bbf_i$'s and $\blambda_j$'s. The joint log-likelihood function of $(\bLambda,\Fb)$ given the observed data $\Yb$ can be written as
\begin{equation}
    L(\bLambda,\Fb|\Yb)\;=\;\sum_{i=1}^N\sum_{j=1}^J l_{ij}(\eta_{ij}\mid Y_{ij}),\label{eq_joint_likleihood}
\end{equation}
where $l_{ij}(\eta)$ is the individual log-likelihood defined as $l_{ij}(\eta)=\log g_{ij}(Y_{ij}\mid\eta)$.

In many applications such as item response theory, the considered factor models also incorporate intercept terms \citep{reckase2009}. A prevalent approach involves introducing an intercept indexed by $j$, specifying $\eta_{ij}$ in \eqref{eq_factor_m} as $\eta_{ij} = \blambda_j^\T\bbf_i + \lambda_{j0}$. Additionally, some other settings include an intercept indexed by $i$, where $\eta_{ij}$ is defined as $\blambda_j^\T\bbf_i + \lambda_{j0} + f_{i0}$. We remark that the setup with both intercepts is equivalent to \eqref{eq_factor_m} by setting latent dimension to be $K+2$ along with $f_{i1}=1$ and $\lambda_{j2} = 1$. In this case, $\lambda_{j1}$ serves as the intercept indexed by $j$ and $f_{i2}$ as the intercept indexed by $i$. Given that the theoretical treatment of intercepts inherently separates from that of the factor structure, we focus our discussion on the model setup \eqref{eq_factor_m} without intercepts in Sections~\ref{sec:setup} and \ref{sec:mainres}, and detailed analysis of models with intercepts is deferred to Section~\ref{sec:intecept} in the Supplementary Material for conciseness.
\begin{remark}\label{remark_random}
{In model \eqref{eq_factor_m}, the latent factors and loadings are treated as model parameters estimated jointly, as both are of great practical interest. Specifically, the latent factors characterize individuals’ underlying features, while the loading parameters provide an embedding of the items into the same latent space, describing how these latent features influence item responses. We note that, in some formulations, the factors are instead modeled as continuous random variables~\citep{andersonintroduction,fan2017sufficient}. Our results can be extended to accommodate this setting; details are provided in Section~\ref{supp_sec_random_factor} of the Supplementary Material.}
\end{remark}

\subsection{Identifiability Conditions}\label{subsec:identification}

Identification is a major problem in latent factor models. For model \eqref{eq_factor_m}, a key identifiability issue involves identifying the factors and loadings from the product $\bTheta := \{\eta_{ij}\}_{J\times N} = \bLambda\Fb^\T$. Specifically, for any invertible matrix $\Gb\in\RR^{K\times K}$, we have $\bTheta =\bLambda\Fb^\T=(\bLambda\Gb)(\Fb\Gb^{-\T})^\T$, which implies that when changing factor $\Fb$ and loading $\bLambda$ to $\Fb\Gb^{-1}$ and $\bLambda\Gb^\T$, the distributions for $\{Y_{ij}\}_{N\times J}$ stay the same.
Such indeterminacy is often referred to as \textit{rotation indeterminacy} in factor analysis literature \citep{andersonintroduction,bai2013principal}.  To characterize the identifiability issue arising from the rotation indeterminacy, we introduce the following definition of \textit{rotational identifiability}.
\begin{definition}\label{def_id}\it
The parameters $(\bLambda,\Fb)$ in model \eqref{eq_factor_m} are called rotationally identifiable if the following holds: for any $(\bLambda,\Fb)$ and $(\bar\bLambda,\bar\Fb)$ such that $\bLambda\Fb^\T=\bar\bLambda\bar\Fb^\T$, $(\bLambda,\Fb)$ and $(\bar\bLambda,\bar\Fb)$ are equivalent up to a reordering of the latent dimensions and a sign change, i.e., $\bar\bLambda = \bLambda\Gb$ and $\bar\Fb=\Fb\Gb$ for some $\Gb\in\cP_K$.
\end{definition}
\begin{remark}\it
Rotational identifiability ensures that given $\bTheta=\bLambda\Fb^\T$, the matrices $\bLambda$ and $\Fb$ are uniquely determined up to a reordering of the latent dimensions and sign changes. Allowing for such reordering and sign changes is a standard practice in factor analysis, as these transformations do not affect the interpretability or structure of the factor loadings and latent variables~\citep{andersonintroduction,bai2013principal}. 
\end{remark}


Throughout the manuscript, we refer identifiability of \eqref{eq_factor_m} as rotational identifiability unless otherwise specified. To ensure rotational identifiability, a minimum of $K^2$ constraints on the loadings and/or factors is needed, as a $K\times K$ invertible matrix $\Gb$ has $K^2$ free parameters~\citep{joreskog1969general,andersonintroduction}. Commonly used identifiability conditions include specifying certain entries in the loading matrix to be zeros, introducing a sparse loading structure, or setting certain factors to be  orthogonal~~\citep{andersonintroduction,bai2012statistical}. 

\begin{figure}[h]
\centering    
        \includegraphics[width=2in]{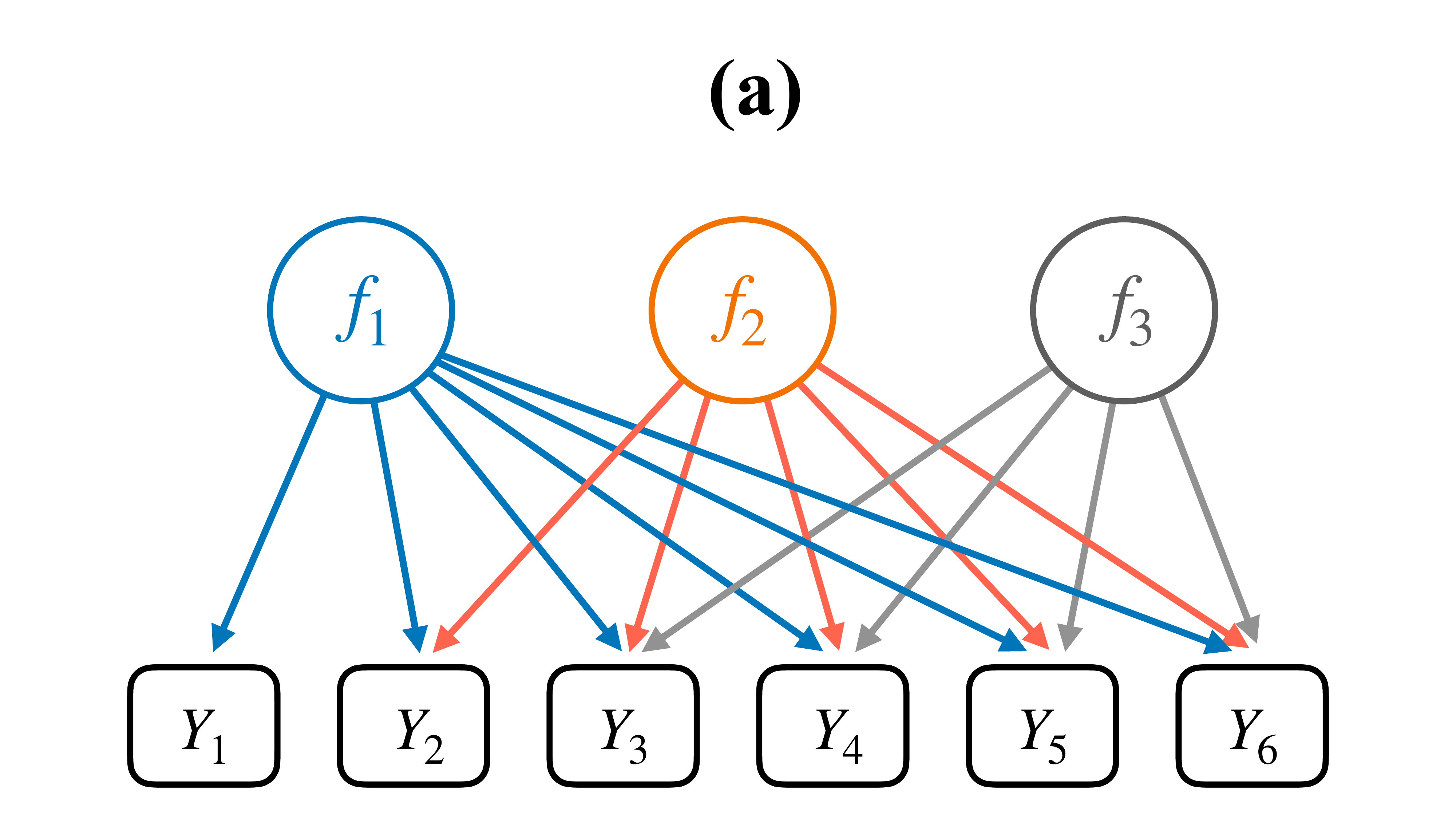}
        \includegraphics[width=2in]{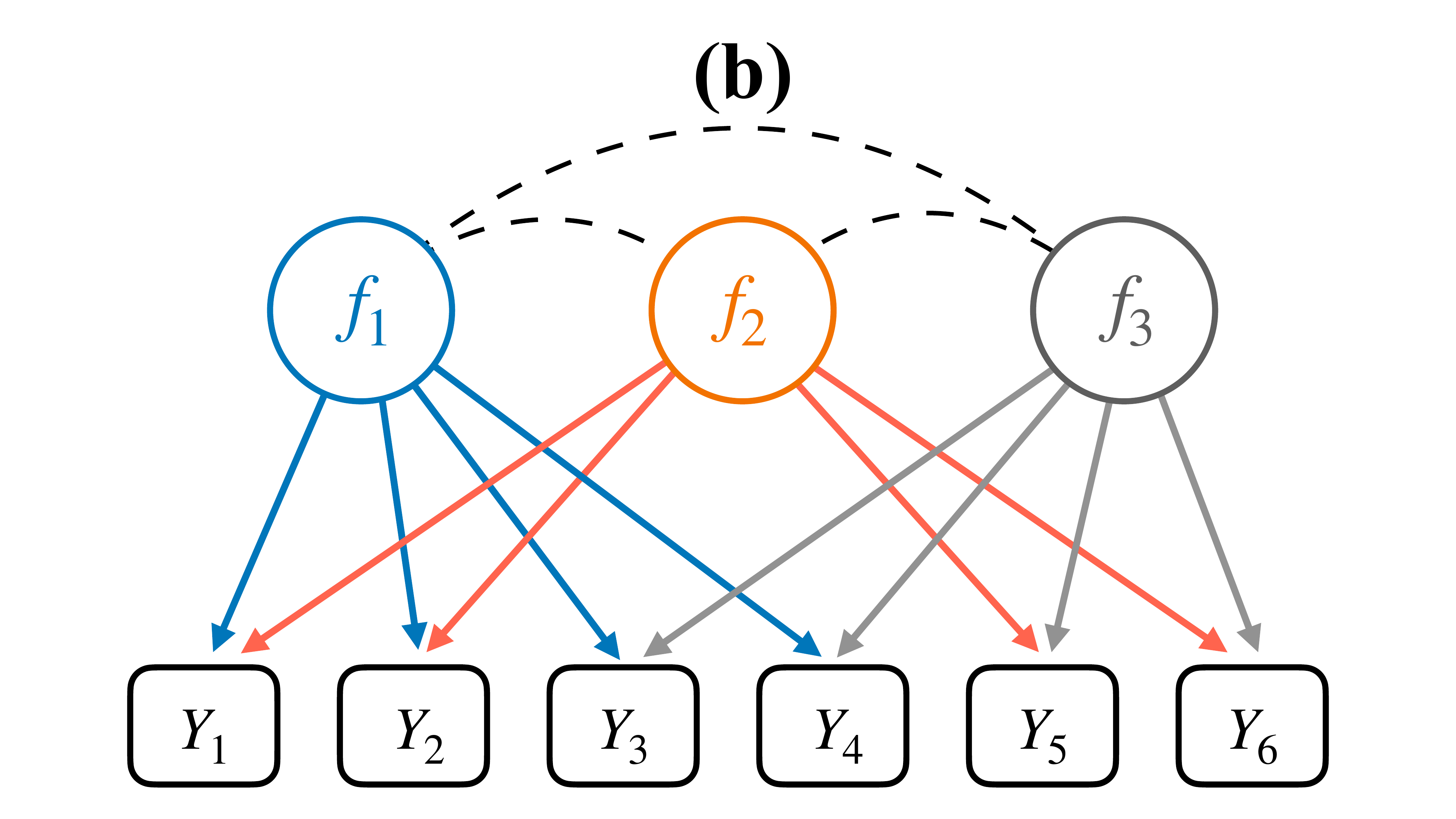}
        \includegraphics[width=2in]{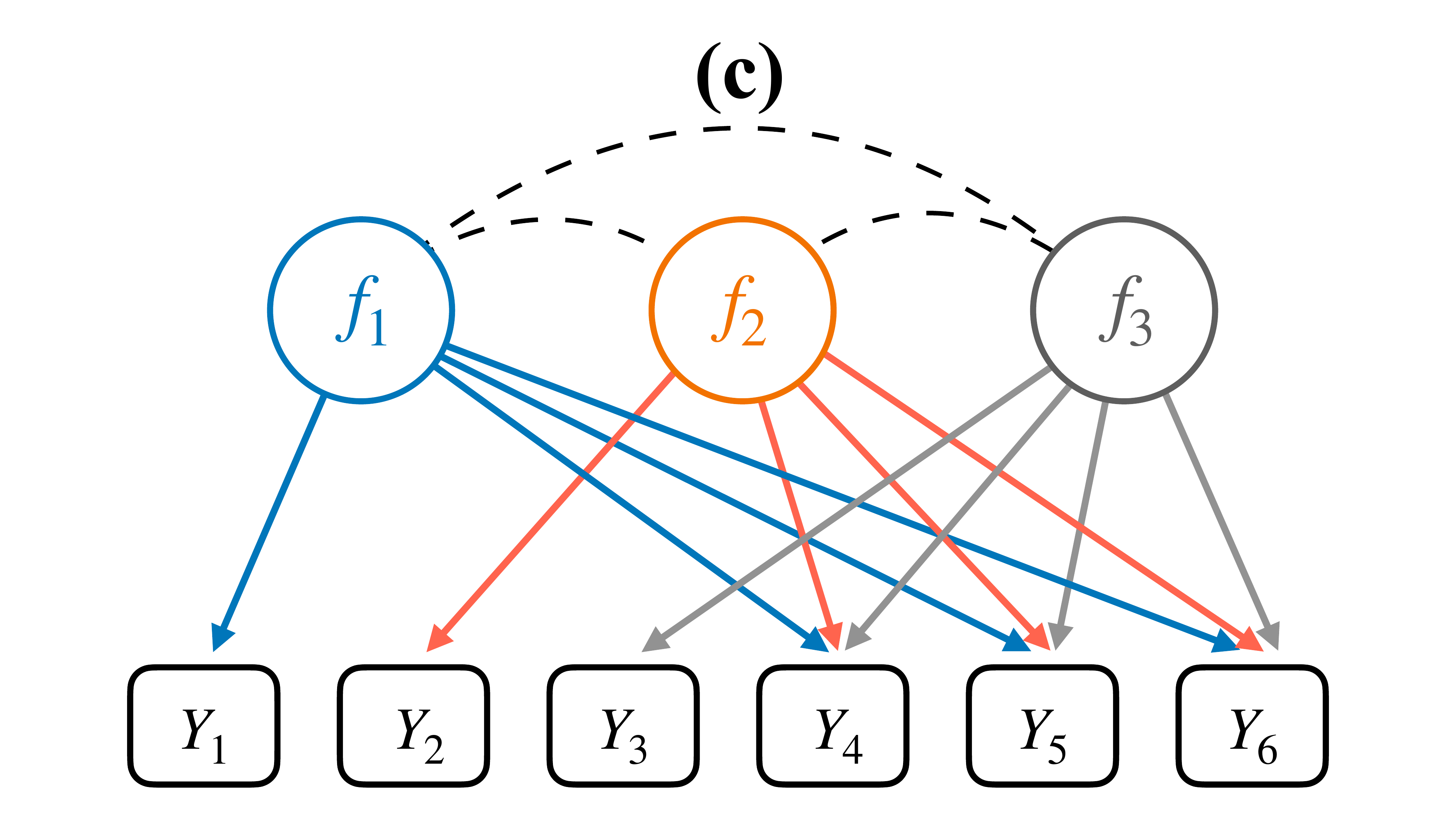}
    \caption{\small Rotationally identifiable examples. The undirected edges among the latent factors $f_1$--$f_3$ indicate possible correlations. Each directed edge represents that a latent factor affects the response variable $Y_j$ for $j\in[6]$. When there is no directed edge from a factor to a response, it indicates the corresponding loading is zero.}
    \label{fig:illus}
\end{figure}
These identifiability conditions can be categorized into two types: orthogonal and oblique (non-orthogonal), depending on whether the factors are assumed to be orthogonal or not. See illustrative examples in Figure~\ref{fig:illus}, where (a) represents an orthogonal case while (b) and (c) are oblique. In the subsequent sections, we present an exhaustive list of minimal conditions under both orthogonal and non-orthogonal settings that guarantee rotational identifiability. {For the rest of the article, we treat these conditions as given. When they are not known in advance, Section~\ref{sec_determine_id} of the Supplementary Material provides a data-driven method for selecting appropriate identifiability conditions.}

\begin{remark}
{ Both linear latent factor model and the generalized latent factor model suffer from rotation indeterminacy, which is the primary source of non-identification in factor analysis~\citep{bai2012statistical,bai2013principal}.}
    Rotational identifiability is a necessary, though not by itself sufficient, condition for full model identification.  Specifically, the full identification of model~\eqref{eq_factor_m} requires that \begin{equation*}\PP(\Yb|\bLambda,\Fb) = \PP(\Yb|\tilde\bLambda,\tilde\Fb)\quad\Rightarrow\quad(\bLambda,\Fb) = (\tilde\bLambda,\tilde\Fb)\text{ up to a signed permutation}\end{equation*} for any parameter pairs $(\bLambda,\Fb) $ and $(\tilde\bLambda,\tilde\Fb)$, where $\PP(\Yb|\bLambda,\Fb)$ denotes the distribution of $\Yb$ under model~\eqref{eq_factor_m}~\citep{newey1994large}. In the subsequent section, we first focus on rotational identifiability and present a series of sufficient conditions that resolve rotation indeterminacy. Once rotational identifiability is established, full model identifiability for \eqref{eq_joint_mle} follows from the identifiability of $\eta$ in $g_{ij}(\cdot|\eta)$; that is, for any $\eta,\tilde\eta$, if $g_{ij}(\cdot|\eta) = g_{ij}(\cdot|\tilde\eta)$, then $\eta = \tilde\eta$. This is guaranteed by Assumption~\ref{assumption:smoothness} in Section~\ref{sec:mainres}, which imposes regularity conditions on the log-likelihood function $-\log g_{ij}(\cdot|\cdot)$. Thus, any of the identifiability conditions below, together with Assumption~\ref{assumption:smoothness}, implies full identifiability of \eqref{eq_factor_m}; see Section~\ref{sup_prove_full_id} of the Supplementary Material. { We emphasize that, although rotational identifiability issues arise in both linear and generalized latent factor models, statistical inference for generalized latent factor models under identification constraints, which is the focus of this work, differs fundamentally from that in the linear case.}
    \end{remark}


\subsubsection{Identifiability under orthogonal factors}\label{sec_id_orthogonal}
We define $\Mb_{\lambda\lambda} = J^{-1} \sum_{j=1}^J \blambda_j \blambda_j^\T$ and $\Mb_{ff} = N^{-1} \sum_{i=1}^N \bbf_i \bbf_i^\T$. In the orthogonal case, the factors are constrained to be orthogonal by setting the off-diagonal entries of  $\Mb_{ff}$ to zero. 
Since $\Mb_{ff}$ is symmetric, additional $K(K+1)/2$ constraints are needed to address the rotation indeterminacy. 
In the following, we study two popular ways to put these constraints to achieve rotational identifiability.  

One approach is to assume factors are orthogonal and normalized, and loadings are orthogonal, which is also called {\it exploratory factor analysis} in the literature \citep{andersonintroduction}. In particular, we consider the following identifiability condition:
\begin{center}
    \textbf{IC}$^{(0)}$: $\Mb_{ff}=\Ib_K$; $\Mb_{\lambda\lambda}$ is diagonal with distinct elements.
\end{center}
Here, \textbf{IC} stands for \textbf{I}dentifiability \textbf{C}ondition and we use superscript $(0)$ to indicate that it serves as a benchmark for comparison with other considered identifiability conditions.
For identification purpose, the diagonal elements of $\Mb_{\lambda\lambda}$ need to be distinct. 
Variants of IC$^{(0)}$ include normalizing the loadings (restricting $\Mb_{\lambda\lambda}=\Ib_K$), or restricting $\Mb_{\lambda\lambda}=\Mb_{ff}$ \citep{Wang2018MaximumLE}. While widely studied in the literature, this condition is short of practical interpretation, as a column-wise orthogonal loading matrix is often considered too restrictive in scientific applications. 

A more common approach in educational and psychological applications to ensure identifiability is to prescribe zeros in the loading matrix $\bLambda$. 
Under the orthogonal factors with $\Mb_{ff} = \Ib_K$, the additional $(K-1)K/2$ constraints are imposed by fixing $(K-1)K/2$ entries in the loading matrix to be zeros. Interestingly, not all configurations of the $(K-1)K/2$ zero entries ensure identifiability. We illustrate this by the two following examples.

\begin{example}\em\label{example_orthogonal_1} 
With $K=3$ and $\Mb_{ff} = \Ib_3$, we need to assign at least 3 zeros in $\bLambda$. However, the zero assignment specified in the following $\bLambda_{[1:2,]} $ does not ensure identifiability:
\begin{equation*}
    \bLambda_{[1:2,]} = \begingroup
    \renewcommand{\arraystretch}{0.5}\begin{pmatrix}
         0&*&*\\ *&0&0
    \end{pmatrix}\endgroup, \qquad  \Gb_{\theta}=\begingroup
    \renewcommand{\arraystretch}{0.5}\begin{pmatrix}
         1&0&0\\ 0&\cos\theta & \sin\theta\\ 0 & -\sin\theta & \cos\theta
    \end{pmatrix}\endgroup.
\end{equation*}
Specifically, for any $\theta$, the above transformation matrix $\Gb_{\theta}$ leaves the factor covariance matrix unchanged, since $\Gb_\theta\Mb_{ff}\Gb_\theta^\T=\Ib_K$, and it preserves the imposed zero pattern because $[\bLambda\Gb_\theta]_{[1:2,]}$ retains the same zero assignment as $\bLambda_{[1:2,]}$. Therefore, $(\bLambda,\Fb)$ is not rotationally identifiable under this constraint.


\end{example} 
\begin{example}\label{example_orthogonal_2} \em
With $K=3$ and $\Mb_{ff} = \Ib_3$, imposing the following zero constraints \begin{equation*}
    \bLambda_{[1:3,]} = \begingroup
    \renewcommand{\arraystretch}{0.5}\begin{pmatrix}
        0&*&*\\ *&0&*\\ *&*&0
    \end{pmatrix}\endgroup \text{ with det}(\bLambda_{[1:3,]})\neq 0,
\end{equation*}does not ensure identifiability either. 
To see this, consider 
    \begin{equation*}
        \bLambda_{[1:3,]} = \begingroup
    \renewcommand{\arraystretch}{0.5}\begin{pmatrix}
        0&1&1\\ 1&0&1\\ 1&2&0
    \end{pmatrix}\endgroup,\quad \Gb = \begingroup
    \renewcommand{\arraystretch}{0.6}\begin{pmatrix}
        0&1/\sqrt{3}&2/\sqrt{6}\\1/\sqrt{2}&-1/\sqrt{3}&-1/\sqrt{6}\\-1/\sqrt{2}&-1/\sqrt{3}&-1/\sqrt{6}
    \end{pmatrix}\endgroup.
    \end{equation*}
    Then $\mathrm{diag}\big([\bLambda\Gb]_{[1:3,]}\big) = \zero_3$ and $\Gb^{-1}\Mb_{ff}\Gb^{-\T} = \Ib_K$, and thus $(\bLambda,\Fb)$ is not rotationally identifiable. For this example, it is interesting to note that the above $\Gb$ matrix is the only non-trivial orthogonal transformation such that 
    $\mathrm{diag}\big([\bLambda\Gb]_{[1:3,]}\big) = \zero_3$, up to a sign change.    
\end{example}

As illustrated by the above examples, an interesting question is to study the minimal sufficient and necessary conditions of assigning zeros in the $\bLambda$ matrix. 
For any $r\in[K]$, we denote $\cA_r\subseteq[J]$ as the set that contains the indices of entries prescribed as zeros in the $r$th column $\bLambda_{[,r]}$ according to the design information, i.e., $\cA_r = \{j:\lambda_{jr} \text{ is fixed as } 0\}$. Note that for $j\notin \cA_r$, $\lambda_{jr}$ can take any value including zero. 

Example~\ref{example_orthogonal_1} shows that, in the orthogonal case, a necessary condition for identifiability is that no two columns of the loading matrix share the same zero pattern, i.e., $\cA_r\neq \cA_{r^{\prime}}$ for any $r\neq r^\prime$. 
To avoid these cases, we consider the following sufficient condition that covers many commonly used identifiability constraints~\citep{dunn1973anote,millsap2001trivial,andersonintroduction}. 
\begin{center}
    \textbf{IC}$^{(1)}$: $\Mb_{ff}=\Ib_K$; up to some column permutation, the loading matrix $\bLambda$ satisfies that, \\for each $r\ge 2$, $|\cA_r|=r-1\text{ and rank}(\bLambda_{[\cA_r,1:r]})=r-1$.
\end{center}
This design can be interpreted as, after certain ordering of the factors, for the $r$ th factor, there are $r-1$ items designed not to load on it.

We show in the following theorem the necessity and sufficiency of the condition {IC}$^{(1)}$.
\begin{theorem}\label{thm_iff_ic1}\it { Identifiability condition IC$^{(1)}$ is sufficient for rotational identifiability. 
Moreover, in the orthogonal case where $\Mb_{ff}=\Ib_K$, 
among loading matrices with exactly $K(K-1)/2$ zero constraints, rotational identifiability holds only if IC$^{(1)}$ is satisfied.
}
\end{theorem}


{Theorem~\ref{thm_iff_ic1} shows that any zero-assignment pattern satisfying IC$^{(1)}$ ensures rotational identifiability in the orthogonal case, and that IC$^{(1)}$ is also necessary when the minimal $(K-1)K/2$ zero constraints are imposed on the loading matrix. We note that although no universally necessary condition for identifiability exists in general, imposing zero constraints on the loading matrix is a common and practically convenient strategy adopted in many applications. In fields such as psychology and education, for example, some testing items are intentionally designed to measure only a subset of factors, which naturally induces zero constraints on their loadings \citep{von2008general,xu2017,gu2020partial}. Our identifiability condition IC$^{(1)}$ provides a complete characterization of these zero-constraint patterns that ensure identifiability in the orthogonal case.}

Practically, one popularly used special case of IC$^{(1)}$ is that $\bLambda_{[1:K,]}$ adopts a lower-triangular structure after proper row and column permutations. Specifically, after a certain re-ordering of the factors, for the $r$th factor, it is assumed that the first $r-1$ manifest items do not depend on this factor, as illustrated in Figure~\ref{fig:illus}(a). It is easy to check this configuration satisfies IC$^{(1)}$ with $\cA_r = \{1,\dots,r-1\}$ for all $r\in\{2,\dots,K\}$, after proper row and column permutations. We summarize it as 
\begin{center}
    \textbf{IC}$^{(2)}$: $\Mb_{ff}=\Ib_K$; $\bLambda_{[1:K,]}$ is lower-triangular with nonzero diagonal elements.
\end{center}
While IC$^{(2)}$ is technically included in IC$^{(1)}$, we highlight it separately here for its widespread use in applications \citep{andersonintroduction,reckase2009}. Note that we require the diagonal elements of $\bLambda_{[1:K,]}$ to be nonzero to meet the rank condition in IC$^{(1)}$. 
A popular variant of this condition  relaxes the assumption of normalizing factors, while instead fixing the diagonal elements of $\bLambda_{[1:K,]}$ to be 1:
\begin{center}
    \textbf{IC}$^{(3)}$: $\Mb_{ff}$ is diagonal; $\bLambda_{[1:K,]}$ is lower-triangular with diagonal elements all being 1.
\end{center}
Under IC$^{(3)}$, we have $\cA_r = \{1,\dots,r-1\}$ for all $r\in\{2,\dots,K\}$. The analysis for these two variants {IC$^{(2)}$} and {IC$^{(3)}$} is closely related, and therefore in this work, we consider IC$^{(3)}$ also as a special case of IC$^{(1)}$.

  \subsubsection{Identifiability under non-orthogonal factors}\label{sec_id_oblique}

Correlated factors are more commonly used than orthogonal ones in scientific applications, particularly in fields such as psychology, education, and social sciences. For instance,  psychological theories often consider different dimensions of intelligence or different types of psychological traits to be related rather than independent  \citep{thurstone1947multiple,brown2015confirmatory,kline2023principles}. This configuration is referred to as the ``oblique'' case in our paper.

To allow correlation among latent factors, we relax restrictions on the off-diagonal entries of $\Mb_{ff}$, while maintaining normalization by setting $\mathrm{diag}(\Mb_{ff}) = \Ib_K$. The remaining $(K-1)K$ constraints are imposed by prescribing zeros in the loading matrix in the following pattern:
\begin{center}
     \textbf{IC}$^{(4)}$: $\mathrm{diag}(\Mb_{ff})=\Ib_K$; for each $r\in[K]$, $|\cA_r|=K-1\text{ and rank}(\bLambda_{[\cA_r,-r]})=K-1$.
\end{center}
Figure~\ref{fig:illus}(b) and (c) represent two special cases of IC$^{(4)}$. The corresponding loading matrices are given as 
\begin{equation*}
    \bLambda_{[1:6,]}=\begingroup
    \renewcommand{\arraystretch}{0.4}\begin{pmatrix}
        * & * & 0\\ * & * & 0\\ * & 0 & *\\ * & 0 & *\\ 0 & * & *\\ 0 & * & *
    \end{pmatrix},\qquad\endgroup\bLambda_{[1:3,]} = \begingroup
    \renewcommand{\arraystretch}{0.5}\begin{pmatrix}
        * & 0 & 0\\ 0 & * & 0\\ 0 & 0 & *
    \end{pmatrix}.\endgroup
\end{equation*}
When assigning zeros to the loading matrix under the restriction $\mathrm{diag}(\Mb_{ff})=\Ib_K$, we can show IC$^{(4)}$ is the necessary and sufficient condition for rotational identifiability.
\begin{theorem}\label{thm_iff_ic4}\it
{ Identifiability condition IC$^{(4)}$ is sufficient for rotational identifiability. Moreover, in the oblique case where $\mathrm{diag}(\Mb_{ff})=\Ib_K$, among loading matrices with exactly $K(K-1)$ zero constraints, rotational identifiability holds only if IC$^{(4)}$ is satisfied.}\end{theorem}


Theorem~\ref{thm_iff_ic4} implies that, to ensure rotation identifiability in the oblique case, for each latent factor, at least $K-1$ items must be specified as not influenced by it. If fewer than $K-1$ items are specified as not influenced by any given factor, the model remains non-identifiable, irrespective of the total number of zeros assigned.

Nevertheless, it is interesting to note that constraining $K-1$ entries to zero in each column ensures that the corresponding column is identifiable up to a scaling, regardless of whether the other columns are identifiable.
To formalize this result, we introduce the definition of \textit{identifiability in direction}: if, for any $(\bLambda,\Fb)$ and $(\bar\bLambda,\bar\Fb)$, $\bLambda\Fb^\T = \bar\bLambda\bar\Fb^\T$ implies that $\sin\big(\bLambda_{[,r]},\bar\bLambda_{[,r]}\big)=0$, then we say the $r$th column loading vector $\bLambda_{[,r]}$ is identifiable in direction. 
We have the following result.
  \begin{proposition}\label{prop_ic4}\it For any $r\in[K]$, if there exists some $\cA_r\subseteq[J]$ of size $K-1$ such that $\mathrm{rank}(\bLambda_{[\cA_r,-r]})=K -1$ and $\bLambda_{[\cA_r,r]}=\zero_{K-1}$, then the $r$th column loading vector is identifiable in direction.
  If the above condition is satisfied for all $r\in[K]$ and $\mathrm{diag}(\Mb_{ff})=\Ib_K$, then rotational identifiability of $(\bLambda,\Fb)$ is guaranteed.
\end{proposition}

In the oblique case, a widely used special case of IC$^{(4)}$ involves specifying an identity submatrix within the loading matrix:
\begin{center}
    \textbf{IC}$^{(5)}$: $\bLambda_{[1:K,]}=\Ib_K$.
\end{center}
Here, no restrictions are placed on the factors. This condition can be interpreted as specifying a ``pure'' item for each latent dimension that loads only on the corresponding latent factor \citep[e.g.,][]{bing2020adaptive,cui2026beyond}.  This ``pure'' item condition is also commonly used in other latent variable models, such as restricted latent class models  \citep{xu2017,xu2018JASA} and mixed membership models \citep{gu2023dimension}. 

\subsection{Maximum Likelihood Estimation under Identifiability Conditions}\label{sec:jmle}

Under the proposed identifiability conditions in Sections ~\ref{sec_id_orthogonal} and \ref{sec_id_oblique}, we consider the maximum likelihood estimation approach to jointly estimate the factors and loadings.
Specifically, for any $S\in\{0,1,\dots,5\}$, the estimator under identifiability condition IC$^{(S)}$ is
\begin{align}\label{eq_joint_mle}
    (\hat\bLambda^{(S)}, \hat\Fb^{(S)}) \quad = \mathop{\arg\min}_{\|\bLambda\|_{\max} \le D,\|\Fb\|_{\max}\le D}-L(\bLambda,\Fb|\Yb), 
     \text{ subject to IC}^{(S)},
\end{align}
where following the literature on joint maximum likelihood estimation for generalized latent factor models \citep[e.g.,][]{chen2019joint,Wang2018MaximumLE,chen2022determining},
we employ a prespecified and sufficiently large positive constant $D$ to regulate the magnitudes of the factor and loading parameters. {Confining the estimation to this compact region is essential for both guaranteeing the existence of a solution to \eqref{eq_joint_mle} and ensuring that the solution satisfies the required regularity conditions}.
Similar to \cite{chen2019joint}, we use an alternating minimization algorithm to solve \eqref{eq_joint_mle}. With initial value for $\hat\Fb^{(0)}$, for each iteration step $t=0,1,\cdots$, we first compute
\begin{align*}
    \tilde\bLambda^{(t+1)}~=~\mathop{\arg\min}_{\|\bLambda\|_{\max}\le D}-L(\bLambda,\hat\Fb^{(t)}|\Yb);\quad  \tilde\Fb^{(t+1)}~=~\mathop{\arg\min}_{\|\Fb\|_{\max}\le D}-L(\tilde\bLambda^{(t+1)},\Fb|\Yb),
\end{align*}
each of which can be solved efficiently by fitting a series of generalized linear models.
 To incorporate the identifiability conditions in each step, 
 we compute a linear transformation $\tilde\Gb^{(t+1)}$ such that $\big(\tilde\bLambda^{(t+1)}\tilde\Gb^{(t+1)},\tilde\Fb^{(t+1)}(\tilde\Gb^{(t+1)})^{-\T}\big)$ satisfy IC$^{(S)}$, and we let $\hat\Fb^{(t+1)} = \tilde\Fb^{(t+1)}(\tilde\Gb^{(t+1)})^{-\T}$ and $\hat\bLambda^{(t+1)} = \tilde\bLambda^{(t+1)}\tilde\Gb^{(t+1)}$ be the update results in the $t$th iteration.
 Due to space limitations, we leave the details for computing $\tilde\Gb^{(t+1)}$ in Section~\ref{sup_sec_express_G} of the Supplementary Material.  

\section{Theoretical Results}\label{sec:mainres}

In this section, we establish the non-asymptotic error bounds and asymptotic normality properties of the maximum likelihood estimators. 
We present their $l_2$ and $l_\infty$ error bounds under the proposed identifiability conditions in Theorems~\ref{thm_consis} and \ref{thm_uniform_consis}, respectively. 
The asymptotic distributions of the loading and latent factor estimators are given in Theorems~\ref{main_thm_loading_asym} and \ref{main_thm_factor_asym}, respectively. The limiting distributions for the corresponding estimators of $\Mb_{ff}$ are presented in Theorem~\ref{thm_mff_limiting_distri}.

Denote the true parameters for the generalized latent factor model \eqref{eq_factor_m} by $(\bLambda^*,\Fb^*)$. 
We further write $\eta_{ij}^*=(\blambda_j^*)^\T\bbf_i^*$ for $i\in[N]$ and $j\in[J]$. 
{ In what follows, we introduce several standard regularity assumptions. Assumption~\ref{assumption: psd covariance} imposes eigenvalue conditions on latent factors and loadings. Assumption~\ref{assumption:smoothness} imposes smoothness conditions on the function $g_{ij}(Y_{ij}\mid \eta)$. Assumption~\ref{assump:scaling} involves scaling between $N$ and $J$. Due to space limit, we defer a detailed discussion of these assumptions to Section~\ref{supp_sec_ill_hessian_discussion} of the Supplementary Material, including how Assumption~\ref{assumption: psd covariance} can be extended to settings in which the relevant eigenvalues depend on $N$ and $J$ and potentially approach zero.}

\begin{assumption}\it
	\label{assumption: psd covariance}
There exists some big constant $D >0$ such that $\|\bLambda^*\|_{\max}\le D/2$ and $\|\Fb^*\|_{\max}\leqslant D/2$. 
 $\bSigma_{\lambda}^* = \lim_{J\rightarrow \infty} \Mb_{\lambda\lambda}^*=\lim_{J\rightarrow \infty} J^{-1}\sum_{j=1}^J\blambda_j^*(\blambda_j^{*})^{\intercal} $  and  $\bSigma_f^* = \lim_{N\rightarrow \infty} \Mb_{ff}^*=\lim_{N \rightarrow \infty}N^{-1}$ $\sum_{i=1}^N\bbf_i^*(\bbf_i^{*})^{\intercal} $ exist and are positive definite. The eigenvalues of $\bSigma_{f}^*\bSigma_{\lambda}^*$ are distinct and nonzero.

	\end{assumption}
 	\begin{assumption}
	\label{assumption:smoothness}\it
 For $i\in[N]$ and $j\in[J]$, {$l_{ij}(\eta) = \log g_{ij}(Y_{ij}|\eta)$} is three times differentiable to $\eta$, with the first, second, and third order derivatives denoted by $l_{ij}^{\prime}(\eta)$, $l_{ij}^{\prime\prime}(\eta)$, and $l_{ij}^{\prime\prime\prime}(\eta)$, respectively. For some constant $M>0$,  $l_{ij}^{\prime} (\eta_{ij}^*)$ is sub-exponential with the sub-exponential norm $\|l_{ij}^{\prime} (\eta_{ij}^*)\|_{\varphi_1} \le M$. Moreover, $-l_{ij}^{\prime\prime}(\eta)>0$ for all $\eta \in \RR$, and there exist constants
$0 < b_L \le b_U $ such that, for all $\eta$ with $|\eta|\le KD^2$, $b_L\le -l_{ij}^{\prime\prime}(\eta)\le b_U$ and $|l_{ij}^{\prime\prime\prime}(\eta)| \leqslant b_U$.
 
   \end{assumption}
   \begin{assumption}\label{assump:scaling}\it
       When $N,J\to\infty$, $\log\big(N\vee J\big)\big\{(\epsilon_J\log N + \epsilon_N\log J)/(N\wedge J)\big\}^{1/2}\to 0$, for some slowly varying $\epsilon_N\to\infty$ as $N\to\infty$, and $\epsilon_J\to\infty$ as $J\to\infty$, {i.e., $\epsilon_N$ and $\epsilon_J$ diverge to infinity at a rate slower than any polynomial order.} 
       Moreover, if $l_{ij}^{\prime}(\eta_{ij})$ is bounded, the above condition can be reduced to $\big\{(\epsilon_J\log N + \epsilon_N\log J)/(N\wedge J)\big\}^{1/2} \to 0$ when $N,J\to\infty$.
   \end{assumption}

{We present the $l_2$ error bound as follows. In light of identification up to a signed permutation, the result holds for the estimators after applying some proper signed permutation matrix. For conciseness, we omit this transformation here and in all subsequent results.}
\begin{theorem}[Average Consistency]\label{thm_consis} \it
    Suppose the true parameters $(\bLambda^*, \Fb^*)$ satisfy the identifiability condition IC$^{(S)}$ for $S\in\{0,1,\dots,5\}$ and Assumptions~\ref{assumption: psd covariance}--\ref{assump:scaling} hold.

    \vspace{-0.1cm}
    For $S=0$ and any constant $\delta>1$, when $N$ and $J$ are large enough, with probability at least $1-(N\wedge J)^{-\delta}$, we have 
    \begin{equation*}
        \frac{1}{J}\sum_{j=1}^J\big\|\hat\blambda_j^{(0)}-\blambda_j^*\big\|^2~\le~ C_\delta\big(N^{-1}+J^{-2}\big),\quad \frac{1}{N}\sum_{i=1}^N\big\|\hat\bbf_i^{(0)}-\bbf_i^*\big\|^2~\le~ C_\delta\big(J^{-1} + N^{-2}\big),
    \end{equation*} where $C_\delta$ is a constant only depending on $\delta$.
    
    \vspace{-0.1cm}
    For $S\in\{1,\dots,5\}$, any $\delta > 1$ and slowly varying $\epsilon_N\to\infty$, when $N$ and $J$ are large enough, with probability at least $1-\exp(-\epsilon_N)-(N\wedge J)^{-\delta}$, we have
    \begin{equation*}
        \frac{1}{J}\sum_{j=1}^J\big\|\hat\blambda_j^{(S)}-\blambda_j^*\big\|^2~\le ~ C_{\delta}(\epsilon_N N^{-1}+J^{-2}),\quad \frac{1}{N}\sum_{i=1}^N\big\|\hat\bbf_i^{(S)}-\bbf_i^*\big\|^2~\le ~C_{\delta}(\epsilon_NN^{-1}+J^{-1}),
    \end{equation*}where $C_\delta$ is a constant only depending on $\delta$.
\end{theorem}
\begin{remark}\it
In order to establish the consistency of $\hat\bLambda^{(S)}$ and $\hat\Fb^{(S)}$, it is necessary for both $N$ and $J$ to go to infinity, a phenomenon referred to as the incidental parameter problem or the Neyman-Scott paradox~\citep{neyman1948consistent,andersen1970asymptotic}. When either $N$ or $J$ is fixed, the joint maximum likelihood estimation approach is inconsistent for the factors and loadings~\citep{haberman1977maximum,chen2020structured}. 
    Our results align with the literature in that, when both $N$ and $J$ diverge to infinity, $\hat\bLambda^{(S)}$ and $\hat\Fb^{(S)}$ consistently estimate the true loadings and factors for any $S\in\{0,1,\dots,5\}$, while estimation errors for both $\hat\bLambda^{(S)}$ and $\hat\Fb^{(S)}$ do not diminish when either $N$ or $J$ is fixed.
\end{remark}
Our non-asymptotic error bounds fill a gap in the literature, where most existing studies primarily focus on asymptotic bounds~\citep{bai2003inferential,bai2012statistical,fan2017sufficient,Wang2018MaximumLE}.
Our results have the following implications.
\begin{enumerate}[leftmargin=0.6cm]
    \item 
{Our theoretical results achieve the oracle rate for factor loadings within a broad range of the asymptotic regimes, where the oracle rate denotes the consistency rate obtained for loading estimation when the true factors are assumed to be {\it known}.} Specifically, for $S\in\{0,1,\dots,5\}$, the consistency rate for $\hat\bLambda^{(S)}$ when the factors $\Fb^*$ are known is $O_p(N^{-1/2})$, { which corresponds to the standard consistency rate in the generalized linear regression setting~\citep{van2000asymptotic}. } Theorem~\ref{thm_consis} implies $J^{-1/2}\big\|\hat\bLambda^{(S)} - \bLambda^{*}\big\|_F = O_p(N^{-1/2}+J^{-1})$, which achieves the oracle rate when $N= O(J^2)$.  
    Similarly, for the factors, the error bounds match the oracle rate when $J= O(N^{2})$ for $S=0$ and $J = O(N)$ for $S\in\{1,2,\dots,5\}$. {Here, the oracle rate is similarly defined to the consistency rate achieved for latent factor estimation when the true loadings are known.}
    
    \item Under identifiability condition IC$^{(0)}$, our error bounds in Theorem~\ref{thm_consis} are sharper than the existing literature. For instance, under similar settings to IC$^{(0)}$, \cite{bai2002determining} and \cite{Wang2018MaximumLE} have obtained consistency rates of $O_p(N^{-1}+ J^{-1})$ for linear and generalized linear models, respectively.
When the true parameters satisfy IC$^{(0)}$, the estimation is closely related to the singular value decomposition of the matrix $\hat\Fb^{(0)}\big(\hat\bLambda^{(0)}\big)^\T $; henceforth, consistency rate can be derived using the perturbation theory for the singular vectors, which readily leads to the consistency rate of $O_p(N^{-1}+J^{-1})$ in the existing literature. 
To derive the sharper non-asymptotic error bound in Theorem~\ref{thm_consis}, we conduct a detailed analysis of the Hessian matrix of the log-likelihood function and its concentration properties; see Lemma~\ref{lemma_hessian_property} in the Supplementary Material. 
\item Under identifiability conditions IC$^{(1)}$--\,IC$^{(5)}$, our results are new and fill a critical gap in the literature. Obtaining consistency results under these conditions is more challenging than that under IC$^{(0)}$, as they are not directly related to the singular decomposition of $\hat\Fb^{(0)}\big(\hat\bLambda^{(0)}\big)^\T$ and the techniques used under the orthogonal case IC$^{(0)}$ cannot be applied. 
        More importantly, as exemplified in Section~\ref{supp_sec_ill_hessian} in the Supplementary Material, the Hessian matrix of the joint log-likelihood function can suffer from diminishing eigenvalues even after incorporating the identifiability constraints. 
        To address these challenges, we develop new technical tools to derive the non-asymptotic error bound for the maximum likelihood estimators under the ill-conditioned Hessian matrix (see more details in Section~\ref{supp_sec_ill_hessian_discussion_technical} of the Supplementary Material). 
\end{enumerate}

The average consistency results allow us to establish corresponding sin-consistency bounds, which characterize the angles between the columns of $\hat\Fb^{(S)}$ and $\Fb^*$, and between $\hat\bLambda^{(S)}$ and $\bLambda^*$. Due to space limit, we present these results in Section~\ref{supp_sec_sin_consis} of the Supplementary Material.

The next theorem shows that the estimators converge to the true parameters uniformly.
\begin{theorem}[Uniform Consistency]\label{thm_uniform_consis}\it
    Suppose that the true parameters $(\bLambda^*,\Fb^*)$ satisfy identifiability condition IC$^{(S)}$ for $S\in \{0,1,\dots,5\}$ and Assumptions~\ref{assumption: psd covariance}--\ref{assump:scaling} hold. Write $\epsilon_{NJ} := (\epsilon_N\log J + \epsilon_J\log N)^{1/2}$. For any $\delta>1$ and any $\epsilon_N,\epsilon_J\to\infty$, when $N$ and $J$ are large enough, with probability at least $1-(N\wedge J)^{-\delta} - \exp(-\epsilon_N\wedge \epsilon_J)$, we have \begin{equation*}
        \max_{1\le j\le J}\big\|\hat\blambda_j^{(S)}-\blambda_j^*\big\|_{\infty}~\le ~C_{\delta} \frac{\epsilon_{NJ}\log J}{\sqrt{N\wedge J}},
        \quad \max_{1\le i\le N}\big\|\hat\bbf_i^{(S)}-\bbf_i^*\big\|_{\infty}~\le ~ C_{\delta}\frac{\epsilon_{NJ}\log N}{\sqrt{N\wedge J}},
    \end{equation*}where $C_{\delta}$ is a constant only depending on $\delta$.
\end{theorem}
Theorem~\ref{thm_uniform_consis} shows the non-asymptotic $l_\infty$ error bound under all identifiability conditions. To the best of our knowledge, we are the first to quantify the maximal deviation for the maximum likelihood estimation of generalized latent factor models under these conditions, particularly the non-orthogonal cases. 
Earlier work obtained asymptotic bounds under orthogonal design for both factors and loadings (IC$^{(0)}$) in linear factor models~\citep{bai2003inferential} and generalized latent factor models~\citep{Wang2018MaximumLE}. 
In \cite{chen2024note}, they derived a finite-sample bound on the maximal entrywise deviation of the estimated parameter matrix in generalized latent factor models. However, those analyses rely on orthogonality and do not extend to the broader identifiability conditions considered here. When $l_{ij}^{\prime}(\eta_{ij}^*)$ is bounded, such as logistic and probit models, the $\log$-terms in Theorem~\ref{thm_uniform_consis}  can be removed and our uniform consistency rate can be simplified to \begin{equation*}
        \max_{1\le j\le J}\big\|\hat\blambda_j^{(S)}-\blambda_j^*\big\|_{\infty}~\le ~ C_{\delta} \frac{\epsilon_{NJ}}{\sqrt{N\wedge J}},
        \quad \max_{1\le i\le N}\big\|\hat\bbf_i^{(S)}-\bbf_i^*\big\|_{\infty}~\le ~ C_{\delta}\frac{\epsilon_{NJ}}{\sqrt{N\wedge J}}.
    \end{equation*}

Next, we present the asymptotic distribution results for the loading estimators. Note that under IC$^{(S)}$ for $S\in\{1,\dots,5\}$, certain loadings $\lambda_{jr}^*$ are set deterministically at $0$ or  $1$ for the itentifiability purpose, and we correspondingly set those $\hat\lambda_{jr}^{(S)}$ as fixed at $0$ or  $1$. Thus we only need to focus on the asymptotic distributions of the remaining free parameters. We denote $\cS_j\subseteq[K]$ as the index set of the free entries in $\blambda_j$ and, when $|\cS_j|>0$, a $|\cS_j|\times K$ selection matrix $\Db_{(j)} = [\Ib_K]_{[\cS_j,]}$. Then the free parameters in $\blambda_j$ are given by $[\blambda_j]_{[\cS_j]} = \Db_{(j)}\blambda_j$.
For instance, under IC$^{(2)}$, the free parameters in $\blambda_j$ are $\Db_{(j)}\blambda_{j}$ with $\Db_{(j)}$ given as 
       $\Db_{(j)} = \left(
        \Ib_j\; \zero_{j\times(K-j)}\right)\text{for }j\le K\text{ and }\Db_{(j)} =\Ib_K\text{ otherwise}$. {Next, we define $\bSigma_{[N],j}^{(S)}$ for $S\in\{0,1,\dots,5\}$, which is the covariance matrix of $\Db_{(j)}(\hat\blambda_j^{(S)} - \blambda_j^*)$ under identifiability condition IC$^{(S)}$, as later shown in Theorem~\ref{main_thm_asymp}. Let
\begin{equation}
    \bSigma_{[N],j}^{(S)} = \left\{\begin{aligned}
        &\bPhi_{[N],j}^*\text{ when }S=0,
        \\&\bPhi_{[N],j}^* + \big(\blambda_j^*\otimes \Ib_K\big)^\T\Xb_{\lambda}^{(S)} \big(\blambda_j^*\otimes \Ib_K\big) + \bOmega_{[N],j}^{(S)}\text{ when }S\in\{1,2,\dots,5\},
    \end{aligned} \right.\label{eq_sigma_lambad_j} 
\end{equation}
where $\otimes$ represents the Kronecker product. Here $\bPhi^*_{[N],j}$ is given as 
\begin{equation}
    \bPhi_{[N],j}^* ~= ~\Big\{-\sum_{i=1}^Nl_{ij}^{\prime\prime}(\eta_{ij}^*)\bbf_i^*(\bbf_i^*)^\T\Big\}^{-1}\Big\{\sum_{i=1}^N\big[l_{ij}^{\prime}(\eta_{ij}^*)\big]^2\bbf_i^*(\bbf_i^*)^\T\Big\}\Big\{-\sum_{i=1}^Nl_{ij}^{\prime\prime}(\eta_{ij}^*)\bbf_i^*(\bbf_i^*)^\T\Big\}^{-1}, \label{eq_sandwich}
\end{equation} 
and for $S\in\{1,2,\dots,5\}$, the terms $\Xb^{(S)}_{\lambda}$ and $\bOmega_{[N],j}^{(S)}$ in   \eqref{eq_sigma_lambad_j} are given as follows.
\begin{enumerate}[label=(\arabic*)]
    \item $S=1$: $\Xb^{(1)}_{\lambda}$ is given in \eqref{eq_expression_Xlambda1} in the Supplementary Material; \\$\bOmega_{[N],j}^{(1)} = \sum_{r=1}^K\sum_{l=1}^{K}1_{(j= \cA_r(l))} \big\{\bPhi_{[N],j}^*\Xb_{\lambda,rl}^{(1)} + \big(\Xb_{\lambda,rl}^{(1)}\big)^\T\bPhi_{[N],j}^*\big\}$ where $1_{(j=\cA_r(l))}$ is the indicator function that equals $1$ if $j$ is the $l$th element in $\cA_r$ and $0$ otherwise, and $\Xb^{(1)}_{\lambda,rl}$ is  given in \eqref{eq_x_lambda_kr_1} in the Supplementary Material.
    \item $S=2$: $\Xb_{\lambda}^{(2)}$ is given in \eqref{eq_expression_Xlambda2} in the Supplementary Material; \\
    $\bOmega_{[N],j}^{(2)} = 1_{(1\le j\le K)}\big(\Ib_K^j\bPhi_{[N],j}^* +\bPhi_{[N],j}^*\Ib_K^j-2\bPhi_{[N],j}^*\big)$ where $\Ib_K^{j}$ is a $K\times K$ diagonal matrix with the first $j$ diagonal elements being 1 and the rest being 0.
    \item $S=3$: $\Xb_{\lambda}^{(3)}$ is given in \eqref{eq_expression_Xlambda3} in the Supplementary Material;\\ $\bOmega_{[N],j}^{(3)} = 1_{(1< j\le K)}\big(\Ib_K^{j-1}\bPhi_{[N],j}^* + \bPhi_{[N],j}^*\Ib_K^{j-1}-2\bPhi_{[N],j}^*\big)$.
    \item $S=4$: $\Xb_{\lambda}^{(4)}$ is given in \eqref{eq_expression_Xlambda4} in the Supplementary Material; \\$\bOmega_{[N],j}^{(4)} = \sum_{r=1}^K\sum_{l=1}^{K-1}1_{(j= \cA_r(l))}\{\bPhi_{[N],j}^{*}\Xb_{\lambda,rl}^{(4)} + (\Xb_{\lambda,rl}^{(4)})^\T\bPhi_{[N],j}^{*}\}$ where $\Xb^{(4)}_{\lambda,rl}$ is  given in \eqref{eq_x_lambda_kr_4} in the Supplementary Material.
    \item $S=5$: $\Xb_{\lambda}^{(5)}$ is given in \eqref{eq_expression_Xlambda5} in the Supplementary Material; $\bOmega_{[N],j}^{(5)} = \zero_{K\times K}$.
\end{enumerate}
Note that given response matrix $\Yb$, for each $S\in\{0,1,\dots,5\}$, $\bSigma_{[N],j}^{(S)}$ depends only on $(\bLambda^*,\Fb^*)$. Therefore, by Theorem~\ref{thm_uniform_consis},  $\bSigma_{[N],j}^{(S)}$ can be consistently estimated by the plug-in estimators $\hat\bSigma_{[N],j}^{(S)}$ that replace  $(\bLambda^*,\Fb^*)$ in $\bSigma_{[N],j}^{(S)}$ with $(\hat\bLambda^{(S)},\hat\Fb^{(S)})$. Specifically, for each $S\in\{0,1,\dots,5\}$, since $\bSigma_{[N],j}^{(S)}$ is a function of $(\bLambda^*,\Fb^*)$ and responses $\Yb$, a plug-in estimator of it can be obtained by replacing the true parameters $(\bLambda^*,\Fb^*)$ with their consistent estimates $(\hat\bLambda^{(S)},\hat\Fb^{(S)})$ under IC$^{(S)}$.
}


\begin{theorem}[Asymptotic Normality for Loading Estimators]\label{main_thm_loading_asym}\it
Suppose the true parameters $(\bLambda^*,\Fb^*)$ satisfy identifiability condition IC$^{(S)}$ for $S\in \{0,1,\dots,5\}$ and Assumptions~\ref{assumption: psd covariance}--\ref{assump:scaling} hold. Assume $\sqrt{N}/J\to 0$ as $N,J\to\infty$. Then, when $|\cS_j|>0$,
\begin{equation}
    \left\{\Db_{(j)}\bSigma_{[N],j} ^{(S)}\Db_{(j)}^\T\right\}^{-1/2}\left\{\Db_{(j)}\big(\hat\blambda_{j}^{(S)} - \blambda_j^*\big)\right\} ~\overset{d}{\to }~\cN(\zero_{|\cS_j|},\Ib_{|\cS_j|})\text{, for any }j\in[J],\label{eq_converge_d_lambda}
\end{equation}
as $N,J\to\infty$, with $\bSigma_{[N],j}^{(S)}$ defined in \eqref{eq_sigma_lambad_j}. Furthermore, the normality results in \eqref{eq_converge_d_lambda} still hold with $\bSigma_{[N],j}^{(S)}$ replaced with the plug-in estimator $\hat\bSigma_{[N],j}^{(S)}$.


\end{theorem}
Our asymptotic results hold both theoretical significance and practical interest. Deriving the limiting distributions under various identifiability conditions presents significant challenges. 
Notably, the smallest eigenvalues of the Hessian matrix of the negative joint log-likelihood function can approach zero, even constrained by the identifiability conditions. These diminishing eigenvalues prevent direct derivation of the limiting distributions. To overcome this challenge, we introduce novel technical tools to establish the asymptotic normality of the loadings, even when the Hessian matrix is ill-conditioned. We highlight that such results can only be expected when the number of model parameters diverges. In this scenario, we can show that the eigenvectors corresponding to the diminishing eigenvalues become asymptotically orthogonal to each of the loading parameters and factors, distributing the error uniformly across all parameters (see Section~\ref{supp_sec_ill_hessian} in the Supplementary Material for a detailed explanation). 

Moreover, our results have strong practical implications. The findings under IC$^{(1)}$ and IC$^{(4)}$ cover a wide range of interpretable and applicable identifiability conditions for both the orthogonal and oblique cases, respectively. These inference results support applications across a broad range of factor and loading designs.
We want to emphasize that the covariance matrices of the loadings estimates in Theorem \ref{main_thm_loading_asym} vary under different identifiability conditions. However, in practice, one popular approach is to directly use $\bPhi_{[N],j}^{*}$ under different conditions \citep{baker2004item}, noting that $\bPhi_{[N],j}^{*}$ corresponds to the Fisher information matrix of the loadings $\blambda_j^*$ with known factors.
Notably, as implied by our theorem, relying solely on $\bPhi_{[N],j}^{*}$ leads to an underestimation of the variance of $\hat\blambda_j^{(S)}$, except in the orthogonal case of IC$^{\,(0)}$.
The covariance matrix $\bSigma_{[N],j}^{(S)}$ given in \eqref{eq_sigma_lambad_j} for $S\ge 1$ includes additional terms that account for errors arising from imposing a prior structure on the loading matrix. 
This indicates that while joint maximum likelihood estimation provides consistent estimates of the true parameters under different identifiability conditions, the asymptotic covariance matrix of the estimates varies across these conditions. Consequently, the common practice of using the same $\bPhi_{[N],j}^{*}$ for estimation under all identifiability conditions is problematic. We further illustrate this issue through a simulation study in Section~\ref{sec:simu}.

\begin{remark}\it
Note that under our assumptions, the following asymptotic equivalence holds:
$\left\|\left\{-\frac{1}{N}\sum_{i=1}^Nl_{ij}^{\prime\prime}(\eta_{ij}^*)\bbf_i^*(\bbf_i^*)^\T\right\}
-\left\{\frac{1}{N}\sum_{i=1}^N\big[l_{ij}^{\prime}(\eta_{ij}^*)\big]^2\bbf_i^*(\bbf_i^*)^\T\right\}\right\|_F\overset{p}{\to} 0$.
Thus, in Theorem~\ref{main_thm_loading_asym}, we can substitute $\bPhi_{[N],j}^*$ with $\big\{-\sum_{i=1}^Nl_{ij}^{\prime\prime}(\eta_{ij}^*)\bbf_i^*(\bbf_i^*)^\T\big\}^{-1}$ or $\big\{\sum_{i=1}^N[l_{ij}^{\prime}(\eta_{ij}^*)]^2\bbf_i^*(\bbf_i^*)^\T\big\}^{-1}$since all three expressions share the same limit when $N\to\infty$, if exists, given by $\lim_{N\to\infty}\big\{-N^{-1}\sum_{i=1}^N\EE[l_{ij}^{\prime\prime}(\eta_{ij}^*)]\bbf_i^*(\bbf_i^*)^\T\big\}^{-1}$.
This corresponds to the Fisher information matrix of the loadings $\blambda_j^*$ when the factors are known.

\end{remark}

\begin{remark}\label{remark_decompose_lambda_two_terms}\it
    Our scaling condition $\sqrt{N}/J\to 0$ allows for $N/J\to c\in[0,\infty]$, making our results flexible for many applications. This condition aligns with the scaling condition that is generally required in factor analysis literature~\citep{bai2003inferential,bai2012statistical,Wang2018MaximumLE}. We remark that this requirement arises from the small order terms when approximating $\hat\blambda^{(S)}_j - \blambda_j^{(S)}$. We show in the proof that $\hat\blambda_{j}^{(S)}-\blambda_j^{(S)}$ can be decomposed into two terms.  The first term is of order $O_p\big(N^{-1/2})$ and is shown to be asymptotically normal. The second term is of order $O_p\big((N\wedge J)^{-1}\big)$, which does not have a tractable limiting distribution, while asymptotically negligible when $\sqrt{N}/J\to 0$. 

\end{remark}


Next, we show the results regarding the limiting distribution of $\hat\bbf_i^{(S)}$. {We introduce $\bSigma_{i,[J]}^{(S)}$ for $S\in\{0,1,\dots,5\}$, which is the covariance matrix of $\hat\bbf_i^{(S)} - \bbf_i^*$ under identifiability condition IC$^{(S)}$, as later shown in Theorem~\ref{main_thm_factor_asym}. \begin{equation}
     \bSigma_{i,[J]}^{(S)} ~=~ \left\{\begin{aligned}&\bPhi_{i,[J]}^*\text{ when }S=0,\\&\bPhi_{i,[J]}^*+\big( \Ib_K\otimes \bbf_i^* \big)^\T\Xb_{\lambda}^{(S)} \big(\Ib_K\otimes \bbf_i^*\big)\text{ when }S\in\{1,2,\dots,5\},\end{aligned}\right.\label{eq_sigma_f_i_exp}
 \end{equation}
 where for each $i\in[N]$, $\bPhi_{i,[J]}^{*}$ is given as
\[
    \bPhi_{i,[J]}^{*}:=\Big\{-\sum_{j=1}^Jl_{ij}^{\prime\prime}(\eta_{ij}^{*})\blambda_j^{*}(\blambda_j^{*})^\T\Big\}^{-1}\Big\{\sum_{j=1}^J\big[l_{ij}^{\prime}(\eta_{ij}^*)\big]^2\blambda_j^{*}(\blambda_j^{*})^\T\Big\}\Big\{-\sum_{j=1}^Jl_{ij}^{\prime\prime}(\eta_{ij}^{*})\blambda_j^{*}(\blambda_j^{*})^\T\Big\}^{-1},\]
and $\Xb_{\lambda}^{(S)}$ has been defined in Theorem~\ref{main_thm_loading_asym} for $S=\{1,2,\dots5\}$. Given response $\Yb$, for each $S\in\{0,1,\dots,5\}$, $\bSigma_{i,[J]}^{(S)}$ depends only on $(\bLambda^*,\Fb^*)$. By Theorem~\ref{thm_uniform_consis},  $\bSigma_{i,[J]}^{(S)}$ can be consistently estimated by the plug-in estimators $\hat\bSigma_{i,[J]}^{(S)}$ that replace  $(\bLambda^*,\Fb^*)$ in $\bSigma_{i,[J]}^{(S)}$ with $(\hat\bLambda^{(S)},\hat\Fb^{(S)})$.}
\begin{theorem}[Asymptotic Normality for Factor Estimators]\label{main_thm_factor_asym}\it
Suppose the true parameters $(\bLambda^*,\Fb^*)$ satisfy identifiability condition IC$^{(S)}$ for $S\in \{0,1,\dots,5\}$ and Assumptions~\ref{assumption: psd covariance}--\ref{assump:scaling} hold. For $S=0$ only, we further require $\sqrt{J}/N \to 0$ as $N,J\to\infty$. Then
 \begin{equation}
     \left(\bSigma_{i,[J]}^{(S)}\right)^{-1/2} \big(\hat\bbf_i^{(S)} - \bbf_i^{*}\big) ~\overset{d}{\to}~ \cN(\zero_K,\Ib_K)\text{, for any }i\in[N],\label{eq_sigma_f_i}
 \end{equation}as $N,J\to\infty$, with $\bSigma_{i,[J]}^{(S)}$ given as \eqref{eq_sigma_f_i_exp}. Furthermore, the normality results in \eqref{eq_converge_d_lambda} still hold with $\bSigma_{i,[J]}^{(S)}$ replaced with the plug-in estimator $\hat\bSigma_{i,[J]}^{(S)}$.

\end{theorem}

\begin{remark}\label{remark_decompose_f_two_terms}\it
    When the true parameters satisfy IC$^{(0)}$, we require scaling condition $\sqrt{J}/N\to 0$ for valid limiting distributions for the estimated factors, similar to Theorem~\ref{main_thm_loading_asym}. In contrast, when the true parameters satisfy IC$^{(S)}$ for $S\in\{1,\dots,5\}$, this scaling condition is not necessary for \eqref{eq_sigma_f_i} to hold. 
    The key difference lies in the decomposition of $\hat\bbf_i^{(S)} - \bbf_i^*$ for different $S$. In particular, when the true parameters satisfy IC$^{(0)}$, $\hat\bbf_i^{(0)} - \bbf_i^*$ consists of two terms. Similar to the discussion in Remark~\ref{remark_decompose_lambda_two_terms}, the first term is of order $O_p(J^{-1/2})$ with a tractable limiting distribution while the second term, of order $O_p\big((N\wedge J)^{-1}\big)$, lacks a tractable limiting distribution but is asymptotically negligible under the condition $\sqrt{J}/N\to0$.
    In contrast, when $S\in\{1,\dots,5\}$, $\hat\bbf_i^{(S)} - \bbf_i^*$ consists of three terms. The first two terms are of orders $O_p(J^{-1/2})$ and $O_p(N^{-1/2})$, respectively. Both terms have tractable limiting distributions and are asymptotically independent. The third term, of order $O_p(N^{-1}+J^{-1})$, is asymptotically negligible compared to the first two terms, regardless of the ratio between $N$ and~$J$.
    
\end{remark}

Now we present the limiting distributions of $\hat\Mb_{ff}^{(S)} = N^{-1}\sum_{i=1}^N \hat\bbf_i^{(S)}\big(\hat\bbf_i^{(S)}\big)^\T$ for $S\in\{3,4,5\}$. Note that under identifiability conditions IC$^{(S)}$ for $S\in \{0,1,2\}$, $\Mb_{ff}$ is fixed as $\Ib_K$ and does not require estimation. 

\begin{theorem}[Asymptotics for factor covariance matrix estimations]\label{thm_mff_limiting_distri}\it
Suppose the true parameters $(\bLambda^*,\Fb^*)$ satisfy identifiability condition IC$^{(S)}$ for $S\in \{3,4,5\}$ and Assumptions~\ref{assumption: psd covariance}--\ref{assump:scaling} hold. Assume $\sqrt{N}/J\to0$ as $N,J\to\infty$. Then
\begin{itemize}
    \item when $S = 3$,
    $\big(\bSigma_{[J]}^{(3)}\big)^{-1/2} \mathrm{vecd}\big(\hat\Mb_{ff}^{(3)} - \Mb_{ff}^* \big)\overset{d}{\to} \cN(\zero_K,\Ib_K)$ as $N,J\to\infty$, where $\bSigma_{[J]}^{(3)}$ is given in \eqref{eq_sigma_mff3} in the Supplementary Material and $\mathrm{vecd}(\Ab)$ denotes the vector consisting of the diagonal elements of a matrix $\Ab\in\RR^{K\times K}$;
\item when $S = 4$, $\big(\bSigma_{[J]}^{(4)}\big)^{-1/2}\mathrm{veck}\big(\hat\Mb_{ff}^{(4)} - \Mb_{ff}^*\big)\overset{d}{\to} \cN(\zero_{K(K-1)/2},\Ib_{K(K-1)/2})$ as $N,J\to\infty$,
where $\bSigma_{[J]}^{(4)}$ is given in \eqref{eq_sigma_mff4} in the Supplementary Material and $\mathrm{veck}(\Ab)$ for $\Ab\in\RR^{K\times K}$ is the $K(K-1)/2$ column vector obtained by vectorizing only the elements above the diagonal of $\Ab$;
\item when $S = 5$,  $\big(\bSigma_{[J]}^{(5)}\big)^{-1/2}\mathrm{vech}\big(\hat\Mb_{ff}^{(5)} - \Mb_{ff}^*\big)\overset{d}{\to} \cN(\zero_{(K+1)K/2},\Ib_{(K+1)K/2})$ as $N,J\to\infty$, where $\bSigma_{[J]}^{(5)}$ is given in \eqref{eq_sigma_mff5} in the Supplementary Material and
    where $\mathrm{vech}(\Ab)$ for $\Ab\in\RR^{K\times K}$ is the $K(K+1)/2$ column vector obtained by vectorizing only the upper-triangular elements $\Ab$.
\end{itemize}
Here the covariance matrices $\bSigma_{[J]}^{(S)}$ can be consistently estimated by the plug-in estimators $\hat\bSigma_{[J]}^{(S)}$ that replace  $(\bLambda^*,\Fb^*)$ in $\bSigma_{[J]}^{(S)}$ with $(\hat\bLambda^{(S)},\hat\Fb^{(S)})$, and the above normality results still hold with $\bSigma_{[J]}^{(S)}$ replaced with $\hat\bSigma_{[J]}^{(S)}$.
\end{theorem}

\begin{remark}\it
Here we impose the scaling condition $\sqrt{N}/J\to 0$. Note that $\hat\Mb_{ff}^{(S)} - \Mb_{ff}^*$ equals $N^{-1}\sum_{i=1}^N\hat\bbf_i^{(S)}(\hat\bbf_i^{(S)} - \bbf_i^*)^\T + N^{-1}\sum_{i=1}^N(\hat\bbf_i^{(S)} - \bbf_i^*)(\bbf_i^*)^\T$. Both summations contain a leading term of order $O_p(N^{-1/2})$, while the remaining terms are of order $O_p\big((N\wedge J)^{-1}\big)$. The condition $\sqrt{N}/J\to 0$ ensures that, in $\hat\Mb_{ff}^{(S)}-\Mb_{ff}^*$, the leading $\sqrt{N}$-order term with a tractable limiting distribution dominates the remaining terms.
\end{remark}



\section{Simulation Studies}\label{sec:simu}

In this section, we present simulation studies to demonstrate our theoretical findings. {The simulations are conducted under linear, logistic, and Poisson latent factor models. We report the results for the logistic latent factor models below, and leave the remaining results to Section~\ref{sec_supp_simu} of the Supplementary Material.} 
We estimate the factors and loadings and conduct statistical inference based on the results established in Section~\ref{sec:mainres}.

We consider three types of loading matrix designs that satisfy IC$^{(2)}$, IC$^{(4)}$ and IC$^{(5)}$, respectively. Due to space limit, the details of the simulation design are left to Section~\ref{supp_sec_simu}.
Under all three designs, we generate the true parameters under the following settings  (1) sample size $N=400, 600, 800, 1000$; (2) test length $J=200, 400, 600$; (3) factor correlation $\tau = 0.3,0.7$.
The alternating minimization algorithm proposed in Section~\ref{sec:jmle} is used for estimation. 
To examine the asymptotic normality, we compute the 95\% Wald intervals for each loading and factor based on Theorem~\ref{main_thm_loading_asym} and Theorem~\ref{main_thm_factor_asym}, respectively. As an alternative, we also consider a naive approach that is commonly used in practice~\citep{baker2004item}. Specifically, this approach uses $\hat\bPhi_{[N],j}^{(S)}$ ($\hat\bPhi_{i,[J]}^{(S)}$) obtained by plugging $\hat\bLambda^{(S)}$ and $\hat\Fb^{(S)}$ into $\bPhi_{[N],j}^*$ ($\bPhi_{i,[J]}^*$) as the covariance estimator for $\hat\blambda_j^{(S)}$ ($\hat\bbf_i^{(S)}$). We also compute the 95\% Wald intervals for both factors and loadings based on naive variance estimates. We generate 200 replications and report the empirical coverage rates for both methods. We summarize the simulation results for loadings under the low correlation setting in Figure~\ref{fig:cover_ic_1}. Other results are left to Section~\ref{sec_supp_simu1} in the Supplementary Material.
{\spacingset{1.3}
\begin{figure}[h]
        \centering
        \includegraphics[width=3in]{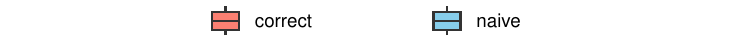}\\
        \includegraphics[width=6in]{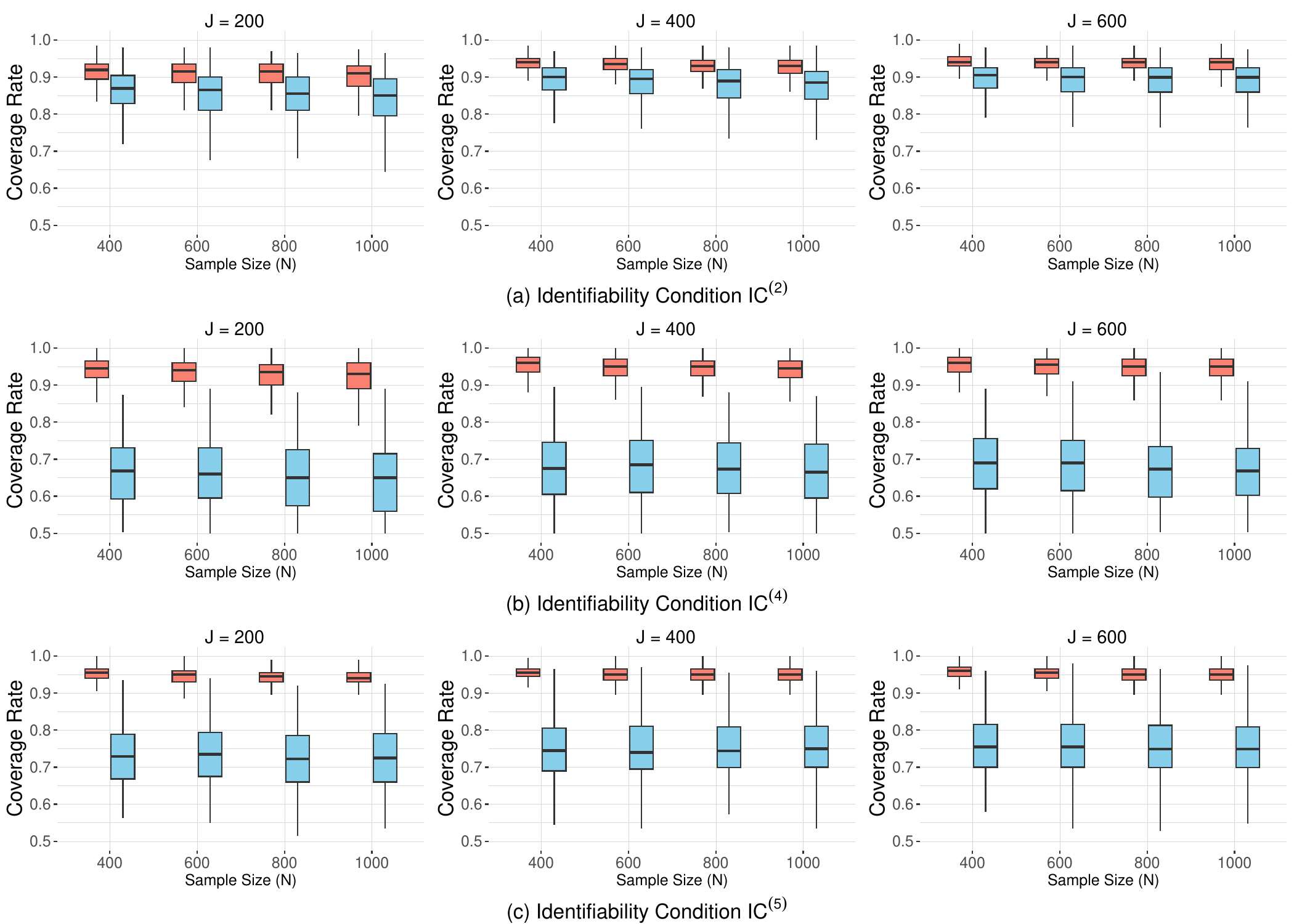}
    \caption{ Empirical coverage of loadings in the low-correlation case ($\tau = 0.3$) under different settings of $N$, $J$ and IC$^{(S)}$ for $S\in\{2,4,5\}$. The coverage rate of the proposed method is colored red; the coverage rate of the naive method is colored blue.}
    \label{fig:cover_ic_1}
\end{figure}
}

Overall, our proposed method demonstrates good coverage results in all scenarios for both factors and loadings. Our proposed method achieves coverage close to 95\% across all setups. In Figure~\ref{fig:cover_ic_1} that displays the low-correlation case, the empirical coverage rates for loadings under different setups remain close to 95\% across all settings, with those under IC$^{(2)}$ and IC$^{(5)}$ exhibiting smaller variations. In contrast, the naive method produces empirical coverage rates significantly below 95\%, indicating substantial underestimation of the variance of the maximum likelihood estimates, regardless of sample size or test length. Similarly, in Figure~\ref{fig:cover_ic_2} in the Supplementary Material, the empirical coverage rates follow a similar pattern in the high-correlation case.  
For estimation of the factors, naive covariance estimation is also problematic, achieving empirical coverage rates of around 80\%, as shown in Figures~\ref{fig:fcover_ic_1} and~\ref{fig:fcover_ic_2} in the Supplementary Material. 
{ The simulation results under both the linear and Poisson factor models display similar empirical coverage patterns for the loadings and the latent factors.}
Overall, our method correctly accounts for these additional variance terms, ensuring more reliable uncertainty quantification and superior performance compared to the naive approach.

\section{Data Application}\label{sec:data}

This section presents a study on a personality dataset based on the International Personality Item Pool (IPIP) NEO personality inventory~\citep{johnson2014measuring}. Originally introduced by~\cite{goldberg1999broad}, the dataset includes 300 items to assess the 30 facet scales of the NEO Personality Inventory~\citep{costa2008revised}, which is designed to measure the Big Five personality factors: ``Neuroticism'', ``Extraversion'', ``Conscientiousness'', ``Agreeableness'', and ``Openness''.

We select a subset of 1,149 individuals collected in Singapore, whose responses to all 300 items are recorded. Responses, measured on a Likert scale, are dichotomized by combining categories $\{1, 2, 3\}$ and $\{4, 5\}$ as in~\cite{chen2020structured}. The 300 items are distributed across five domains, with 60 items per domain designed primarily to assess a single personality trait, while possibly also loading on other domains. 
Based on the previous study, we select one item from each domain restricted to measuring only the personality trait associated with this domain. The selected items are presented in Section~\ref{sec_supp_simu3} in the Supplementary Material. We further assume that these personality traits are normalized, but allow them to be correlated. Then it can be checked that IC$^{(4)}$ is satisfied. We fit a logistic factor model featuring five latent dimensions and an intercept indexed by $j$, known as the multidimensional 2-parameter logistic model (M2PL) \citep{reckase2009}. The estimation is obtained following a similar procedure as in
Section~\ref{sec:jmle}, detailed in Section~\ref{sec_sup_single_intercept} in the Supplementary Material.

 {\spacingset{1.3}
    \begin{figure}[h]
\centering    
        \includegraphics[width=5.4in, height=1.6in]{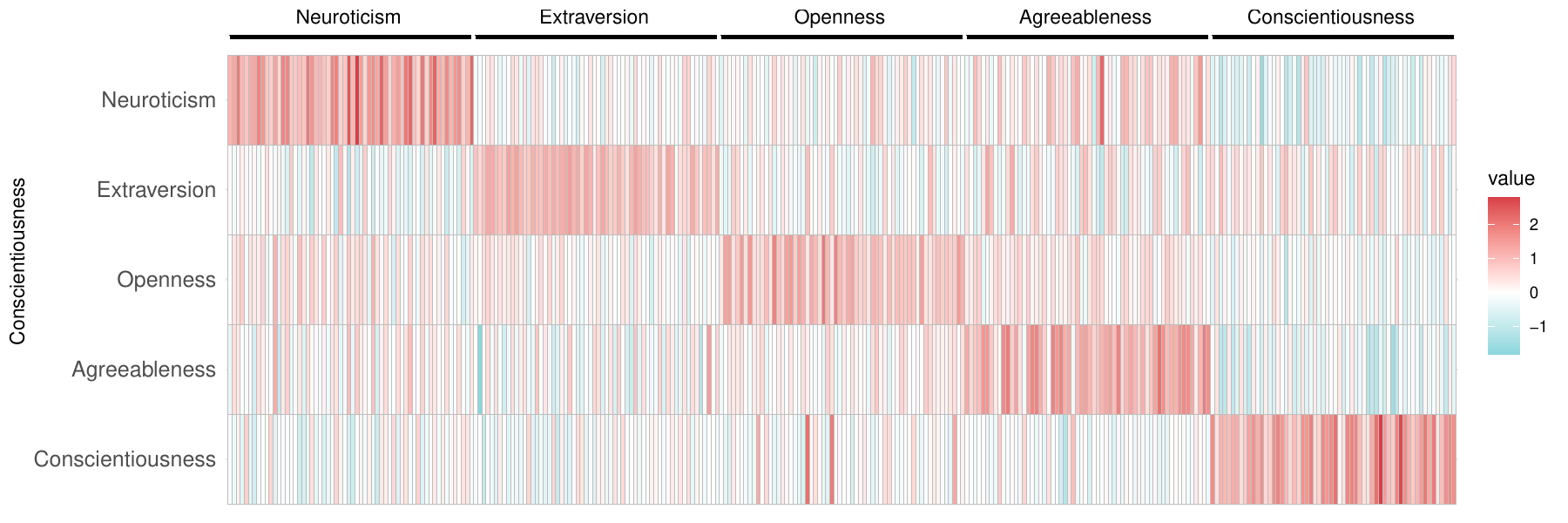}
  \caption{\small Heatmap for the (transposed) estimated loading matrix for the IPIP-NEO dataset. Each row contains estimated loadings in a specific personality domain. The columns are grouped and labeled by the item pools in each personality domain.}\label{fig:loadings}
\end{figure}
 }

\begin{figure}[htbp!]
\centering    
        \includegraphics[width=5.4in]{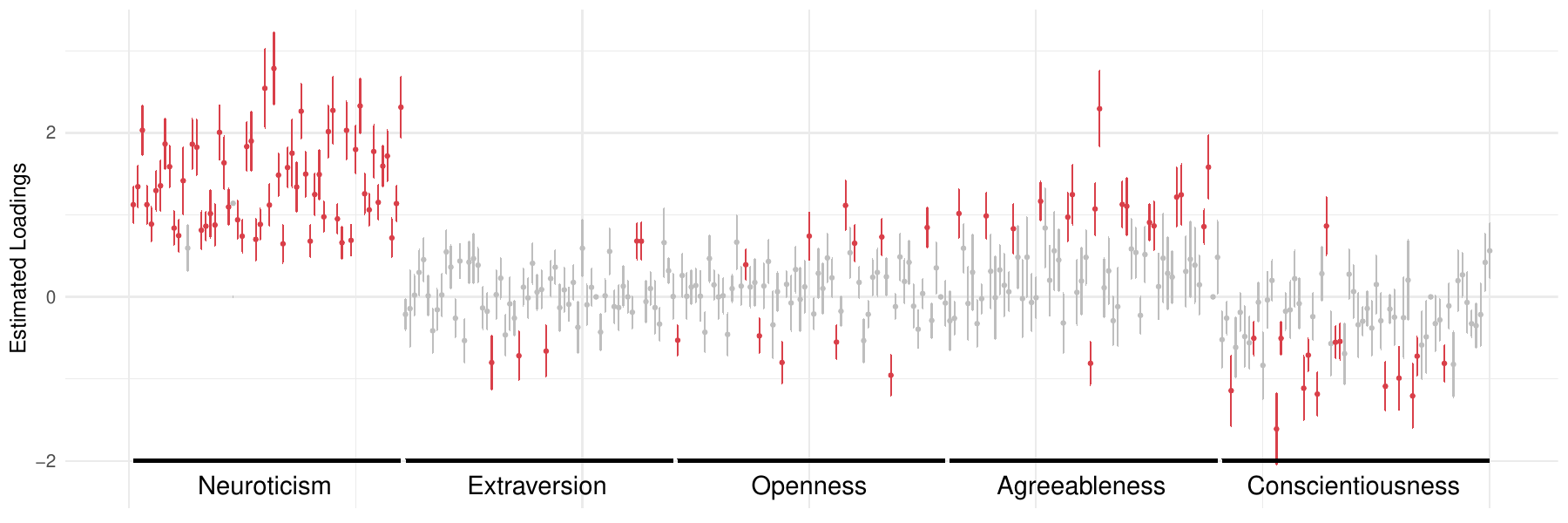}
   \caption{\small Confidence intervals for the first column of the loadings matrix corresponding to the ``Neuroticism'' facet. Red intervals represent significant loadings after Bonferroni correction.} \label{fig:ci_loading1}
\end{figure}

Figure~\ref{fig:loadings} presents the estimated loadings, where they are grouped into five item pools. From the heatmap, it is observed that each group is primarily associated with one personality domain, together with some cross-domain impacts, consistent with existing findings~\citep{gow2005goldberg}. For instance, the first factor shows several large loadings on the ``Agreeableness'' domain, along with a couple of negative loadings on the ``Conscientiousness'' domain. Using  Theorem~\ref{main_thm_asymp_intj} in the Supplementary Material, we construct 95\% Wald intervals and compute $p$-values to test whether each loading is nonzero. Figure~\ref{fig:ci_loading1} illustrates the results of the first column in the loading matrix, where significant loadings are highlighted in red based on a Bonferroni-corrected threshold at the $0.05$ significance level. 
The results for other columns are omitted in Section~\ref{sec_supp_simu2} in the Supplementary Material. We identify significant impacts of each factor on nearly all items in its associated item pool and on approximately 60 other items outside its pool.
Notably, in Figure~\ref{fig:ci_loading1}, one item from the ``Agreeableness'' pool—corresponding to the question ``Love a good fight''—shows a large loading on the ``Neuroticism'' factor. This suggests that the ``Neuroticism'' factor strongly influences responses to this item as well.

Next, we compute $\hat\Mb_{ff}^{(4)}$ and calculate the $p$-values (using Theorem~\ref{thm_mff_limiting_distri_intj}  in the Supplementary Material) with Bonferroni correction to test whether each off-diagonal entry is zero. Since $\hat\Mb_{ff}^{(4)}$ is symmetric, we display only its lower-triangular entries, highlighting those that indicate significant correlations with $p$-values below the threshold of 0.005.
\begin{equation*}\begingroup\renewcommand{\arraystretch}{0.5}
   \begin{matrix}
       \text{Neuroticism. }\\\text{Extraversion. }\\\text{Openness. }\\\text{Agreeableness. }\\\text{Conscientiousness. }
   \end{matrix} \begin{bmatrix}
 1&  &  &  &  \\
  0.017 &  1 & & &  \\
 \bm{-0.238} & -0.232&  1 & & \\
 -0.186 & -0.071 &  \bm{0.343} &  1 & \\
  0.039 &  0.090 & -0.115 &  0.079 &  1 \\
\end{bmatrix}\hspace{2.8cm}\endgroup
\end{equation*}
We observe a strong negative correlation between the ``Neuroticism'' factor and the ``Openness'' factor, with a $p$-value of $5.3\times 10^{-4}$, and a strong positive correlation between the ``Agreeableness'' factor and the ``Openness'' factor, with a $p$-value of $4.4\times 10^{-3}$. These results confirm the existence of correlations between these personality traits~\citep{park2020meta}.

\section{Conclusion}\label{sec:end}
In this article, we present a statistical inference framework for maximum likelihood estimation of generalized latent factor models under a series of easily satisfied identifiability conditions that have been commonly applied in practice. We present necessary and sufficient conditions for prescribing zeros in the loading matrix for both orthogonal and oblique designs. We establish consistency and asymptotic normality for maximum likelihood estimators across a wide range of identifiability conditions. Our contributions significantly advance the factor analysis literature by offering guidelines on statistical inference involving correlated factors and the design of the loadings. We validate our findings through simulation studies and an application to the International Personality Item Pool dataset.

Our work opens up several possibilities for future research. In this study, we have assumed that the number of latent factors is known and fixed. It remains to explore the scenarios where the number of latent dimensions diverges with $N$ and $J$. Another related direction is to develop inference procedures for determining the number of latent dimensions. Furthermore, given our theoretical results of the limiting distributions for all loading parameters, another potential direction involves developing multiple testing procedures to further investigate the loading matrix structure. {Third, the identifiability conditions in our paper focus on restrictions on the second-moment matrices of factors and loadings. A very interesting direction is to study other identifiability conditions that involve higher moments~\citep{Rohe2020VintageFA}.} {In addition, our paper focuses on minimal identifiability conditions for resolving rotational indeterminacy. An important direction for future research is to study over-identified settings. In deciding which constraints to impose under such settings, likelihood-based methods such as goodness-of-fit tests and information criteria may be useful, and our work provides a first step toward understanding the behavior of the likelihood under identifiability constraints.}


{
\setlength{\bibsep}{0pt} 
 \small
 \bibstyle{agsm}
\bibliography{reference}}

@article{Wang2018MaximumLE,
  title={Maximum likelihood estimation and inference for high dimensional generalized factor models with application to factor-augmented regressions},
  author={Wang, Fa},
  journal={Journal of econometrics},
  volume={229},
  number={1},
  pages={180--200},
  year={2022},
  publisher={Elsevier}
}

@article{gu2023dimension,
  title={Dimension-grouped mixed membership models for multivariate categorical data},
  author={Gu, Yuqi and Erosheva, Elena E and Xu, Gongjun and Dunson, David B},
  journal={Journal of machine learning research},
  volume={24},
  number={88},
  pages={1--49},
  year={2023}
}

@book{kline2023principles,
  title={Principles and practice of structural equation modeling},
  author={Kline, Rex B},
  year={2023},
  publisher={Guilford publications}
}

@book{brown2015confirmatory,
  title={Confirmatory factor analysis for applied research},
  author={Brown, Timothy A},
  year={2015},
  publisher={Guilford publications}
}

@article{rijmen2009efficient,
  title={Efficient full information maximum likelihood estimation for multidimensional IRT models},
  author={Rijmen, Frank},
  journal={ETS Research Report Series},
  volume={2009},
  number={1},
  pages={i--31},
  year={2009},
  publisher={Wiley Online Library}
}

@book{ullman2012structural,
  title={Structural equation modeling},
  author={Ullman, Jodie B and Bentler, Peter M},
  journal={Handbook of Psychology, Second Edition},
  volume={2},
  year={2012},
  publisher={Wiley Online Library}
}

@book{thurstone1947multiple,
  title={Multiple-factor analysis: a development and expansion of the vectors of mind},  
author={Thurstone, LL },
  year={1947},
  publisher={University of Chicago Press}
}

@article{bai2012statistical,
author = {Jushan Bai and Kunpeng Li},
title = {{Statistical analysis of factor models of high dimension}},
volume = {40},
journal = {The Annals of Statistics},
number = {1},
publisher = {Institute of Mathematical Statistics},
pages = {436 -- 465},

  year={2012}
}

@article{johnson2014measuring,
  title={Measuring thirty facets of the Five Factor Model with a 120-item public domain inventory: Development of the {IPIP}-{NEO}-120},
  author={Johnson, John A},
  journal={Journal of Research in Personality},
  volume={51},
  pages={78--89},
  year={2014},
  publisher={Elsevier}
}

@article{costa2008revised,
  title={The revised neo personality inventory ({NEO-PI-R})},
  author={Costa, Paul T and McCrae, Robert R},
  journal={The SAGE handbook of personality theory and assessment},
  volume={2},
  number={2},
  pages={179--198},
  year={2008}
}

@article{goldberg1999broad,
  title={A broad-bandwidth, public domain, personality inventory measuring the lower-level facets of several five-factor models},
  author={Goldberg, Lewis R and others},
  journal={Personality psychology in Europe},
  volume={7},
  number={1},
  pages={7--28},
  year={1999},
  publisher={Tilburg Netherland}
}

@article{yu2015useful,
  title={A useful variant of the {D}avis--{K}ahan theorem for statisticians},
  author={Yu, Yi and Wang, Tengyao and Samworth, Richard J},
  journal={Biometrika},
  volume={102},
  number={2},
  pages={315--323},
  year={2015},
  publisher={Oxford University Press}
}

@book{van2000asymptotic,
  title={Asymptotic statistics},
  author={van der Vaart, Aad W},
  volume={3},
  year={2000},
  publisher={Cambridge University Press}
}

@article{fan2011high,
  title={High dimensional covariance matrix estimation in approximate factor models},
  author={Fan, Jianqing and Liao, Yuan and Mincheva, Martina},
  journal={The Annals of Statistics},
  volume={39},
  number={6},
  pages={3320},
  year={2011}
}

@article{chen2020constrained,
  title={Constrained factor models for high-dimensional matrix-variate time series},
  author={Chen, Elynn Y and Tsay, Ruey S and Chen, Rong},
  journal={Journal of the American Statistical Association},
  volume={115},
  number={530},
  pages={775–793},
  year={2020},
  publisher={Taylor \& Francis}
}

@article{holland1990sampling,
  title={On the sampling theory roundations of item response theory models},
  author={Holland, Paul W},
  journal={Psychometrika},
  volume={55},
  number={4},
  pages={577--601},
  year={1990},
  publisher={Springer-Verlag}
}

@article{sweeting1980uniform,
  title={Uniform asymptotic normality of the maximum likelihood estimator},
  author={Sweeting, Trevor J},
  journal={The Annals of Statistics},
  pages={1375--1381},
  year={1980},
  publisher={JSTOR}
}

@article{chen2020structured,
  title={Structured latent factor analysis for large-scale data: Identifiability, estimability, and their implications},
  author={Chen, Yunxiao and Li, Xiaoou and Zhang, Siliang},
  journal={Journal of the American Statistical Association},
  volume={115},
  number={532},
  pages={1756--1770},
  year={2020},
  publisher={Taylor \& Francis}
}

@article{wang2025robust,
  title={Robust matrix completion with heavy-tailed noise},
  author={Wang, Bingyan and Fan, Jianqing},
  journal={Journal of the American Statistical Association},
  volume={120},
  number={550},
  pages={922--934},
  year={2025},
  publisher={Taylor \& Francis}
}

@article{von2008general,
  title={A general diagnostic model applied to language testing data},
  author={Von Davier, Matthias},
  journal={British Journal of Mathematical and Statistical Psychology},
  volume={61},
  number={2},
  pages={287--307},
  year={2008},
  publisher={Wiley Online Library}
}

@article{chen2020noisy,
  title={Noisy matrix completion: Understanding statistical guarantees for convex relaxation via nonconvex optimization},
  author={Chen, Yuxin and Chi, Yuejie and Fan, Jianqing and Ma, Cong and Yan, Yuling},
  journal={SIAM Journal on Optimization},
  volume={30},
  number={4},
  pages={3098--3121},
  year={2020},
  publisher={SIAM}
}

@article{cui2026beyond,
  title={Beyond Vintage Rotation: Bias-Free Sparse Representation Learning with Oracle Inference},
  author={Cui, Chengyu and Chen, Yunxiao and Ouyang, Jing and Xu, Gongjun},
  journal={arXiv preprint arXiv:2602.22590},
  year={2026}
}

@article{choi2024matrix,
  title={Matrix completion when missing is not at random and its applications in causal panel data models},
  author={Choi, Jungjun and Yuan, Ming},
  journal={Journal of the American Statistical Association},
  pages={1--15},
  year={2024},
  publisher={Taylor \& Francis}
}

@article{chen2019inference,
  title={Inference and uncertainty quantification for noisy matrix completion},
  author={Chen, Yuxin and Fan, Jianqing and Ma, Cong and Yan, Yuling},
  journal={Proceedings of the National Academy of Sciences},
  volume={116},
  number={46},
  pages={22931--22937},
  year={2019},
  publisher={National Academy of Sciences}
}

@article{bing2025optimal,
  title={Optimal vintage factor analysis with deflation varimax},
  author={Bing, Xin and He, Xin and Jin, Dian and Zhang, Yuqian},
  journal={The Annals of Statistics},
  volume={53},
  number={4},
  pages={1639--1666},
  year={2025},
  publisher={Institute of Mathematical Statistics}
}

@article{browne2001overview,
  title={An overview of analytic rotation in exploratory factor analysis},
  author={Browne, Michael W},
  journal={Multivariate behavioral research},
  volume={36},
  number={1},
  pages={111--150},
  year={2001},
  publisher={Taylor \& Francis}
}

@article{kaiser1958varimax,
  title={The varimax criterion for analytic rotation in factor analysis},
  author={Kaiser, Henry F},
  journal={Psychometrika},
  volume={23},
  number={3},
  pages={187--200},
  year={1958},
  publisher={Springer}
}

@article{jennrich2006rotation,
  title={Rotation to simple loadings using component loss functions: The oblique case},
  author={Jennrich, Robert I},
  journal={Psychometrika},
  volume={71},
  number={1},
  pages={173--191},
  year={2006},
  publisher={Springer}
}

@article{jennrich1966rotation,
  title={Rotation for simple loadings},
  author={Jennrich, Robert I and Sampson, PF},
  journal={Psychometrika},
  volume={31},
  number={3},
  pages={313--323},
  year={1966},
  publisher={Springer-Verlag}
}

@article{Rohe2020VintageFA,
  title={Vintage Factor Analysis with Varimax Performs Statistical Inference},
  author={Karl Rohe and Muzhe Zeng},
  journal={Journal of the Royal Statistical Society Series B: Statistical Methodology},
  volume={85},
  number={4},
  pages={1037-1060},
  year={2023}
}

@article{wang2025data,
  title={Data thinning for Poisson factor models and its applications},
  author={Wang, Zhijing and Xu, Peirong and Zhao, Hongyu and Wang, Tao},
  journal={Journal of the American Statistical Association},
  number={just-accepted},
  pages={1--30},
  year={2025},
  publisher={Taylor \& Francis}
}

@article{newey1994large,
  title={Large sample estimation and hypothesis testing},
  author={Newey, Whitney K and McFadden, Daniel},
  journal={Handbook of Econometrics},
  volume={4},
  pages={2111--2245},
  year={1994},
  publisher={Elsevier}
}

@incollection{ross2013arbitrage,
  title={The arbitrage theory of capital asset pricing},
  author={Ross, Stephen A},
  booktitle={Handbook of the fundamentals of financial decision making: Part I},
  pages={11--30},
  year={2013},
  publisher={World Scientific}
}

@article{bai2013principal,
  title={Principal components estimation and identification of static factors},
  author={Bai, Jushan and Ng, Serena},
  journal={Journal of Econometrics},
  volume={176},
  number={1},
  pages={18--29},
  year={2013},
  publisher={Elsevier}
}

@article{preacher2002exploratory,
  title={Exploratory factor analysis in behavior genetics research: Factor recovery with small sample sizes},
  author={Preacher, Kristopher J and MacCallum, Robert C},
  journal={Behavior genetics},
  volume={32},
  pages={153--161},
  year={2002},
  publisher={Springer}
}

@article{maccallum1999sample,
  title={Sample size in factor analysis.},
  author={MacCallum, Robert C and Widaman, Keith F and Zhang, Shaobo and Hong, Sehee},
  journal={Psychological methods},
  volume={4},
  number={1},
  pages={84},
  year={1999},
  publisher={American Psychological Association}
}

@article{bai2003inferential,
  title={Inferential theory for factor models of large dimensions},
  author={Bai, Jushan},
  journal={Econometrica},
  volume={71},
  number={1},
  pages={135--171},
  year={2003},
  publisher={Wiley Online Library}
}

@article{chen2019joint,
  title={Joint maximum likelihood estimation for high-dimensional exploratory item factor analysis},
  author={Chen, Yunxiao and Li, Xiaoou and Zhang, Siliang},
  journal={Psychometrika},
  volume={84},
  pages={124--146},
  year={2019},
  publisher={Springer}
}

@book{skrondal2004generalized,
  title={Generalized latent variable modeling: Multilevel, longitudinal, and structural equation models},
  author={Skrondal, Anders and Rabe-Hesketh, Sophia},
  year={2004},
  publisher={Chapman and Hall/CRC}
}

@article{stock2002forecasting,
  title={Forecasting using principal components from a large number of predictors},
  author={Stock, James H and Watson, Mark W},
  journal={Journal of the American statistical association},
  volume={97},
  number={460},
  pages={1167--1179},
  year={2002},
  publisher={Taylor \& Francis}
}

@book{fabrigar2012exploratory,
  title={Exploratory factor analysis},
  author={Fabrigar, Leandre R and Wegener, Duane T},
  year={2012},
  publisher={Oxford University Press}
}

@article{gu2020partial,
  title={Partial identifiability of restricted latent class models},
  author={Gu, Yuqi and Xu, Gongjun},
  journal={The Annals of Statistics},
  volume={48},
  number={4},
  pages={2082--2107},
  year={2020},
  publisher={JSTOR}
}

@article{chen2025item,
  title={Item response theory—A statistical framework for educational and psychological measurement},
  author={Chen, Yunxiao and Li, Xiaoou and Liu, Jingchen and Ying, Zhiliang},
  journal={Statistical Science},
  volume={40},
  number={2},
  pages={167--194},
  year={2025},
  publisher={Institute of Mathematical Statistics}
}

@book{mulaik2009foundations,
  title={Foundations of factor analysis},
  author={Mulaik, Stanley A},
  year={2009},
  publisher={CRC press}
}

@article{Liu2023RotationTS,
  title={Rotation to Sparse Loadings Using ${L}^p$ Losses and Related Inference Problems},
  author={Xinyi Liu and Gabriel Wallin and Yunxiao Chen and Irini Moustaki},
  journal={Psychometrika},
  year={2023},
  volume={88},
  pages={527-553}
}

@article{hendrickson1964promax,
  title={Promax: A quick method for rotation to oblique simple structure},
  author={Hendrickson, Alan E and White, Paul Owen},
  journal={British Journal of Statistical Psychology},
  volume={17},
  number={1},
  pages={65--70},
  year={1964},
  publisher={Wiley Online Library}
}

@book{andersonintroduction,
  title={An introduction to multivariate statistical analysis},
  author={Anderson, Theodore Wilbur},
   year={2003},
  publisher={Wiley New York}
}

@article{bai2006confidence,
  title={Confidence intervals for diffusion index forecasts and inference for factor-augmented regressions},
  author={Bai, Jushan and Ng, Serena},
  journal={Econometrica},
  volume={74},
  number={4},
  pages={1133--1150},
  year={2006},
  publisher={Wiley Online Library}
}

@article{bing2020adaptive,
  title={Adaptive estimation in structured factor models with applications to overlapping clustering},
  author={Bing, Xin and Bunea, Florentina and Ning, Yang and Wegkamp, Marten},
  journal={The Annals of Statistics},
  pages={2055–2081},
  volume={48},
  year={2020}
}

@article{jennrich1969asymptotic,
  title={Asymptotic properties of non-linear least squares estimators},
  author={Jennrich, Robert I},
  journal={The Annals of Mathematical Statistics},
  volume={40},
  number={2},
  pages={633--643},
  year={1969},
  publisher={JSTOR}
}

@article{wu1981asymptotic,
  title={Asymptotic theory of nonlinear least squares estimation},
  author={Wu, Chien-Fu},
  journal={The Annals of Statistics},
  volume={9},
  number={3},
  pages={501--513},
  year={1981},
  publisher={Institute of Mathematical Statistics}
}

@article{thompson2004exploratory,
  title={Exploratory and Confirmatory Factor Analysis: Understanding Concepts and Applications},
  author={Thompson, B},
  journal={American Psychological Association},
  year={2004}
}

@book{reckase2009,
  title={Multidimensional Item Response Theory},
  author={M.D. Reckase},
  year={2009},
  publisher={Springer New York, NY}
}

@article{neyman1948consistent,
  title={Consistent estimates based on partially consistent observations},
  author={Neyman, Jerzy and Scott, Elizabeth L},
  journal={Econometrica: Journal of the Econometric Society},
  pages={1--32},
  year={1948},
  publisher={JSTOR}
}

@article{bock1981marginal,
  title={Marginal maximum likelihood estimation of item parameters: Application of an EM algorithm},
  author={Bock, R Darrell and Aitkin, Murray},
  journal={Psychometrika},
  volume={46},
  number={4},
  pages={443--459},
  year={1981},
  publisher={Springer}
}

@article{bai2002determining,
  title={Determining the number of factors in approximate factor models},
  author={Bai, Jushan and Ng, Serena},
  journal={Econometrica},
  volume={70},
  number={1},
  pages={191--221},
  year={2002},
  publisher={Wiley Online Library}
}

@book{baker2004item,
  title={Item response theory: Parameter estimation techniques},
  author={Baker, Frank B and Kim, Seock-Ho},
  year={2004},
  publisher={CRC press}
}

@book{billingsley2017probability,
  title={Probability and measure},
  author={Billingsley, Patrick},
  year={2017},
  publisher={John Wiley \& Sons}
}

@article{gow2005goldberg,
  title={Goldberg’s ‘{IPIP}’Big-Five factor markers: Internal consistency and concurrent validation in Scotland},
  author={Gow, Alan J and Whiteman, Martha C and Pattie, Alison and Deary, Ian J},
  journal={Personality and Individual Differences},
  volume={39},
  number={2},
  pages={317--329},
  year={2005},
  publisher={Elsevier}
}

@article{fan2017sufficient,
  title={Sufficient forecasting using factor models},
  author={Fan, Jianqing and Xue, Lingzhou and Yao, Jiawei},
  journal={Journal of econometrics},
  volume={201},
  number={2},
  pages={292--306},
  year={2017},
  publisher={Elsevier}
}

@article{chen2021nonlinear,
  title={Nonlinear factor models for network and panel data},
  author={Chen, Mingli and Fern{\'a}ndez-Val, Iv{\'a}n and Weidner, Martin},
  journal={Journal of Econometrics},
  volume={220},
  number={2},
  pages={296--324},
  year={2021},
  publisher={Elsevier}
}

@article{joreskog1969general,
  title={A general approach to confirmatory maximum likelihood factor analysis},
  author={J{\"o}reskog, Karl G},
  journal={Psychometrika},
  volume={34},
  number={2},
  pages={183--202},
  year={1969},
  publisher={Springer}
}

@article{dunn1973anote,
  title={A NOTE ON A SUFFICIENCY CONDITION FOR UNIQUENESS OF
A RESTRICTED FACTOR MATRIX. },
  author={James E. Dunn},
  journal={Psychometrika},
  volume={38},
  number={1},
  pages={141--143},
  year={1973},
  publisher={Springer}
}

@article{benitez2020perform,
  title={How to perform and report an impactful analysis using partial least squares: Guidelines for confirmatory and explanatory IS research},
  author={Benitez, Jose and Henseler, J{\"o}rg and Castillo, Ana and Schuberth, Florian},
  journal={Information \& management},
  volume={57},
  number={2},
  pages={103168},
  year={2020},
  publisher={Elsevier}
}

@article{henseler2020using,
  title={Using confirmatory composite analysis to assess emergent variables in business research},
  author={Henseler, J{\"o}rg and Schuberth, Florian},
  journal={Journal of Business Research},
  volume={120},
  pages={147--156},
  year={2020},
  publisher={Elsevier}
}

@article{xu2017,
author = "Xu, Gongjun",
journal = "Annals of Statistics",
number = "2",
pages = "675--707",
publisher = "The Institute of Mathematical Statistics",
title = "Identifiability of restricted latent class models with binary responses",
volume = "45",
year = "2017"
}

@article{xu2018JASA,
author = {Gongjun Xu and Zhuoran Shang},
title = {Identifying Latent Structures in Restricted Latent Class Models},
journal = {Journal of the American Statistical Association},
volume = {113},
number = {523},
pages = {1284-1295},
year  = {2018},
publisher = {Taylor & Francis},
}

@article{millsap2001trivial,
  title={When trivial constraints are not trivial: The choice of uniqueness constraints in confirmatory factor analysis},
  author={Millsap, Roger E},
  journal={Structural Equation Modeling},
  volume={8},
  number={1},
  pages={1--17},
  year={2001},
  publisher={Taylor \& Francis}
}

@article{bandeira2016sharp,
  title={Sharp nonasymptotic bounds on the norm of random matrices with independent entries},
  author={Bandeira, Afonso S and Van Handel, Ramon},
journal={Annals of Probability},
  volume={44},
  number={4},
  pages={2479--2506},
  year={2016},
  publisher={Taylor \& Francis}
}

@article{vershynin2010introduction,
  title={Introduction to the non-asymptotic analysis of random matrices},
  author={Vershynin, Roman},
  journal={arXiv preprint arXiv:1011.3027},
  year={2010}
}

@article{haberman1977maximum,
  title={Maximum likelihood estimates in exponential response models},
  author={Haberman, Shelby J},
  journal={The Annals of statistics},
  volume={5},
  number={5},
  pages={815--841},
  year={1977},
  publisher={Institute of Mathematical Statistics}
}

@article{andersen1970asymptotic,
  title={Asymptotic properties of conditional maximum-likelihood estimators},
  author={Andersen, Erling Bernhard},
  journal={Journal of the Royal Statistical Society Series B: Statistical Methodology},
  volume={32},
  number={2},
  pages={283--301},
  year={1970},
  publisher={Oxford University Press}
}

@article{chen2022determining,
  title={Determining the number of factors in high-dimensional generalized latent factor models},
  author={Chen, Yunxiao and Li, Xiaoou},
  journal={Biometrika},
  volume={109},
  number={3},
  pages={769--782},
  year={2022},
  publisher={Oxford University Press}
}

@article{cui2026convexity,
  title={Convexity in Disguise: A Theoretical Framework for Nonconvex Low-Rank Matrix Estimation},
  author={Cui, Chengyu and Xu, Gongjun},
  journal={arXiv preprint arXiv:2605.05446},
  year={2026}
}

@article{chen2024note,
  title={A note on entrywise consistency for mixed-data matrix completion},
  author={Chen, Yunxiao and Li, Xiaoou},
  journal={Journal of Machine Learning Research},
  volume={25},
  number={343},
  pages={1--66},
  year={2024}
}

@article{park2020meta,
  title={Meta-analytic five-factor model personality intercorrelations: Eeny, meeny, miney, moe, how, which, why, and where to go.},
  author={Park, HyeSoo Hailey and Wiernik, Brenton M and Oh, In-Sue and Gonzalez-Mul{\'e}, Erik and Ones, Deniz S and Lee, Youngduk},
  journal={Journal of Applied Psychology},
  volume={105},
  number={12},
  pages={1490},
  year={2020},
  publisher={American Psychological Association}
}

\newtheorem{suppTheorem}{Theorem}[section]  
\renewcommand{\thesuppTheorem}{\thesection.\arabic{suppTheorem}}  
\renewcommand{\theequation}{S\arabic{equation}} 
\setcounter{equation}{0}  
\renewcommand{\thefigure}{S\arabic{figure}} 
\setcounter{figure}{0} 

\vspace{-2cm}
\newpage
\begin{center}
{\bf\large Supplementary Material for ``Identifiability and  Inference for Generalized Latent Factor Models''}
\end{center}
\appendix
\startcontents[appendix]

\section*{Content}
\printcontents[appendix]{}{1}{\setcounter{tocdepth}{1}}


\section{Preliminaries}\label{sec:supp_prlim}

In this section, we clarify some notation used throughout the subsequent theoretical proofs. We use the notation $A\lesssim B$ or $B\gtrsim A$ to indicate that $A$ is asymptotically bounded above by $B$, i.e., there exists a positive constant $C$ such that $A\le CB$ for sufficiently large values of relevant parameters. 
We define $\bm{e}_x^{(y)}$ as the $y$-dimensional vector with the $x$-th entry equal to 1 and all other entries equal to 0. When $y=K$, we omit the superscript and denote the indicator vector as $\be_x$. For any matrix $\Ab = \{a_{ij}\}_{n\times m}$, let $\|\Ab\|=\sup_{\|\bx\|=1}\|\Ab\bx\|$ be the operator norm, $\|\Ab\|_{1} = \max_{j=1,\ldots,m} \sum_{i=1}^n |a_{ij}|$ be the maximum absolute column sum and $\| \Ab\|_{\infty} = \max_{i=1,\ldots,n} \sum_{j=1}^m |a_{ij}|$ be the maximum absolute row sum. Let $\sigma_{\max}(\Ab)$ and $\sigma_{\min}(\Ab)$ denote the largest and smallest singular value of $\Ab$, respectively. For matrix $\Ab \in \mathbb{R}^{n \times m}$, $\mathrm{vec}(\Ab)$ is the $nm$ vector obtained by stacking the columns of the matrix $\Ab$ on top of one another. For square matrix $\Ab\in\RR^{n\times n}$, we define some operators as follows:  $\mathrm{ndiag}(\Ab)$ preserves only the non-diagonal elements in $\Ab$, $\text{lower}(\Ab)$ preserves only the elements below the diagonal of $\Ab$; $\text{nupper}(\Ab)$ preserves the elements on and below the diagonal of $\Ab$. Define $J_K = \{1,2,\cdots,JK\}$ and $N_K = \{1,2,\cdots,NK\}$. We use $\zero$ without a subscript when there is no confusion.

We define $\btheta=(\blambda_1^\T,\cdots,\blambda_J^\T,\bbf_1^\T,\cdots,\bbf_N^\T)^\T$ as the vectorized model parameters. 
We will use $\btheta^{(S)}$ or $(\bLambda^{(S)},\Fb^{(S)})$ to denote the parameters such that (i) they satisfy IC$^{(S)}$; (ii) $\bLambda^{(S)}(\Fb^{(S)})^\T = \bLambda^* (\Fb^*)^\T$. As will be demonstrated in Section~\ref{sec_supp_identification}, if the true parameters $(\bLambda^*, \Fb^*)$ satisfy IC$^{(S)}$, $\bLambda^{(S)} = \bLambda^* $ and $\Fb^{(S)} = \Fb^*$ up to a signed permutation for any $S\in \{0,1,\cdots,5\}$. Then throughout the Supplementary Material, we will write $(\bLambda^{(S)},\Fb^{(S)})$ as the true parameters. Since $(\blambda_j^{(S)})^\T \bbf_i^{(S)} = (\blambda_j^*)^\T\bbf_i^*$ does not change with $S$, we will use $\eta_{ij}^*$ to denote the product for simplicity. We denote the feasible set in \eqref{eq_joint_mle} as $\cB(D) = \{\btheta:\|\btheta\|_{\infty}\le D\}$ and assume that $D$ is large enough such that $\btheta^{(S)} \in \cB(D/2)$ for all $S\in\{0,1,\cdots,5\}$.
Define first order derivative of the negative log-likelihood as $\bS_L(\btheta)=-\partial_{\btheta}L(\btheta)$ and the Hessian matrix as $\Hb_L(\btheta)=-\partial_{\btheta\btheta}^2L(\btheta)$. By Assumption~\ref{assumption: psd covariance}, the sub-exponential norm of $l_{ij}^{\prime}(\eta_{ij}^*)$ can be bounded by $M$. We take $M$ large enough such that $\EE[|l^{\prime}_{ij}(\eta^*_{ij})|^\xi]\le M$ for some $\xi> 4$. 

\vspace{-0.5cm}
\section{Simulation Details and Additional Empirical Results}\label{sec_supp_simu}
Section~\ref{supp_sec_simu} provides additional details on the simulation setups used in Section~\ref{sec:simu}. Section~\ref{sec_supp_simu1} reports further simulation results. Section~\ref{sec_supp_simu3} lists the items selected to satisfy the identifiability condition used in Section~\ref{sec:data}. Finally, Section~\ref{sec_supp_simu2} presents additional real data results for Section~\ref{sec:data}.

\subsection{Simulation Setup}\label{supp_sec_simu}
We provide details of the simulation setups considered in Section~\ref{sec:simu} of the main text. The true parameters for these designs are denoted as $(\bLambda^{(2)}, \Fb^{(2)})$, $(\bLambda^{(4)}, \Fb^{(4)})$, and $(\bLambda^{(5)}, \Fb^{(5)})$. The corresponding loading structures are specified in equation \eqref{simulation-lambda},  where ``$*$''  indicates an entry that is generated randomly as follows. 
\begin{equation}
\label{simulation-lambda}
    \bLambda^{(2)} = \begingroup
    \renewcommand{\arraystretch}{0.45}\begin{pmatrix}
        * & 0 & 0\\ * & * & 0\\ * & * & *\\\vdots & \vdots & \vdots
    \end{pmatrix},\endgroup \quad \bLambda^{(4)}=\begingroup
    \renewcommand{\arraystretch}{0.45}\begin{pmatrix}
        * & * & 0\\ * & * & 0\\ * & 0 & *\\ * & 0 & *\\ 0 & * & *\\ 0 & * & * \\\vdots & \vdots & \vdots
    \end{pmatrix},\quad\endgroup\bLambda^{(5)} = \begingroup
    \renewcommand{\arraystretch}{0.45}\begin{pmatrix}
        1 & 0 & 0\\ 0 & 1 & 0\\ 0 & 0 & 1\\\vdots & \vdots & \vdots
    \end{pmatrix}.\endgroup
\end{equation}
We denote $\lambda_{jr}^{(S)}$ as the $(j,r)$-th entry in $\bLambda^{(S)}$ for $S\in\{2,4,5\}$. For parameters simulated under IC$^{(2)}$, we fix $\lambda_{21}^{(2)}, \lambda_{31}^{(2)},\lambda_{32}^{(2)}$ as zeros and other entries in $\bLambda_{{[1:3,]}}^{(2)}$ are generated i.i.d. from $\cN(0,1)$. We simulate $\blambda_j^{(2)}$ i.i.d. from $\cN(\zero_3,\bSigma_\tau)$ for $j\ge 4$. Here, $\bSigma
 _\tau\in\RR^{3\times 3}$ has the $(l,h)$th entry set as $\tau^{|l-h|}$ for $l,h\in\{1,2,3\}$, where $\tau$ is specified later. Factors are first simulated i.i.d. from $\cN(\zero_3,\Ib_3)$ and then transformed to satisfy $N^{-1}(\Fb^{(2)})^\T\Fb^{(2)} = \Ib_3$. 
 For the design IC$^{(4)}$, the construct of $\bLambda^{(4)}$ satisfies IC$^{(4)}$ by letting $\cA_1=\{5,6\}$, $\cA_2=\{3,4\}$ and $\cA_3=\{1,2\}$. For parameters simulated under IC$^{(4)}$, we fix $\lambda_{13}^{(4)}, \lambda_{23}^{(4)},\lambda_{32}^{(4)}, \lambda_{42}^{(4)},\lambda_{51}^{(4)}, \lambda_{61}^{(4)}$ as zeros and simulate all other entries in $\bLambda^{(4)}$ i.i.d. from $\cN(0,1)$. Factors are first simulated i.i.d. from $\cN(\zero_3,\bSigma_{\tau})$ and then transformed to satisfy $\mathrm{diag}\{N^{-1}(\Fb^{(4)})^\T\Fb^{(4)}\} = \Ib_3$. For parameters simulated under IC$^{(5)}$, we fix $[\bLambda^{(5)}]_{[1:3,]} = \Ib_{3}$ and simulate all other entries in $\bLambda^{(5)}$ i.i.d. from $\cN(0,1)$. Factors $\bbf_i^{(5)}$s are simulated i.i.d. from $\cN(\zero_3,\bSigma_{\tau})$.

\subsection{Additional Simulation Results}\label{sec_supp_simu1}
Section~\ref{sec:simu} of the main text presents the simulation results for the loadings estimates under the low correlation case for a logistic latent factor model. In this subsection, we provide additional results. Under the setup of the logistic latent factor model, for loadings estimates under the high correlation case ($\tau = 0.7$), Figure~\ref{fig:cover_ic_2} shows the empirical coverage rates. For latent factor estimates, Figure~\ref{fig:fcover_ic_1} presents the empirical coverage rates under the low correlation case ($\tau = 0.3$), while Figure~\ref{fig:fcover_ic_2} presents the empirical coverage rates under the high correlation case ($\tau = 0.7$). {We also conduct simulation studies beyond the logistic latent factor model under the setup of Section \ref{supp_sec_simu}. The results of the empirical coverages of loading parameters and latent factors for the linear latent factor model are presented in Figures~\ref{fig:linear_cover_ic_2}--\ref{fig:linear_fcover_ic_2}.  The results of the empirical coverages of loading parameters and latent factors for the Poisson latent factor model are presented in Figures~\ref{fig:poisson_cover_ic_1}--\ref{fig:poisson_fcover_ic_2}. }

{\spacingset{1}
\begin{figure}
        \centering\includegraphics[width=3in]{legend.pdf}\\\includegraphics[width=5in]{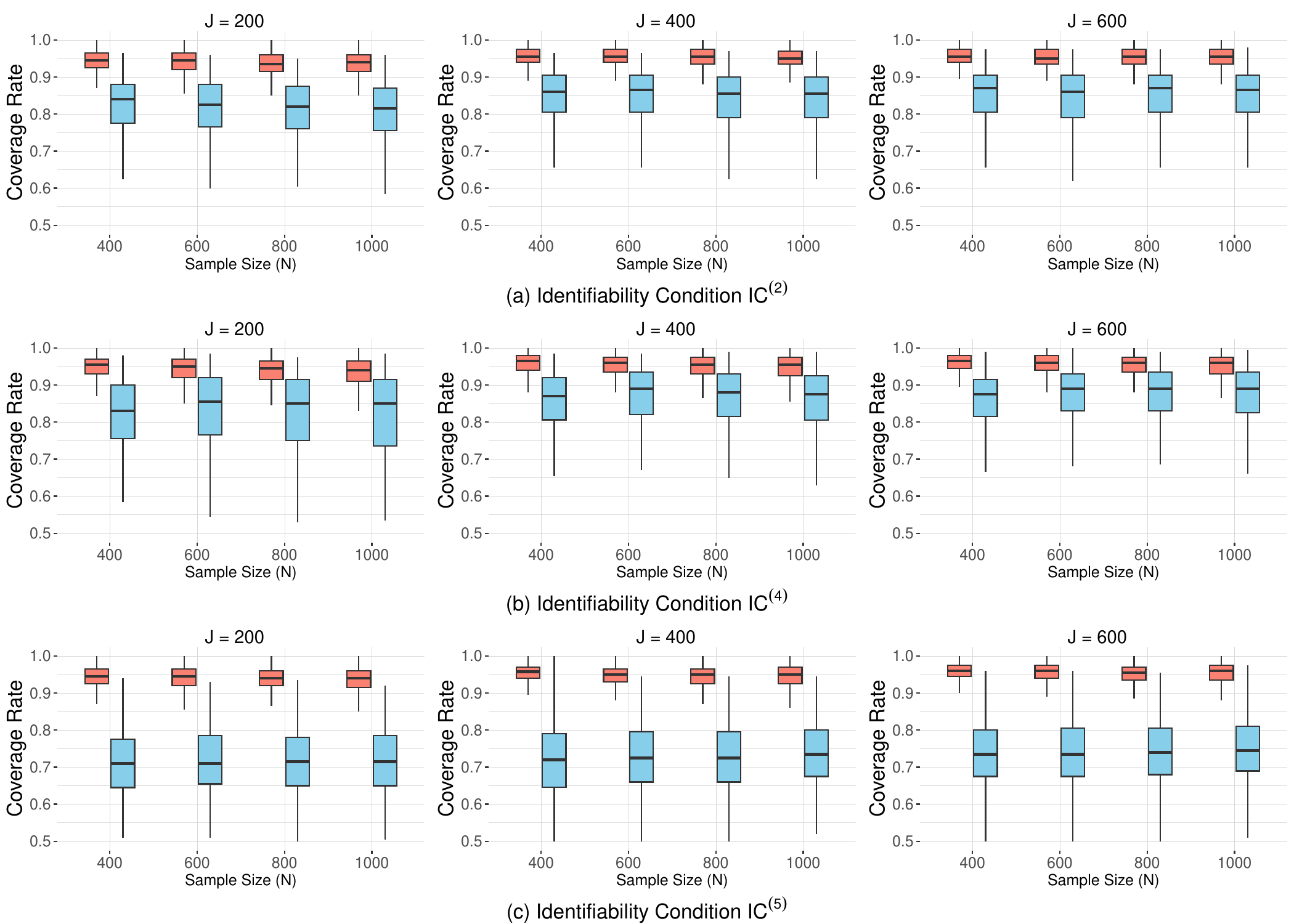}
    \caption{\small Empirical coverage of loadings in the logistic latent factor model and high-correlation case ($\tau = 0.7$) under different settings of $N$, $J$, and IC$^{(S)}$ for $S\in\{2,4,5\}$. The coverage rate of the proposed method is colored red; the coverage rate of the naive method is colored blue.}
    \label{fig:cover_ic_2}
\end{figure}

\begin{figure}
        \centering\includegraphics[width=3in]{legend.pdf}\\\includegraphics[width=5in]{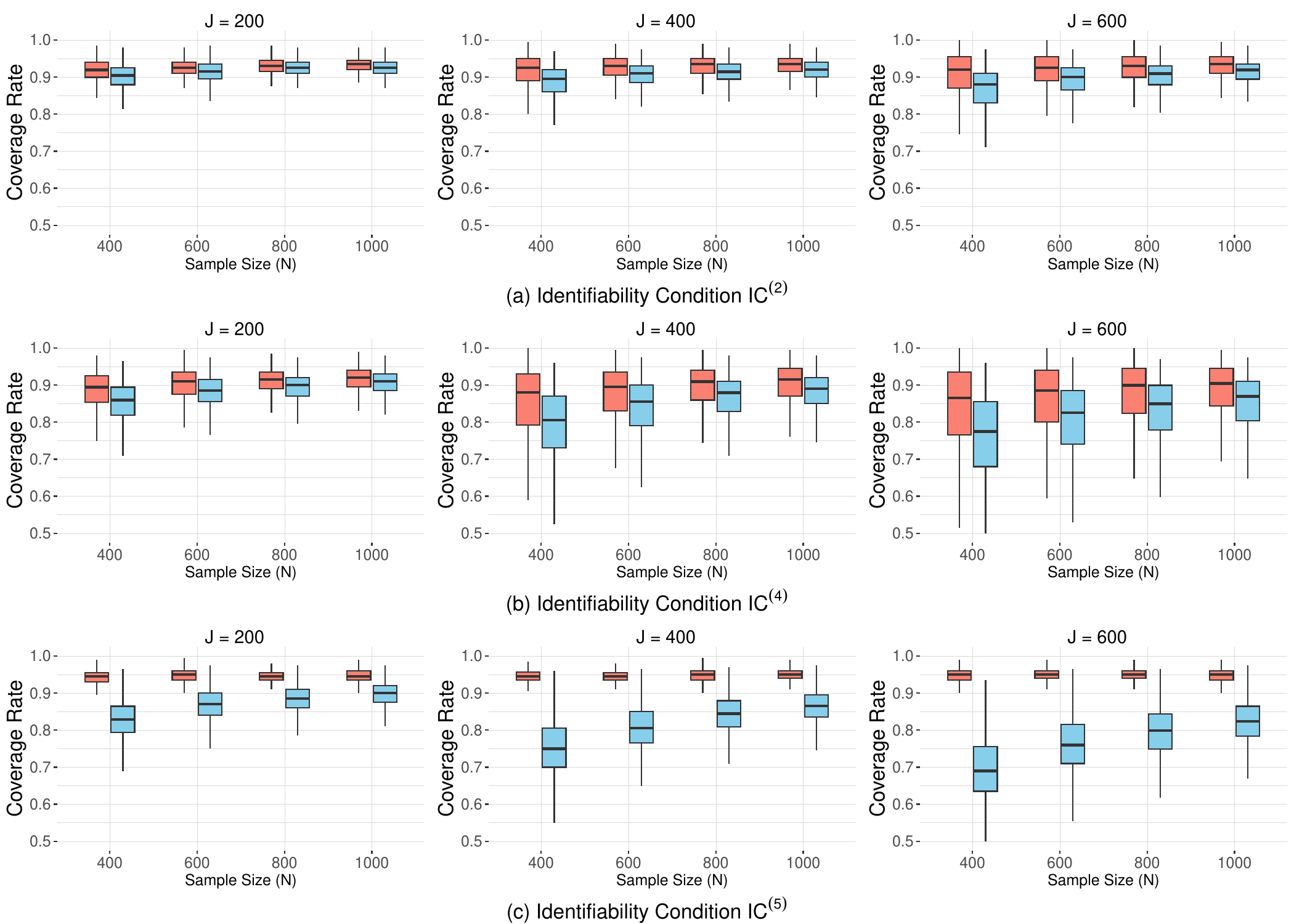}
    \caption{\small Empirical coverage of factors in the logistic latent factor model and low-correlation case ($\tau = 0.3$) under different settings of $N$, $J$, and IC$^{(S)}$ for $S\in\{2,4,5\}$. The coverage rate of the proposed method is colored red; the coverage rate of the naive method is colored blue.}
    \label{fig:fcover_ic_1}
\end{figure}
\begin{figure}
        \centering\includegraphics[width=3in]{legend.pdf}\\\includegraphics[width=5in]{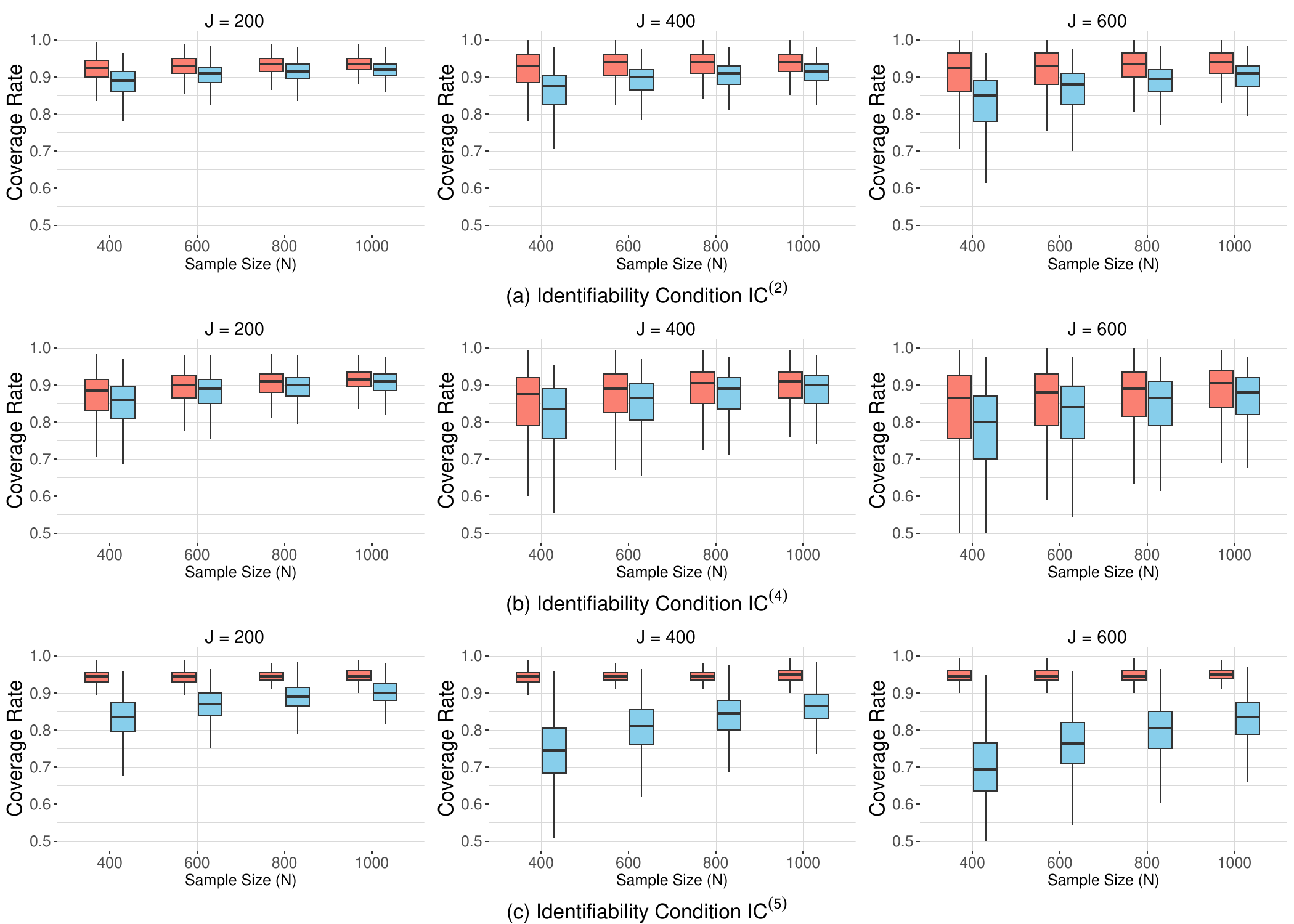}
    \caption{\small Empirical coverage of factors in the logistic latent factor model and high-correlation case ($\tau = 0.7$) under different settings of $N$, $J,$ and IC$^{(S)}$ for $S\in\{2,4,5\}$. The coverage rate of the proposed method is colored red; the coverage rate of the naive method is colored blue.}
    \label{fig:fcover_ic_2}
\end{figure}

\begin{figure}
        \centering\includegraphics[width=3in]{legend.pdf}\\\includegraphics[width=5in]{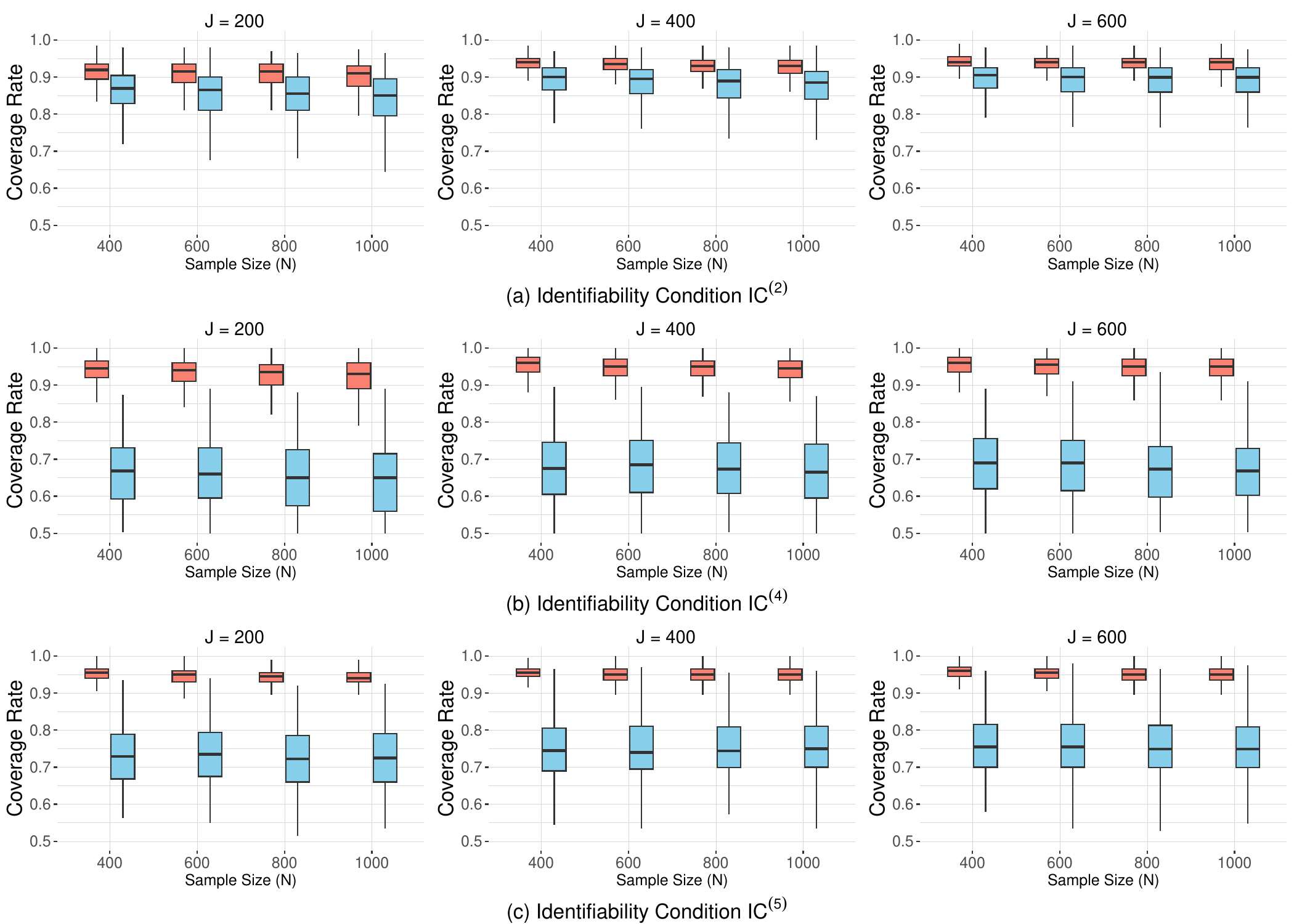}
    \caption{\small Empirical coverage of loadings in the linear latent factor model and low-correlation case ($\tau = 0.3$) under different settings of $N$, $J$, and IC$^{(S)}$ for $S\in\{2,4,5\}$. The coverage rate of the proposed method is colored red; the coverage rate of the naive method is colored blue.}
    \label{fig:linear_cover_ic_2}
\end{figure}
\begin{figure}
        \centering\includegraphics[width=3in]{legend.pdf}\\\includegraphics[width=5in]{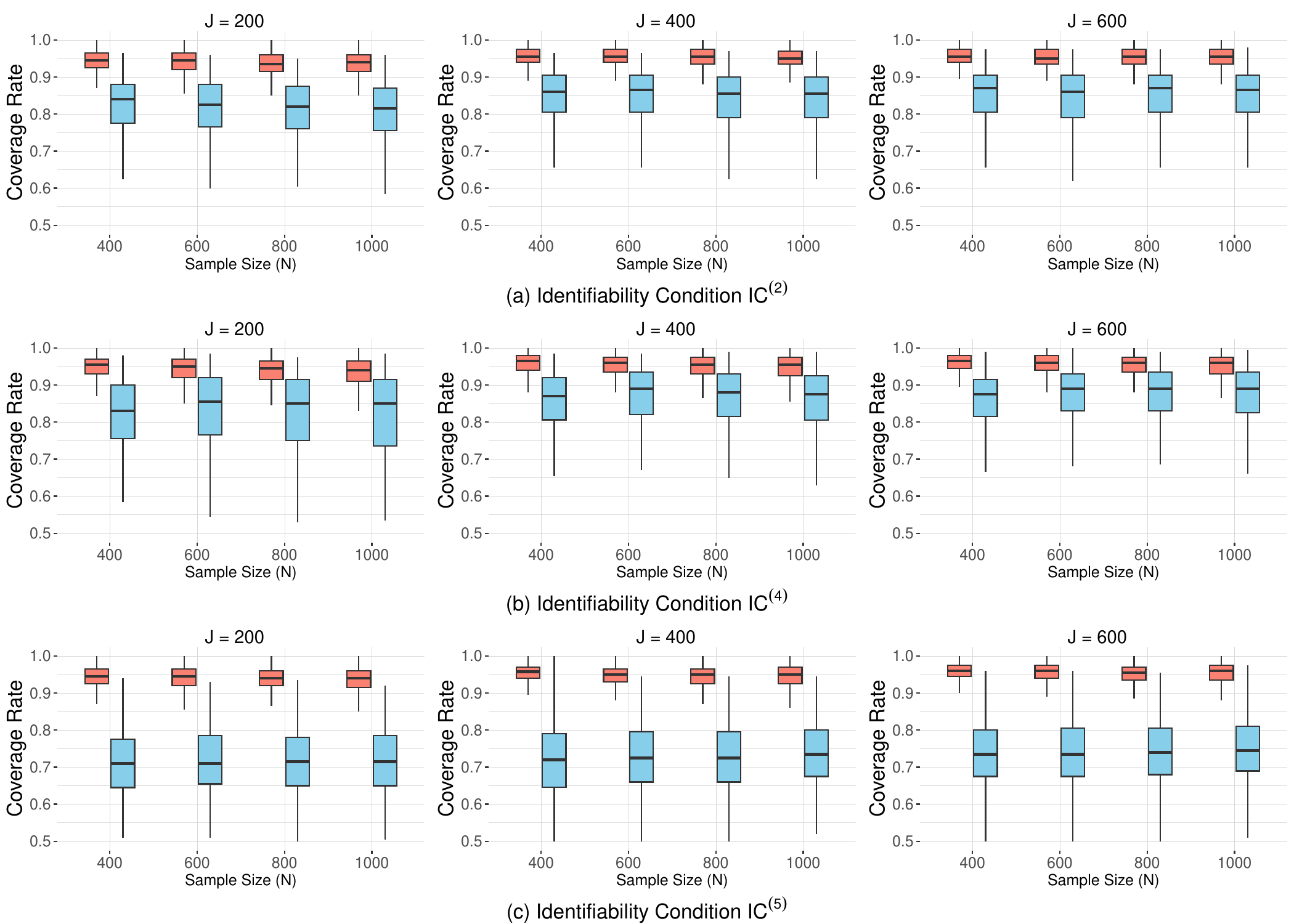}
    \caption{small Empirical coverage of loadings in the linear latent factor model and high-correlation case ($\tau = 0.7$) under different settings of $N$, $J$, and IC$^{(S)}$ for $S\in\{2,4,5\}$. The coverage rate of the proposed method is colored red; the coverage rate of the naive method is colored blue.}
    \label{fig:linear_coverage_cover_ic_2}
\end{figure}

\begin{figure}
        \centering\includegraphics[width=3in]{legend.pdf}\\\includegraphics[width=5in]{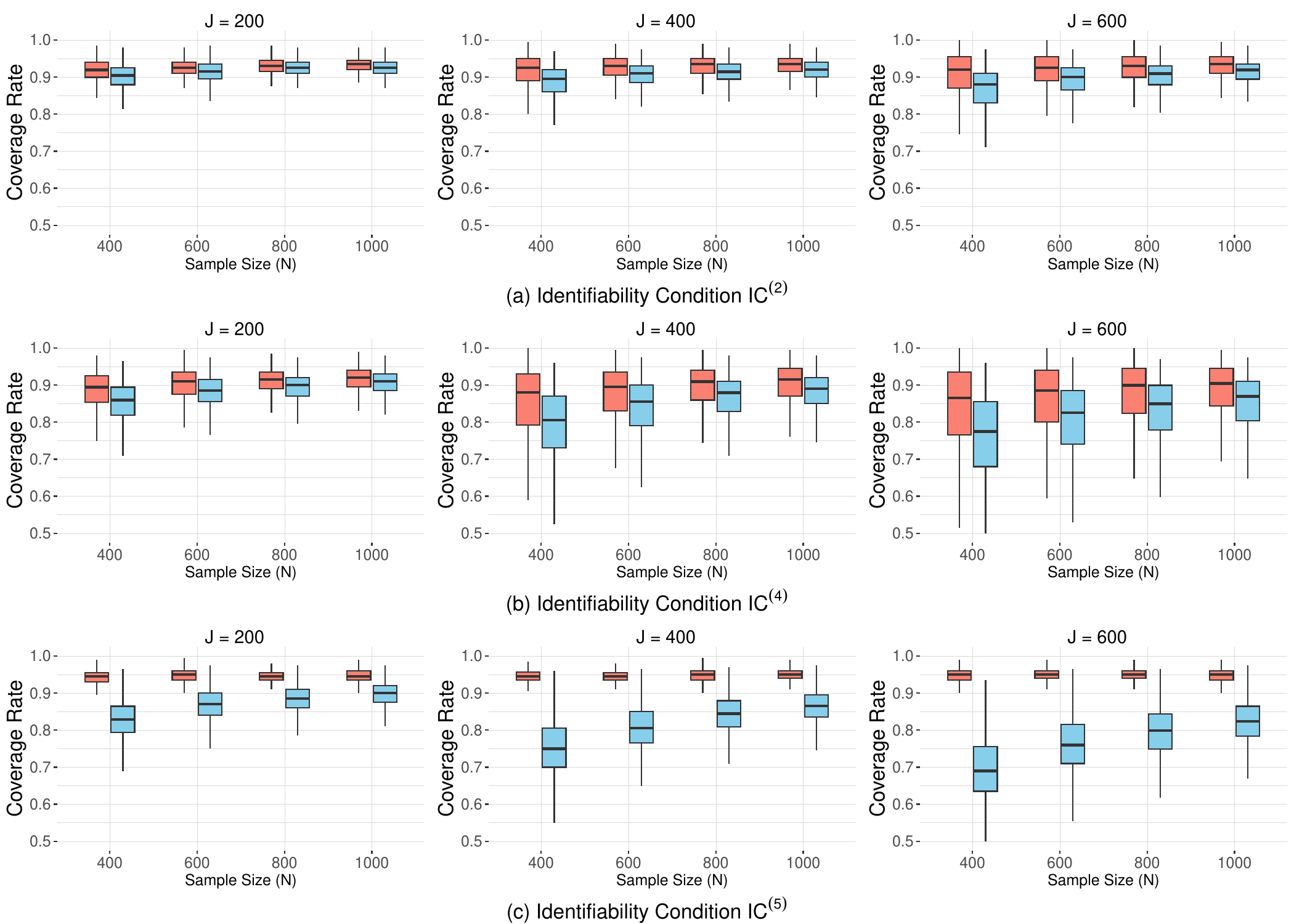}
    \caption{\small Empirical coverage of factors in the linear latent factor model and low-correlation case ($\tau = 0.3$) under different settings of $N$, $J$, and IC$^{(S)}$ for $S\in\{2,4,5\}$. The coverage rate of the proposed method is colored red; the coverage rate of the naive method is colored blue.}
    \label{fig:linear_fcover_ic_1}
\end{figure}
\begin{figure}
        \centering\includegraphics[width=3in]{legend.pdf}\\\includegraphics[width=5in]{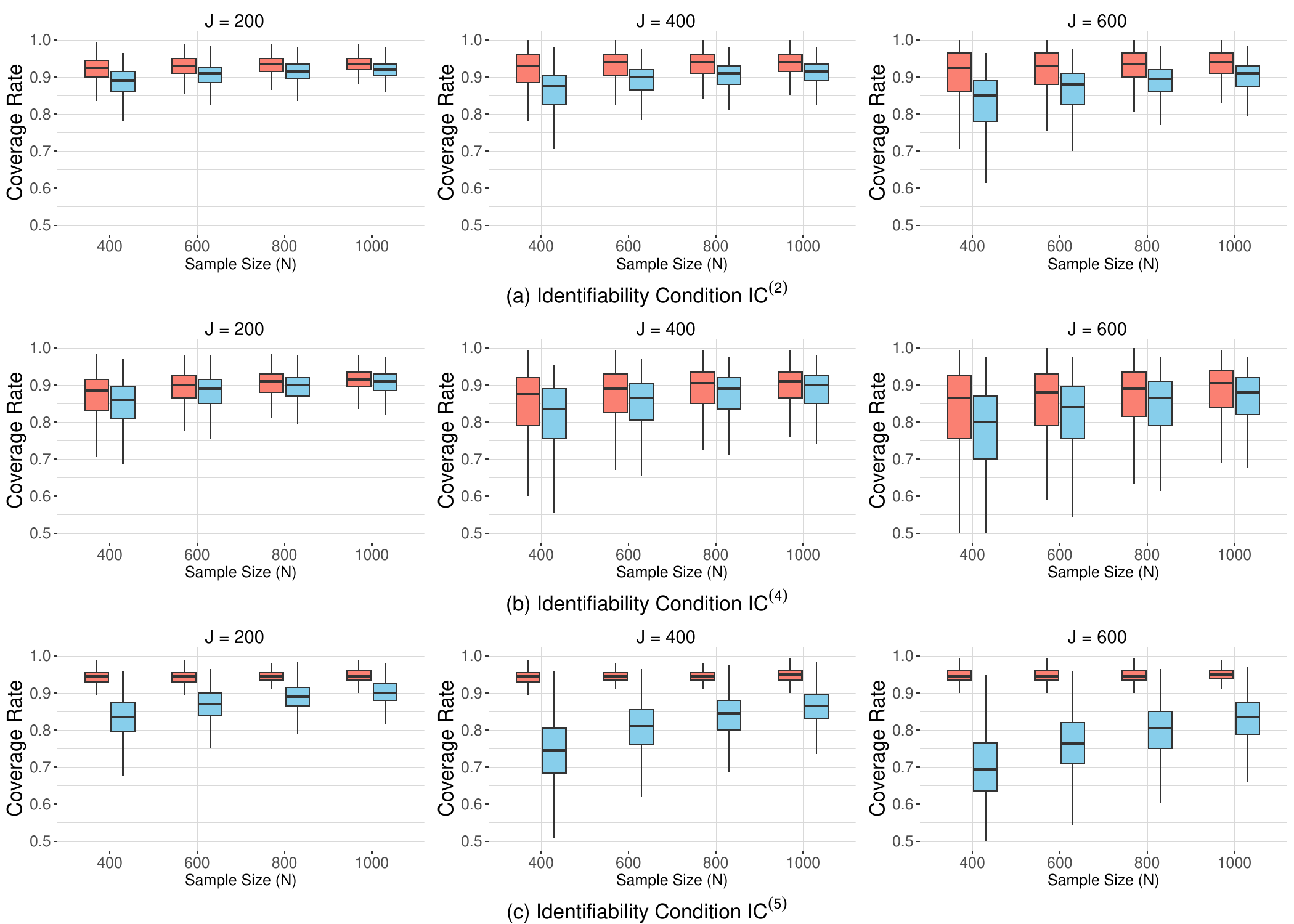}
    \caption{\small Empirical coverage of factors in the linear latent factor model and high-correlation case ($\tau = 0.7$) under different settings of $N$, $J,$ and IC$^{(S)}$ for $S\in\{2,4,5\}$. The coverage rate of the proposed method is colored red; the coverage rate of the naive method is colored blue.}
    \label{fig:linear_fcover_ic_2}
\end{figure}

\begin{figure}
        \centering\includegraphics[width=3in]{legend.pdf}\\\includegraphics[width=5in]{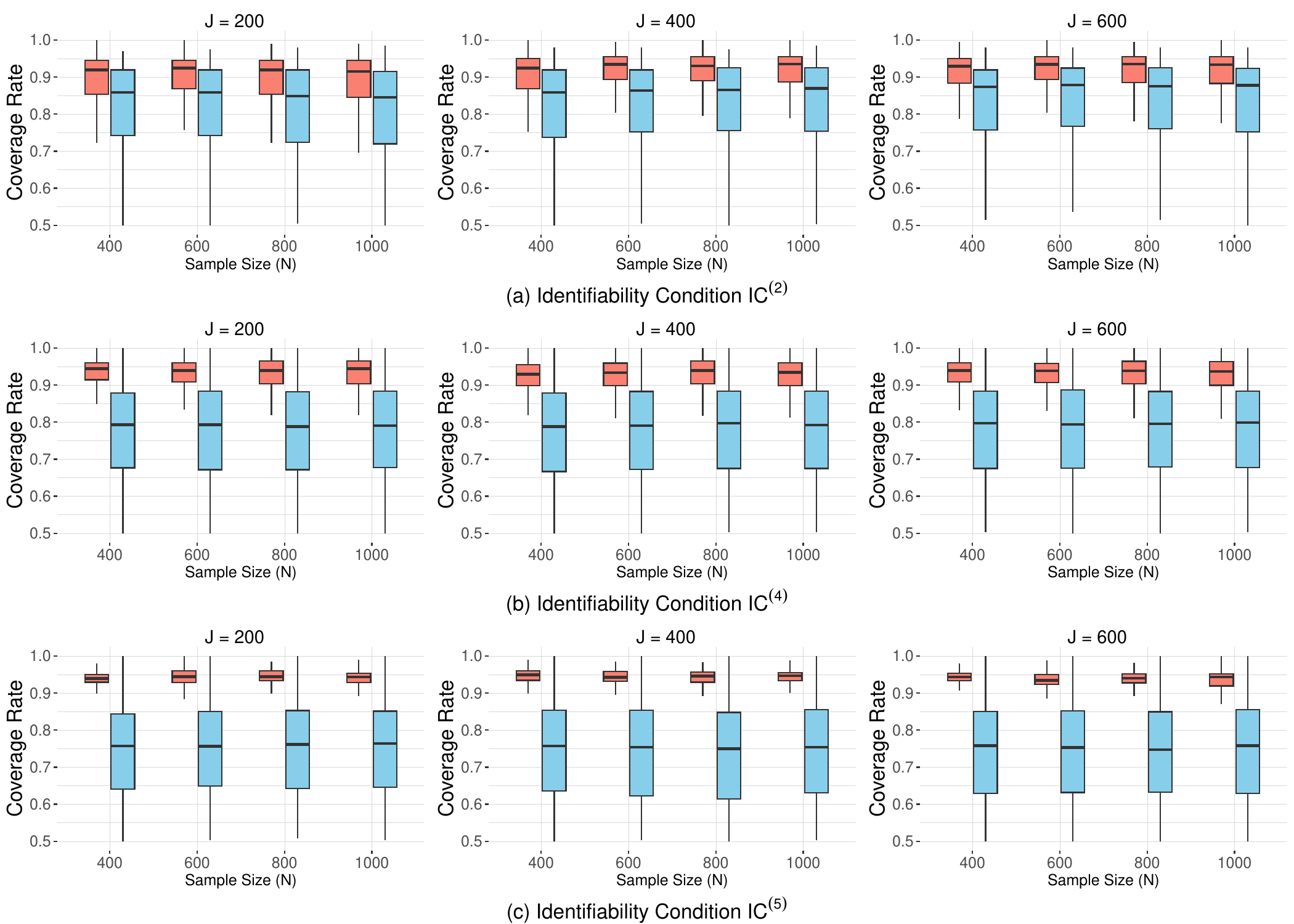}
    \caption{\small Empirical coverage of loadings in the Poisson latent factor model and low-correlation case ($\tau = 0.3$) under different settings of $N$, $J$, and IC$^{(S)}$ for $S\in\{2,4,5\}$. The coverage rate of the proposed method is colored red; the coverage rate of the naive method is colored blue.}
    \label{fig:poisson_cover_ic_1}
\end{figure}
\begin{figure}
        \centering\includegraphics[width=3in]{legend.pdf}\\\includegraphics[width=5in]{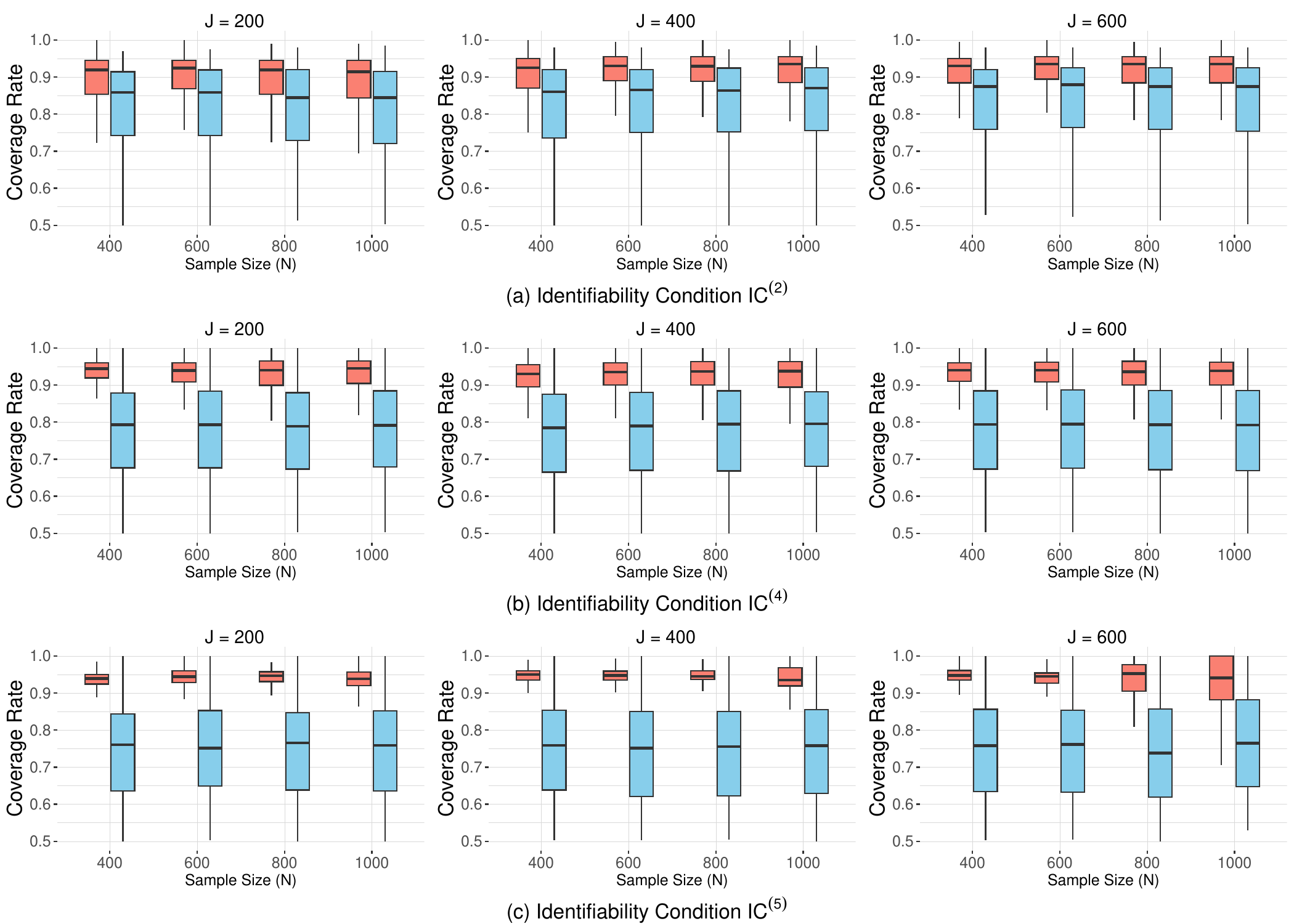}
    \caption{\small Empirical coverage of loadings in the Poisson latent factor model and high-correlation case ($\tau = 0.7$) under different settings of $N$, $J$, and IC$^{(S)}$ for $S\in\{2,4,5\}$. The coverage rate of the proposed method is colored red; the coverage rate of the naive method is colored blue.}
    \label{fig:poisson_cover_ic_2}
\end{figure}

\begin{figure}
        \centering\includegraphics[width=3in]{legend.pdf}\\\includegraphics[width=5in]{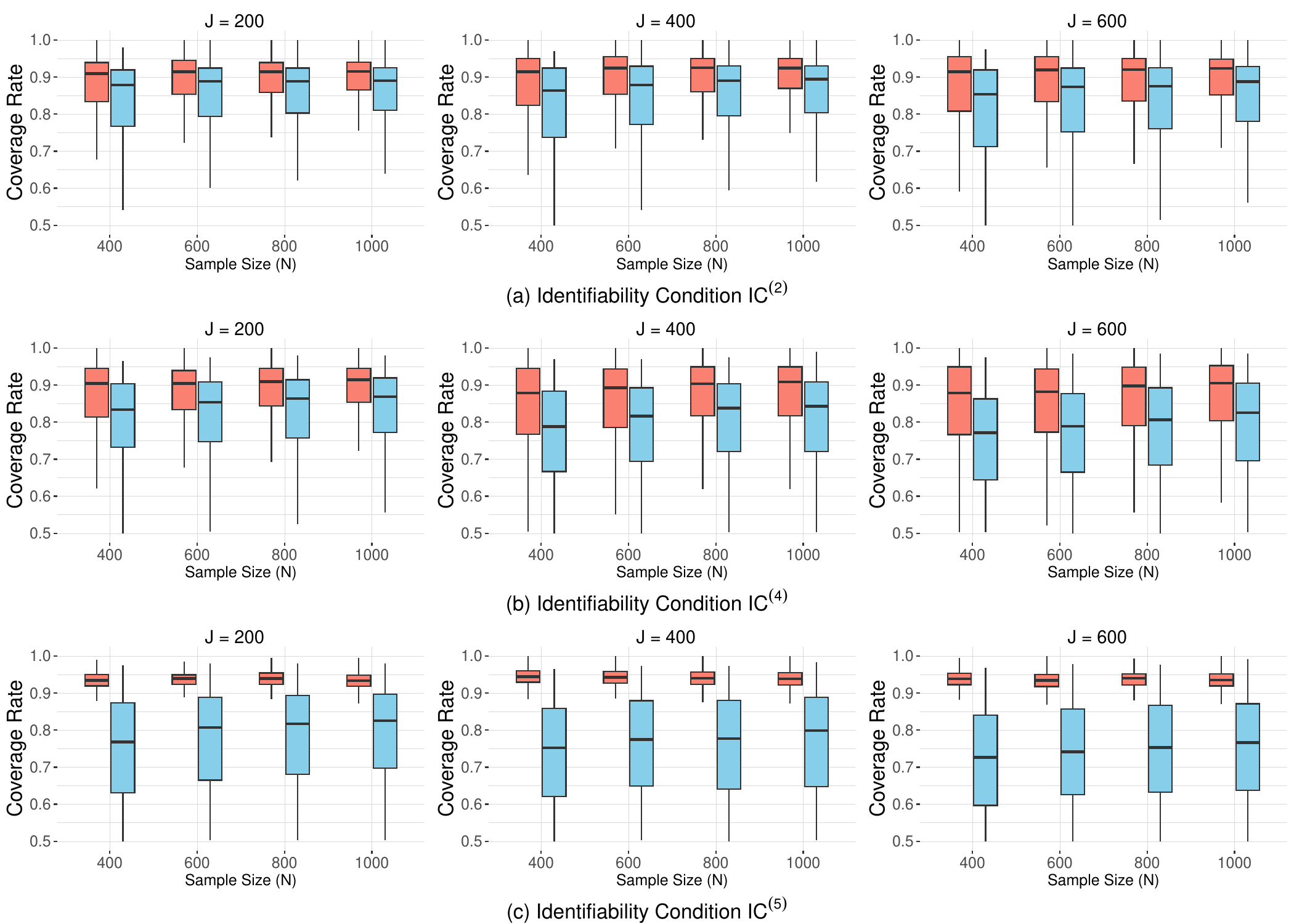}
    \caption{\small Empirical coverage of factors in the Poisson latent factor model and low-correlation case ($\tau = 0.3$) under different settings of $N$, $J$, and IC$^{(S)}$ for $S\in\{2,4,5\}$. The coverage rate of the proposed method is colored red; the coverage rate of the naive method is colored blue.}
    \label{fig:poisson_fcover_ic_1}
\end{figure}
}{\spacingset{1}
\begin{figure}
        \centering\includegraphics[width=3in]{legend.pdf}\\\includegraphics[width=5in]{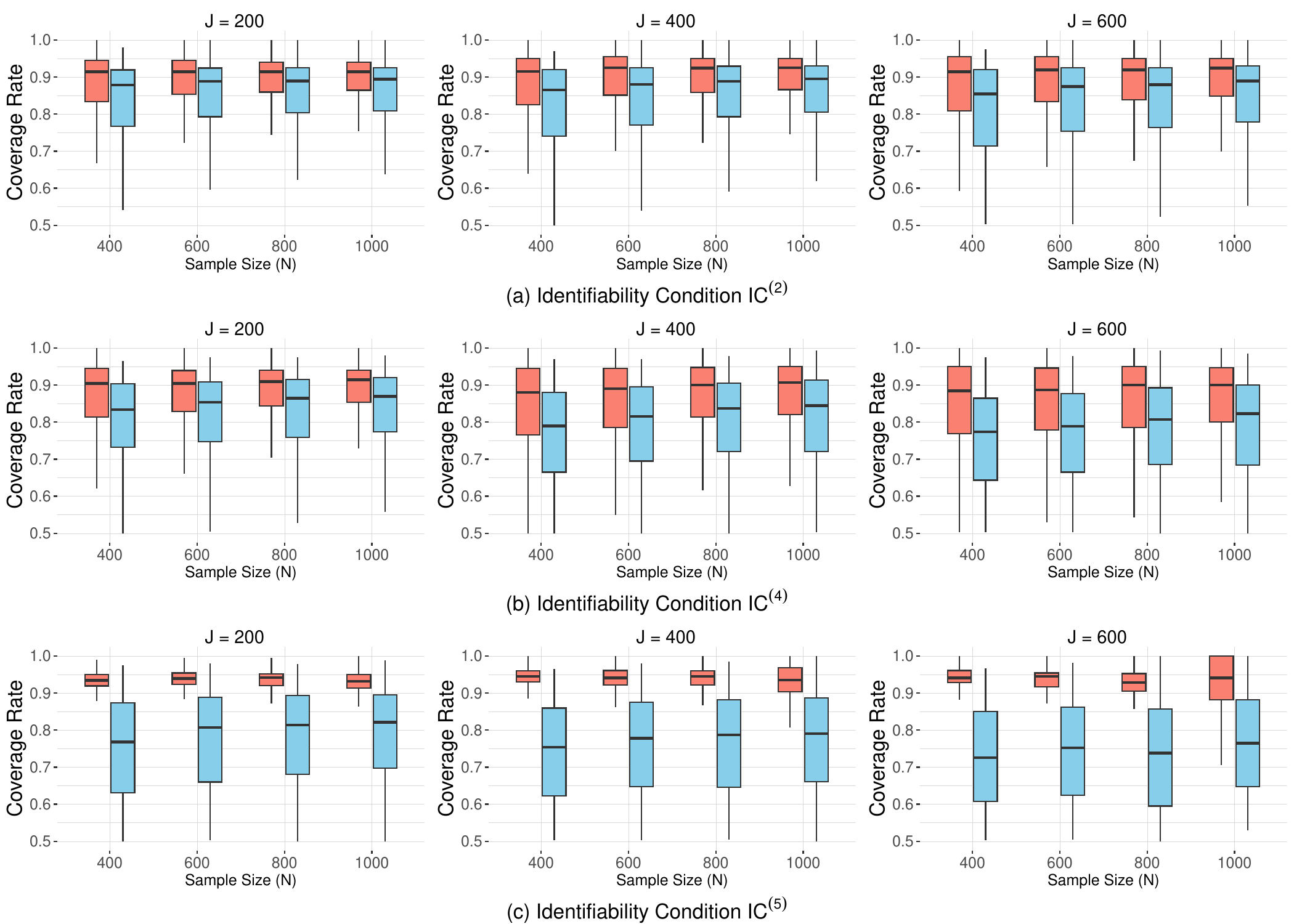}
    \caption{\small Empirical coverage of factors in the Poisson latent factor model and high-correlation case ($\tau = 0.7$) under different settings of $N$, $J,$ and IC$^{(S)}$ for $S\in\{2,4,5\}$. The coverage rate of the proposed method is colored red; the coverage rate of the naive method is colored blue.}
    \label{fig:poisson_fcover_ic_2}
\end{figure}
}

\subsection{Item Code for the Selected Items}\label{sec_supp_simu3}
The following table records the item code for the items that were selected in Section~\ref{sec:data} in the main text to satisfy the identifiability condition in the real data analysis. Each item is fixed to load only on its corresponding personality trait.
\begin{table}[H]
\centering
\begin{tabular}{|c|c|c|l|}
\hline
\textbf{Factor}& \textbf{Key (position)} & \textbf{Facet} & \textbf{Item} \\ \hline
N & N5 (111) & Immoderation & Go on binges \\ \hline
E & E1 (212) & Friendliness & Avoid contacts with others \\ \hline
O & O5 (293) & Intellect & Avoid difficult reading material \\ \hline
A & A5 (294) & Modesty & Make myself the center of attention \\ \hline
C & C5 (235) & Self-Discipline & Need a push to get started \\ \hline
\end{tabular}
\caption{Item codes and descriptions for the selected items}
\label{tab:trait_descriptions}
\end{table}

\subsection{Confidence Intervals for Loadings of the Remaining Dimensions}\label{sec_supp_simu2}
\begin{figure}[h]
\centering    
\includegraphics[width=5.4in]{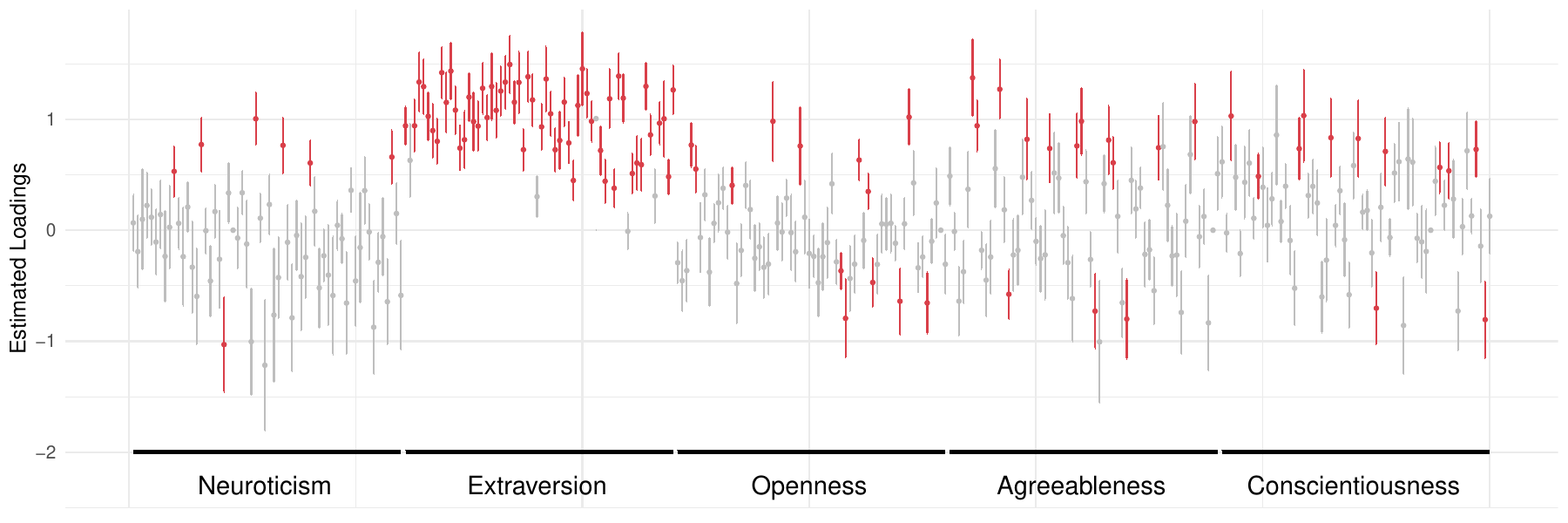}
    \caption{\small Confidence intervals for entries in the second column of loadings matrix corresponding to the ``Extraversion'' facet. Intervals in red represent significant loadings after adjustment with Bonferroni correction.}\label{fig:l2}
\end{figure}

\begin{figure}[h]
\centering    
\includegraphics[width=5.4in]{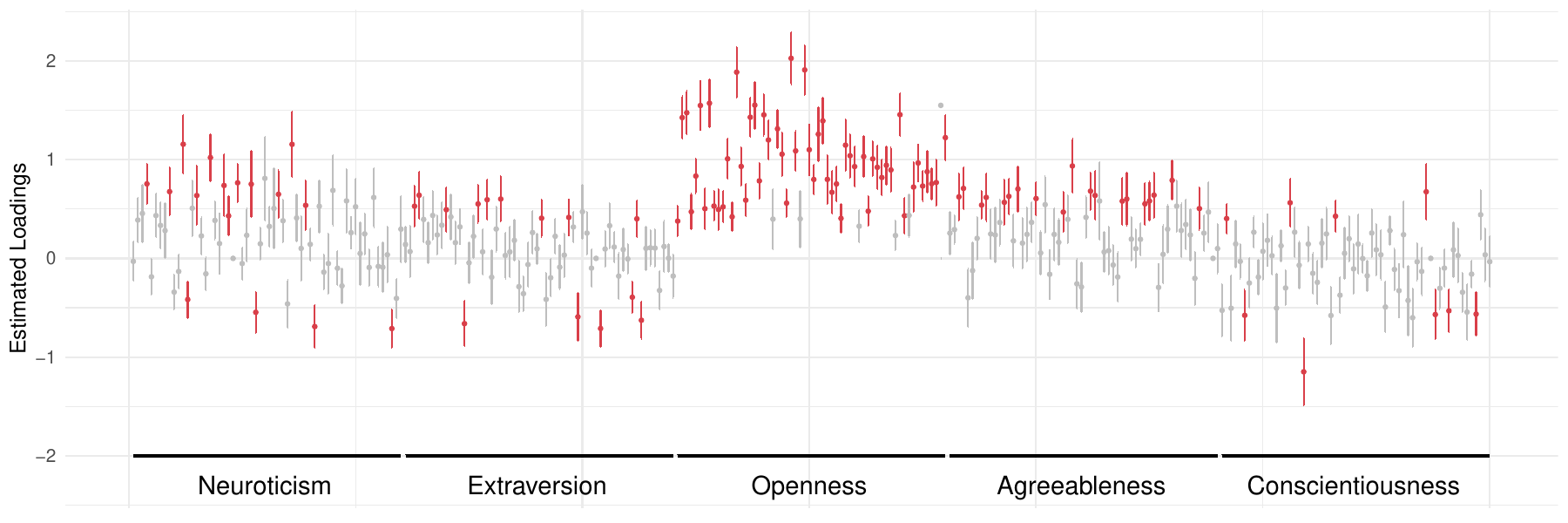}
    \caption{\small Confidence intervals for entries in the second column of loadings matrix corresponding to the ``Openness'' facet. Intervals in red represent significant loadings after adjustment with Bonferroni correction.}\label{fig:l3}
\end{figure}

\begin{figure}[h]
\centering    
\includegraphics[width=5.4in]{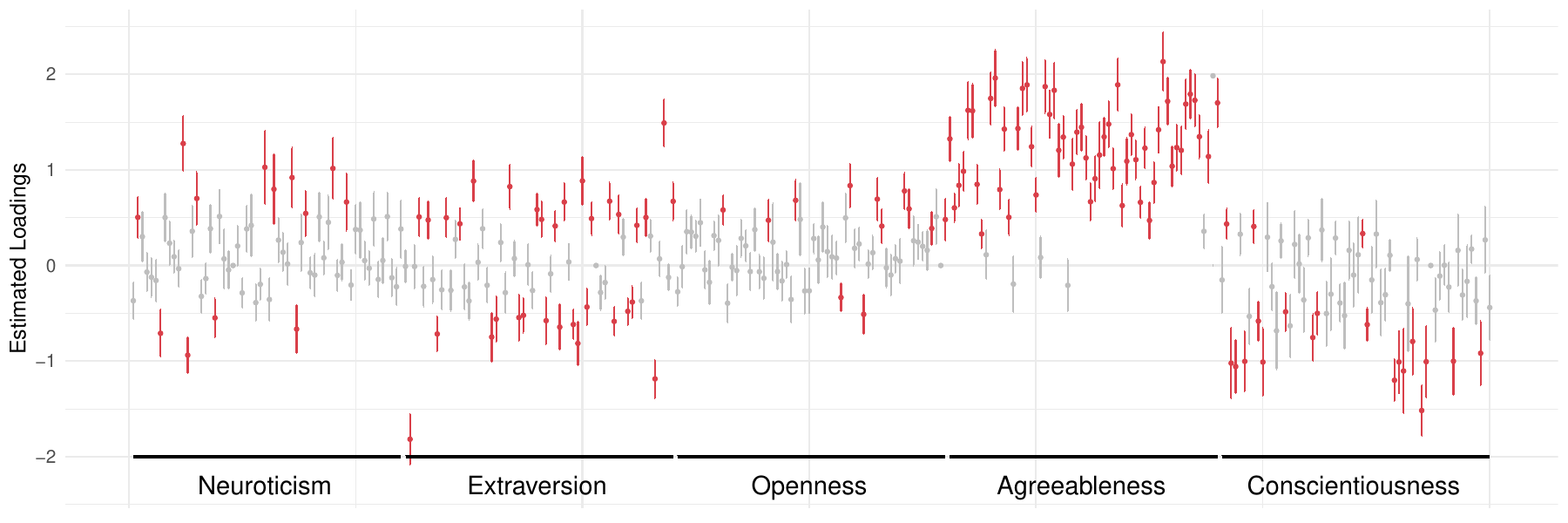}
    \caption{\small Confidence intervals for entries in the second column of loadings matrix corresponding to the ``Agreeableness'' facet. Intervals in red represent significant loadings after adjustment with Bonferroni correction.}\label{fig:l4}
\end{figure}

\begin{figure}[h]
\centering    
        \includegraphics[width=5.4in]{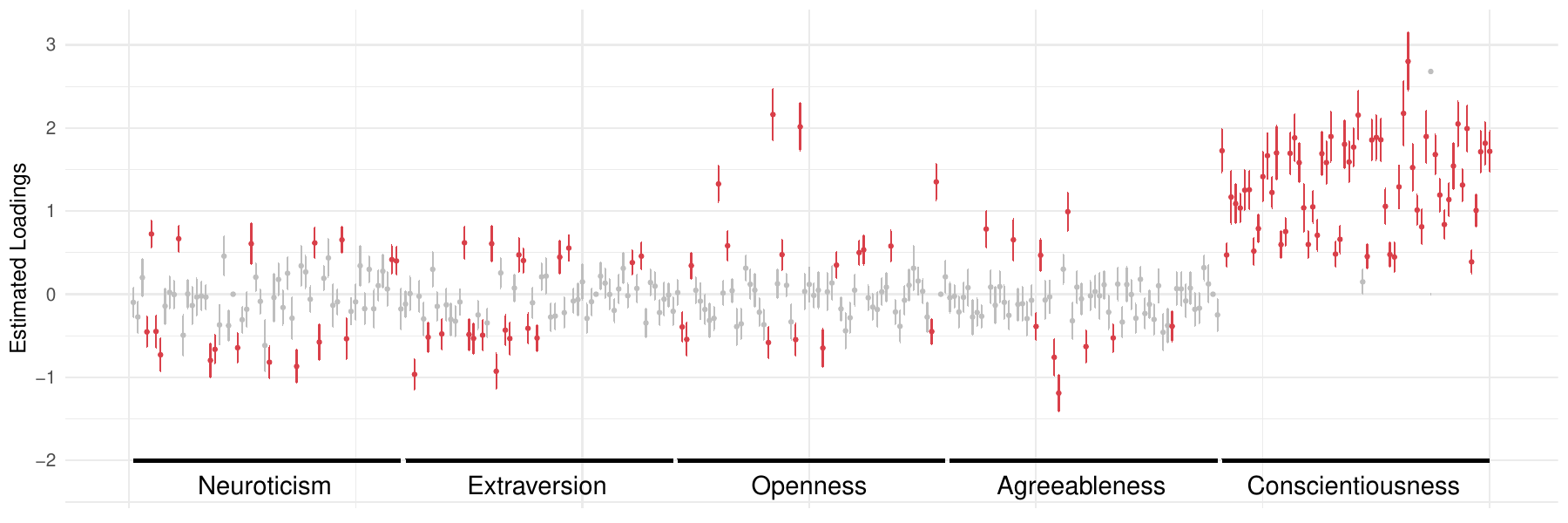}
\caption{\small Confidence intervals for entries in the second column of loadings matrix corresponding to the ``Conscientiousness'' facet. Intervals in red represent significant loadings after adjustment with Bonferroni correction.}    \label{fig:l5}
\end{figure}

In Section~\ref{sec:data} of the main text, we present the confidence intervals for the first column in the loading matrix corresponding to the ``Neuroticism'' facet. The confidence intervals for the rest of the columns are presented in Figures~\ref{fig:l2}--\ref{fig:l5}.


\newpage

\section{Additional Consistency Results}\label{supp_sec_sin_consis}
Estimation consistency can also be evaluated using the angle between the column vectors $\hat\Fb^{(S)}_{[,r]}$ and $\Fb^*_{[,r]}$ \big(or $\hat\bLambda^{(S)}_{[,r]}$ and $\bLambda^*_{[,r]}$\big), for $r\in[K]$, specifically measured by the sine of this angle. In particular, the sine of the angle between any two vectors $\bx,\by\in\RR^n$ is defined as $$\sin\langle \bx,\by\rangle=\left\{1-\frac{(\bx^\T\by)^2}{\|\bx\|^2\|\by\|^2}\right\}^{1/2},$$
and sin-consistency for $\hat\Fb_{[,r]}^{(S)}$ \big(or $\hat\bLambda_{[,r]}^{(S)}$\big) is established if $\sin\langle \hat\Fb_{[,r]}^{(S)},\Fb_{[,r]}^*\rangle\overset{p}{\to} 0$ \big(or $\sin\langle \hat\bLambda_{[,r]}^{(S)},\bLambda_{[,r]}^*\rangle\overset{p}{\to} 0$\big) when $N,J\to\infty$.
\cite{chen2020structured} recently proved the sin-consistency of factor estimation, 
and as discussed in Remark~\ref{remark_chen_ident}, their analysis requires a diverging number of loading entries to be fixed as zeros, with their proportions converging to a positive constant. Based on Theorem~\ref{thm_consis}, we can establish sin-consistency for both factors and loadings under weaker requirements and in a nonasymptotic setting, requiring only a fixed number of specified zero entries in the loading matrix. Our results are summarized as follows.
\begin{corollary}[Sin-consistency]\it
    \label{prop_consis_ic4} 
Suppose that the true parameters $(\bLambda^*,\Fb^*)$ satisfy the identifiability condition IC$^{(S)}$ for $S\in\{0,1,\dots,5\}$ and  Assumptions~\ref{assumption: psd covariance}--\ref{assump:scaling} hold. 

\vspace{0.1cm}
For $S=0$ and any constant $\delta>1$, when $N$ and $J$ are large enough, with probability at least $1-(N\wedge J)^{-\delta}$, we have  for any $r\in[K]$,
    \begin{equation*}
        \sin\big\langle \hat\bLambda_{[,r]}^{(0)},\bLambda_{[,r]}^*\big\rangle\le C_{\delta}\big( N^{-1/2} + J^{-1}\big),\quad \sin\big\langle \hat\Fb_{[,r]}^{(0)},\Fb_{[,r]}^*\big\rangle\le C_{\delta}(J^{-1/2} + N^{-1}),
    \end{equation*}
    where $C_\delta$ is a constant only depending on $\delta$. 
    
    \vspace{0.1cm}
    For $S\in\{1,\dots,5\}$ and any constant $\delta>1$, when $N$ and $J$ are large enough, with probability at least $1-\exp(-\epsilon_N)-(N\wedge J)^{-\delta}$, we have
    \begin{equation*}
        \sin\big\langle \hat\bLambda_{[,r]}^{(S)},\bLambda_{[,r]}^*\big\rangle\le C_{\delta}\big( \sqrt{\epsilon_N}N^{-1/2} + J^{-1}\big),\quad \sin\big\langle \hat\Fb_{[,r]}^{(S)},\Fb_{[,r]}^*\big\rangle\le C_{\delta}\big(J^{-1/2} + \sqrt{\epsilon_N}N^{-1/2}\big),
    \end{equation*}
for any $r\in[K]$, where $C_\delta$ is a constant only depending on $\delta$.
\end{corollary}

\section{Results for the Generalized Latent Factor Model with Intercepts}\label{sec:intecept}
In this section, we extend the theoretical results to generalized latent factor models with intercepts. Incorporating intercepts $f_{i0}$ index by $i$ and $\lambda_{j0}$ indexed by $j$, model \eqref{eq_factor_m} can be reformulated as 
\begin{equation}
    g_{ij}(Y_{ij} = y_{ij}\mid\eta_{ij})\text{, where }\eta_{ij}=\lambda_{j0} + f_{i0} + \blambda_j^\T\bbf_i.
    \label{eq_factor_m_intij}
\end{equation}
Note that \eqref{eq_factor_m} is a special case of \eqref{eq_factor_m_intij} in which all intercepts are set to zero. Another commonly used special case sets $f_{i0} = 0$, leading to the following form\begin{equation}
    g_{ij}(Y_{ij} = y_{ij}\mid\eta_{ij})\text{, where }\eta_{ij}=\lambda_{j0} + \blambda_j^\T\bbf_i. 
    \label{eq_factor_m_intj}
\end{equation}
In the following, we present the analysis for model \eqref{eq_factor_m_intij} and \eqref{eq_factor_m_intj} in Sections~\ref{sec_sup_double_intercept} and~\ref{sec_sup_single_intercept}, respectively. Specifically, we propose additional constraints to resolve the identifiability issues introduced by the intercepts. Then we use a similar maximum likelihood estimation approach to jointly estimate the model parameters as in Section~\ref{sec:jmle}, and derive the theoretical properties for the estimation. 
\subsection{Analysis for Model \eqref{eq_factor_m_intij}}\label{sec_sup_double_intercept}
We begin by discussing the model identifiability for \eqref{eq_factor_m_intij}. Denote $\bbeta_{\lambda0} = (\lambda_{10},\cdots,\lambda_{J0})^\T$ and $\bbeta_{f0} = (f_{10},\cdots,f_{N0})^\T$. In addition to the identifiability issues as in Section~\ref{subsec:identification}, due to the inclusion of the intercepts $\bbeta_{\lambda0}$ and $\bbeta_{f0}$, we need additional restrictions for model identifiability. Following~\citet{chen2019joint, chen2020structured}, we consider the following additional condition.
\begin{center}
    \textbf{ICC}: $N^{-1}\Fb^\T\one_N = \zero_{K}\;\text{and}\; J^{-1}\bLambda^\T\one_J=\zero_K$ with  either  $N^{-1}\bbeta_{f0}^\T\one_N = 0\text{ or }J^{-1}\bbeta_{\lambda0}^\T\one_J=0$,
\end{center}
where $\one_n$ denotes all-one vector of dimension $n$.
We label this condition by ICC to distinguish it from identifiability conditions IC$^{(S)}$  in Section~\ref{subsec:identification}. 
Because of symmetry, ICC requires either $N^{-1}\bbeta_{f0}^\T\one_N = 0$ or $J^{-1}\bbeta_{\lambda0}^\T\one_J = 0$. For convenience and without loss of generality, we assume $N^{-1}\bbeta_{f0}^\T\one_N = 0$ whenever we refer to ICC below.

\begin{remark}\it
Without ICC, the model parameters in \eqref{eq_factor_m_intij} are not identifiable. To illustrate this, suppose the factors are not constrained to be centered, i.e., $N^{-1}\Fb^\T\one_N$ is unrestricted. Then, for any $(\bLambda,\Fb,\bbeta_{\lambda0},\bbeta_{f0})$, the following holds
$    \bLambda\Fb^\T + \bbeta_{\lambda0}\one_N^\T + \one_J\bbeta_{f0}^\T=\bar\bLambda\bar\Fb^\T+\bar\bbeta_{\lambda0}\one_N^\T + \one_J\bar\bbeta_{f0}^\T,
$
    where $\bar\bLambda = \bLambda\Gb$, $\bar \Fb = (\Fb - \one_N\bmu^\T)\Gb^{-\T}$, $\bar\bbeta_{\lambda0} = \bbeta_{\lambda0} + \bLambda\bmu$ and $\bar\bbeta_{f0} = \bbeta_{f0}$ for any $\bmu\in\RR^{K}$ and invertible $\Gb\in\RR^{K\times K}$. This implies that $(\bLambda,\Fb,\bbeta_{\lambda0},\bbeta_{f0})$ and $(\bar\bLambda,\bar\Fb,\bar\bbeta_{\lambda0},\bar\bbeta_{f0})$ lead to identical distributions for $\{Y_{ij}\}_{N\times J}$, making the model parameters unidentifiable. Imposing additional identifiability conditions discussed in Section~\ref{subsec:identification} can resolve the rotational indeterminacy only but cannot fully eliminate the above indeterminacy issue. Similarly, the model parameters are not identifiable if the loadings are not constrained to be centered, or if both intercepts $\bbeta_{f0}$ and $\bbeta_{\lambda0}$ are not constrained to be centered.
\end{remark}


Together with any identifiability condition IC$^{(S)}$ in Section~\ref{subsec:identification}, ICC guarantees the model parameters in \eqref{eq_factor_m_intij} are identifiable.
Estimation of model \eqref{eq_factor_m_intij} follows a similar procedure as in Section~\ref{sec:jmle}. The estimators for model \eqref{eq_factor_m_intij} under identifiability condition IC$^{(S)}$ for $S\in\{0,1,\dots,5\}$ and ICC are obtained by 
\begin{align}
    (\hat\bLambda^{(S)},\hat\Fb^{(S)},\hat\bbeta_{\lambda0}^{(S)},\hat\bbeta_{f0}^{(S)}) = &\,\mathop{\arg\max}_{{\|\bLambda\|_{\max},\|\Fb\|_{\max},\|\bbeta_{\lambda0}\|_{\max},\|\bbeta_{f0}\|_{\max}\le D}}\sum_{i=1}^N\sum_{j=1}^J l_{ij}(\blambda_j^\T\bbf_i+\lambda_{j0}+f_{i0}\mid Y_{ij}),\label{eq_joint_mle_intij}\\&\text{ s.t. IC}^{(S)}\text{ and ICC}.\nonumber
\end{align}

Next, we present the theoretical properties of the above maximum likelihood estimators. We denote the true parameters for \eqref{eq_factor_m_intij} as $(\bLambda^*,\Fb^*,\bbeta_{\lambda0}^*,\bbeta_{f0}^*)$. To establish these properties, we make slight modifications to Assumptions~\ref{assumption: psd covariance} and~\ref{assumption:smoothness} to account for the presence of intercepts. 
\newenvironment{assumptionPrime}[1]{
    \renewcommand{\theassumption}{#1$^{\prime}$} 
    \begin{assumption}
}{
    \end{assumption}
    \renewcommand{\theassumption}{\arabic{assumption}} 
}
The modifications are denoted by Assumption~\ref{assump:psd covariance int} and Assumption~\ref{assumption:smoothness int} given below.
\begin{assumptionPrime}{1}\label{assump:psd covariance int}
There exists $D >0$ such that, $\big\|\bLambda^*\big\|_{\max}\le D/2$, $\big\|\Fb^*\big\|_{\max}\le D/2$, $\big\|\bbeta_{\lambda0}^*\big\|_{\infty}\le D/2$ and $\big\|\bbeta_{f0}^*\big\|_{\infty}\le D/2$. 
 $\bSigma_{\lambda}^* = \lim_{J\rightarrow \infty} \Mb_{\lambda\lambda}=\lim_{J\rightarrow \infty} J^{-1}\sum_{j=1}^J\blambda_j^{c*}(\blambda_j^{c*})^{\intercal} $  and  $\bSigma_f^* = \lim_{N\rightarrow \infty} \Mb_{ff}=\lim_{N \rightarrow \infty} $ $N^{-1}\sum_{i=1}^N\bbf_i^{c*}(\bbf_i^{c*})^{\intercal} $ exist and are positive definite. The eigenvalues of $\bSigma_{f}^*\bSigma_{\lambda}^*$ are distinct and nonzero.
	\end{assumptionPrime}
 	\begin{assumptionPrime}{2}
	\label{assumption:smoothness int}
 The individual log-likelihood function $l_{ij}(\eta)$ is three times differentiable to $\eta$, with the first, second, and third order derivatives denoted by $l_{ij}^{\prime}(\eta)$, $l_{ij}^{\prime\prime}(\eta)$, and $l_{ij}^{\prime\prime\prime}(\eta)$. For some constant $M>0$,  $l_{ij}^{\prime} (\eta_{ij}^*)$ is sub-exponential with the sub-exponential norm $\|l_{ij}^{\prime} (\eta_{ij}^*)\|_{\varphi_1} \le M$. 
 For any $\eta\le (K + 2)D^2$, there exists $0<b_L<b_U$ such that $b_L\le -l_{ij}^{\prime\prime}(\eta)\le b_U$ and $|l_{ij}^{\prime\prime\prime}(\eta)| \leqslant b_U$.
 
   \end{assumptionPrime}
   Here $b_L$ and $b_U$ may be chosen differently from those in Assumption~\ref{assumption:smoothness}; however, for simplicity, we use the same notation for $b_L$ and $b_U$.
   The following results show the non-asymptotic error bounds for the estimators in \eqref{eq_joint_mle_intij}.
\begin{suppTheorem}\it
    For model \eqref{eq_factor_m_intij}, suppose the true parameters $(\bLambda^*,\Fb^*,\bbeta^*_{\lambda0},\bbeta_{f0}^*)$ satisfy identifiability condition ICC and IC$^{(S)}$ for $S\in\{0,1,\dots,5\}$. Under Assumptions~\ref{assump:psd covariance int}, \ref{assumption:smoothness int}, and \ref{assump:scaling}, the results in Theorems~\ref{thm_consis} and \ref{thm_uniform_consis} hold for $\hat\bLambda^{(S)}$ and $\hat\Fb^{(S)}$ obtained in \eqref{eq_joint_mle_intij}. For $\hat\bbeta_{\lambda0}^{(S)}$ and $\hat\bbeta_{f0}^{(S)}$ in \eqref{eq_joint_mle_intij}, for any $\delta>1$, when $N,J$ are large enough, with probability at least $1-(N\wedge J)^{-\delta}$, \begin{equation*}
        \frac{1}{J}\sum_{j=1}^J\big(\hat\lambda_{j0}^{(S)} - \lambda_{j0}^*\big)^2\le C_{\delta}\big( N^{-1}+J^{-2}\big),\quad \frac{1}{N}\sum_{i=1}^N\big(\hat f_{i0}^{(S)} - f_{i0}^*\big)^2\le C_{\delta}\big(N^{-2} + J^{-1}\big),
    \end{equation*}
    and with probability at least $1 - (N\wedge J)^{-\delta} - \exp\{-(\epsilon_N\wedge \epsilon_J)\}$,
    \begin{equation*}
        \max_{1\le j\le J}\big|\hat\lambda_{j0}^{(S)} - \lambda_{j0}^*\big| \le C_{\delta}\frac{\epsilon_{NJ}\log N}{\sqrt{N\wedge J}},\quad \max_{1\le i\le N}\big|\hat f_{i0}^{(S)} - f_{i0}^*\big| \le C_{\delta}\frac{\epsilon_{NJ}\log J}{\sqrt{N\wedge J}},
    \end{equation*}
    where $C_{\delta}$ is some constant only depending on $\delta$, 
 and $\epsilon_N$, $\epsilon_J$ and $\epsilon_{NJ}$ are defined as in Theorem~\ref{thm_uniform_consis}.
\label{thm_consis_int}
\end{suppTheorem}

Next, we present the results for limiting distributions for the estimators in \eqref{eq_joint_mle_intij}.
\begin{suppTheorem}\label{main_thm_asymp}
   \it
  For model \eqref{eq_factor_m_intij}, suppose the true parameters $(\bLambda^*,\Fb^*,\bbeta^*_{\lambda0},\bbeta_{f0}^*)$ satisfy identifiability condition IC$^{(S)}$ for $S\in\{0,1,\dots,5\}$ and ICC, and that Assumptions~\ref{assump:psd covariance int}, \ref{assumption:smoothness int}, and \ref{assump:scaling} hold . 
  For $\hat\lambda_{j0}^{(S)}$ and $\hat\blambda_j^{(S)}$ in \eqref{eq_joint_mle_intij}, if $\sqrt{N}/J\to 0 $ as $N,J\to\infty$, we have
    \begin{equation}
      \big(\bSigma_{[N],-j}^{(S)}\big)^{-1/2}\begingroup
    \renewcommand{\arraystretch}{0.7}\begin{pmatrix}
         \hat\lambda_{j0}^{(S)}-\lambda_{j0}^*\\\Db_{(j)}\big( \hat\blambda_j^{(S)} - \blambda_j^*\big)
      \end{pmatrix}\endgroup \overset{d}{\to}\cN\left(\zero_{|\cS_j|+1},\Ib_{|\cS_j|+1}\right),\text{ for any }j\in[J]\label{eq_asym_int_lambda}
  \end{equation}
   as $N,J\to\infty$ with $\cS_j$ as specified in Theorem~\ref{main_thm_loading_asym}. For $\hat f_{i0}^{(S)}$ and $\hat \bbf_i^{(S)}$ in \eqref{eq_joint_mle_intij}, if $\sqrt{J}/N \to 0$ as $N,J\to\infty$ when $S=0$, we have
   \begin{equation}
      \big(\bSigma_{-i,[J]}^{(S)}\big)^{-1/2}\begingroup
    \renewcommand{\arraystretch}{0.7}\begin{pmatrix}\hat f_{i0}^{(S)}-f_{i0}^*\\\hat\bbf_i^{(S)} - \bbf_i^*\\\end{pmatrix}\endgroup \overset{d}{\to}\cN\left(\zero_{K+1},\Ib_{K+1}\right),\text{ for any }i\in[N],\label{eq_asym_int_f}
  \end{equation}as $N,J\to\infty$.
  Expressions for $\bSigma_{[N],-j}^{(S)}$ and $\bSigma_{-i,[J]}^{(S)}$ are given in \eqref{eq_expression_int_gfm_cov_lambda} and \eqref{eq_expression_int_gfm_cov_f} in the Supplementary Material. The covariance matrices $\bSigma_{[N],-j}^{(S)}$ and $\bSigma_{-i,[J]}^{(S)}$ can be consistently estimated by the plug-in estimators $\hat\bSigma_{[N],-j}^{(S)}$ and $\hat\bSigma_{-i,[J]}^{(S)}$ that replace $(\bLambda^*,\Fb^*,\bbeta_{\lambda0}^{*},\bbeta_{f0}^{*})$ in $\bSigma_{[J]}^{(S)}$ with $(\hat\bLambda^{(S)},\hat\Fb^{(S)},\hat\bbeta_{\lambda0}^{(S)},\hat\bbeta_{f0}^{(S)})$, and the above normality results of \eqref{eq_asym_int_lambda} and \eqref{eq_asym_int_f} still hold with $\bSigma_{[N],-j}^{(S)}$ and $\bSigma_{-i,[J]}^{(S)}$ replaced with $\hat\bSigma_{[N],-j}^{(S)}$ and $\hat\bSigma_{-i,[J]}^{(S)}$, respectively.

 \end{suppTheorem}
Section~\ref{sec_sketch_proof} outlines the proofs for these two theorems. The non-asymptotic error bounds presented in Theorems~\ref{thm_consis} and \ref{thm_uniform_consis} remain valid for $\big(\hat\bLambda^{(S)},\hat\Fb^{(S)}\big)$ in \eqref{eq_joint_mle_intij}. 
The limiting distributions of $\hat\blambda_j^{(S)}$s and $\hat\bbf_i^{(S)}$s in \eqref{eq_joint_mle_intij} differ from those in Theorems~\ref{main_thm_loading_asym} and \ref{main_thm_factor_asym} due to the more complex model setup involving intercepts.

\subsection{Analysis for Model \eqref{eq_factor_m_intj}}\label{sec_sup_single_intercept}
In this part, we present the theoretical results for the generalized latent factor model with one intercept indexed by $j$:
\begin{equation*}
    g_{ij}(Y_{ij} = y_{ij}\mid\eta_{ij})\text{, where }\eta_{ij}=\lambda_{j0} + \blambda_j^\T\bbf_i.
\end{equation*}
Similar to the discussion of model \eqref{eq_factor_m_intij} in Section~\ref{sec_sup_double_intercept}, we impose the following condition to resolve the indeterminacy introduced by $\bbeta_{\lambda0} = (\lambda_{10},\cdots,\lambda_{J0})^\T$.
\begin{center}
    \textbf{ICC}$^{\prime}$: $N^{-1}\Fb^\T\one_N = \zero_K$.
\end{center}
The rotational indeterminacy is resolved using IC$^{(S)}$ for some $S\in\{0,1,\cdots,5\}$.
When the true parameters $(\bLambda,\Fb,\bbeta_{\lambda0})$ satisfy IC$^{(S)}$ and ICC$^{\prime}$, the estimation can similarly be obtained through a constrained maximum likelihood estimation, given as
\begin{align}
    (\hat\bLambda^{(S)},\hat\Fb^{(S)},\hat\bbeta_{\lambda0}^{(S)}) = &\,\mathop{\arg\max}_{\substack{\|\bLambda\|_{\max}\le D,\|\Fb\|_{\max}\le D\\\|\bbeta_{\lambda0}\|_{\max}\le D}}\sum_{i=1}^N\sum_{j=1}^J l_{ij}(\blambda_j^\T\bbf_i+\lambda_{j0}\mid Y_{ij}),\label{eq_joint_mle_intj}\\&\text{ s.t. IC}^{(S)}\text{ and ICC}^{\prime}.\nonumber
\end{align}
For this estimation, we have the following results.
\begin{suppTheorem}\it
    For model \eqref{eq_factor_m_intj}, suppose Assumptions~\ref{assump:psd covariance int}, \ref{assumption:smoothness int}, and \ref{assump:scaling} hold and the true parameters $(\bLambda^*,\Fb^*,\bbeta^*_{\lambda0})$ satisfy idenfitication condition IC$^{(S)}$ and ICC$\, ^{\prime}$ for $S\in\{0,1,\cdots,5\}$. Under , the results in Theorem~\ref{thm_consis} and \ref{thm_uniform_consis} hold for $\hat\bLambda^{(S)}$ and $\hat\Fb^{(S)}$ obtained in \eqref{eq_joint_mle_intj}. For $\hat\bbeta_{\lambda0}^{(S)}$ in \eqref{eq_joint_mle_intj}, for any $\delta>1$, when $N,J$ are large enough, with probability at least $-1(N\wedge J)^{-\delta}$, we have\begin{equation*}
        \frac{1}{J}\sum_{j=1}^J\big(\hat\lambda_{j0}^{(S)} - \lambda_{j0}^*\big)^2\le C_{\delta}\big( N^{-1}+J^{-2}\big),
    \end{equation*}
    where $C_{\delta}$ is some constant only depending on $\delta$ and with probability at least $1 - (N\wedge J)^{-\delta} - \exp(-\epsilon_N\wedge \epsilon_J)$ 
    \begin{equation*}
        \max_{1\le j\le J}\big|\hat\lambda_{j0}^{(S)} - \lambda_{j0}^*\big| \le C_{\delta}\frac{\epsilon_{NJ}\log N}{\sqrt{N\wedge J}}.
    \end{equation*}
    where $\epsilon_N$, $\epsilon_J$ and $\epsilon_{NJ}$ are defined in Theorem~\ref{thm_uniform_consis}.
\label{thm_consis_intj}
\end{suppTheorem}
For this asymptotic distribution, we have the following results.
\begin{suppTheorem}\label{main_thm_asymp_intj}\it
   
  For model \eqref{eq_factor_m_intj}, suppose Assumptions~\ref{assump:psd covariance int}, \ref{assumption:smoothness int}, and \ref{assump:scaling} hold and the true parameters $(\bLambda^*,\Fb^*,\bbeta^*_{\lambda0})$ satisfy identifiability condition IC$^{(S)}$ for $S\in\{0,1,\cdots,5\}$ and ICC$\, ^{\prime}$. For estimation \eqref{eq_joint_mle_intij}, if $\sqrt{N}/J\to 0$ as $N,J\to\infty$, we have
    \begin{equation*}
      \big(\bSigma_{[N],-j}^{(S)}\big)^{-1/2}\begin{pmatrix}\hat\lambda_{j0}-\lambda_{j0}\\\hat\blambda_j^{(S)} - \blambda_j^*\\\end{pmatrix} \overset{d}{\to}\cN\left(\zero,\Ib_{|\cS_j|+1}\right),\text{ for any }j\in[J]
  \end{equation*}
   as $N,J\to\infty$ where $\cS_j$ is as specified in Theorem~\ref{main_thm_loading_asym}. If $\sqrt{J}/N \to 0$ when $S=0$, we have
   \begin{equation*}
      \big(\bSigma_{i,[J]}^{(S)}\big)^{-1/2}\left(\hat\bbf_i^{(S)} - \bbf_i^*\right)\overset{d}{\to}\cN\left(\zero,\Ib_{K+1}\right),\text{ for any }i\in[N],
  \end{equation*}as $N,J\to\infty$. 
  Here $\bSigma_{[N],-j}^{(S)}$ is introduced in Theorem~\ref{main_thm_asymp}. The covariance matrices can be consistently estimated by replacing $(\bLambda^*,\Fb^*,\bbeta_{\lambda0}^{*})$ in $\bSigma_{[N],-j}^{(S)}$ and $\bSigma_{-i,[J]}^{(S)}$ with $(\hat\bLambda^{(S)},\hat\Fb^{(S)},\hat\bbeta_{\lambda0}^{(S)})$.

 \end{suppTheorem}
 The proofs of these two theorems can be obtained similar to the proofs in Section~\ref{se:integfm proof}.
Note that the asymptotic distributions for $\Big\{ \big(\bSigma_{i,[J]}^{(S)}\big)^{-1/2}\left(\hat\bbf_i^{(S)} - \bbf_i^*\right)\Big\}_{i=1}^N$ are identical to those in Theorem~\ref{main_thm_factor_asym}. Following the proof in Section~\ref{prove_thm_mff_limiting_dist}, we have the following limiting distributions for $\hat\Mb_{ff}$.
\begin{suppTheorem} \label{thm_mff_limiting_distri_intj}\it
Suppose the true parameters $(\bLambda^*,\Fb^*,\bbeta_{\lambda0})$ satisfy identifiability condition IC$^{(S)}$ for $S\in \{0,1,\cdots,5\}$ and ICC$\, ^{\prime}$, and that Assumptions~\ref{assump:psd covariance int}, \ref{assumption:smoothness int}, and \ref{assump:scaling} hold. If $\sqrt{N}/J\to0$ as $N,J\to\infty$, we have
\begin{itemize}
\item when $S\in\{0,1,2\}$, $\hat\Mb_{ff}^{(S)} =\Mb_{ff}^{(S)} = \Ib_K$ by the identifiability constraints.
    \item when $S = 3$,
    $\big(\bSigma_{[J]}^{(3)}\big)^{-1/2} \mathrm{vec}\big(\mathrm{diag}\big(\hat\Mb_{ff}^{(3)} - \Mb_{ff}^*\big) \big)\overset{d}{\to} \cN(\zero_K,\Ib_K)$;
\item when $S = 4$, $\big(\bSigma_{[J]}^{(4)}\big)^{-1/2}\mathrm{veck}\big(\hat\Mb_{ff}^{(4)} - \Mb_{ff}^*\big)\overset{d}{\to} \cN(\zero_{K(K-1)/2},\Ib_{K(K-1)/2})$,
where $\mathrm{veck}(\Ab)$ for $\Ab\in\RR^{K\times K}$ is the $K(K-1)/2$ column vector obtained by vectorizing only the elements above the diagonal of $\Ab$;
\item when $S = 5$,  $\big(\bSigma_{[J]}^{(5)}\big)^{-1/2}\mathrm{vech}\big(\hat\Mb_{ff}^{(5)} - \Mb_{ff}^*\big)\overset{d}{\to} \cN(\zero_{(K+1)K/2},\Ib_{(K+1)K/2})$,
    where $\mathrm{vech}(\Ab)$ for $\Ab\in\RR^{K\times K}$ is the $K(K+1)/2$ column vector obtained by vectorizing only the upper-triangular elements $\Ab$.
\end{itemize}
Here $\bSigma_{[J]}^{(S)}$s for $S\in\{3,4,5\}$ is the same as in Theorem~\ref{thm_mff_limiting_distri}. 
\end{suppTheorem}

\section{Results for Generalized Latent Factor Model with Random Factors}\label{supp_sec_random_factor}
In this section, we extend the analysis to generalized latent
factor models where latent factors are treated as random effects. Specifically, we assume $\{\bbf_i\}_{i=1}^N$ are independent random vectors with zero mean and covariance matrix $\bSigma$. 
Under this specification, identifiability conditions IC$^{(0)}$--IC$^{(4)}$, which involve restrictions on $\Mb_{ff} = N^{-1}\sum_{i=1}^N\bbf_i\bbf_i^\T$, cannot be directly imposed as $\Mb_{ff}$ is also random. Identifiability condition IC$^{(5)}$ still applies as it involves the loading parameters only.  We therefore impose the corresponding restriction on the population covariance $\bSigma$ instead. We denote the adapted conditions by $\overline{\mathrm{IC}}^{(0)}$--$\overline{\mathrm{IC}}^{(4)}$, presented in Table~\ref{tab:identifiability_conditions}. We omit the identifiability condition IC$^{(5)}$ as it does not involve any constraint on the latent factors and thus needs no modification.



\begin{table}[h!]
\centering
\renewcommand{\arraystretch}{1.3}
\begin{tabular}{p{0.18\linewidth} p{0.75\linewidth}}
\toprule
\textbf{Condition} & \textbf{Description} \\
\midrule
$\overline{\mathrm{IC}}^{(0)}$ &
$\bSigma = \Ib_K$; $\Mb_{\lambda\lambda}$ is diagonal with distinct elements. \\

$\overline{\mathrm{IC}}^{(1)}$ &
$\bSigma = \Ib_K$; up to some column permutation, the loading matrix $\bLambda$ satisfies that for each $r \ge 2$, 
$|\cA_r| = r-1$ and $\operatorname{rank}(\bLambda_{[\cA_r, 1:r]}) = r-1$. \\

$\overline{\mathrm{IC}}^{(2)}$ &
$\bSigma = \Ib_K$; $\bLambda_{[1:K,\,]}$ is lower triangular with nonzero diagonal elements. \\

$\overline{\mathrm{IC}}^{(3)}$ &
$\bSigma$ is diagonal; $\bLambda_{[1:K,\,]}$ is lower triangular with diagonal elements all equal to 1. \\

$\overline{\mathrm{IC}}^{(4)}$ &
$\mathrm{diag}(\bSigma) = \Ib_K$; for each $r \in [K]$, 
$|\cA_r| = K-1$ and $\operatorname{rank}(\bLambda_{[\cA_r, -r]}) = K-1$. \\
\bottomrule
\end{tabular}
\caption{Modified identifiability conditions $\overline{\mathrm{IC}}^{(0)}$--$\overline{\mathrm{IC}}^{(4)}$ under random latent factors.}
\label{tab:identifiability_conditions}
\end{table}


Regarding the estimation under $\overline{\mathrm{IC}}^{(S)}$ for $S\in\{0,1,\dots,4\}$, we follow the same procedure described in Section~\ref{sec:jmle} to obtain the estimator $(\hat\bLambda^{(S)},\hat\Fb^{(S)})$ under IC$^{(S)}$, which serves as a good approximation for $\overline{\mathrm{IC}}^{(S)}$. We show that, under mild regularity conditions, the estimators $(\hat\bLambda^{(S)},\hat\Fb^{(S)})$ obtained from \eqref{eq_joint_mle} consistently still recover the loading parameters and realizations of the latent factors under the random factor specification.

\begin{remark}
    \it A popular estimation approach for generalized latent factor models with latent factors treated as random variables is the marginal maximum likelihood estimation~\citep{bock1981marginal,holland1990sampling,skrondal2004generalized}, in which the latent factors are equipped with a prior distribution and then integrated out of the likelihood. 
    Despite its popularity, this method has several drawbacks. First, it requires specifying a prior distribution for the latent factors, which may affect estimation consistency~\citep{chen2019joint}. Moreover, it can be computationally challenging due to the integrals involved, particularly in the large-scale setting considered here, where both $N$ and $J$ diverge~\citep{rijmen2009efficient}. Theoretical analysis of the marginal estimator also relies critically on properties of the marginal likelihood, which differ fundamentally from the joint likelihood considered here. For these reasons, we focus on the estimator in \eqref{eq_joint_mle} and leave the marginal approach for future investigation.
\end{remark}

When the factors are modeled as random, Assumption~\ref{assumption: psd covariance} cannot be imposed directly, since it is formulated in terms of a fixed (non-random) second moment matrix. We therefore introduce a modified version of Assumption~\ref{assumption: psd covariance} that is tailored to the setting with random latent factors. Define the empirical second-moment matrix $\Mb_{ff} = N^{-1} \sum_{i=1}^N \bbf_i \bbf_i^\T$ with singular value decomposition $\Mb_{ff} = \Ub_f\bUpsilon_f\Ub_f^\T$. Let $\sigma$-field $\cG^{(S)}$ be $\sigma\big(\bar\bbf_1^{(S)},\dots,\bar\bbf_N^{(S)}\big)$ where for each $i\in[N]$,
\begin{equation*}
    \bar\bbf_i^{(S)} := \left\{\begin{aligned}
        &(\Mb_{ff})^{-1/2}\bbf_i\text{ for }S\in\{0,1,2\}\\&\Ub_f^\T\bbf_i\text{ for }S=3\\
        &\{\mathrm{diag}(\Mb_{ff})\}^{-1/2}\bbf_i\text{ for }S=4
    \end{aligned}\right.
\end{equation*}
Here, $(\Mb_{ff})^{-1/2}$ is given as $\Ub_f\bUpsilon_f^{-1/2}\Ub_f^\T$. Then one can verify $\big\{\bar\bbf_i^{(S)}\big\}_{i\in[N]}$ always satisfy the identifiability condition IC$^{(S)}$ regard the latent factor part. Correspondingly, for each $j\in[J]$, let $\bar\blambda_j^{(S)} = (\Mb_{ff})^{1/2}\blambda_j^*$ for $S\in\{0,1,2\}$, $\bar\blambda_j^{(S)} = \Ub_f^\T\blambda_j^*$ when $S=3$, and $\bar\blambda_j^{(S)} = \{\mathrm{diag}(\Mb_{ff})\}^{1/2}\blambda_j^*$ when $S=4$. Here, $\blambda_j^*$ denotes the true loading parameters. 
Now we can introduce the following assumption for the randomness of the latent factors.
\setcounter{assumption}{0} 
\renewcommand{\theassumption}{R\arabic{assumption}}

\begin{assumption}
\label{assump_random_factor}\it 
For the true loading parameters $\bLambda^*$, assume that for some big constant $D >0$, $\|\bLambda^*\|_{\max}\le D/2$ and
 $\bSigma_{\lambda}^* = \lim_{J\rightarrow \infty} \Mb_{\lambda\lambda}^*=\lim_{J\rightarrow \infty} J^{-1}\sum_{j=1}^J\blambda_j^*(\blambda_j^{*})^{\intercal} $ exists and is positive definite. 
Let $\{\bbf_i\}_{i=1}^N$ be independent random vectors in $\mathbb{R}^K$ with mean zero and positive definite covariance $\bSigma\in\RR^{K\times K}$. Assume that there exists some constant $C$ such that
 for any $j\in[J]$,
\begin{equation*}
    \left\|\EE\left[\Big\{\sum_{i=1}^Nl_{ij}^{\prime\prime}\big(\bar\blambda_j^{(S)}{}^\T\bar\bbf_i^{(S)}\big)\bar\bbf_i^{(S)}\bar\bbf_i^{(S)}{}^\T\Big\}^{-1}\sum_{i=1}^Nl_{ij}^{\prime}\big(\bar\blambda_j^{(S)}{}^\T\bar\bbf_i^{(S)}\big)\bar\bbf_i^{(S)}(\bar\blambda_j^{(S)} - \blambda_j^*)^\T\Big|\cG^{(S)}\right]\right\| \le C(N\wedge J)^{-1},
\end{equation*}
and for each $i\in[N]$,
\begin{equation*}
    \left\|\EE\left[\Big\{\sum_{j=1}^Jl_{ij}^{\prime\prime}\big(\bar\blambda_j^*{}^\T\bar\bbf_i\big)\bar\blambda_j^{(S)}\bar\blambda_j^{(S)}{}{}^\T\Big\}^{-1}\sum_{j=1}^Jl_{ij}^{\prime}\big(\bar\blambda_j^{(S)}{}^\T\bar\bbf_i^{(S)}\big)\bar\blambda_j^{(S)}(\bar\bbf_i^{(S)} - \bbf_i)^\T\Big|\cG^{(S)}\right]\right\| \le C(N\wedge J)^{-1},
\end{equation*}with $S\in\{0,1,2,3,4\}$.
In addition, we assume a uniform boundedness condition: for some sufficiently large constant $D>0$, $\PP(\max_{i\in[N]}\|\bbf_i\|_{\infty}\le D/2)\to 1$ as $N\to\infty$. Finally, the eigenvalues of $\bSigma\bSigma_{\lambda}^*$ are distinct and nonzero.

\end{assumption}

   Assumption~\ref{assump_random_factor} imposes a weak form of dependence between the randomness of the latent factors and the responses. This condition is mild and commonly adopted in the literature of linear factor models~\citep{fan2011high, fan2017sufficient, chen2020constrained}. Specifically, in the linear factor model $Y_{ij} = \blambda_j^\T\bbf_i + \epsilon_{ij}$, where $l^{\prime}_{ij}(\cdot) = \epsilon_{ij}$ denotes the noise term, many existing works either assume independence between $\epsilon_{ij}$ and $\bbf_i$ or impose a weak dependence condition~\citep{bai2003inferential}. Assumption~\ref{assump_random_factor} can be verified to hold under these settings. Moreover, it requires that the latent factors are bounded with high probability, which aligns with the boundedness assumption introduced under the settings where latent factors are treated as fixed parameters (Assumption~\ref{assumption: psd covariance}).


Now we are ready to present the consistency and asymptotic distributions for the estimator \eqref{eq_joint_mle}. For each $i\in[N]$, denote $\bbf_i^*$ as the realization of the random vector $\bbf_i$, and the matrix form as $\Fb^* = (\bbf_1^*,\cdots,\bbf_i^*)^\T$.
\begin{suppTheorem}\it
    For model \eqref{eq_factor_m} where the latent factors $\{\bbf_i\}_{i=1}^N$ are random vectors with zero mean and covariance $\bSigma$, suppose the loading parameters $\bLambda^*$ and covariance $\bSigma$ satisfy identifiability condition $\overline{\mathrm{IC}}^{(S)}$ for $S\in\{0,1,\dots,4\}$. Under Assumptions~\ref{assump_random_factor}, \ref{assumption:smoothness}, and \ref{assump:scaling}, for the estimator $(\hat\bLambda^{(S)},\hat\Fb^{(S)})$ obtained in \eqref{eq_joint_mle}, we have
    \begin{equation*}
        \frac{1}{J}\sum_{j=1}^J\big\|\hat\blambda_j^{(S)}-\blambda_j^*\big\|^2 = O_p \big(N^{-1}+J^{-2}\big),\quad \frac{1}{N}\sum_{i=1}^N\big\|\hat\bbf_i^{(S)}-\bbf_i^*\big\|^2 = O_p\big(J^{-1} + N^{-1}\big),
    \end{equation*}
and
    \begin{equation*}
        \max_{1\le j\le J}\big\|\hat\blambda_j^{(S)}-\blambda_j^*\big\|_{\infty} = O_p\Big( \frac{\epsilon_{NJ}\log J}{\sqrt{N\wedge J}}\Big),
        \quad \max_{1\le i\le N}\big\|\hat\bbf_i^{(S)}-\bbf_i^*\big\|_{\infty} = O_p\Big(\frac{\epsilon_{NJ}\log N}{\sqrt{N\wedge J}}\Big),
    \end{equation*}
    where $\epsilon_{NJ} := (\epsilon_N\log J + \epsilon_J\log N)^{1/2}$ with $\epsilon_N$ and $\epsilon_J$ given in Assumption~\ref{assump:scaling}.
\label{thm_consis_random}
\end{suppTheorem}

Next, we present the results for limiting distributions for the estimators in \eqref{eq_joint_mle}. Let $\Sbb_K\in\{0,1\}^{K^2\times K^2}$ denote the commutation matrix satisfying $\Sbb_K\mathrm{vec}(\Ab) = \mathrm{vec}(\Ab^\T)$ for any $\Ab\in\RR^{K\times K}$. Define $\circ$ as the Hadamard product, representing the element-wise multiplication of matrices. The following asymptotic distributions are obtained conditioned on $\cG^{(S)}$ for $S\in\{0,1,2,3,4\}$.
\begin{suppTheorem}\label{main_thm_asymp_random}
   \it
  For model \eqref{eq_factor_m} where the latent factors $\{\bbf_i\}_{i=1}^N$ are random vectors with zero mean and covariance $\bSigma$, suppose the loading parameters $\bLambda^*$ and covariance $\bSigma$ satisfy identifiability condition $\overline{\mathrm{IC}}^{(S)}$ for $S\in\{0,1,\dots,4\}$, and Assumptions~\ref{assump_random_factor}, \ref{assumption:smoothness}, and \ref{assump:scaling} hold.
  For $\hat\blambda_j^{(S)}$ in \eqref{eq_joint_mle}, when $|\cS_j|>0$,
\begin{equation}
    \left\{\Db_{(j)}\bar{\bSigma}_{[N],j} ^{(S)}\Db_{(j)}^\T\right\}^{-1/2}\left\{\Db_{(j)}\big(\hat\blambda_{j}^{(S)} - \blambda_j^*\big)\right\} \overset{d}{\to }\cN(\zero_{|\cS_j|},\Ib_{|\cS_j|})\text{, for any }j\in[J],\label{eq_converge_d_lambda_random}
\end{equation}
   as $N,J\to\infty$ with $\cS_j$ as specified in Theorem~\ref{main_thm_loading_asym}. Here,
   \begin{equation}
       \bar{\bSigma}_{[N],j} ^{(S)} =\left\{\begin{aligned} &{\bSigma}_{[N],j} ^{(S)} + (2N)^{-1}(\blambda_j^*{}^\T\otimes \Ib_K)(\Ib_{K} + \Sbb_K)(\blambda_j^*\otimes \Ib_K)\text{, when }S\in\{0,1,2,3\}\\&{\bSigma}_{[N],j} ^{(S)} + N^{-1}(\blambda_j^*\blambda_j^*{}^\T)\circ\bSigma\circ\bSigma\text{, when }S=4\end{aligned}\right.
   \end{equation}
   with ${\bSigma}_{[N],j} ^{(S)}$ given in \eqref{eq_sigma_lambad_j}, depending only on $(\Yb,\bLambda^*,\Fb^*)$ with $\Fb^*$ denoting the realization of $\Fb$.
   For $\hat \bbf_i^{(S)}$ in \eqref{eq_joint_mle}, we have 
   \begin{equation}
      \big(\bar\bSigma_{i,[J]}^{(S)}\big)^{-1/2}\big(\hat\bbf_i^{(S)} - \bbf_i^*\big)\overset{d}{\to}\cN\left(\zero_{K},\Ib_{K}\right),\text{ for any }i\in[N],\label{eq_asym_int_f_random}
  \end{equation}as $N,J\to\infty$. Here, 
  \begin{equation*}
      \bar\bSigma_{i,[J]}^{(S)} = \left\{\begin{aligned} &\bSigma_{i,[J]}^{(S)} + (2N)^{-1}(\bbf_i^*{}^\T\otimes \Ib_K)(\Ib_{K} + \Sbb_K)(\bbf_i^*\otimes \Ib_K)\text{, when }S\in\{0,1,2,3\}\\&\bSigma_{i,[J]}^{(S)} + N^{-1}(\bbf_i^*\bbf_i^*{}^\T)\circ\bSigma\circ\bSigma\text{, when }S = 4\end{aligned}\right.
  \end{equation*}with $\bSigma_{i,[J]}^{(S)}$ given in \eqref{eq_sigma_f_i}, depending only on $(\Yb,\bLambda^*,\Fb^*)$. The covariance matrices can be consistently estimated by replacing $\bLambda^*,\Fb^*$, and $\bSigma$ in $\bar\bSigma_{[N],j}^{(S)}$ and $\bar\bSigma_{i,[J]}^{(S)}$ with their consistent estimates $\hat\bLambda^{(S)},\hat\Fb^{(S)}$, and $\hat\Mb_{ff}^{(S)}$, respectively. Here, $\hat\Mb_{ff}^{(S)} := N^{-1}\sum_{i=1}^N\hat\bbf_i^{(S)}\{\hat\bbf_i^{(S)}\}^\T$. Denote the estimated covariance matrix $\hat{\bar\bSigma_{[N],j}^{(S)}}$ and $\hat{\bar\bSigma_{i,[J]}^{(S)}}$. Then \eqref{eq_converge_d_lambda_random} and \eqref{eq_asym_int_f_random} continue hold when replacing $\bar\bSigma_{[N],j}^{(S)}$ with $\hat{\bar\bSigma_{[N], -j}^{(S)}}$ and $\bar\bSigma_{i,[J]}^{(S)}$ with $\hat{\bar\bSigma_{-i,[J]}^{(S)}}$, respectively.

 \end{suppTheorem}
Section~\ref{sec_sketch_proof_random} outlines the proofs for these two theorems. 
Our results show that when the latent factors are treated as random vectors, the constrained maximum likelihood estimator in \eqref{eq_joint_mle} remains consistent for both the true loading parameters and the realizations of the latent factors. For the asymptotic distributions, the randomness of the latent factors introduces an additional variance component. In what follows, we illustrate through a simulation study that the extra term in $\bar\bSigma_{[N],j}^{(S)}$ and $\bar\bSigma_{i,[J]}^{(S)}$ is much smaller than their counterpart $\bSigma_{[N],j}^{(S)}$ and $\bSigma_{i,[J]}^{(S)}$. This implies that our theoretical results established in Theorems~\ref{main_thm_loading_asym} and Theorem~\ref{main_thm_factor_asym} are robust to the model specification in the sense that imposing restrictions on their realizations provides a good approximation to imposing the corresponding restrictions on their population covariance. 

We consider simulation settings that satisfy IC$^{(2)}$ and IC$^{(4)}$ as described in Section~\ref{sec:simu}. The setup is identical to that described in Section~\ref{supp_sec_simu}, except that the latent factors are generated without the normalization step. Specifically, under IC$^{(2)}$, the latent factors are simulated i.i.d. from $\cN(\zero_3,\Ib_3)$; under IC$^{(4)}$, they are simulated i.i.d. from $\cN(\zero_3,\bSigma_{\tau})$ with $\bSigma_{\tau}$ specified in Section~\ref{supp_sec_simu}. The setups are consistent with identifiability conditions $\overline{\mathrm{IC}}^{(2)}$ and $\overline{\mathrm{IC}}^{(4)}$, respectively. The empirical coverage rates for both the loading parameters and the latent factors are reported in Figures~\ref{fig:random_low_loading}--\ref{fig:random_high_factor}. We compare the coverage of (1) the naive method (in blue); (2) results under fixed-factor setting (red) where the asymptotic distributions are given in Theorems~\ref{main_thm_loading_asym} and~\ref{main_thm_factor_asym}; (3) results under random-factor setting (in green) where the asymptotic distributions are given in Theorem~\ref{main_thm_asymp_random}. The results show that, when the latent factors are random, treating them as fixed parameters leads to a slight reduction in coverage under $\overline{\mathrm{IC}}^{(2)}$, and the impact is nearly negligible under $\overline{\mathrm{IC}}^{(4)}$.
{\spacingset{1}
\begin{figure}
\centering 
        \includegraphics[width=5in]{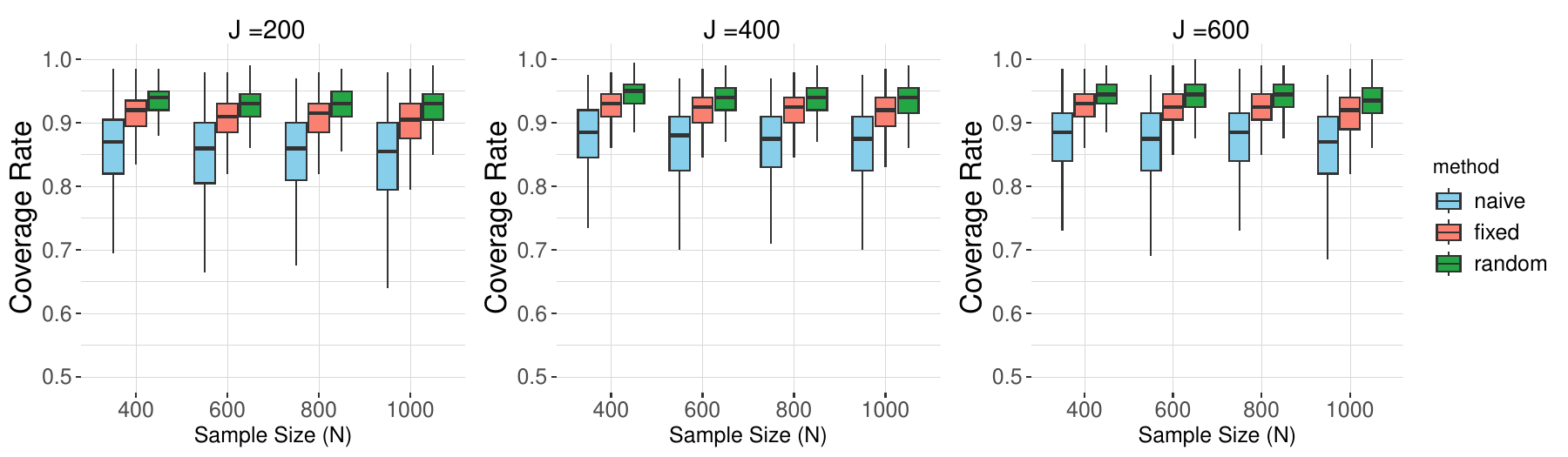}\\
        {\scriptsize (a) Identifiability Condition $\overline{\mathrm{IC}}^{(2)}$}\\\includegraphics[width=5in]{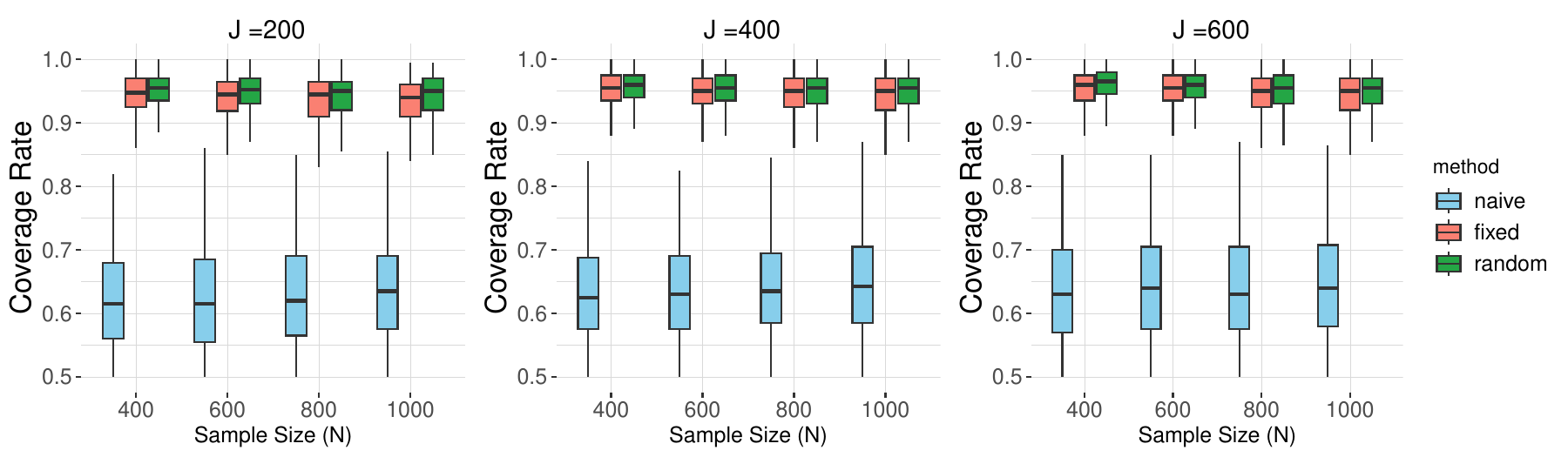}\\
        {\scriptsize (b) Identifiability Condition $\overline{\mathrm{IC}}^{(4)}$}
    \caption{\small Empirical coverage of loadings in the low-dimensional case $(\tau = 0.3)$ under different settings of $N$, $J$, and sparsity for $S\in\{2,4,5\}$. The naive method is shown in blue, the fixed-factor approximation is shown in red, and the random-factor method is shown in green.}
    \label{fig:random_low_loading}
\end{figure}
\begin{figure}
\centering 
        \includegraphics[width=5in]{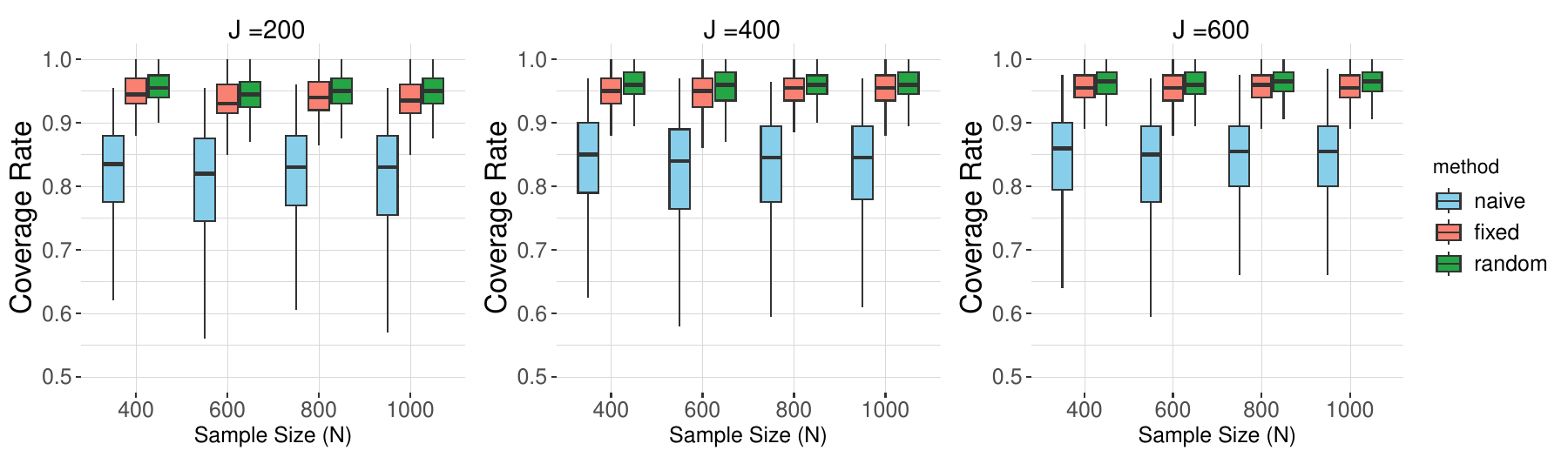}\\
        {\scriptsize (a) Identifiability Condition $\overline{\mathrm{IC}}^{(2)}$}\\\includegraphics[width=5in]{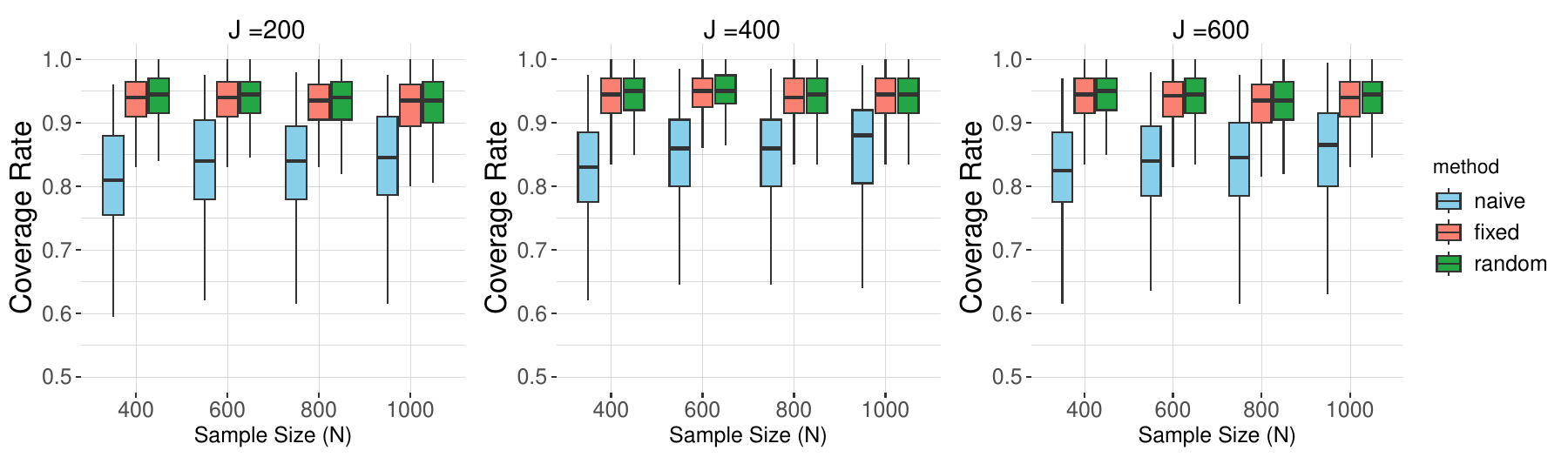}\\
        {\scriptsize (b) Identifiability Condition $\overline{\mathrm{IC}}^{(4)}$}
    \caption{\small Empirical coverage of loadings in the high-correlation case $(\tau = 0.7)$ under different settings of $N$, $J$, and $\overline{\mathrm{IC}}^{(S)}$ $S\in\{2,4\}$. The naive method is shown in blue, the fixed-factor approximation is shown in red, and the random-factor method is shown in green.}
    \label{fig:random_high_loading}
\end{figure}}

{\spacingset{1}
\begin{figure}
\centering 
        \includegraphics[width=5in]{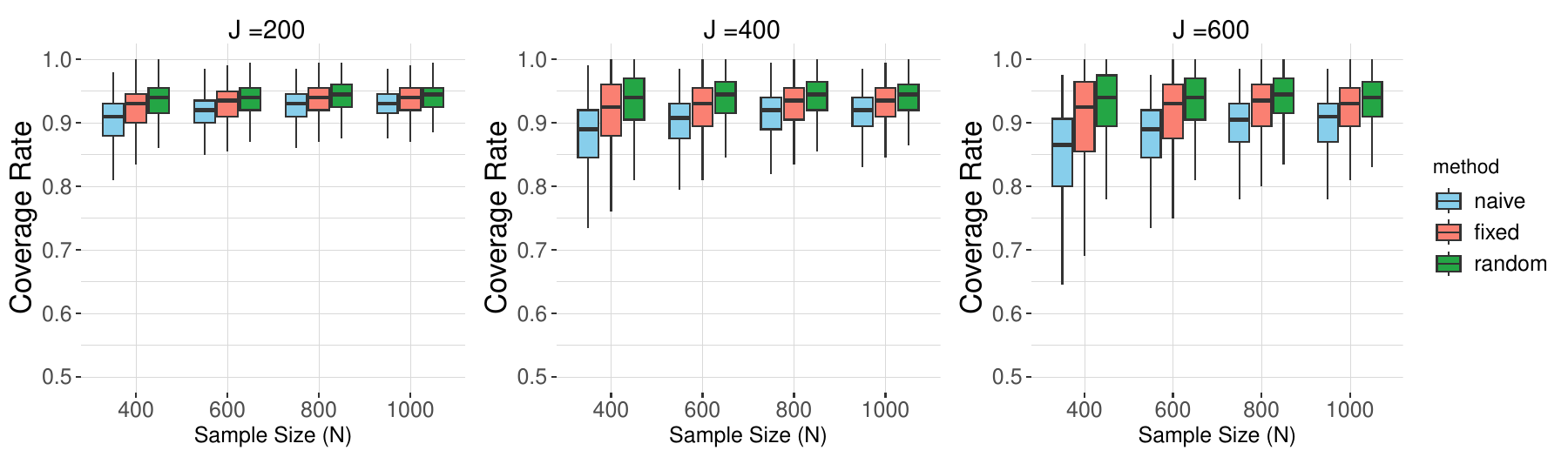}\\
        {\scriptsize (a) Identifiability Condition $\overline{\mathrm{IC}}^{(2)}$}\\\includegraphics[width=5in]{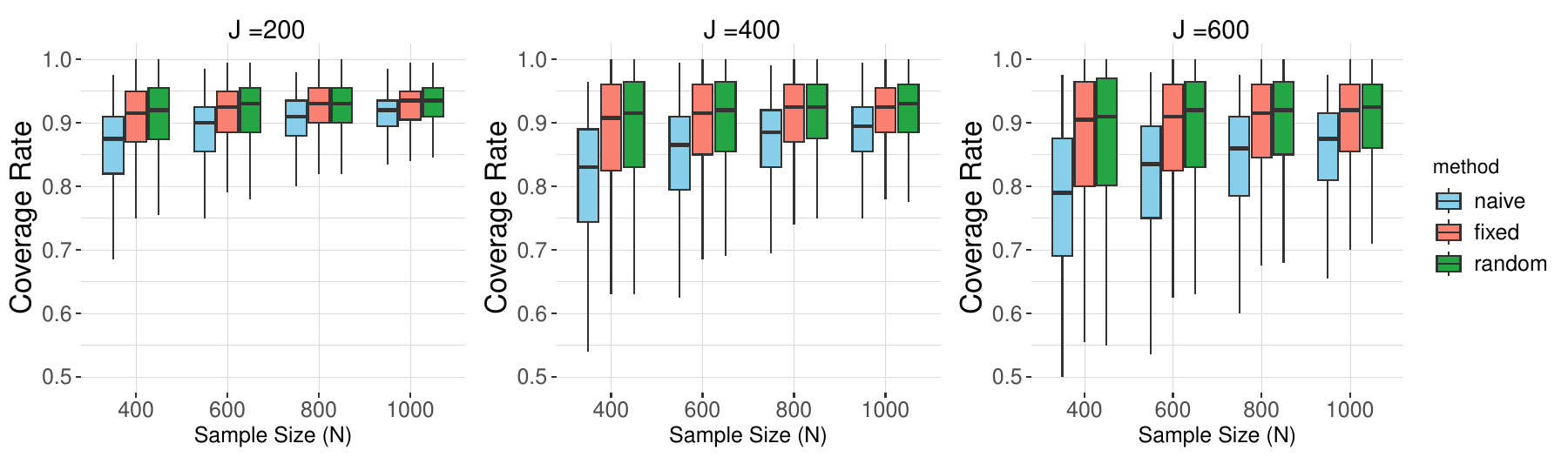}\\
        {\scriptsize (b) Identifiability Condition $\overline{\mathrm{IC}}^{(4)}$}
    \caption{\small Empirical coverage of latent factors in the low-correlation case $(\tau = 0.3)$ under different settings of $N$, $J$, and $\overline{\mathrm{IC}}^{(S)}$ for $S\in\{2,4,5\}$. The naive method is shown in blue, the fixed-factor approximation is shown in red, and the random-factor method is shown in green.}
    \label{fig:random_low_factor}
\end{figure}
\begin{figure}
\centering 
        \includegraphics[width=5in]{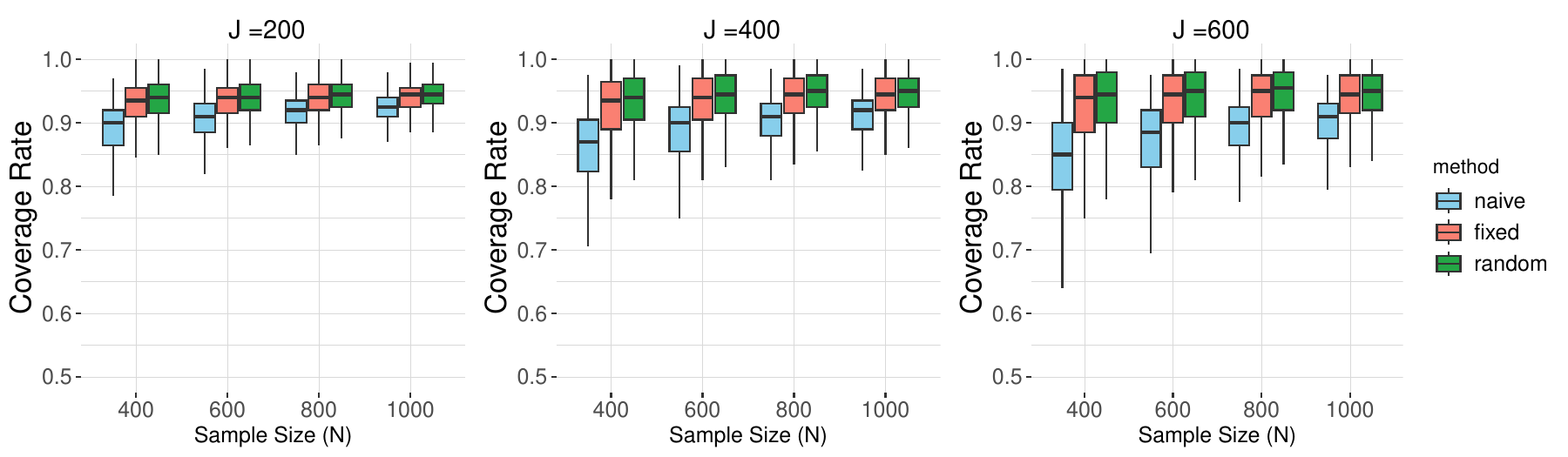}\\
        {\scriptsize (a) Identifiability Condition $\overline{\mathrm{IC}}^{(2)}$}\\\includegraphics[width=5in]{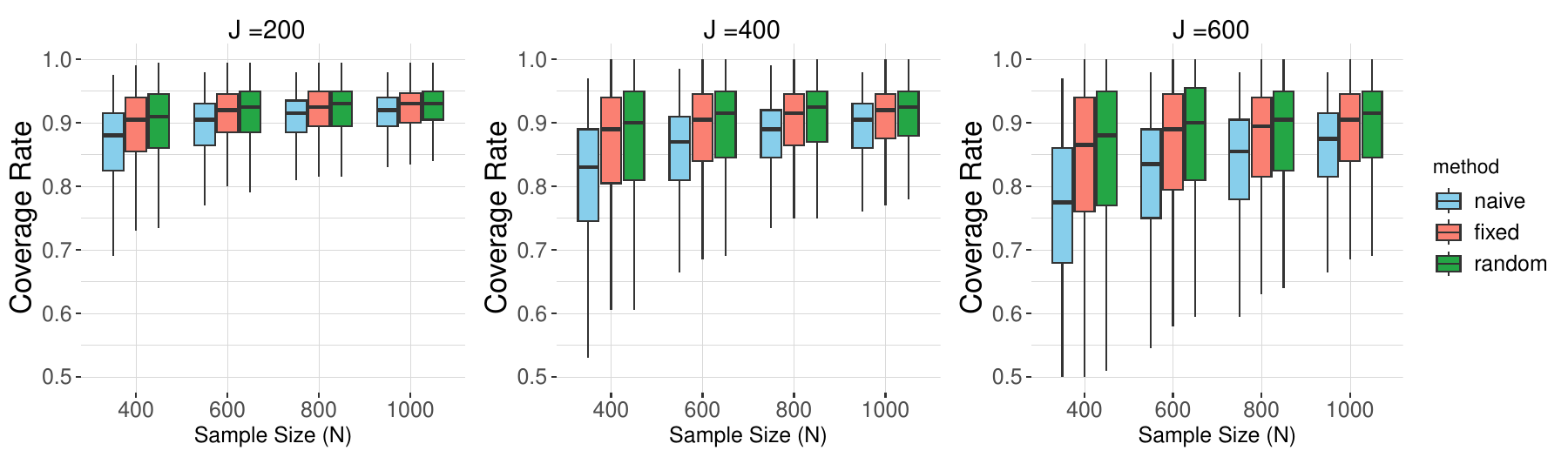}\\
        {\scriptsize (b) Identifiability Condition $\overline{\mathrm{IC}}^{(4)}$}
    \caption{\small Empirical coverage of latent factors in the high-correlation case $(\tau = 0.7)$ under different settings of $N$, $J$, and $\overline{\mathrm{IC}}^{(S)}$ for $S\in\{2,4,5\}$. The naive method is shown in blue, the fixed-factor approximation is shown in red, and the random-factor method is shown in green.}
    \label{fig:random_high_factor}
\end{figure}}

\section{Selecting Identifiability Condition}\label{sec_determine_id}


 In this section, we discuss the practical approaches to selecting an identifiability condition and show how our inferential results can be used to evaluate whether a selected condition is appropriate.
In practice, the specific identifiability condition can be selected based on practitioners’ scientific or domain knowledge about the underlying data structure. When such knowledge is not available, practitioners typically apply an exploratory factor analysis (EFA)~\citep{fabrigar2012exploratory}. Below, we outline a practical approach in the EFA for selecting an appropriate identifiability condition, and apply our asymptotic results to evaluate this selection. The idea is to choose a constraint set on the loading matrix that satisfies the zero assignment pattern required by IC$^{(4)}$. Here, we focus on IC$^{(4)}$ as it allows various patterns of zero constraints on the loading matrix and factor correlations.
Additionally, we can impose that the latent factors are normalized by $\mathrm{diag}(\Mb_{ff}) = \Ib_K$ without loss of generality.
Under the constraint set and $\mathrm{diag}(\Mb_{ff}) = \Ib_K$, we can obtain an estimator via \eqref{eq_joint_mle} under IC$^{(4)}$. Using the derived asymptotic distributions, we then select an alternative constraint set that also satisfies IC$^{(4)}$ to assess if this choice of identifiability condition is appropriate. 
The presence of factor correlations can further be assessed using Theorem~\ref{thm_mff_limiting_distri}. We describe the details as follows.

 Following the EFA framework~\citep{mulaik2009foundations,fabrigar2012exploratory,chen2025item}, we begin by obtaining an initial estimate under some arbitrary but mathematically convenient constraints. Here, we adopt the identifiability condition IC$^{(0)}$. A rotation method is then applied to produce a sparse loading matrix. Several rotation techniques are available, including varimax rotation~\citep{kaiser1958varimax,Rohe2020VintageFA}, promax rotation~\citep{hendrickson1964promax}, and methods based on the $l_p$-loss~\citep{Liu2023RotationTS}. Among these, varimax rotation has long been one of the most popular approaches in exploratory analysis, and recent studies demonstrate that, under mild regularity conditions, it can consistently recover the underlying factor loading structure~\citep{kaiser1958varimax,Rohe2020VintageFA}. Promax rotation, an oblique generalization of varimax, accommodates correlated factors and remains one of the most widely used methods in exploratory factor analysis~\citep{browne2001overview}. A more recent study~\citep{Liu2023RotationTS} proposes a rotation method based on the $l_p$-loss, showing that it can effectively reduce bias in the estimated loading matrix and, in many cases, establish the consistency of the resulting estimator. These results indicate that the rotated loading matrix can effectively signal potential zero entries through coefficients whose estimated values are near zero. 

Denote the rotated loading matrix as $\hat\bLambda^0$. After obtaining a sparse loading matrix, we select, for each column $k\in[K]$, a set of $K-1$ indices denoted by $\cD_k$, that approximately correspond to the smallest entries in the $k$th column of $\hat\bLambda^0$ and satisfy
\begin{equation}
    det\big(\hat\bLambda^0_{[\cD_k,-k]}\big)>\delta_0,\label{eq_rank_constraints}
\end{equation}for some pre-specified threshold $\delta_0$.
Here, $\cD_k$ does not necessarily contain the absolute smallest entries, since the rank condition required for identifiability must also be taken into account. 
The resulting identifiability condition takes the form IC$^{(4)}$ with
\begin{equation}
    \mathrm{diag}(\Mb_{ff}) = \Ib_K\text{ and for each }k\in[K],~ \bLambda_{[\cD_{k},k]} = \zero_{K-1},\label{eq_id_d_rot}
\end{equation} where the rank condition is approximated by \eqref{eq_rank_constraints}. Under condition \eqref{eq_id_d_rot}, we obtain an estimator using the framework developed in Section~\ref{sec:jmle} and conduct statistical inference on the free loadings and latent factors.

With the derived asymptotic distributions in Section~\ref{sec:mainres}, we can further assess whether this choice of $\{\cD_k\}_{k\in[K]}$ is appropriate.
Specifically, we repeat the selection step to choose, for each column $k\in[K]$, another set of $K-1$ indices from the complement $[J]\setminus\cD_k$ that approximately correspond to the smallest entries in $\hat\bLambda^0_{[-\cD_k,k]}$. Let $\cD_k^{\prime}$ denote the resulting index set. Using $\{\cD_k^{\prime}\}_{k\in[K]}$, we can construct another identifiability condition of the IC$^{(4)}$-type, obtain another estimator and perform statistical inference with Theorems~\ref{main_thm_loading_asym}. Specifically, we consider the hypothesis test
 \[
    H_0 : \bLambda_{[\cD_k, k]} = 0 \quad \text{for all } k \in [K].
    \]
If we accept $H_0$, we conclude that our initial selection indeed yields a faithful identifiability condition. There are $(K-1)K$ individual tests in $H_0$, and we apply the Benjamini–Hochberg procedure with significance level $\alpha = 0.05$.

We conduct numerical studies to evaluate the performance of the proposed approach. Throughout, we set the latent dimension $K$ as $3$. The latent factors are first simulated i.i.d. from $\cN(\zero_3,\bSigma_{\tau})$ and then transformed to satisfy $\mathrm{diag}\{N^{-1}(\Fb^{(4)})^\T\Fb^{(4)}\} = \Ib_3$. Here, $\bSigma
 _\tau\in\RR^{3\times 3}$ has the $(l,h)$th entry set as $\tau^{|l-h|}$ for $l,h\in\{1,2,3\}$ and  $\tau = 0.3$. For the loading matrix, we generate $s \times J$ as simple loadings, where each row contains exactly one nonzero entry drawn from $Unif(1,2)$. The simple rows are distributed evenly across the three columns. The remaining $(1-s)\times J$ rows are generated from $\cN(\zero_3,\Ib_3)$. Simple loadings are common in exploratory factor analysis and are known to facilitate consistent recovery under rotation methods~\citep{browne2001overview,andersonintroduction,Liu2023RotationTS}. To avoid generating entries that are small but nonzero, we truncate any element with magnitude below $0.2$ a zero. In all simulations, we take $\delta_0 $ in \eqref{eq_rank_constraints} as $ 0.3$.

\begin{figure}[h]
        \centering\includegraphics[width=6in]{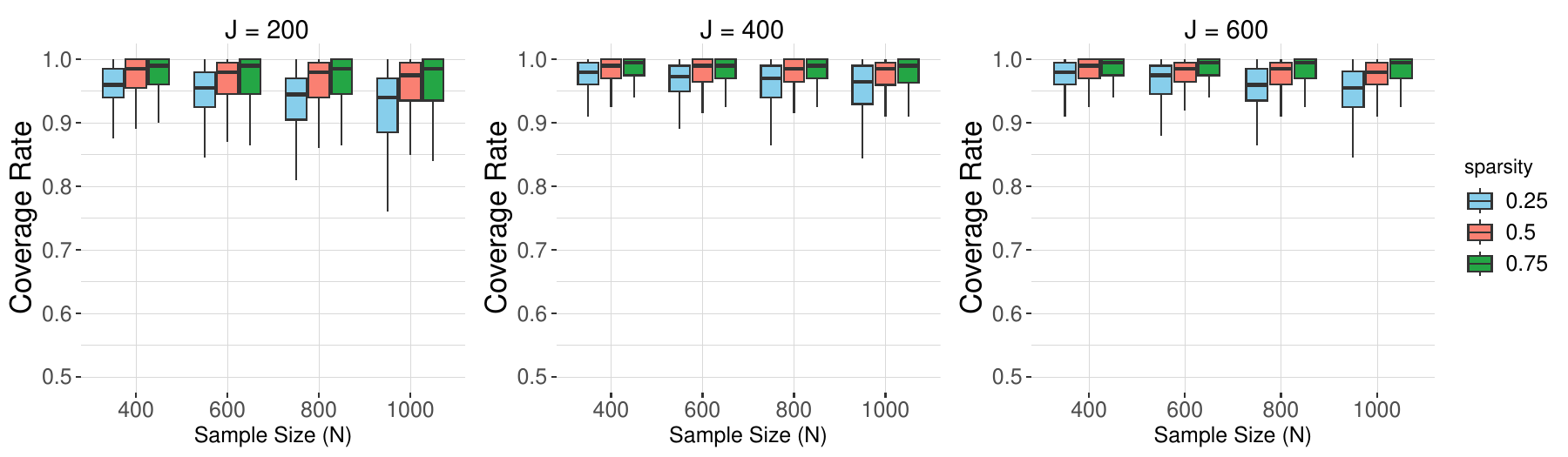}
{\scriptsize Empirical coverage for loading parameters}\\
        \includegraphics[width=6in]{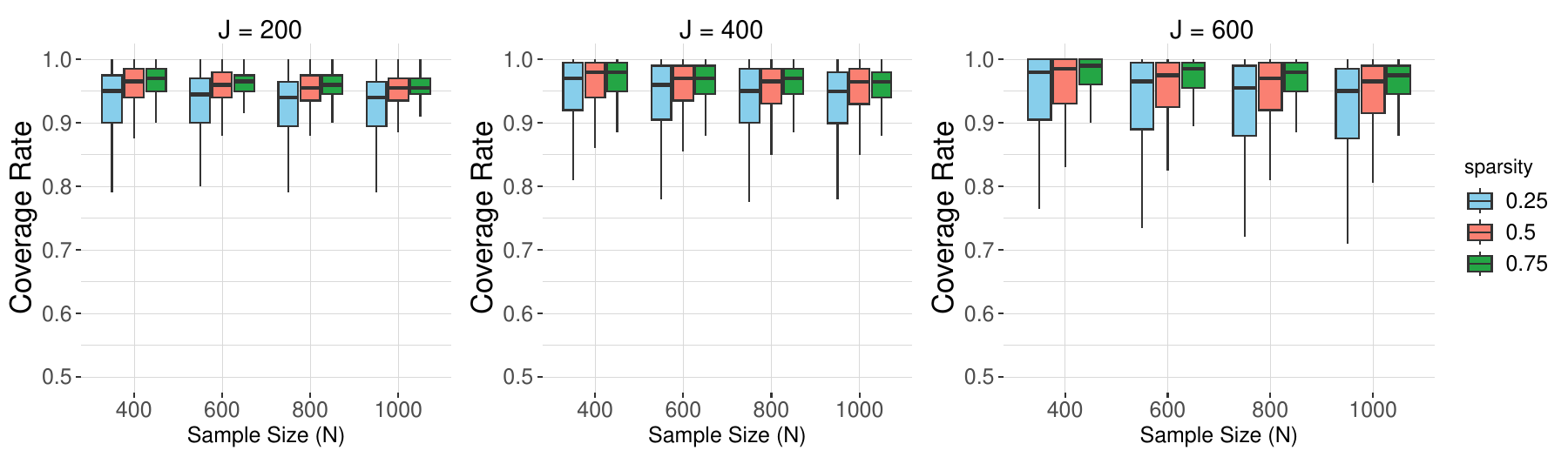}
        {\scriptsize Empirical coverage for factor parameters}\\
    \caption{Empirical coverage of loadings and latent factors with the identifiability condition determined by the proposed data-driven method under different settings of $N$, $J$, and sparsity level.}
    \label{fig:select}
\end{figure}

 The manipulated settings includes  (1) sample size $N=400, 600, 800, 1000$; (2) test length $J=200, 400, 600$; (3) proportion of simple rows $s = 0.25,0.5,0.75$. A total of 200 replications are generated under each setup. We compute the empirical coverage rates for both factors and loading parameters similar to Section~\ref{sec:simu}. The results are presented in Figure~\ref{fig:select}. We observe good coverage for both the loadings and the latent factors, demonstrating that the rotation method successfully selects a valid identifiability condition, even in scenarios with only a few nonzero loadings ($s = 0.25$). Across all settings considered above, the null hypothesis $H_0$ is accepted in more than 99\% of the replications, suggesting that the selection procedure based on EFA can reliably identify appropriate constraint sets in these scenarios.

\begin{remark}
    { The effectiveness of the data-driven approach depends on whether the initial loading estimator $\hat\bLambda^0$, obtained by a suitable rotation method, can faithfully recover the unknown sparsity pattern. For rotation-based methods, many existing results can handle sparsity patterns in which each row has at most one nonzero entry~\citep{jennrich1966rotation,jennrich2006rotation,Liu2023RotationTS}. Remarkably, several recent works have established statistical guarantees under more complicated sparsity patterns~\citep{Rohe2020VintageFA,bing2025optimal,cui2026beyond}. But these theories typically rely on additional constraints on the loading matrix. It remains an open problem whether rotation methods can consistently recover more general unknown sparsity patterns. Our current inference theory provides useful insight into this problem and offers potential tools for addressing it by deriving a stochastic expansion for $\hat\bLambda^{(S)}$ (see Section~\ref{sec:proof limit}) under different identifiability conditions $S\in\{0,1,\dots,5\}$. We expect that these results can provide a useful foundation for future work on the theoretical analysis of such problems.}
\end{remark}


\section{Proofs of Identifiability}\label{sec_supp_identification}

In Section~\ref{supp_sec_prove_thm_iff_ic1}, we prove Theorem~\ref{thm_iff_ic1}, Theorem~\ref{thm_iff_ic4} and Proposition~\ref{prop_consis_ic4}. Section~\ref{sup_sec_express_suff} presents the sufficiency of all the identifiability conditions in Section~\ref{subsec:identification} in the main text, and Section~\ref{sup_sec_express_G} gives the expressions for $\tilde\Gb^{(t+1)}$ that were omitted in Section~\ref{sec:jmle}. Before we start, define $\Gb_K^{ij}(\theta)$ as the Givens matrix to be a $K$-by-$K$ identity matrix with elements in rows and columns $i$ and $j$ modified to create a rotation by an angle $\theta$ in the $i$-$j$ plane. The matrix rotates vectors in the $i$-$j$ subspace, leaving other elements unchanged. In particular, 
\begin{equation*}
    \Gb_{K}^{ij}(\theta)=
    \begingroup
    \renewcommand{\arraystretch}{0.8}
    \begin{pmatrix}
            \Ib_{i-1}\\&\cos\theta&&\sin\theta\\&&\Ib_{j-i}\\ &-\sin\theta&&\cos\theta&\\ &&&&\Ib_{K-j-1}
    \end{pmatrix}
    \endgroup
    \text{, for }i<j.
\end{equation*}
where the blank spaces in $\Gb_{K}^{ij}(\theta)$ denote zero elements. 
\subsection{Proofs of Theorem~\ref{thm_iff_ic1}, Theorem~\ref{thm_iff_ic4} and Proposition~\ref{prop_ic4}}\label{supp_sec_prove_thm_iff_ic1}
\noindent\textbf{Proof of Theorem~\ref{thm_iff_ic1}}

\textbf{Sufficiency. } We show that if $(\bLambda,\Fb)$ and $(\bar\bLambda,\bar\Fb)$ both satisfy this condition and $\bLambda\Fb^\T = \bar\bLambda\bar\Fb^\T$, we must have $\bar\bLambda = \bLambda\Gb$ and $\bar\Fb = \Fb(\Gb)^{-\T}$ with $\Gb$ being some signed permutation. Since $\bLambda$ and $\Fb$ are full rank, by $\bLambda\Fb^\T = \bar\bLambda\bar\Fb^\T$ we know that there exists some invertible matrix $\Gb$ such that $\bar\bLambda = \bLambda\Gb$ and $\bar\Fb = \Fb(\Gb)^{-\T}$. By $N^{-1}\bar\Fb^\T \bar\Fb = N^{-1}\Fb^\T \Fb = \Ib_K$, we know $\Gb$ must be orthogonal ($\Gb\Gb^\T= \Ib_K$). Write $\Gb = (\bg_1,\cdots,\bg_K)$. By (i) $|\cA_K| = K-1$; (ii) $\bLambda_{[\cA_K,K]} = \zero_{K-1}$; (iii) $\bar\bLambda_{[\cA_K,K]} = \bLambda_{[\cA_K,]}\bg_K = \zero_{K-1}$; (iv) $\mathrm{rank}(\bLambda_{[\cA_K,1:K]}) = K-1$, we know $\bg_K = \pm\be_K$. Suppose we have determined $\bg_{r} = \pm\be_r$ for $r = k+1,\cdots, K$. For $\bg_k$ we have $\bLambda_{[\cA_k,k]} = \zero_{k-1}$ and $\bar\bLambda_{[\cA_k,k]} = \bLambda_{[\cA_k,]}\bg_k = \zero_{k-1}$. Since $\bg_k$ should be orthogonal to $\bg_{k+1},\cdots,\bg_{K}$, we know $[\bg_{k}]_{[k+1:K]}=\zero_{K-k}$, and thus $ \bLambda_{[\cA_k,1:k]}[\bg_k]_{[1:k]} = \zero_{k-1}$, which together with $\mathrm{rank}(\bLambda_{[\cA_k,1:k]}) = k-1$ implies that $\bg_k = \pm\be_k$. Therefore we proved that $\Gb = \mathrm{diag}(\pm1,\cdots,\pm1)$.

\textbf{Necessity.} We then construct counterexamples for cases where the zero-assignment or rank condition is not satisfied. Suppose for each $r$, $\bLambda_{[\cA_r,r]}$ is specified as $\zero_{k_r}$ where $k_r=|\cA_r|$. For convenience, we assume $k_1\le k_2\le \cdots\le k_K $. Since we are assigning the minimal $K(K-1)/2$ to the loading matrix, $k_1+\cdots+k_K = (K-1)K/2$.
\begin{itemize}
    \item We claim that if the minimal $K(K-1)/2$ zeros are not assigned as in IC$^{(1)}$, there exists $r\neq r^{\prime}\in[K]$ such that $k_r=k_{r^{\prime}}$. Suppose not, then $k_1<k_2<\cdots<k_K$, which implies that $(K-1)K/2=k_1+\cdots+k_K\ge K\times k_1 + (K-1)K/2$. Therefore we must have $k_1 = 0$ and $k_r=k_1+r-1=r-1$, which implies IC$^{(1)}$ is satisfied. 

Then the counterexample is constructed such that, for $r< r^{\prime}$ with $k_r=k_{r^{\prime}}$, $\bLambda_{[\cA_r,]} = \bLambda_{[\cA_{r^{\prime}},]}$. Then by the identifiability condition that $\bLambda_{[\cA_r,r]} =\zero_{k_r}$ and $\bLambda_{[\cA_{r^{\prime}},r^{\prime}]} =\zero_{k_r}$, we know
\begin{equation*}
    \bLambda_{[\cA_r,r]} = \bLambda_{[\cA_{r^{\prime}},r]}=\bLambda_{[\cA_r,r^{\prime}]} = \bLambda_{[\cA_{r^{\prime}},r^{\prime}]} =\zero_{k_r}.
\end{equation*}
Under this setting, the rank condition can still be satisfied by letting $\mathrm{rank}\big(\bLambda_{[\cA_r,1:r]}\big)= k_r$. Observe that $\big[\bLambda\Gb_{K}^{rr^{\prime}}(\theta)\big]_{[\cA_l,l]}=\zero_{k_l}$ for any $\theta$ and $l\in[K]$ and we know this setting does not guarantee identification.
 \item Next we show the necessity of the rank condition. Still, we permute the column such that $|\cA_r|= r-1$. If $\mathrm{rank}(\bLambda_{[\cA_r,1:r]})<r-1$ for some $r\ge 2$, we construct a counterexample as $\bLambda_{[\cA_{r-1},r]} = \zero_{r-1}$ and \begin{equation*}\bLambda_{[\cA_r,]} = \begin{pmatrix}\zero\\\bLambda_{[\cA_{r-1},]}\end{pmatrix}\end{equation*}
Observe that $\big[\bLambda\Gb_{K}^{r-1\, r}(\theta)\big]_{[\cA_l,l]}=\zero_{l-1}$ for any $\theta$ and $l\in[K]$ and we know without the rank condition identification cannot be guaranteed.
\end{itemize}

    \noindent\textbf{Proof of Theorem~\ref{thm_iff_ic4}}\label{supp_sec_prove_thm_iff_ic4}
    
    \textbf{Sufficiency.} Sufficiency is directly implied by Proposition~\ref{prop_ic4}, the proof of which is left to the subsequent section.
    
\textbf{Necessity.} 
We prove the necessity of $|\cA_r|=K-1$ and $\mathrm{rank}(\bLambda_{[\cA_r,1:K]}) = K-1$, respectively.
\begin{itemize}
    \item \textbf{Necessity of $|\cA_r|=K-1$.}
Suppose without loss of generality that $|\cA_1|<K-1$. Then for linear equations $\bLambda_{[\cA_r,2:K]}\bx = 0$, there exists a nontrivial solution $\bx \in \RR^{K-1}$ ($\bx\neq \zero_{K-1}$). It is easy to see for any nonzero $g_2,\cdots,g_K$, let 
\begin{equation*}
    \Gb = \begingroup\renewcommand{\arraystretch}{0.6}\begin{pmatrix}
            1&\zero^\T\\\bx&\begin{matrix}
                g_2&&0\\&\ddots\\0&&g_{K}\end{matrix}
        \end{pmatrix}\endgroup.
\end{equation*}
Then $\bLambda\Gb$ still adopts the structure, no matter how other zeros are assigned. Note that 
\begin{equation*}
    \Gb^{-1} = \begingroup\renewcommand{\arraystretch}{0.6}\begin{pmatrix}
            1&\zero^\T\\\begin{matrix}
                -g_2^{-1}x_2\\\vdots\\-g_K^{-1}x_K
            \end{matrix}&\begin{matrix}
                g_2&&0\\&\ddots\\0&&g_{K}\end{matrix}
        \end{pmatrix}\endgroup.
\end{equation*}
Clearly $\big[\Gb^{-1}\Mb_{ff}\Gb^{-\T}\big]_{[1,1]} = 1$ with $[\Mb_{ff}]_{11}=1$. Next, we let 
    \begin{equation*}
        g_r = \left\{\big(-x_r\be_1+\be_r\big)^\T\Mb_{ff}\big(-x_r\be_1+\be_r\big)\right\}^{1/2}>0,
    \end{equation*}
    and therefore for every $r\ge 2$
    \begin{equation*}
        \big[\Gb^{-1}\Mb_{ff}\Gb^{-\T}\big]_{[r,r]} = \big(-g_r^{-1}x_r\be_1+g_r^{-1}\be_r\big)^\T\Mb_{ff}\big(-g_r^{-1}x_r\be_1+g_r^{-1}\be_r\big) = 1,
    \end{equation*}
    which gives $\mathrm{diag}(\Gb^{-1}\Mb_{ff}\Gb^{-\T}) = \Ib_K$.
    \item \textbf{Necessity of $\mathrm{rank}(\bLambda_{[\cA_r,1:K]}) = K-1$.} Suppose without loss of generality that $\mathrm{rank}(\bLambda_{[\cA_r,1:K]})<K-1$. Similarly, for linear equations $\bLambda_{[\cA_r,2:K]}\bx = 0$, there exists a nontrivial solution $\bx \in \RR^{K-1}$ ($\bx\neq \zero_{K-1}$). The counterexample can be similarly constructed.
    \end{itemize}

\noindent\textbf{Proof of Proposition~\ref{prop_ic4}}

By rank decomposition, we know there exists some $\Gb\in\RR^{K\times K}$ such that $\bar\bLambda=\bLambda\Gb$.
        For any $r\in[K]$, by the identifiability condition, we have $\bar\bLambda_{[\cA_r,r]}=\bLambda_{[\cA_r,]}\Gb_{[,r]}=\zero$. Further with the condition that $\mathrm{rank}(\bLambda_{[\cA_r,-r]})=K-1$ and $\bLambda_{[\cA_r,r]}=\zero$, we know that $\bLambda_{[\cA_r,-r]}\Gb_{[-r,r]}=\zero$, which implies that $\Gb_{[,r]}=\nu_r\be_r$ for some $\nu_r\neq 0$. Then $\bar\bLambda_{[,r]}=\nu_r\bLambda_{[,r]}$, that is $\sin\big(\bLambda_{[,k]},\bar\bLambda_{[,k]}\big)=0$.
        
        When this condition is satisfied for all $r\in[K]$, we know $\Gb = \mathrm{diag}(\nu_1,\cdots,\nu_K)$. Then 
    \begin{equation*}
        \Ib_K=\mathrm{diag}\big(\bar\Mb_{ff}\big)=\mathrm{diag}\big(\Gb^{-1}\Mb_{ff}\Gb^{-\T}\big)=\begin{pmatrix}
            \nu_1^2&&\zero\\&\ddots\\\zero&&\nu_{K}^2
        \end{pmatrix},
    \end{equation*}
    which implies that all diagonal elements of $\Gb$ are $\pm1$.
    
    \begin{remark}\label{remark_chen_ident} \it
 In a confirmatory manner, \cite{chen2020structured} studied the structural identifiability of latent factors, 
   which defines the model identifiability in an asymptotic sense ($J\to\infty$).  Their identifiability condition requires a diverging number of zero entries in each column of the loading matrix, with their proportions converging to some positive constants. In contrast,  our identifiability conditions for generalized latent factor models apply to any non-asymptotic setting, and we only require $K(K-1)$ entries in total specified as zeros, offering greater practical flexibility.
\end{remark}

\subsection{Sufficiency of Other Identifiability Conditions}\label{sup_sec_express_suff}

In this subsection, we show that IC$^{(0)}$, IC$^{(2)}$, IC$^{(3)}$, and IC$^{(5)}$ are sufficient to guarantee rotational identifiability. Specifically, for any $(\bLambda,\Fb)$ and $(\bar\bLambda,\bar\Fb)$ such that $\bLambda\Fb^\T = \bar\bLambda\bar\Fb^\T$, we know there is some invertible matrix $\Gb$ such that $(\bar\bLambda, \bar\Fb) = (\bLambda\Gb, \Fb\Gb^{-\T})$. Then, we show the sufficiency by showing that when both $(\bLambda,\Fb)$ and $(\bar\bLambda,\bar\Fb)$ satisfy IC$^{(S)}$, $\Gb$ can only be a signed permutation. To this end, we express $\Gb$ in terms of $(\bLambda,\Fb)$ for general $(\bLambda,\Fb)$ that does not necessarily satisfy IC$^{(S)}$. Furthermore, $\tilde\Gb^{(n+1)}$, which we omitted in Section~\ref{sec:jmle},  obtained analogously by replacing $(\bLambda,\Fb)$ with $(\tilde\bLambda^{(n+1)}, \tilde\Fb^{(n+1)})$. 

For IC$^{(0)}$, we know by $N^{-1}\Fb^\T\Fb = N^{-1}\bar\Fb^\T\bar\Fb = \Ib_K$ that $\Gb^{-\T}\Gb^\T =\Ib_K$, which implies that $\Gb$ must be orthogonal. Next, with $J^{-1}\bLambda^\T\bLambda$ and $J^{-1}\bar\bLambda^\T\bar\bLambda = \Gb^\T(J^{-1}\bLambda^\T\bLambda)\Gb$ both being diagonal with distinct elements, one can easily check that $\Gb$ must also be diagonal. Therefore, $\Gb$ can only be a signed permutation.

For IC$^{(2)}$, it is a special case of IC$^{(1)}$, and by Theorem~\ref{thm_iff_ic1}, it is sufficient.

For IC$^{(3)}$, let $\Mb_{ff} = N^{-1}\Fb^\T\Fb$ and $\bar\Mb_{ff} = N^{-1}\bar\Fb^\T\bar\Fb$. Since $\Mb_{ff}$, $\bar\Mb_{ff}$ are diagonal, and $\Fb$, $\bar\Fb$ are full rank, we know $\Mb_{ff}^{1/2}$ and $\bar\Mb_{ff}^{1/2}$ are well defined with positive diagonal entries. We note that $(\bLambda\Mb_{ff}^{1/2},\Fb\Mb_{ff}^{-1/2})$ and $(\bar\bLambda\bar\Mb_{ff}^{1/2},\bar\Fb\bar\Mb_{ff}^{-1/2})$ satisfy IC$^{(2)}$ with $\bLambda\Fb^\T = \bar\bLambda\bar\Fb^\T$. Then $\bLambda\Mb_{ff}^{1/2} = \bar\bLambda\bar\Mb_{ff}^{1/2}\Gb$ for some signed permutation $\Gb$. Namely, $\bLambda = \bar\bLambda\Gb'$ for some diagonal matrix $\Gb'$. Since $\bLambda_{[1:K,]}$ and $\bar\bLambda_{[1:K,]}$ are  lower-triangular with nonzero diagonal elements, we know $\Gb = \Ib_K$.

For IC$^{(5)}$, with $\bLambda_{[1:K,]}=\bar\bLambda_{[1:K,]} = \Ib_K$, we know immediately from $\bLambda = \bar\bLambda\Gb$ that $\Gb = \Ib_K$.

\subsection{Expression for Transformation \texorpdfstring{$\tilde\Gb^{(t+1)}$}{tildeG}}\label{sup_sec_express_G}
Next, we discuss how to obtain $\tilde\Gb^{(t+1)}$ from $(\tilde\bLambda^{(t+1)},\tilde\Fb^{(t)})$ under each identifiability condition $S\in\{0,1,\dots,5\}$. For brevity, we write $(\tilde\bLambda,\tilde\Fb) $ as $ (\tilde\bLambda^{(t+1)},\tilde\Fb^{(t)})$ and $\tilde\Gb$ as $\tilde\Gb^{(t+1)}$.
\begin{itemize}
    \item For IC$^{(0)}$, we have $\tilde\Gb=\tilde\Mb_{ff}^{1/2}\tilde\Ub$ where $\tilde\Ub$ is the matrix of left singular vectors of $\tilde\Mb_{ff}^{1/2}\tilde\Mb_{\lambda\lambda}\tilde\Mb_{ff}^{1/2}$ for $\tilde\Mb_{ff} = N^{-1}\tilde\Fb^\T\tilde\Fb$ and $\tilde\Mb_{\lambda\lambda} = J^{-1}\tilde\bLambda^\T\tilde\bLambda$. 
    \item For IC$^{(1)}$, we have $\tilde\Gb = \tilde\Mb_{ff}^{1/2}\Gb_1$ where $\Gb_1 = (\bg_1,\dots,\bg_K)$ with $\bg_K,\bg_{K-1},\cdots, \bg_{1}$ being being constructed recursively backward with $\|\bg_r\|=1$ and
    \begin{equation*}
        \bg_r\perp \mathrm{row}\big([\tilde\bLambda\tilde\Mb_{ff}^{1/2}]_{[\cA_r,]}\big)\cup \{\bg_{r+1},\dots, \bg_{K}\}.
    \end{equation*}
    \item For IC$^{(2)}$, it is a special case of IC$^{(1)}$ and thus $\tilde\bG$ can be determined as above.
    \item For IC$^{(3)}$, we have $\tilde \Gb = \tilde\bLambda_{[1:K,]}^{-1}\Lb_1$ where $\Lb_1$ is the $L$-matrix from  the LDL decomposition of $\tilde\bLambda_{[1:K,]}\tilde\Mb_{ff}^*\bLambda_{[1:K,]}^\T$, namely, $\tilde\bLambda_{[1:K,]}\tilde\Mb_{ff}\tilde\bLambda_{[1:K,]}^\T = \Lb_1\Db\Lb_1^\T$ for some diagonal matrix $\Db$ and the diagonal elements of $\Lb_1$ are $1$. 
   \item For IC$^{(4)}$, we have $\tilde\Gb=\tilde\Mb_{ff}^{1/2}\Hb\big\{\mathrm{diag}[(\Hb^\T\Hb)^{-1}]\big\}^{1/2}$ where $\Hb_{[,r]}$ is unit vector spanning the null space of $\tilde\bLambda_{[\cA_r,]}\tilde\Mb_{ff}^{1/2}$.
    \item For IC$^{(5)}$, we have $\tilde\Gb = (\tilde\bLambda_{[1:K,
   ] })^{-1}$.
\end{itemize}

For IC$^{(2)}$--IC$^{(5)}$, $\tilde\Gb$ is well-defined under certain rank conditions on $\tilde\bLambda$. These rank conditions do not depend on the particular identifiability condition satisfied by $\tilde\bLambda$. Thus, under the assumption that the true loading $\bLambda^*$ satisfies the corresponding identifiability condition and $\tilde\bLambda$ is close to $\bLambda^*\Gb$ for some invertible $\Gb$, $\tilde\bLambda$ satisfies the required rank conditions and thus $\tilde\Gb$ can be constructed as above. 

\subsection{Proof of Full Model Identifiability}\label{sup_prove_full_id}

Recall that our definition for full model identifiability is
$$\PP(\Yb|\bLambda,\Fb) = \PP(\Yb|\tilde\bLambda,\tilde\Fb)\quad\Rightarrow\quad(\bLambda,\Fb) = (\bLambda^{\prime},\Fb^{\prime})\text{ up to a signed permutation}$$ for any parameter pairs $(\bLambda,\Fb) $ and $(\bLambda^{\prime},\Fb^{\prime})$. Let $\eta_{ij} = (\bLambda\Fb^\T)_{[j,i]}$ be the parameters underlying $Y_{ij}$. Define $Q_{ij}(\eta) = \EE_{\eta_{ij}}[l_{ij}(\eta)]$. Then 
\begin{align*}
\partial_{\eta}Q_{ij}(\eta)|_{\eta =\eta_{ij}}&= \mathbb{E}_{\eta_{ij}}\big[\partial_{\eta} l_{ij}(\eta; Y_{ij} )\big]|_{\eta = \eta_{ij}}
\\&= \int \partial_{\eta} \log g_{ij}(Y_{ij} \mid \eta_{ij})\, g_{ij}(y \mid \eta_{ij})\, dy \\
&= \int \partial_{\eta} p(y \mid \eta_{ij})\, dy \\
&= \partial_{\eta} \int p(y \mid \eta_{ij})\, dy  \\
&= 0.
\end{align*}
Therefore, $\eta_{ij}$ is the maximizer of $Q_{ij}(\eta)$. With $l^{\prime\prime}(\eta)<0$ and linearity of expectation,
$Q_{ij}(\eta) $ is also concave on
${\RR}$. Specifically, for any $\eta_1,\eta_2 \in {\RR}$ and
$t \in [0,1]$,
\begin{align*}
  Q_{ij}\bigl(t\eta_1 + (1-t)\eta_2\bigr)
  &= \EE_{\eta_{ij}}\Bigl[
       l_{ij}\bigl(t\eta_1 + (1-t)\eta_2; Y_{ij}\bigr)
     \Bigr] \\
  &\ge \EE_{\eta_{ij}}\Bigl[
         t \,l_{ij}(\eta_1;Y) + (1-t)\,l_{ij}(\eta_2;Y)
         + \frac{b_L(\eta_1,\eta_2)}{2} t(1-t)(\eta_1-\eta_2)^2
       \Bigr] \\
  &= t\,Q_{ij}(\eta_1) + (1-t)\,Q_{ij}(\eta_2)
     + \frac{b_L(\eta_1,\eta_2)}{2} t(1-t)(\eta_1-\eta_2)^2,
\end{align*}
for some $b_L(\eta_1,\eta_2)<0$ depending only on $\eta_1$ and $\eta_2$.
Thus $Q_{ij}$ is concave on $\RR$. Consequently, $\eta_{ij}$ is the unique maximizer. With $\PP(\Yb|\bLambda,\Fb) = \PP(\Yb|\tilde\bLambda,\tilde\Fb)$, we know $g_{ij}(\cdot|\eta_{ij}) = g_{ij}(\cdot|\tilde\eta_{ij})$ almost everywhere, where $\tilde\eta_{ij} = (\tilde\bLambda\tilde\Fb)_{[j,i]}$. Therefore, $Q(\tilde\eta_{ij}) = Q(\eta_{ij})$. By the uniqueness of $\eta_{ij}$, we conclude that $\eta_{ij} = \tilde\eta_{ij}$ for each $i\in[N]$ and $j\in[J]$; that is $\bLambda\Fb^\T = \tilde\bLambda\tilde\Fb^\T$. With  the identifiability results established above, we have proved that the the full model identification is achieved.


\section{Proofs of Theorem~\ref{thm_consis}, Theorem~\ref{thm_uniform_consis}, and Corollary~\ref{prop_consis_ic4}}\label{sec:supp_prove_consistency}
In this section, we derive the theoretical results relating to the consistency in the Main text. Specifically, the proofs for average consistency (Theorem~\ref{thm_consis}) and uniform consistency (Theorem~\ref{thm_uniform_consis}) under different identifiability conditions are presented in Sections~\ref{sec:supp_prove_consistency_ic0}--\ref{supp_prove_consistency_ic5}. The proof for sin-consistency (Corollary~\ref{prop_consis_ic4}) is presented in Section~\ref{supp_prove_sin_consistency}. We start with discussions on the Assumptions~\ref{assumption: psd covariance}--\ref{assump:scaling}, and the technical details in the proof.

\subsection{Preliminaries}\label{supp_sec_ill_hessian}
\subsubsection{Discussion on the Assumptions}\label{supp_sec_ill_hessian_discussion}
Assumption~\ref{assumption: psd covariance} is similar to Assumptions A and B in~\cite{bai2003inferential}, or Assumptions A and C in \cite{bai2012statistical} under linear factor models. In this work, the factors $\bbf_i$ are treated as fixed parameters, consenting to our joint estimation framework. 
     The boundedness assumption for the model parameters is also commonly assumed under nonlinear models~\citep{jennrich1969asymptotic, wu1981asymptotic}. We also restrict the eigenvalues of $\bSigma_{f}^*\bSigma_{\lambda}^*$ to be distinct, which is standard in literature~\citep{bai2003inferential,fan2017sufficient}. 
     Since $D$ is fixed, this assumption guarantees a nonzero eigen-gap for $\bSigma_{f}^* \bSigma_{\lambda}^*$. Adequate separation between the eigenvalues is essential for ensuring that the eigenvectors of $\bSigma_{f}^* \bSigma_{\lambda}^*$ are identifiable and can therefore be stably recovered under small perturbations~\citep{yu2015useful}. { The minimal eigenvalue assumptions can be further relaxed by allowing $\lambda_{\min}(\bSigma_{f}^*)$ and $\lambda_{\min}(\bSigma_{\lambda}^*)$ to depend on $N$ and $J$ and vanish. Specifically, let $\sigma^* = \lambda_{\min}(\bSigma_{f}^*)\wedge \lambda_{\min}(\bSigma_{\lambda}^*)$ and assume that Assumption~\ref{assump:scaling} strengthens to
\begin{equation*}
\log(N\vee J) \left\{ \frac{\epsilon_J \log N + \epsilon_N \log J}{\sigma_*^2 (N\wedge J)} \right\}^{1/2} \to 0,\text{ as }N,J\to\infty. 
\end{equation*}
In this setting, the equivalence argument in the subsequent proof can be similarly derived, and thus we can derive Theorem~\ref{thm_consis} with all convergence rates divided by a factor of $1/(\sigma_*)^2$, and Theorem~\ref{thm_uniform_consis} with the convergence rates divided by a factor of $1/\sigma_*$.}

Assumption~\ref{assumption:smoothness} imposes smoothness requirement on the individual log-likelihood function $l_{ij}(\cdot)$ and is commonly considered in the literature \citep{bai2003inferential,Wang2018MaximumLE}. The sub-exponential assumption for $l_{ij}^{\prime}(\eta^*)$ is crucial for deriving the non-asymptotic error bound for the estimators.
The rest of Assumption~\ref{assumption:smoothness} guarantees the concavity of $l_{ij}(\eta^*_{ij})$ when the true parameters are inside a compact set. It can be easily verified that a wide range of generalized latent factor models satisfy this assumption, including linear models, logistic models, Probit models, and Poisson models.

Assumption~\ref{assump:scaling} is a mild condition and allows many combinations of scaling for $N,J$. Here $\epsilon_N$ and $\epsilon_J$ are assumed to slowly diverge to infinity with $N$ and $J$, respectively, controlling the tail probability of $l_\infty$~error of the estimation (see Theorem~\ref{thm_uniform_consis}). In particular, $\epsilon_N$ $(\epsilon_J)$ can diverge at an arbitrary slow rate, e.g., $\log N$ or $\log \log N$ $(\log J$ or $\log \log J)$. Moreover, when $l_{ij}^{\prime}(\eta)$ is bounded, as in logistic and probit models, the conditions simplify to $N\epsilon_N^{-1}\gg \log J$ and $J\epsilon_J^{-1}\gg \log N$ for some slowly diverging $\epsilon_N$ and $\epsilon_J$.

\subsubsection{Challenges in Constrained Estimation}\label{supp_sec_ill_hessian_challenge}
In this part, we present a significant challenge in the direct analysis of constrained joint maximum likelihood estimation. A central result in our analysis is Lemma~\ref{lemma_hessian_property}, which summarizes the critical properties of the Hessian matrix within a neighborhood of $\btheta^{(0)}$. This lemma establishes, through (i), that when augmented with a small-order term constructed under IC$^{(0)}$, the scaled Hessian matrix $\Hb_L$ is strictly positive definite with the addition of an augmentation matrix $\Hb_P$ in a small region around $\btheta^{(0)}$. Furthermore, properties (viii) and (ix) demonstrate that its inverse can be effectively approximated by considering only the diagonal blocks. These results are non-trivial and require intricate analysis of the Hessian structure and the imposed identifiability constraints. Notably, the construction of such an augmentation to the Hessian matrix is unique to the specified identifiability conditions and cannot be generalized to other scenarios. In the following, we demonstrate that, in the case of a one-dimensional factor model, the Hessian matrix becomes inherently ill-conditioned at $\btheta^{(5)}$.
In this case $\Hb_L(\btheta)$ can be written as:
\begin{equation*}
    \Hb_L(\btheta) = \begin{pmatrix}
        \Hb_{L\lambda\lambda}(\btheta)&\Hb_{L\lambda f}(\btheta)\\\Hb_{f\lambda}(\btheta)&\Hb_{Lff}(\btheta)
    \end{pmatrix},
\end{equation*}
where $\Hb_{L\lambda\lambda}(\btheta)$ and $\Hb_{Lff}(\btheta)$ are diagonal with
\begin{equation*}
    \big[\Hb_{L\lambda\lambda}(\btheta)\big]_{[j,j]} = -\sum_{i=1}^N l_{ij}^{\prime\prime}(\eta_{ij}) f_i^2\text{, for }j\in[J],\; \big[\Hb_{Lff}(\btheta)\big]_{[i,i]} = -\sum_{j=1}^J l_{ij}^{\prime\prime}(\eta_{ij}) \lambda_j^2\text{, for }i\in[N],
\end{equation*}
and $\Hb_{L\lambda f}(\btheta)=\Hb_{Lf\lambda}(\btheta)$ with 
\begin{equation*}
    \big[\Hb_{L\lambda f}(\btheta)\big]_{[j,i]} = - l_{ij}^{\prime\prime}(\eta_{ij}) \lambda_jf_i-l_{ij}^{\prime}(\eta_{ij})\text{, for }i\in[N]\text{ and }j\in[J].
\end{equation*}
Define
\begin{equation*}
    \Db_m = \begin{pmatrix}
        N\Ib_{J}&\zero\\\zero&J\Ib_N
    \end{pmatrix}.
\end{equation*}
Here we use $\Db_m$ to scale the Hessian matrix such that the diagonal blocks of \\$\Db_m^{-1/2}\Hb_{L}(\btheta)\Db_m^{-1/2}$ can be bounded by some constant. Let
\begin{equation}
    \bv = \big(J^{-1/2}\lambda_1^{(5)},\cdots,J^{-1/2}\lambda_J^{(5)},-N^{-1/2}f_1^{(5)},\cdots,-N^{-1/2}f_N^{(5)}\big)^\T.
\end{equation}
We have $\|\bv\| \le D$ and 
\begin{align*}
    \big[\Db_m^{-1/2}\Hb_L(\btheta^{(5)})\Db_m^{-1/2}\bv\big]_{[j]} = (NJ)^{-1/2} \frac{1}{\sqrt N}\sum_{i=1}^N l_{ij}^{\prime}(\eta_{ij}^*) f_i^{(5)}\text{, for }j=1,\cdots, J;\\\big[\Db_m^{-1/2}\Hb_L(\btheta^{(5)})\Db_m^{-1/2}\bv\big]_{[J+i]} = -(NJ)^{-1/2} \frac{1}{\sqrt J}\sum_{j=1}^J l_{ij}^{\prime}(\eta_{ij}^*) \lambda_j^{(5)}\text{, for }i= 1,\cdots, N.
\end{align*}
Since $l_{ij}^{\prime}(\eta_{ij}^*)$ is sub-exponential, we know $\big|\sum_{i=1}^N l_{ij}^{\prime}(\eta_{ij}^*) f_i^{(5)}\big| = O_p(\sqrt{N})$, $\big|\sum_{j=1}^J l_{ij}^{\prime}(\eta_{ij}^*) \lambda_j^{(5)}\big|=O_p(\sqrt{J})$ and $\big|\sum_{i=1}^N\sum_{j=1}^J l_{ij}^{\prime}(\eta_{ij}^*) f_i^{(5)} \lambda_j^{(5)}\big|=O_p\big((NJ)^{1/2}\big)$ by the Bernstein-type probability bound.  This implies that 
\begin{equation*}
    \left|\bv^\T \Db_m^{-1/2}\Hb_L(\btheta^{(5)})\Db_m^{-1/2}\bv\right| = 2(NJ)^{-1}\left|\sum_{i=1}^N\sum_{j=1}^J l_{ij}^{\prime}(\eta_{ij}^*) f_i^{(5)} \lambda_j^{(5)}\right| = O_p\big((NJ)^{-1/2}\big).
\end{equation*}
Under IC$^{(5)}$ in the one-dimensional case, $\lambda_1^{(5)}$ is fixed as $1$. In this case, the Hessian matrix of the constrained negative joint log-likelihood can be expressed as $[\Hb_L(\btheta)]_{[2:N+J,2:N+J]}$. However, we note that by Assumptions~\ref{assumption: psd covariance} and \ref{assumption:smoothness},
\begin{align*}
    &\,\Big|\bv_{[2:N+J]}^\T[\Db_m^{-1/2}\Hb_L(\btheta^{(5)})\Db_m^{-1/2}]_{[2:N+J,2:N+J]}\bv_{[2:N+J]}\Big|\\ \le &\,\Big|\bv^\T\Db_m^{-1/2}\Hb_L(\btheta^{(5)})\Db_m^{-1/2}\bv\Big|\\ &\,+ \frac{1}{NJ}\Big|\sum_{i=1}^Nl_{i1}^{\prime\prime}(\eta_{i1}^*)\big(f_i^{(5)}\lambda_1^{(5)}\big)^2\Big| + \frac{2}{NJ}\Big|\sum_{i=1}^N\big\{l_{i1}^{\prime}(\eta_{i1}^*)-l_{i1}^{\prime\prime}(\eta_{i1}^*)\lambda_1^{(5)}f_i^{(5)}\big\}\lambda_1^{(5)}f_i^{(5)}\Big|\\ \le  &\, O_p\big((NJ)^{-1/2} + J^{-1}\big).
\end{align*}
Then we conclude that the Hessian matrix in this case is ill-posed:
\begin{equation}
    \lambda_{\min}\big\{[\Db_m^{-1/2}\Hb_L(\btheta^{(5)})\Db_m^{-1/2}]_{[2:N+J,2:N+J]}\big\} \le O_p\Big(\frac{1}{\sqrt{J(N\wedge J)}}\Big).\label{eq_first_equation}
\end{equation}
Nonetheless, consistency and limiting distributions of our estimation are still valid. This is because there are only a fixed number of diminishing eigenvalues and the eigenvectors corresponding to the eigenvalues become asymptotically orthogonal to all model parameters. Specifically, the projection of $\bv_{[2:N+J]}$ onto $\lambda_{j}$ is of order $O(J^{-1/2})$, and its projection onto $f_i$ is of order $O(N^{-1/2})$.
As a result, the effects of the ill-posedness are uniformly distributed across an increasing number of parameters, which grows to infinity. To navigate this conundrum, we adopt an alternative approach to derive our asymptotic results. In particular, we will show in Section~\ref{sec:supp_prove_consistency_ic1} that the estimations of constrained MLE under different identifiability constraints can be obtained from each other, enabling the derivation of our theoretical results under different identifiability conditions.

\subsubsection{Overview of the Technical Details}\label{supp_sec_ill_hessian_discussion_technical}
{To address the challenge present in Section~\ref{supp_sec_ill_hessian_challenge}, we employ the guarantees established for the estimator under IC$^{(0)}$, and then devise a mapping that carries this estimator $(\hat\bLambda^{(0)},\hat\Fb^{(0)})$ into the maximum likelihood estimation under every other identifiability condition. While the mapping can be constructed from the identifiability condition, proving the equivalence between the original constrained estimator and the mapped estimator from $(\hat\bLambda^{(0)},\hat\Fb^{(0)})$ is technically involved. Specifically, in Section~\ref{sec:supp_prove_consistency_ic0}, we first show that the constrained estimator $(\hat\bLambda^{(0)},\hat\Fb^{(0)})$ is a strict interior point of the feasible set $\{(\bLambda,\Fb):\|\bLambda\|_{\max}\le D,\|\Fb\|_{\max}\le D\}$ for \eqref{eq_joint_mle}, which in turn allows us to derive the first-order conditions for the constrained estimators (see step 2 in Section~\ref{sec:supp_prove_consistency_ic0}). In Section~\ref{sec:supp_prove_consistency_ic1}, we then derive sharp entrywise error bounds for $(\hat\bLambda^{(0)},\hat\Fb^{(0)})$. Taken together, we can further show that, with probability approaching 1, the mapped estimator coincides with the original constrained estimator under each identifiability condition, and we establish our theoretical results based on this equivalence.}


Furthermore, we remark that the difference in our non-asymptotic bounds reflects the intrinsic difficulty of the problem, not a limitation of the proof technique. First, the $l_2$ error bound for the estimation under IC$^{(1)}$--IC$^{(5)}$ includes a term related to $\epsilon_N$, diverging to infinity at an arbitrarily slow rate as $N\to\infty$ (e.g., $\log N$ or $\log\log N$), which is needed to bound the additional errors under the more complex non-orthogonal designs. Second, under IC$^{(1)}$–IC$^{(5)}$, the convergence rate for the factor estimates cannot be better than $O_p(N^{-1}+J^{-1})$, as Theorem \ref{main_thm_factor_asym} below shows that their asymptotic variances are also of that order, $O_p(N^{-1}+J^{-1})$. This is also demonstrated in the simulation results for the factors under IC$^{(4)}$ and IC$^{(5)}$ in Figures~\ref{fig:fcover_ic_1} and~\ref{fig:fcover_ic_2} in the Supplementary Material.

   \subsection[Consistency under IC$^{(0)}$]{Consistency under IC$^{(0)}$}\label{sec:supp_prove_consistency_ic0}
The proof consists of three steps:
\begin{itemize}
    \item\textbf{Step 1.} Let 
    \begin{align*}
        \cE_1(C) =&\, \left\{J^{-1}\sum_{j=1}^J\big\|\hat\blambda_j^{(0)}-\blambda_j^{(0)}\big\|^2\le C(N\wedge J)^{-1}\right.\\&\qquad\left.\text{ and }{N}^{-1}\sum_{i=1}^N\big\|\hat\bbf_i^{(0)}-\bbf_i^{(0)}\big\|^2\le C(N\wedge J)^{-1}\right\}.
    \end{align*}
    We prove that for any $\delta>0$, there exists some constant $C_{\delta}$ depending only on $\delta$ such that $P(\cE_1(C_{\delta}))>1-(NJ)^{\delta} - (N+J)^{-\delta}$. This rate does not match the rate in Theorem~\ref{thm_consis}, but we can therefore focus on parameters in $\{\btheta:J^{-1}\sum_{j=1}^J\big\|\blambda_j-\blambda_j^{(0)}\big\|^2+{N}^{-1}\sum_{i=1}^N\big\|\bbf_i-\bbf_i^{(0)}\big\|^2\le c_{NJ}\}$ for some diminishing $c_{NJ}$ that will be specified later.
\item\textbf{Step 2.} We derive the first order optimality condition for constrained  MLE under IC$^{(0)}$. In particular, we show that, conditioned on $\cE_1(C_{\delta})$ in step 1, \begin{equation*}
         P\big(\cE_S\mid\cE_1(C_{\delta})\big)\ge 1-N^{-\delta}-J^{-\delta}- 5(NJ)^{-\delta} - 5(N+J)^{-\delta}-2\exp(-C\epsilon_N)-2\exp(-C\epsilon_J),
     \end{equation*}
     where $\cE_S=\big\{\bS_L(\hat\btheta^{(0)})=\zero\big\}$. 
\item \textbf{Step 3.} We prove the average (Theorem~\ref{thm_consis}) and uniform consistency results (Theorem~\ref{thm_uniform_consis}) by expanding $\bS_L(\hat\btheta^{(0)})=\zero$. 
\end{itemize}
\noindent\textbf{Step 1.} 
In this step, we will use the following lemma, the proof of which is left to Section~\ref{prove_lemma_concentration_l_mat}.
\begin{lemma}\it
    Let $\Lb_{\eta} = \big\{l_{ij}^{\prime}(\eta_{ij}^*)\big\}_{N\times J}$ be an $N\times J$ matrix with the $(i,j)$th entry being $l_{ij}^{\prime}(\eta_{ij}^*)$. Denote its $i$th row and $j$th column as $\Lb_{\eta,i} = [\Lb_{\eta}]_{[i,]}$ and $\Lb_{\eta,j} = [\Lb_{\eta}]_{[,j]}$, respectively. Under Assumption~\ref{assumption:smoothness}, for any constant $\delta>0$, we have 
    \begin{enumerate}[label = (\roman*)]
    \item with probability at least $1-J^{-\delta}$, for any $i\in[N]$,
    \begin{equation*}\big\|[\Lb_{\eta}]_{[i,]}\big\|\le C_{\delta}\sqrt{J},
    \end{equation*}
    and with probability at least $1-(NJ)^{-\delta}$,
    \begin{equation*}\max_{1\le i\le N}\big\|[\Lb_{\eta}]_{[i,]}\big\|\le C_{\delta}\sqrt{J\log N},
    \end{equation*}
    where $C_{\delta}$ is some constant only depending on $\delta$;
    \item with probability at least $1-N^{-\delta}$, for any $j\in[J]$,
    \begin{equation*}\big\|[\Lb_{\eta}]_{[,j]}\big\|\le C_{\delta}\sqrt{N},
    \end{equation*}
    and with probability at least $1-(NJ)^{-\delta}$,
    \begin{equation*}\max_{1\le j\le J}\big\|[\Lb_{\eta}]_{[,j]}\big\|\le C_{\delta}\sqrt{N\log J},
    \end{equation*}
    where $C_{\delta}$ is some constant only depending on $\delta$;
        \item with probability at least $1- (N+J)^{-\delta} - (NJ)^{-\delta}$,
    \begin{equation*}\big\|\Lb_{\eta}\big\|\le C_{\delta}\sqrt{N\vee J}.
    \end{equation*}
    where $C_{\delta}$ is some constant only depending on $\delta$.
    \end{enumerate}
     
    \label{lemma_concentration_l_mat}
\end{lemma}
Note that by definition we have $\sum_{i=1}^N\sum_{j=1}^Jl_{ij}\big(\eta_{ij}^*\big)=\sum_{i=1}^N\sum_{j=1}^Jl_{ij}\big((\blambda_j^{(0)})^\T\bbf_i^{(0)}\big)\le$ \\$\sum_{i=1}^N\sum_{j=1}^Jl_{ij}\big((\hat\blambda_j^{(0)})^\T\hat\bbf_i^{(0)}\big)$. Write $\hat\eta_{ij}=(\hat\blambda_j^{(0)})^\T\hat\bbf_i^{(0)}$. Expand each individual log-likelihood to the true parameters to get
    \begin{align*}
        \sum_{i=1}^N\sum_{j=1}^Jl_{ij}\big(\eta_{ij}^*\big)\le\sum_{i=1}^N\sum_{j=1}^Jl_{ij}\big(\eta_{ij}^*\big)+\sum_{i=1}^N\sum_{j=1}^Jl_{ij}^{\prime}(\eta_{ij}^*)(\hat\eta_{ij}-\eta_{ij}^*)+\sum_{i=1}^N\sum_{j=1}^Jl_{ij}^{\prime\prime}(\tilde\eta_{ij})(\hat\eta_{ij}-\eta_{ij}^*)^2.
    \end{align*}
    Here $\tilde\eta_{ij}$ is between $\eta_{ij}^*$ and $\hat\eta_{ij}$, and therefore $l_{ij}^{\prime\prime}(\tilde\eta_{ij})\le -b_L$ by Assumption~\ref{assumption:smoothness}.
By Assumption~\ref{assumption:smoothness} and (iii) of Lemma~\ref{lemma_concentration_l_mat},
    \begin{align*}
        \sum_{i=1}^N\sum_{j=1}^J&\,l_{ij}^{\prime}(\eta_{ij}^*)(\hat\eta_{ij}-\eta_{ij}^*)\\=&\,\mathrm{Trace}\left[\big\{l_{ij}^{\prime}(\eta_{ij}^*)\big\}_{N\times J}\big\{\hat\bLambda^{(0)}\big(\hat\Fb^{(0)}\big)^\T-\bLambda^{(0)}\big(\Fb^{(0)}\big)^\T\big\}\right]\\\le &\,\mathrm{rank}\big[\hat\bLambda^{(0)}\big(\hat\Fb^{(0)}\big)^\T-\bLambda^{(0)}\big(\Fb^{(0)}\big)^\T\big]\big\|\hat\bLambda^{(0)}\big(\hat\Fb^{(0)}\big)^\T-\bLambda^{(0)}\big(\Fb^{(0)}\big)^\T\big\|_F\big\|\big\{l_{ij}^{\prime}(\eta_{ij}^*)\big\}_{N\times J}\big\|\\\le &\,2C_{\delta}K\sqrt{N\vee J}{\Big(\sum_{i=1}^N\sum_{j=1}^J(\hat\eta_{ij}-\eta_{ij}^*)^2\Big)^{1/2}},
    \end{align*}
    with probability at least $1-(NJ)^{-\delta} - (N+J)^{-\delta}$ where $C_\delta$ is a constant that depends only on $\delta$ but not on $N$ or $J$. In the following, we will omit the dependence on $\delta$ and write `$\lesssim$' instead for simplicity.
    Here for the second inequality we used $\mathrm{Trace}(\Ab\Bb^\T) \le \mathrm{rank}(\Bb)\|\Ab\|\|\Bb\|_F$.
    Combine these altogether and we have, with probability at least $1-(NJ)^{-\delta} - (N+J)^{-\delta}$,
    \begin{equation}
        \big\|\hat\bLambda^{(0)}\big(\hat\Fb^{(0)}\big)^\T-\bLambda^{(0)}\big(\Fb^{(0)}\big)^\T\big\|_F=\Big(\sum_{i=1}^N\sum_{j=1}^J(\hat\eta_{ij}-\eta_{ij}^*)^2\Big)^{1/2}\lesssim\sqrt{N\vee J}.
    \end{equation}
    
    If we write the singular value decomposition of $(NJ)^{-1/2}\bLambda^{(0)}\big(\Fb^{(0)}\big)^\T$ as $\Ub^*\bUpsilon^*(\Vb^*)^\T$ and $(NJ)^{-1/2}\hat\bLambda^{(0)}\big(\hat\Fb^{(0)}\big)^\T$ as $\hat\Ub^*\hat\bUpsilon^*(\hat\Vb^*)^\T$. Then by the identifiability condition, know $\bLambda^{(0)}=\sqrt{J}\Ub^*\bUpsilon^*$, $\Fb^{(0)}=\sqrt N\Vb^*$, $\hat\bLambda^{(0)}=\sqrt J\hat\Ub^*\hat\bUpsilon^*$ and $\hat\Fb^{(0)}=\sqrt N\hat\Vb^*$ up to a signed permutation. Since $\bSigma_{f}^*\bSigma_{\lambda}^*$ has distinct eigenvalues, the diagonals of $\bUpsilon^*$ are different for large $N,J$. By \cite{yu2015useful} we know, with probability at least $1-(NJ)^{-\delta} - (N+J)^{-\delta}$,
    \begin{equation*}
        \big\|\hat\Ub^*-\Ub^*\big\|\lesssim \frac{1}{\sqrt{N\wedge J}}, \;\big\|\hat\Vb^*-\Vb^*\big\|\lesssim \frac{1}{\sqrt{N\wedge J}},\text{ and }\big\|\hat\bUpsilon^*-\bUpsilon^*\big\|\lesssim \frac{1}{\sqrt{N\wedge J}}.
    \end{equation*}
    Therefore we have $N^{-1/2}\|\hat\Fb^{(0)}-\Fb^{(0)}\|\lesssim (N\wedge J)^{-1/2}$ and 
    \begin{align*}
        J^{-1/2}\|\hat\bLambda^{(0)}-\bLambda^{(0)}\|&\,\le \big\|\hat\Ub^*-\Ub^*\big\|\big\|\hat\bUpsilon^*-\bUpsilon^*\big\|+\big\|\Ub^*\big\|\big\|\hat\bUpsilon^*-\bUpsilon^*\big\|+\big\|\hat\Ub^*-\Ub^*\big\|\big\|\bUpsilon^*\big\|\\&\,\lesssim \frac{1}{\sqrt{N\wedge J}},
    \end{align*}
     with probability at least $1-(NJ)^{-\delta} - (N+J)^{-\delta}$.
     
\noindent\textbf{Step 2.}
Suppose for the rest of the proof that $\cE_1(C)$ holds for some constant $C>0$, where we omit the dependence on $\delta$ for simplicity. Define \begin{equation*}
    \Db_{m}=\begin{pmatrix}
            N\Ib_{JK}&\zero\\\zero&J\Ib_{NK}
        \end{pmatrix}.
\end{equation*}
Define $\Hb_P$ as 
\begin{equation*}
    \Hb_P(\btheta) = cNJ\begin{pmatrix}
        \sum_{l=1}^K\sum_{h>l}^K\bv_{\lambda,lh}(\btheta)\bv_{\lambda,lh}(\btheta)^\T&\zero\\\zero&\sum_{l=1}^K\sum_{h\le l}\bv_{f,lh}(\btheta)\bv_{f,lh}(\btheta)^\T
    \end{pmatrix},
\end{equation*}
where $$\bv_{f,rr}(\btheta)=2N^{-1}\left(f_{1r}\be_r^\T,\cdots,f_{Nr}\be_r^\T\right)^\T\text{ for }r\in[K];$$
$$\bv_{f,lh}(\btheta)=\left(N^{-1}\big(f_{1l}\be_h+f_{1h}\be_l\big)^\T,\cdots,N^{-1}\big(f_{Nl}\be_h+f_{Nh}\be_l\big)^\T\right)^\T\text{ for }1\le h\le l\le K;$$
$$\bv_{\lambda,lh}(\btheta)=\left(J^{-1}\big(\lambda_{1l}\be_h+\lambda_{1h}\be_l\big)^\T,\cdots,J^{-1}\big(\lambda_{Jl}\be_h+\lambda_{Jh}\be_l\big)^\T\right)^\T\text{ for }1\le l<h\le K.$$
Next, we present two useful lemmas for the second-order derivatives and first-order derivatives for the log-likelihood function, respectively. Define $\bS_L^* = \bS_{L}(\btheta^{(0)})$. First, we show that by augmenting $\Hb_P$ to the Hessian matrix $\Hb_L = -\partial_{\btheta\btheta}^2L(\btheta)$, the summation is strictly positive definite and its inverse can be properly bounded. Specifically, we have the following lemma, whose proof is left to Section~\ref{supp_sec_proof_hessian_property_1}.
\begin{lemma}\label{lemma_hessian_property}\it
Fix a $\tilde\btheta\in\cB(D)$ with $N^{-1}\sum_{i=1}^N\|\tilde\bbf_i-\bbf_i^{(0)}\|^2+J^{-1}\sum_{j=1}^J\|\tilde\blambda_j-\blambda_j^{(0)}\|^2\le c_{NJ}$ for some $c_{NJ} \to 0$ when $N,J\to\infty$. Let 
\begin{equation*}
    \tilde\Hb(\tilde\btheta)=\int_0^1\Big\{\Hb_L\big(\btheta^{(0)}+s(\tilde\btheta-\btheta^{(0)})\big)+c\Hb_P\big(\btheta^{(0)}+s(\tilde\btheta-\btheta^{(0)})\big)\Big\}ds,
\end{equation*}
with $\Hb_L=-\partial^2_{\btheta\btheta}L(\btheta)$ and $c$ being some positive constant. 
    Suppose Assumptions~\ref{assumption: psd covariance} and \ref{assumption:smoothness} hold. Then, there exists a positive constant $c$ that such for any constant $\delta>0$ and some constant $C_\delta$ that depends only on $\delta$, when $N$ and $J$ are large enough,
    \begin{enumerate}[label = (\roman*)]
        \item $\PP\Big\{\lambda_{\min}(\Db_{m}^{-1/2}\tilde\Hb(\tilde\btheta)\Db_{m}^{-1/2})>\gamma\Big\}\ge1 - (N+J)^{-\delta} - (NJ)^{-\delta}$ with $\gamma = Cb_L\min\{\lambda_{\min}(\bSigma_f^*),$ $\lambda_{\min}(\bSigma_\lambda^*)\}$ for some constant $C$;
    \item  with probability at least $1 - 2(NJ)^{-\delta} - (N+J)^{-\delta}$, $\big\|[\{\tilde\Hb(\tilde\btheta)\}^{-1}\Db_m]_{[J_K,]}\big\|_{\infty}\le C_{\delta}\log J$;
    \item  with probability at least $1 - 2(NJ)^{-\delta} - (N+J)^{-\delta}$, $\big\|[\{\tilde\Hb(\tilde\btheta)\}^{-1}\Db_m]_{[JK+N_K,]}\big\|_{\infty}\le C_{\delta}\log N$.
    \item with probability at least $1-2N^{-\delta}-J^{-\delta}-6(NJ)^{-\delta} - 2(N+J)^{-\delta}$,\\
    $\big\|\big[\{\tilde\Hb(\tilde\btheta)\}^{-1}\big]_{[J_K,J_K]}\big\|\le C_{\delta} N^{-1}$; $\big\|\big[\{\tilde\Hb(\tilde\btheta)\}^{-1}\big]_{[J_K,JK+N_K]}\big\|\le C_{\delta} (NJ)^{-1/2}$;
    \\ $\big\|\big[\{\tilde\Hb(\tilde\btheta)\}^{-1}\bS_{L}^*\big]_{[J_K,]}\big\|\le C_{\delta}\left( \sqrt{J/N}+\sqrt{c_{NJ}}+\sqrt{\log (NJ)/N} + 1/\sqrt{J}\right)$.
    \item with probability at least $1-\exp(-C\epsilon_N) - 2N^{-\delta}-2J^{-\delta} - 6(NJ)^{-\delta} - 2(N+J)^{-\delta}$\\
    $\big\|\big[\{\tilde\Hb(\tilde\btheta)\}^{-1}\bS_{L}^*\big]_{[(j-1)K+1:jK,]}\big\|\le C_{\delta}\left( \sqrt{\epsilon_N/N}+\sqrt{c_{NJ}/J}+\sqrt{\log (NJ)/(NJ)} + 1/{J}\right)$.
    \item with probability at least $1-N^{-\delta}-2J^{-\delta}-6(NJ)^{-\delta} - 2(N+J)^{-\delta}$,\\
    $\big\|\big[\{\tilde\Hb(\tilde\btheta)\}^{-1}\big]_{[JK+N_K,J_K]}\big\|\le C_{\delta}(NJ)^{-1/2}$; $\big\|\big[\{\tilde\Hb(\tilde\btheta)\}^{-1}\big]_{[JK+N_K,JK+N_K]}\big\|\le C_{\delta}J^{-1}$;\\
    $\big\|\big[\{\tilde\Hb(\tilde\btheta)\}^{-1}\bS_{L}^*\big]_{[JK+N_K,]}\big\|\le C_{\delta}\left( \sqrt{N/J}+\sqrt{c_{NJ}}+\sqrt{\log (NJ)/J} + 1/\sqrt{N}\right)$. 
    \item with probability at least $1-\exp(-C\epsilon_J) - 2N^{-\delta}-2J^{-\delta} - 6(NJ)^{-\delta} - 2(N+J)^{-\delta}$\\
    $\big\|\big[\{\tilde\Hb(\tilde\btheta)\}^{-1}\bS_{L}^*\big]_{[JK+(i-1)K+1:JK+iK,]}\big\|\le C_{\delta}\left(\sqrt{\epsilon_J/J}+\sqrt{c_{NJ}/N}+\sqrt{\log (NJ)/(NJ)} + 1/{N}\right)$.
    \end{enumerate}
    Furthermore, write 
\begin{equation*}\Hb_{L\lambda\lambda}(\tilde\btheta) = \big[\Hb_L(\tilde\btheta)\big]_{[J_K,J_K]} ,\quad \Hb_{Lff}(\tilde\btheta) = \big[\Hb_L(\tilde\btheta)\big]_{[JK+N_K,JK+N_K]} .\end{equation*}
    \begin{enumerate}[label = (\roman*)]\setcounter{enumi}{7}
        \item with probability at least $1 - N^{-\delta} - 3(NJ)^{-\delta} - (N+J)^{-\delta}$, for any $j\in[J]$, \begin{equation*}\Big\|\big[\Db_m^{1/2}\{\tilde\Hb(\tilde\btheta)\}^{-1}\Db_m^{1/2}\big]_{[(j-1)K+1:jK,]} - N\big[\{\Hb_{L\lambda\lambda}(\tilde\btheta)\}^{-1}\big]_{[(j-1)K+1:jK,]}\Big\|\le C_{\delta} J^{-1/2};\end{equation*} 
    \item with probability at least $1 - J^{-\delta} - 3(NJ)^{-\delta} - (N+J)^{-\delta}$, for any $i\in[N]$, we have\begin{equation*}
        \Big\|\big[\Db_m^{1/2}\{\tilde\Hb(\tilde\btheta)\}^{-1}\Db_m^{1/2}\big]_{[JK+(i-1)K+1:JK+iK,]} - J[\{\Hb_{Lff}(\tilde\btheta)\}^{-1}]_{[(i-1)K+1:iK,]}\Big\|\le  C_{\delta}N^{-1/2};\end{equation*}
    \end{enumerate}
\end{lemma}
\begin{remark} \it
    This lemma summarizes key properties of the negative joint log-likelihood function at a local regime of $\btheta^{(0)}$. In particular, we showed by (i) that with penalty $P$, the scaled Hessian matrix is strictly positive definite, ensuring local invertibility. Then (ii)--(ix) are used to carve the inverse of the Hessian matrix in the measure of the spectral norm and $l_{\infty}$ norm. Note that the penalty $P$ is different from the Lagrange multiplier, where it takes the square of the Frobenius norm for theoretical analysis.
    
    This set of properties stands in contrast to the ill-posed Hessian matrix discussed in Section~\ref{supp_sec_ill_hessian}. We attribute this refined behavior to the fact that IC$^{~(0)}$ induces proper directions that maintain non-negligible overlapping with the eigenvectors of the Hessian of the negative joint log-likelihood function.
\end{remark}
Lemma~\ref{lemma_hessian_property} trivially leads to the corollaries below.
\begin{corollary}\label{cor_lemma_hessian_property}\it
     Define $P(\btheta) = NJ\big\|N^{-1}\sum_{i=1}^N\bbf_i\bbf_i^\T-\Ib_K\big\|_F^2/2+NJ\big\|\text{ndiag}\big(J^{-1}\sum_{j=1}^J\blambda_j\blambda_j^\T\big)\big\|_F^2/2$ and $\Hb^*(\btheta) = \Hb_L(\btheta) + c\partial^2_{\btheta\btheta}P(\btheta)$.
    Under all conditions in Lemma~\ref{lemma_hessian_property}, (i)--(ix) hold when replacing $\tilde\Hb(\tilde\btheta)$ with $\tilde\Hb^*(\tilde\btheta) :=\int_0^1 \Hb^*\big\{\btheta^{(0)} + s(\tilde\btheta-\btheta^*)\big\}ds$.
\end{corollary}
\begin{corollary}\label{cor_lemma_hessian_property2}\it
    Under all conditions in Lemma~\ref{lemma_hessian_property}, (i)--(ix) hold when replacing $\tilde\Hb(\tilde\btheta)$ with $\Hb(\tilde\btheta) = \Hb_L(\tilde\btheta) + \Hb_P(\tilde\btheta)$.
\end{corollary}
\begin{corollary}
    \label{cor_lemma_hessian_property_bound_l_prime}\it
    When $l_{ij}^{\prime}(\eta)$ is bounded, all the above results hold, and $\log J$ in (ii) and $\log N$ in (iii) can be replaced by a constant not depending on $N$ or $J$.
\end{corollary}
The proofs of the corollaries are left to Sections~\ref{supp_sec_prove_cor_lemma_hessian_property}-- \ref{supp_sec_prove_cor_lemma_hessian_property_bound_l_prime}. For simplicity, we refer to statements (i)–(ix) of Corollaries~\ref{lemma_hessian_property} and~\ref{cor_lemma_hessian_property2} as the corresponding statements (i)–(ix) of Lemma~\ref{lemma_hessian_property}, with $\tilde\Hb(\tilde\btheta)$ replaced by $\tilde\Hb^*(\tilde\btheta)$ and $\Hb(\tilde\btheta)$, respectively.

The following lemma provides first-order concentration in the $l_2$-norm and $l_{\infty}$-norm.
\begin{lemma}\it Suppose Assumption~\ref{assumption:smoothness} holds.\begin{enumerate}[label = (\roman*)]
    \item Suppose $\epsilon_{N}\to \infty$ as $N\to\infty$. Then for any fixed sequence $\{a_i\}_{i=1}^N$ with $\max_{1\le i\le N}|a_i|\le A$ and some constant $C$, with probability at least $1-\exp(-C\epsilon_N)$,
    \begin{align*}
        \forall j\in[J]\;\Big| \sum_{i=1}^N a_i l_{ij}^{\prime}(\eta_{ij}^*) \Big|&\, \lesssim \sqrt{\epsilon_NN};\\ \max_{1\le j\le J}\Big| \sum_{i=1}^N a_i l_{ij}^{\prime}(\eta_{ij}^*) \Big|&\, \lesssim \sqrt{\epsilon_NN\log J}.
    \end{align*}
    \item Suppose $\epsilon_{J}\to \infty$ as $J\to\infty$. Then for any fixed sequence $\{a_j\}_{j=1}^J$ with $\max_{1\le j\le J}|a_j|\le A$ and some constant $C$, with probability at least $1-\exp(-C\epsilon_J)$,
    \begin{align*}
        \forall i\in[N],\;\left| \sum_{j=1}^J a_j l_{ij}^{\prime}(\eta_{ij}^*) \right| &\,\lesssim \sqrt{\epsilon_JJ};\\ \max_{1\le i\le N}\left| \sum_{j=1}^J a_j l_{ij}^{\prime}(\eta_{ij}^*) \right| &\,\lesssim \sqrt{\epsilon_JJ\log N}.
    \end{align*}
    \item Suppose $\epsilon_{NJ}\to \infty$ as $N,J\to\infty$. For any fixed sequence $\{a_{ij}\}_{i=1,j=1}^{N,J}$ with $\max_{1\le i\le N,1\le j\le J}|a_{ij}|\le A$ and some constant $C$, with probability at least $1-\exp(-C\epsilon_{NJ})$,
    \begin{align*}
        \left| \sum_{i=1}^N\sum_{j=1}^J a_{ij} l_{ij}^{\prime}(\eta_{ij}^*) \right| &\,\lesssim \sqrt{NJ\epsilon_{NJ}},\\ \max_{1\le i\le N}\max_{1\le j\le J}\left|l_{ij}^{\prime}(\eta_{ij}^*)\right|&\,\lesssim\sqrt{\epsilon_{NJ}\log (NJ)}
    \end{align*}
    \end{enumerate}\label{lemma_concentration}
\end{lemma}
The proof of this lemma follows the Bernstein-type probability bound, and is left to Section~\ref{prove_lemma_concentation}. The following is a corollary of Lemma~\ref{lemma_concentration}, and its proof is left to Section~\ref{sup_sec_prove_coro_first_order_concern}.
\begin{corollary}\label{coro_first_order_concern}\it
    Suppose Assumptions~\ref{assumption: psd covariance} and \ref{assumption:smoothness} hold. On the true parameters $\btheta^{*}$, for any $\delta>0$,
    \begin{enumerate}[label = (\roman*)]
        \item with probability at least $1-J^{-\delta}$,
    \begin{equation*}
        \Big\|N^{-1/2}\big[\bS_L(\btheta^*)\big]_{J_K}\Big\|\le C_{\delta} \sqrt{J},
    \end{equation*}
    where $C_{\delta}$ is some constant only depending on $\delta$;
    \item 
    with probability at least $1- N^{-\delta}$,
    \begin{equation*}
        \Big\|J^{-1/2}\big[\bS_L(\btheta^*)\big]_{JK + N_K}\Big\|\le C_{\delta} \sqrt{N},
    \end{equation*}
    where $C_{\delta}$ is some constant only depending on $\delta$;
    \item with probability at least $1-J^{-\delta} - N^{-\delta}$,
    \begin{equation*}
        \Big\|\Db_m^{-1/2}\bS_L(\btheta^*)\Big\|\le C_{\delta} \sqrt{N+J},
    \end{equation*}
    where $C_{\delta}$ is some constant only depending on $\delta$.
    \end{enumerate}
    Here, we use the superscript $*$ to denote the true parameters, as the above results hold irrespective of the identifiability conditions satisfied by the true parameters.
\end{corollary}

Define $\bS_P(\btheta) = c\partial_{\btheta}P(\btheta)$ for a constant $c$ as specified in Lemma~\ref{lemma_hessian_property} (Corollary~\ref{cor_lemma_hessian_property}). 
Consider the following optimization problem:
\begin{align}
    \mathop{\arg\min}_{\btheta\in\cB(D)}\Big\|&\Db_{m}^{-1/2}\big\{\bS_L(\btheta)+\bS_P(\btheta)\big\}\Big\|_{\infty}^2 \label{eq_joint_mle2}\\
    \text{ subject to }&\,N^{-1}\sum_{i=1}^N\|\bbf_i-\bbf_i^{(0)}\|^2+J^{-1}\sum_{j=1}^J\|\blambda_j-\blambda_j^{(0)}\|^2\le c_{NJ}\nonumber.
\end{align}
Here we take $c_{NJ}\gg \log^2(NJ)/(N\wedge J)$ to ensure that $\hat\btheta^{(0)}$ necessarily lies within this regime.
The solution is denoted by $\tilde\btheta^{(0)}$. Write $\bS(\btheta) = \bS_L(\btheta) + \bS_P(\btheta)$. 
By the mean value theorem, we have 
    \begin{align}
        \bS(\tilde\btheta^{(0)})-\bS(\btheta^{(0)})=\tilde\Hb^*(\tilde\btheta)\big(\tilde\btheta^{(0)}-\btheta^{(0)}\big)\label{eq_mvt},
    \end{align}where $\tilde\Hb^*(\tilde\btheta)$ is as defined in Corollary~\ref{cor_lemma_hessian_property}.
     Multiply both side with $\big\{\tilde\Hb^*(\tilde\btheta)\big\}^{-1}$ and by (ii) of Corollary~\ref{cor_lemma_hessian_property}, with probability at least $1 - 2(NJ)^{-\delta} - (N+J)^{-\delta}$
    \begin{align*}
        \max_{1\le j\le J}\big\|\tilde\blambda_j^{(0)}-\blambda_j^{(0)}\big\|_{\infty}=&\,\Big\|\big[\{\tilde\Hb^*(\tilde\btheta)\}^{-1}\Db_m\Db_{m}^{-1}\big\{\bS(\tilde\btheta^{(0)})-\bS(\btheta^{(0)})\big\}\big]\Big\|_{\infty}\\\le&\, \Big\|[\{\tilde\Hb^*(\tilde\btheta)\}^{-1}\Db_m]_{[J_K,]}\Big\|_{\infty}\Big\|\Db_{m}^{-1}\big\{\bS(\tilde\btheta^{(0)})-\bS(\btheta^{(0)})\big\}\Big\|_{\infty}\\\lesssim &\, \frac{\log J}{\sqrt{N\wedge J}}\Big\|\Db_{m}^{-1/2}\big\{\bS(\tilde\btheta^{(0)})-\bS(\btheta^{(0)})\big\}\Big\|_{\infty}.
    \end{align*} Here we omit the constant $C_\delta$ for simplicity.
    Since that $\|\Db_{m}^{-1/2}\bS_L(\btheta^{(0)})\|_{\infty}\lesssim(\epsilon_N\log J+\epsilon_J\log N)^{1/2}$ with probability at least $1-\exp(-C\epsilon_N)-\exp(-C\epsilon_J)$ by (i) and (ii) of Lemma~\ref{lemma_concentration} and that $\big\|\Db_m^{-1/2}\bS(\tilde\btheta^{(0)})\big\|_{\infty}\le\big\|\Db_m^{-1/2}\bS(\btheta^{(0)})\big\|_{\infty}$ by definition of \eqref{eq_joint_mle2}, we conclude with Assumption~\ref{assump:scaling} that 
    \begin{align}
        \max_{1\le j\le J}\big\|\tilde\blambda_j^{(0)}-\blambda_j^{(0)}\big\|_{\infty}\lesssim \log J\sqrt{\frac{\epsilon_N\log J+\epsilon_J\log N}{N\wedge J}}\to0,\label{eq_infinite_bound_lambda_raw}
    \end{align}
    with probability at least $1 - 2(NJ)^{-\delta} - (N+J)^{-\delta}-\exp(-C\epsilon_N)-\exp(-C\epsilon_J)$ by union bound.
    Similarly, we have
    \begin{equation}
         \max_{1\le i\le N}\big\|\tilde\bbf_i^{(0)}-\bbf_i^{(0)}\big\|_{\infty}\lesssim \log N\sqrt{\frac{\epsilon_N\log J+\epsilon_J\log N}{N\wedge J}}\to0,\label{eq_infinite_bound_f_raw}
    \end{equation}
    with probability at least $1 - 2(NJ)^{-\delta} - (N+J)^{-\delta}-\exp(-C\epsilon_N)-\exp(-C\epsilon_J)$.
 Next multiplying $\Db_m^{1/2}\big\{\tilde\Hb^*(\tilde\btheta)\big\}^{-1}$ from the left to both sides in \eqref{eq_mvt}, we have 
 \begin{align}
        \Db_m^{1/2}\big(\tilde\btheta^{(0)}-\btheta^{(0)}\big) =&\, \big\{\Db_m^{-1/2}\tilde\Hb^*(\tilde\btheta)\Db_m^{-1/2}\big\}^{-1}\Db_m^{-1/2}\Big(\bS(\tilde\btheta^{(0)})-\bS(\btheta^{(0)})\Big).\label{eq_bound_ave_theta_raw}
    \end{align}
    By (i) of Corollary~\ref{cor_lemma_hessian_property}, we know that $\lambda_{\min}\big(\Db_m^{-1/2}\tilde\Hb^*(\tilde\btheta)\Db_m^{-1/2}\big)$ can be lower bounded by some constant $\gamma$ with probability at least $1-(N+J)^{-\delta} - (NJ)^{-\delta}$, we have
    \begin{equation}
        \big\|\Db_m^{1/2}\big(\tilde\btheta^{(0)}-\btheta^{(0)}\big)\big\| \le \gamma^{-1}\Big\|\Db_m^{-1/2}\Big(\bS(\tilde\btheta^{(0)})-\bS(\btheta^{(0)})\Big)\Big\|.\label{eq_bound_ave_theta}
    \end{equation}
 Since that $\|\Db_{m}^{-1/2}\bS_L(\btheta^{(0)})\|\lesssim\sqrt{N+J}$ with probability at least $1-N^{-\delta}-J^{-\delta}$ by Corollary~\ref{coro_first_order_concern} and that $\big\|\Db_m^{-1/2}\bS(\tilde\btheta^{(0)})\big\|\le\sqrt{NK+JK}\big\|\Db_m^{-1/2}\bS(\tilde\btheta^{(0)})\big\|_{\infty}$ by definition of \eqref{eq_joint_mle2},
     we further bound \eqref{eq_bound_ave_theta} as
     \begin{equation*}
         \big\|\Db_m^{1/2}\big(\tilde\btheta^{(0)}-\btheta^{(0)}\big)\big\|^2\lesssim (N+J),
     \end{equation*}
     with probability at least $1-N^{-\delta}-J^{-\delta} - (N+J)^{-\delta} - (NJ)^{-\delta}$, in which case
     \begin{equation}
         \frac{1}{N}\sum_{i=1}^N\big\|\tilde\bbf_i^{(0)}-\bbf_i^{(0)}\big\|^2+\frac{1}{J}\sum_{j=1}^J\big\|\tilde\blambda_j^{(0)}-\blambda_j^{(0)}\big\|^2\lesssim\frac{(\epsilon_N\log J + \epsilon_J\log N)^2}{N\wedge J}\ll c_{NJ}.\label{eq_interior_point_condition_3}
     \end{equation}
     When \eqref{eq_infinite_bound_lambda_raw}, \eqref{eq_infinite_bound_f_raw} and \eqref{eq_interior_point_condition_3} hold, we know $\tilde\btheta^{(0)}$ is an interior point of \eqref{eq_joint_mle2}, that is,
     \begin{align*}
         &\,P\left(\big\{\tilde\btheta^{(0)}\text{ is an interior solution of }\eqref{eq_joint_mle2} \big\}\mid\cE_1(C)\right)\\&\,\ge 1-N^{-\delta}-J^{-\delta}- 5(NJ)^{-\delta} - 5(N+J)^{-\delta}-2\exp(-C\epsilon_N)-2\exp(-C\epsilon_J),
     \end{align*} When $\tilde\btheta^{(0)}$ is an interior solution, we have  $\partial_{\btheta}\big\|\Db_m^{-1/2}\bS(\tilde\btheta^{(0)})\big\|_{\infty}^2=\zero$.
     Therefore by the Hardy-Littlewood derivative to maximal function, we must have $\bS(\tilde\btheta^{(0)})$ $=\zero$. Then $\tilde\btheta^{(0)}$ must maximize the penalized log-likelihood function $L(\btheta) - cP(\btheta)$ inside the region specified by \eqref{eq_joint_mle2}. We claim that $P(\tilde\btheta^{(0)})=0$. Suppose not. Then since $\hat\btheta^{(0)}$ is inside the region under $\cE_1(C)$ when $N,J$ are large enough, $\hat\btheta^{(0)}$ maximizes the likelihood function inside this region among all $\btheta$ that satisfy IC$^{(0)}$. So if $P(\tilde\btheta^{(0)})>0$, by $L(\tilde\btheta^{(0)}) - cP(\tilde\btheta^{(0)})\ge L(\hat\btheta^{(0)}) - cP(\hat\btheta^{(0)}) = L(\hat\btheta^{(0)})$, we know $L(\tilde\btheta^{(0)})>L(\hat\btheta^{(0)})$. Note we have assumed $\big\|\btheta^{(0)}\big\|\le D/2$. Then by \eqref{eq_infinite_bound_lambda_raw} and \eqref{eq_infinite_bound_f_raw}, we know when $N$ and $J$ are large enough, with high probability (specified in \eqref{eq_infinite_bound_lambda_raw} and \eqref{eq_infinite_bound_f_raw}), $\big\|\tilde\btheta^{(0)}\big\|\le 3D/4$.
Next, when $N^{-1}\sum_{i=1}^N\|\tilde\bbf_i^{(0)}-\bbf_i^{(0)}\|^2+J^{-1}\sum_{j=1}^J\|\tilde\blambda_j^{(0)}-\blambda_j^{(0)}\|^2\le \big(\epsilon_N\log J+\epsilon_J\log N\big)^2/(N\wedge J)$, 
\begin{align*}
    \Big\|N^{-1}\sum_{i=1}^N\tilde\bbf_i^{(0)}\big(\tilde\bbf_i^{(0)}\big)^\T-\Ib_K\Big\|^2&\,=\Big\|N^{-1}\sum_{i=1}^N\tilde\bbf_i^{(0)}\big(\tilde\bbf_i^{(0)}\big)^\T-N^{-1}\sum_{i=1}^N\bbf_i^{(0)}\big(\tilde\bbf_i^{(0)}\big)^\T\Big\|^2\\\le &\,2\Big\|N^{-1}\sum_{i=1}^N\bbf_i^{(0)}\big(\tilde\bbf_i^{(0)} - \bbf_i^{(0)}\big)^\T\Big\|^2 \\&\,+ \Big\|N^{-1}\sum_{i=1}^N\big(\tilde\bbf_i^{(0)} - \bbf_i^{(0)}\big)\big(\tilde\bbf_i^{(0)} - \bbf_i^{(0)}\big)^\T\Big\|\\\le &\,2\Big\{N^{-1}\sum_{i=1}^N\|\bbf_i^{(0)}\|^2\Big\}\Big\{N^{-1}\sum_{i=1}^N\big\|\tilde\bbf_i^{(0)} - \bbf_i^{(0)}\big\|^2\Big\} \\&\,+ \Big\|N^{-1}\sum_{i=1}^N\big(\tilde\bbf_i^{(0)} - \bbf_i^{(0)}\big)\big(\tilde\bbf_i^{(0)} - \bbf_i^{(0)}\big)^\T\Big\|\\\lesssim  &\, \frac{(\epsilon_N\log J + \epsilon_J\log N)^2}{N\wedge J}.
\end{align*}
Similarly, $\big\|\text{ndiag}\{J^{-1}\sum_{j=1}^J\tilde\blambda^{(0)}(\tilde\blambda_j^{(0)})^\T\}\big\|^2 \lesssim \big(\epsilon_N\log J+\epsilon_J\log N\big)^2/(N\wedge J)$. Next, we apply the following lemma, whose proof is left to Section~\ref{prove_lemma_for_convergenceG}.
\begin{lemma}\it
    Suppose $\Ab,\Bb\in\mathbb{R}^{K\times K}$ are positive-definite matrices and the eigen-gap of $\Ab\Bb$ is positive. If symmetric matices $\hat\Ab$ and $\hat\Bb $ are consistent estimates of $\Ab$ and $\Bb $ with rate $\nu\to0$.
    \begin{equation*}
        \|\hat\Ab-\Ab\|\lesssim\nu,\quad\|\hat\Bb -\Bb \|\lesssim\nu.
    \end{equation*}
     Suppose $\Gb$ satisfies $
        \Gb^\T\Ab \Gb=\Ib_K$ and $\Gb^{-1}\Bb \Gb^{-\T} $ is diagonal, and correspondingly $\hat \Gb$ satisfies $
         \hat \Gb^\T\hat\Ab\hat \Gb=\Ib_K$ and $\hat \Gb^{-1}\hat\Bb \hat \Gb^{-\T}$ is  diagonal.
    Then
    \begin{equation*}
        \|\hat \Gb- \Gb\|\lesssim \nu.
    \end{equation*}\label{lemma_for_convergenceG}
\end{lemma}
Then we can find $\tilde\Gb^{\delta}$ such that $\|\tilde\Gb^{\delta}-\Ib_K\| \lesssim \big(\epsilon_N\log J+\epsilon_J\log N\big)/\sqrt{N\wedge J}$ and $\tilde\btheta^{\delta} := \{(\tilde\Gb^{\delta})^{-\T}\tilde\blambda_j^{(0)},\tilde\Gb^{\delta}\tilde\bbf_i^{(0)}\}_{i,j}$ satisfy IC$^{(0)}$. Furthermore, we have $$\max_{1\le i\le N}\big\|\tilde\Gb^{\delta}\tilde\bbf_i^{(0)}\big\|_{\infty}\le\max_{1\le i\le N}\big\|\tilde\bbf_i^{(0)}\big\|_{\infty}+\max_{1\le i\le N}\big\|\tilde\Gb^{\delta}-\Ib_K\big\|_{\infty}\big\|\tilde\bbf_i^{(0)}\big\|_{\infty}\le D,$$ $$\max_{1\le j\le J}\big\|(\tilde\Gb^{\delta})^{-\T}\tilde\blambda_j^{(0)}\big\|_{\infty}\le \max_{1\le j\le J}\big\|\tilde\blambda_j^{(0)}\big\|_{\infty}+\max_{1\le j\le J}\big\|(\tilde\Gb^{\delta})^{-\T}-\Ib_K\big\|_{\infty} \big\|\tilde\blambda_j^{(0)}\big\|_{\infty}\le D$$ and 
     \begin{align*}
    &\,N^{-1}\sum_{i=1}^N\big\|\tilde\Gb^{\delta}\tilde\bbf_i^{(0)}-\bbf_i^{(0)}\big\|^2+J^{-1}\sum_{j=1}^J\big\|(\tilde\Gb^{\delta})^{-\T}\tilde\blambda_j^{(0)}-\blambda_j^{(0)}\big\|^2 \\\le &\,N^{-1}\sum_{i=1}^N\big\|\tilde\bbf_i^{(0)}-\bbf_i^{(0)}\big\|^2+N^{-1}\sum_{i=1}^N\big\|\tilde\bbf_i^{(0)}\big\|^2\big\|\tilde\Gb^{\delta}-\Ib_K\big\|^2\\&\,+J^{-1}\sum_{j=1}^J\big\|\tilde\blambda_j^{(0)}-\blambda_j^{(0)}\big\|^2+J^{-1}\sum_{j=1}^J\big\|\tilde\blambda_j^{(0)}\big\|^2\big\|(\tilde\Gb^{\delta})^{-\T}-\Ib_K\big\|^2\\\lesssim &\,  \frac{(\epsilon_N\log J + \epsilon_J\log N)^2}{N\wedge J} \ll c_{NJ},
\end{align*}where $N,J$ are large enough. Here we used $\|(\tilde\Gb^{\delta})^{-1}-\Ib_K\| \le \|(\tilde\Gb^{\delta})^{-1}\|\|\tilde\Gb^{\delta}-\Ib_K\|\le \|\tilde\Gb^{\delta}-\Ib_K\|/(1 - \|\tilde\Gb^{\delta}-\Ib_K\|)\lesssim (\epsilon_N\log J+\epsilon_J\log N)/\sqrt{N\wedge J}$, which can be made arbitrarily small as $N$ and $J\to\infty$.

Therefore $\big(\tilde\Fb^{(0)}(\tilde\Gb^{\delta})^{\T},\tilde\bLambda^{(0)}(\tilde\Gb^{\delta})^{-1}\big)$ is inside the region specified by \eqref{eq_joint_mle2} when  \eqref{eq_infinite_bound_lambda_raw}, \eqref{eq_infinite_bound_f_raw} and \eqref{eq_interior_point_condition_3} hold.
     But 
     \begin{align*}
         &\,-L\big(\tilde\Fb^{(0)}(\tilde\Gb^{\delta})^{\T},\tilde\bLambda^{(0)}(\tilde\Gb^{\delta})^{-1}\big) + cP\big(\tilde\Fb^{(0)}(\tilde\Gb^{\delta})^{\T},\tilde\bLambda^{(0)}(\tilde\Gb^{\delta})^{-1}\big)\\ =&\, -L(\tilde\btheta^{(0)}) 
         \\<&\, -L(\tilde\btheta^{(0)})+cP(\tilde\btheta^{(0)}),
     \end{align*}which contradicts that $\tilde\btheta^{(0)}$ maximizes the penalized log-likelihood function. Therefore $P(\tilde\btheta^{(0)})=0$, and subsequently $\tilde\btheta^{(0)}$ maximizes the likelihood function inside this region among all $\btheta$ that satisfy IC$^{(0)}$. It is easy to see that $P(\hat\btheta^{(0)}) = 0$ implies $\partial_{\btheta}P(\hat\btheta^{(0)}) = 0$.
     Therefore $\tilde\btheta^{(0)}=\hat\btheta^{(0)}$ and \begin{equation}
         \zero = \bS\big(\hat\btheta^{(0)}\big) = \bS_L\big(\hat\btheta^{(0)}\big).\label{eq_first_order_condition}
     \end{equation}
Consequently \begin{equation*}
         P\big(\cE_S\mid\cE_1(C)\big)\ge 1-N^{-\delta}-J^{-\delta}- 5(NJ)^{-\delta} - 5(N+J)^{-\delta}-2\exp(-C\epsilon_N)-2\exp(-C\epsilon_J).
     \end{equation*}
We summarize this result in the following lemma.
\begin{lemma}
    Under Assumptions~\ref{assumption: psd covariance}--\ref{assump:scaling}, for the estimator $\hat\btheta^{(0)}$
 obtained in \eqref{eq_joint_mle} under identifiability condition IC$^{(0)}$, for any constant $\delta>0$, when $N$ and $J$ are large enough, with probability at $1-(N\wedge J)^{-\delta}$, the first order condition holds
 \begin{equation*}
     \bS_{L}(\hat\btheta^{(0)}) = \zero
 \end{equation*}
 \end{lemma}
\noindent{\textbf{Step 3.}} 
Now suppose $\cE_S$ and $\cE_1(C)$ hold. Define 
\begin{equation*}
    \bS_{L\lambda}(\btheta) = \big[\bS_{L}(\btheta)\big]_{J_K},\quad \bS_{Lf}(\btheta) = \big[\bS_{L}(\btheta)\big]_{JK + N_K}.
\end{equation*}
Plugging in $\bS(\hat\btheta^{(0)}) = \zero$ and $\bS_P(\btheta^{(0)})=\zero$ into \eqref{eq_mvt} to get
     \begin{align}
        \hat\btheta^{(0)}-\btheta^{(0)} =&\, -\big\{\tilde\Hb^*(\tilde\btheta)\big\}^{-1}\bS_L(\btheta^{(0)}) .\label{eq_bound_ave_theta_without_sca}
    \end{align}
    By (iv) of Corollary~\ref{cor_lemma_hessian_property}, we know
    \begin{equation*}\Big\|\big[\big\{\tilde\Hb(\tilde\btheta)\big\}^{-1}\bS_L(\btheta^{(0)})\big]_{J_K}\Big\|\lesssim\sqrt{J/N}+\sqrt{c_{NJ}}+\sqrt{\log (NJ)/N} + 1/\sqrt{J},\end{equation*} with probability at least $1-2N^{-\delta}-J^{-\delta} - 6(NJ)^{-\delta} - 2(N+J)^{-\delta}$. Here we omit the constant $C_\delta$ for simplicity.
     Note we have assumed $\log(NJ)/J\to 0$ in Assumption~\ref{assump:scaling}. Plugging the bound to \eqref{eq_bound_ave_theta_without_sca} with $c_{NJ} \asymp (N\wedge J)^{-1}$, with probability at least $1-2N^{-\delta}-J^{-\delta} - 6(NJ)^{-\delta} - 2(N+J)^{-\delta}$, we have
     \begin{equation}
         J^{-1/2}\big\|[\hat\btheta^{(0)}-\btheta^{(0)}]_{J_K}\big\|\le C_{\delta}\left( \frac{1}{\sqrt{N}} + \frac{1}{J}\right).\label{eq_avera_blambda}
     \end{equation}where $C_{\delta}$ is some constant depending only on $\delta$.
     Similarly, with probability at least $1-N^{-\delta}-2J^{-\delta} - 6(NJ)^{-\delta} - 2(N+J)^{-\delta}$, we have
     \begin{equation}
         N^{-1/2}\big\|[\hat\btheta^{(0)}-\btheta^{(0)}]_{JK + N_K}\big\|\le C_{\delta}\left( \frac{1}{\sqrt{J}}+ \frac{1}{N}\right).\label{eq_avera_f}
     \end{equation}where $C_{\delta}$ is some constant depending only on $\delta$.
     Therefore we obtained the average consistency results.
     
     Followed by \eqref{eq_infinite_bound_lambda_raw}, \eqref{eq_infinite_bound_f_raw} and $\hat\btheta^{(0)} = \tilde\btheta^{(0)}$, we obtained that with probability at least $1- 4(NJ)^{-\delta} - 2(N+J)^{-\delta}-2\exp(-C\epsilon_N)-2\exp(-C\epsilon_J)$, 
\begin{align}
        \max_{1\le j\le J}\big\|\hat\blambda_j^{(0)}-\blambda_j^{(0)}\big\|\le C_{\delta}\log J\sqrt{\frac{\epsilon_N\log J+\epsilon_J\log N}{N\wedge J}};\label{eq_max_blambda}\\\max_{1\le i\le N}\big\|\hat\bbf_i^{(0)}-\bbf_i^{(0)}\big\|\le C_{\delta}\log N\sqrt{\frac{\epsilon_N\log J+\epsilon_J\log N}{N\wedge J}},\label{eq_max_f}
        \end{align}where $C_{\delta}$ is some constant depending only on $\delta > 0$. 

Note that tails probabilities for \eqref{eq_avera_blambda}--\eqref{eq_max_f} consist of several terms. Replacing $\delta$ with $\delta^{\prime} - 1$ with $\delta^{\prime} >1$, we see that when $N$ and $J$ are large enough, the coefficients before these terms can be any small. Therefore we obtain the $l_2$ error bounds states in Theorem~\ref{thm_consis} and $l_{\infty}$ error bounds in Theorem~\ref{thm_uniform_consis}.
        
        In the following, we will omit this $C_{\delta}$ for simplicity. Finally, when $l_{ij}^{\prime}(\eta)$ is bounded, by Corollary~\ref{cor_lemma_hessian_property_bound_l_prime}, we know the right sides of \eqref{eq_max_blambda} and \eqref{eq_max_f} can be replaced by $$C_{\delta}\sqrt{\frac{\epsilon_N\log J+\epsilon_J\log N}{N\wedge J}}.$$

\subsection[Consistency under IC$^{(1)}$]{Consistency under IC$^{(1)}$}\label{sec:supp_prove_consistency_ic1}

In the previous section, we established a non-asymptotic bound for the maximum likelihood estimation under IC$^{(0)}$, relying on a detailed analysis of the augmented Hessian matrix, as described in Lemma~\ref{lemma_hessian_property}. 
However, this construction involves significant effort and may not generalize to other identifiability conditions. We adopt a different approach to address these limitations to prove the consistency results under alternative identifiability conditions.

Suppose the true parameters $(\bLambda^*,\Fb^*)$ satisfy IC$^{(S)}$ for some $S\in\{1,\cdots,5\}$. By definition, we can write them as $(\bLambda^{(S)},\Fb^{(S)})$. By Assumption~\ref{assumption: psd covariance}, the eigenvalues of $\bSigma_f^{(S)}\bSigma_\lambda^{(S)}$ are distinct. Then we know that there exists a unique $\Gb^{(S)}$ (up to a signed permutation) such that $(\bLambda^{(S)}(\Gb^{(S)})^{-1},\Fb^{(S)}(\Gb^{(S)})^{\T})$ satisfy IC$^{(0)}$. In such case $\btheta^{(0)} = \big( \bLambda^{(S)}(\Gb^{(S)})^{-1},\Fb^{(S)}(\Gb^{(S)})^{\T}\big)$ can be viewed as a variant of the true parameters that satisfy IC$^{(0)}$, and $\btheta^{(0)}\in\cB(D/2)$ by selecting $D$ large enough. Therefore, according to the results in Section~\ref{sec:supp_prove_consistency_ic0}, $\hat\btheta^{(0)}$, obtained in \eqref{eq_joint_mle} by setting $S=0$, is an interior point in $\cB(D)$ with probability approaching 1 and maximizes the joint likelihood function inside $\cB(D)$, regardless of the identifiability condition $S$ that $\btheta^{(S)}$ satisfy. We will prove in Section~\ref{sec:supp_prove_consistency_ic1}--\ref{supp_prove_consistency_ic5} that there exists $\hat\Gb^{(S)}$, which can be obtained as detailed in Section~\ref{sup_sec_express_G}, such that
\begin{itemize}
    \item $\big(\hat\bLambda^{(0)}\hat\Gb^{(S)},\hat\Fb^{(0)}(\hat\Gb^{(S)})^{-\T}\big)$ satisfy identifiability condition IC$^{(S)}$;
    \item $\|\hat\Gb^{(S)}-\Gb^{(S)}\|\overset{p}{\to} 0$;
\end{itemize}
then $\big(\hat\bLambda^{(0)}\hat\Gb^{(S)},\hat\Fb^{(0)}(\hat\Gb^{(S)})^{-\T}\big)$ is inside $\cB(D)$ with probability approaching 1, and therefore by definition $\hat\bLambda^{(S)}=\hat\bLambda^{(0)}\hat\Gb^{(S)},\hat\Fb^{(S)}=\hat\Fb^{(0)}(\hat\Gb^{(S)})^{-\T}$, up to a signed permutation. Therefore we prove our results by showing that $\big(\hat\bLambda^{(0)}\hat\Gb^{(S)},\hat\Fb^{(0)}(\hat\Gb^{(S)})^{-\T}\big)$ consistently estimate $\big(\bLambda^{(0)}\Gb^{(S)},\Fb^{(0)}(\Gb^{(S)})^{-\T}\big)$.

We first show individual consistency of the MLE by bounding $\hat\blambda_j^{(0)} - \blambda_{j}^{(0)}$ for all $j\in[N]$ and $\hat\bbf_i^{(0)} - \bbf_i^{(0)}$ for $i\in[N]$. Suppose $\cE_1(C)$ and $\cE_S$ hold. Take rows $(j-1)K+1$ through $jK$ of \eqref{eq_bound_ave_theta_without_sca} to obtain
\begin{equation*}
    \hat\blambda_j^{(0)} - \blambda_j^{(0)} = -\big[\{\tilde\Hb^*(\tilde\btheta)\}^{-1}\bS_L(\btheta^{(0)})\big]_{[(j-1)K+1:jK]}
\end{equation*}
By (vi) of Corollary~\ref{cor_lemma_hessian_property} and plugging in $c_{NJ} \asymp 1/(N\wedge J)$, we get, with probability at least $1 - \exp(-C\epsilon_N) - 2N^{-\delta} - 2J^{-\delta} - 6(NJ)^{-\delta} - 2(N+J)^{-\delta}$,
\begin{equation*}
    \big[\{\tilde\Hb^*(\tilde\btheta)\}^{-1}\bS_L(\btheta^{(0)})\big]_{[(j-1)K+1:jK]}\lesssim \sqrt{\epsilon_N/N}+\sqrt{\log(NJ)/(NJ)}+J^{-1}
\end{equation*}
Still, we omit the constant that depends on $\delta$.
With $\log(NJ)/J\to 0$, we have that, for each $j\in[J]$, with probability at least $1 - \exp(-C\epsilon_N) - N^{-\delta} - 4J^{-\delta} - 2(NJ)^{-\delta} - (N+J)^{-\delta}$,
\begin{equation}
    \big\|\hat\blambda_j^{(0)}-\blambda_j^{(0)}\big\|\lesssim \sqrt{\frac{\epsilon_N}{N}}+\frac{1}{J}.\label{eq_derive_accurate_consis}
\end{equation}
By symmetry we have for each $i\in[N]$, with probability at least $1 - \exp(-C\epsilon_J) - 2N^{-\delta} - 2J^{-\delta} - 6(NJ)^{-\delta} - 2(N+J)^{-\delta}$,\begin{equation}\big\|\hat\bbf_i^{(0)}-\bbf_i^{(0)}\big\|\lesssim\sqrt{\frac{\epsilon_J}{J}}+\frac{1}{N}.\label{eq_derive_accurate_consis_f}\end{equation}

Next, we proceed to prove the results under IC$^{(1)}$. From now we suppose that $\cE_1$ holds, and \eqref{eq_derive_accurate_consis} holds for all $j\in \cA_r$ for any $r\in[K]$, i.e., $\big\|\hat\blambda_j^{(0)}-\blambda_j^{(0)}\big\|\lesssim \sqrt{\epsilon_N/N} + 1/J$ for all $j$ such that $\blambda_j$ is involved in the identifiability conditions. 

We first present the following technique lemma, the proof of which is left to Section~\ref{prove_lemma_linear_equation_perturb}
    \begin{lemma}\it \label{lemma_linear_equation_perturb}
        Suppose $\Hb\in\RR^{K-1\times K}$ and $\mathrm{rank}(\Hb) = K-1$. Denote the solutions of $\Hb\bx = 0$ as $\{\bx_h\nu:\nu\in\RR\}$. Suppose we have $\tilde\Hb\in\RR^{K-1\times K}$ such that $\|\Hb - \tilde\Hb\|\le \epsilon$. Denote the solutions of $\tilde\Hb\bx = 0$ as $\{\tilde\bx_h\nu:\nu\in\RR\}$ When $\epsilon$ is small enough, we have
        \begin{equation*}
            \sin\big\langle\bx_h,\tilde\bx_h\rangle \lesssim \epsilon.
        \end{equation*}
    \end{lemma}

 For estimation under IC$^{(1)}$, denote $\Gb^{(1)} = \big(\bg_1^{(1)},\cdots,\bg_K^{(1)}\big)^\T$, where for $r=K,K-1,\cdots,1$,
 \begin{equation*}
     \bg_r^{(1)} \in \text{span}\left\{\mathrm{row}\big(\bLambda_{[\cA_r,]}^{(0)}\big),\bg_{r+1}^{(1)},\cdots,\bg_K^{(1)}\right\}^\perp,\;\|\bg_r^{(1)}\| = 1.
 \end{equation*}
 Next $\hat\Gb^{(1)} = \big(\bg_1^{(1)},\cdots,\bg_K^{(1)}\big)^\T$ is similarly defined as:
 \begin{equation*}
     \hat\bg_r^{(1)} \in \text{span}\left\{\mathrm{row}\big(\hat\bLambda_{[\cA_r,]}^{(0)}\big),\hat\bg_{r+1}^{(1)},\cdots,\hat\bg_K^{(1)}\right\}^\perp,\;\|\hat\bg_r^{(1)}\| = 1.
 \end{equation*}
For $r=K$ we know $\bLambda_{[\cA_K,]}^{(0)}\bg_K^{(1)} = 0$ and $\hat\bLambda_{[\cA_K,]}^{(0)}\hat\bg_K^{(1)} = 0$. Also, since $\Gb^{(1)}$ and $\hat\Gb^{(1)}$ are orthogonal by construct, $\Ib_K = (\Gb^{(1)})^{-1}N^{-1}(\Fb^{(0)})^\T\Fb^{(0)}(\Gb^{(1)})^{-\T} = (\hat\Gb^{(1)})^{-1}N^{-1}(\hat\Fb^{(0)})^\T\hat\Fb^{(0)}(\hat\Gb^{(1)})^{-\T}$. Therefore $(\bLambda^{(0)}\Gb^{(1)},\Fb^{(0)}(\Gb^{(1)})^{-\T})$ and $(\hat\bLambda^{(0)}\hat\Gb^{(1)},\hat\Fb^{(0)}(\hat\Gb^{(1)})^{-\T})$ satisfy IC$^{(1)}$. By \eqref{eq_derive_accurate_consis}, we know $\big\|\hat\bLambda_{[\cA_r,]}^{(0)} - \bLambda_{[\cA_r,]}^{(0)}\big\|\lesssim (\epsilon_N/N)^{1/2}+1/J$ for any $r\in[K]$, and therefore by Lemma~\ref{lemma_linear_equation_perturb}, $\sin\langle \hat\bg_K^{(1)} , \bg_K^{(1)}\rangle \lesssim (\epsilon_N/N)^{1/2}+1/J$. Finally, since $\|\hat\bg_K^{(1)}\| =\| \bg_K^{(1)}\|=1$, $\big\|\hat\bg_K^{(1)} - \bg_K^{(1)}\big\|^2 = 2-2(\hat\bg_K^{(1)})^\T  \bg_K^{(1)}=2\sin\langle \hat\bg_K^{(1)} , \bg_K^{(1)}\rangle\lesssim (\epsilon_N/N)^{1/2}+1/J$. For remaining $r = 1,\cdots, K-1$,  \begin{equation*}
    \left\{\begin{aligned}
       0&\,= \bLambda_{[\cA_r,]}^{(0)}\bg_r^{(1)} \\ 0&\,=(\bg^{(1)}_l)^\T\bg_r^{(1)},\,l>r
    \end{aligned}\right.\quad \left\{\begin{aligned}
       0&\,= \hat\bLambda_{[\cA_r,]}^{(0)}\hat\bg_r^{(1)} \\ 0&\,=(\hat\bg^{(1)}_l)^\T\hat\bg_r^{(1)},\,l>r
    \end{aligned}\right.
\end{equation*}
Apply Lemma~\ref{lemma_linear_equation_perturb} recursively for each $r\in[K]$, we know similarly $\big\|\hat\bg_K^{(1)} - \bg_K^{(1)}\big\|\lesssim (\epsilon_N/N)^{1/2}+1/J$, which implies $\big\|\hat\Gb^{(1)} - \Gb^{(1)}\big\|\lesssim (\epsilon_N/N)^{1/2}+1/J$. So, $\bLambda^{(1)}=\bLambda^{(0)}\Gb^{(1)}$ and $\Fb^{(1)}=\Fb^{(0)}(\Gb^{(1)})^{-\T}=\Fb^{(0)}\Gb^{(1)}$, with at most a sign change for the columns of $\hat\Gb^{(1)}$. Similarly $\hat\bLambda^{(1)}=\hat\bLambda^{(0)}\hat\Gb^{(1)}$ and $\hat\Fb^{(1)}=\hat\Fb^{(0)}\hat\Gb^{(1)}$. Then by \eqref{eq_derive_accurate_consis} and $\|\blambda_{j}^{(0)}\|\le \sqrt{KD}$,
   \begin{align*}
       \frac{1}{J}\sum_{j=1}^J\big\|\hat\blambda_j^{(1)}-\blambda_j^{(1)}\big\|^2\le&\, \big\|\hat\Gb^{(1)}-\Gb^{(1)}\big\|^2\frac{1}{J}\sum_{j=1}^J\|\blambda_{j}^{(0)}\|^2+\frac{1}{J}\sum_{j=1}^J\big\|\hat\blambda_j^{(0)}-\blambda_j^{(0)}\big\|^2\|\Gb^{(1)}\|\\&\,+\frac{1}{J}\sum_{j=1}^J\big\|\hat\blambda_j^{(1)}-\blambda_j^{(1)}\big\|^2\big\|\hat\Gb^{(1)}-\Gb^{(1)}\big\|^2\\\lesssim &\,{\frac{\epsilon_N}{N}} + \frac{1}{J^2}.
   \end{align*}
For $\hat\Fb^{(1)}$, by \eqref{eq_derive_accurate_consis_f}, we have
\begin{align*}
       \frac{1}{N}\sum_{i=1}^N\big\|\hat\bbf_i^{(1)}-\bbf_i^{(1)}\big\|^2\le&\, \big\|\hat\Gb^{(1)}-\Gb^{(1)}\big\|^2\frac{1}{N}\sum_{i=1}^N\|\bbf_i^{(0)}\|^2+\frac{1}{N}\sum_{i=1}^N\big\|\hat\bbf_i^{(0)}-\bbf_i^{(0)}\big\|^2\|\Gb^{(1)}\|\\&\,+\frac{1}{N}\sum_{i=1}^N\big\|\hat\bbf_i^{(1)}-\bbf_i^{(1)}\big\|^2\big\|\hat\Gb^{(1)}-\Gb^{(1)}\big\|^2\\\lesssim &\,\frac{\epsilon_N}{N} + \frac{1}{J}.
   \end{align*}
For uniform consistency, by \eqref{eq_max_blambda}, we have by taking $\epsilon_N = \log N$ that
\begin{align*}
       \max_{1\le j\le J}\big\|\hat\blambda_j^{(1)}-\blambda_j^{(1)}\big\|\le&\, \max_{1\le j\le J}\big\|\hat\Gb^{(1)}-\Gb^{(1)}\big\|\|\blambda_{j}^{(0)}\|+\max_{1\le j\le J}\big\|\hat\blambda_j^{(0)}-\blambda_j^{(0)}\big\|\|\Gb^{(1)}\|\\&\,+\max_{1\le j\le J}\big\|\hat\blambda_j^{(1)}-\blambda_j^{(1)}\big\|\big\|\hat\Gb^{(1)}-\Gb^{(1)}\big\|\\\lesssim &\,\log J\sqrt{\frac{\epsilon_N\log J+\epsilon_J\log N}{N\wedge J}},
   \end{align*}
   and by \eqref{eq_max_f}, we have
   \begin{align*}
       \max_{1\le i\le N}\big\|\hat\bbf_i^{(1)}-\bbf_i^{(1)}\big\|\le&\, \max_{1\le i\le N}\big\|\hat\Gb^{(1)}-\Gb^{(1)}\big\|\|\bbf_i^{(0)}\|+\max_{1\le i\le N}\big\|\hat\bbf_i^{(0)}-\bbf_i^{(0)}\big\|\|\Gb^{(1)}\|\\&\,+\max_{1\le i\le N}\big\|\hat\bbf_i^{(1)}-\bbf_i^{(1)}\big\|\big\|\hat\Gb^{(1)}-\Gb^{(1)}\big\|\\\lesssim &\,\log N\sqrt{\frac{\epsilon_N\log J+\epsilon_J\log N}{N\wedge J}}.
   \end{align*}

Proof for consistency under IC$^{(2)}$ as it is a special case of IC$^{(1)}$. When $l_{ij}^{\prime}(\eta)$ is bounded, the infinity bound can be similarly improved as in Section~\ref{sec:supp_prove_consistency_ic0}.
\subsection[Consistency under IC$^{(3)}$]{Consistency under IC$^{(3)}$}
   For estimation under IC$^{(3)}$, similar to our previous analysis, we define $\Gb^{(3)}$ and $\hat\Gb^{(3)}$ as follows:
   \begin{align*}
       \Gb^{(3)}_{[,1]}=&\frac{\blambda_{1}^{(0)}}{\|\blambda_{1}^{(0)}\|^2}\\\Gb^{(3)}_{[,r]}=&\frac{\blambda_{r}^{(0)}-\Gb^{(3)}_{[,1:r-1]}(\Db_r^{(3)})^{-1}(\Gb^{(3)}_{[,1:r-1]})^\T\blambda_{r}^{(0)}}{\|\blambda_{r}^{(0)}\|^2-(\blambda_{r}^{(0)})^\T\Gb^{(3)}_{[,1:r-1]}(\Db_r^{(3)})^{-1}(\Gb^{(3)}_{[,1:r-1]})^\T\blambda_{r}^{(0)}},\;1<r\le K,
   \end{align*}
   and
\begin{align*}
       \hat\Gb^{(3)}_{[,1]}=&\frac{\hat\blambda_{1}^{(0)}}{\|\hat\blambda_{1}^{(0)}\|^2}\\\hat\Gb^{(3)}_{[,r]}=&\frac{\hat\blambda_{r}^{(0)}-\hat\Gb^{(3)}_{[,1:r-1]}(\hat\Db_r^{(3)})^{-1}(\hat\Gb^{(3)}_{[,1:r-1]})^\T\hat\blambda_{r}^{(0)}}{\|\hat\blambda_{r}^{(0)}\|^2-(\hat\blambda_{r}^{(0)})^\T\hat\Gb^{(3)}_{[,1:r-1]}(\hat\Db_r^{(3)})^{-1}(\hat\Gb^{(3)}_{[,1:r-1]})^\T\hat\blambda_{r}^{(0)}},\;1<r\le K.
   \end{align*}
   Here $\Db_r^{(3)} = \mathrm{diag}(\|\Gb_{[,1]}^{(3)}\|^2,\cdots,\|\Gb_{[,r-1]}^{(3)}\|^2)$ and $\hat\Db_r^{(3)} = \mathrm{diag}(\|\hat\Gb_{[,1]}^{(3)}\|^2,\cdots,\|\hat\Gb_{[,r-1]}^{(3)}\|^2)$. It is easy to check that $\big(\bLambda^{(0)}\Gb^{(3)},\Fb^{(0)}(\Gb^{(3)})^{-\T}\big)$ and $\big(\hat\bLambda^{(0)}\hat\Gb^{(3)},\hat\Fb^{(0)}(\hat\Gb^{(3)})^{-\T}\big)$ satisfy IC$^{(3)}$.
   It can also be checked recursively that $\big\|\hat\Gb_{[,r]} - \Gb_{[,r]}\big\| \lesssim (\epsilon_N/N)^{1/2}+1/J$. Then $\|\hat\Gb^{(3)}-\Gb^{(3)}\|\lesssim (\epsilon_N/N)^{1/2}+1/J$ and $\hat\Gb^{(3)}$ is invertible when $N$ is large w.h.p.. With
   \begin{align*}
       \big\|(\hat\Gb^{(3)})^{-1}-(\Gb^{(3)})^{-1}\big\|\le&\,\big\|(\Gb^{(3)})^{-1}\big\|\big\|\hat\Gb^{(3)}-\Gb^{(3)}\big\|\big\|(\hat\Gb^{(3)})^{-1}\big\|\\\le&\,\big\|(\Gb^{(3)})^{-1}\big\|^2\big\|\hat\Gb^{(3)}-\Gb^{(3)}\big\|\\&\,+\big\|(\Gb^{(3)})^{-1}\big\|\big\|\hat\Gb^{(3)}-\Gb^{(3)}\big\|\big\|(\hat\Gb^{(3)})^{-1}-(\Gb^{(3)})^{-1}\big\|,
   \end{align*}
   we know $\big\|(\hat\Gb^{(3)})^{-1}-(\Gb^{(3)})^{-1}\big\|\le 2\big\|(\Gb^{(3)})^{-1}\big\|^2\big\|\hat\Gb^{(3)}-\Gb^{(3)}\big\|\lesssim (\epsilon_N/N)^{1/2}+1/J$, and therefore consistency results can be similarly derived as in Section~\ref{sec:supp_prove_consistency_ic1}.
\subsection[Consistency under IC$^{(4)}$]{Consistency under IC$^{(4)}$}
Write $\bLambda_{r}^{(4)}=\bLambda_{[\cA_r,]}^{(4)}\in\RR^{(K-1)\times K}$, $\hat\bLambda_{r}^{(4)}=\hat\bLambda_{[\cA_r,]}^{(4)}\in\RR^{(K-1)\times K}$.  Define $\bx_r^{(4)}$ as the solution to $\bLambda_{r}^{(0)}\bx=\zero$ and $\hat\bx_r^{(4)}$ as the solution to $\hat\bLambda_{r}^{(0)}\bx=\zero$. 
Note that $\mathrm{rank}(\bLambda^{(0)}_r)=\mathrm{rank}(\bLambda_r^{(4)})=K-1$. By lemma~\ref{lemma_linear_equation_perturb},
\begin{equation*}
    \sin\langle\bx_r^{(4)},\hat\bx_r^{(4)}\rangle=\sqrt{1-\frac{\big\{\big(\bx_r^{(4)}\big)^\T\hat\bx_r^{(4)}\big\}^2}{\big\|\bx_r^{(4)}\big\|^2\big\|\hat\bx_r^{(4)}\big\|^2}}\lesssim (\epsilon_N/N)^{1/2}+1/J.
\end{equation*}
Write $\Hb^{(4)}=\big(\bx_1^{(4)},\cdots,\bx_{K}^{(4)}\big)$ and $\hat\Hb^{(4)}=\big(\hat\bx_1^{(4)},\cdots,\hat\bx_{K}^{(4)}\big)$. Define
 \begin{equation*}
       \Gb^{(4)}=\Hb^{(4)}\Big[\mathrm{diag}\big\{(\Hb^{(4)})^{-1}(\Hb^{(4)})^{-\T}\big\}\Big]^{1/2},\;\hat\Gb^{(4)}=\hat\Hb^{(4)}\Big[\mathrm{diag}\big\{(\hat\Hb^{(4)})^{-1}(\hat\Hb^{(4)})^{-\T}\big\}\Big]^{1/2}.
   \end{equation*}
   Note that $\Gb^{(4)}$ and $\hat\Gb^{(4)}$ are invariant with the scale of $\bx_r^{(4)}$ and $\hat\bx_r^{(4)}$. Without loss of generality, we take $\|\bx_r^{(4)}\|=\|\hat\bx_r^{(4)}\|=1$. Then $\|\bx_r^{(4)}-\hat\bx_r^{(4)}\|\lesssim (\epsilon_N/N)^{1/2}+1/J$.
  Then $\big\|\hat\Hb^{(4)} - \Hb^{(4)}\big\| \lesssim (\epsilon_N/N)^{1/2}+1/J$ and it can be checked that $\big\|\hat\Gb^{(4)}-\Gb^{(4)}\big\|\lesssim (\epsilon_N/N)^{1/2}+1/J$ and therefore $\hat\Gb^{(4)}$ is invertible when $N$ is large w.h.p. And by the following two facts: (1) for any diagonal matrix $\Db$, $\bLambda^{(0)}\Hb^{(4)}\Db$ and $\hat\bLambda^{(0)}\hat\Hb^{(4)}\Db$ satisfy the restrictions for loading parameters in IC$^{(4)}$; (2) $\mathrm{diag}\big((\Gb^{(4)})^{-1}(\Gb^{(4)})^{-\T}\big)=\mathrm{diag}\big((\hat\Gb^{(4)})^{-1}(\hat\Gb^{(4)})^{-\T}\big)=\Ib_K$, we conclude that $\bLambda^{(4)}=\bLambda^{(0)}\Gb^{(4)}$, $\Fb^{(4)}=\Fb^{(0)}(\Gb^{(4)})^{-\T}$, $\hat\bLambda^{(4)}=\hat\bLambda^{(0)}\hat\Gb^{(4)}$ and $\hat\Fb^{(4)}=\hat\Fb^{(0)}(\hat\Gb^{(4)})^{-\T}$ up to a sign change.
Consistency rates follow similarly as in Section~\ref{sec:supp_prove_consistency_ic1}.

\subsection[Consistency under IC$^{(5)}$]{Consistency under IC$^{(5)}$}\label{supp_prove_consistency_ic5}
   For estimation under IC$^{(5)}$, define $\Gb^{(5)}$ and $\hat\Gb^{(5)}$ as follows:
   \begin{equation*}
       \Gb^{(5)}=\big(\bLambda_{[1:K,]}^{(0)}\big)^{-1},\;\hat\Gb^{(5)}=\big(\hat\bLambda_{[1:K,]}^{(0)}\big)^{-1}.
   \end{equation*}
Here $\hat\Gb^{(5)}$ is well-defined when $N$ is large by individual consistency. Then $\bLambda_{[1:K,]}^{(0)}\Gb^{(5)}=\hat\bLambda_{[1:K,]}^{(0)}\hat\Gb^{(5)}=\Ib_K$, which implies that $\bLambda^{(5)}=\bLambda^{(0)}\Gb^{(5)}$, $\Fb^{(5)}=\Fb^{(0)}(\Gb^{(5)})^{-\T}$, $\hat\bLambda^{(5)}=\hat\bLambda^{(0)}\hat\Gb^{(5)}$ and $\hat\Fb^{(5)}=\hat\Fb^{(0)}(\hat\Gb^{(5)})^{-\T}$. Note that 
   \begin{align*}
       \big\|\hat\Gb^{(5)}-\Gb^{(5)}\big\|\le&\,\big\|\big(\bLambda_{[1:K,]}^{(0)}\big)^{-1}\big\|\big\|\bLambda_{[1:K,]}^{(0)}-\hat\bLambda_{[1:K,]}^{(0)}\big\|\big\|\big(\hat\bLambda_{[1:K,]}^{(0)}\big)^{-1}\big\|\\\le&\,\big\|\big(\bLambda_{[1:K,]}^{(0)}\big)^{-1}\big\|^2\big\|\bLambda_{[1:K,]}^{(0)}-\hat\bLambda_{[1:K,]}^{(0)}\big\|\\&\,+\big\|\big(\bLambda_{[1:K,]}^{(0)}\big)^{-1}\big\|\big\|\bLambda_{[1:K,]}^{(0)}-\hat\bLambda_{[1:K,]}^{(0)}\big\|\big\|\hat\Gb^{(5)}-\Gb^{(5)}\big\|
       \\\lesssim &\,\sqrt{\frac{\epsilon_N}{N}}+\frac{1}{J}+\frac{\big\|\hat\Gb^{(5)}-\Gb^{(5)}\big\|}{\sqrt{N}},
   \end{align*}
   which implies that $\big\|\hat\Gb^{(5)}-\Gb^{(5)}\big\|\lesssim\sqrt{\epsilon_N/N} + 1/J$. We can derive the results for $\hat\btheta^{(5)}$ as in Section~\ref{sec:supp_prove_consistency_ic1}.

\subsection{Sin-consistency}\label{supp_prove_sin_consistency}
Note that if $\ba,\bb\in\RR^{N}$, with $\|\ba\|_{\infty}\le M$, $\|\ba\|\ge mN^{1/2}$, $\|\bb\|_{\infty}\le M$ for some constant $m$ and $M$, and $N^{-1/2}\|\ba-\bb\|\le c$, then
\begin{equation*}
    { \sin\langle \ba,\bb\rangle = \frac{\|(\Ib_N-\Pb_{\bb})\ba\|}{\|\ba\|}\le \frac{\|\ba-\bb\|}{\|\ba\|}\le cm^{-1}.}
\end{equation*}
Then sin-consistency follows from previous consistency results and $N^{-1}\sum_{i=1}^N(f_{ir}^{(S)})^2 \ge $\\$\lambda_{\min}(\Mb_{\lambda\lambda}^{(S)})$ and $J^{-1}\sum_{j=1}^J(\lambda_{jr}^{(S)})^2 \ge \lambda_{\min}(\Mb_{ff}^{(S)})$ for any $r\in[K]$.

   \section{Proofs of Theorems~\ref{main_thm_loading_asym},~\ref{main_thm_factor_asym} and~\ref{thm_mff_limiting_distri}}
   
   \label{sec:proof limit}
In this section, we prove the theorems concerning the limiting distributions, including Theorems~\ref{main_thm_loading_asym},~\ref{main_thm_factor_asym} and~\ref{thm_mff_limiting_distri}. The results for $\hat\blambda^{(S)}$s and $\hat\bbf_i^{(S)}$s with $S\in\{0,1,\cdots,5\}$ are proved in Sections~\ref{sec:use_clt}--\ref{sec_limit_ic5}. The proof for Theorem~\ref{thm_mff_limiting_distri} is presented in Section~\ref{prove_thm_mff_limiting_dist}. We provide a roadmap for expressing $\Xb_{\lambda}^{(S)}$ and $\Xb_{\lambda,rk}^{(S)}$, which will be detailed in the following sections. In particular, 
\begin{itemize}
    \item $\Xb_{\lambda}^{(1)}$ is given in \eqref{eq_expression_Xlambda1} and $\Xb_{\lambda,kr}^{(1)}$ is given in \eqref{eq_x_lambda_kr_1}.
    \item $\Xb_{\lambda}^{(2)}$ is given in \eqref{eq_expression_Xlambda2}.
    \item $\Xb_{\lambda}^{(3)}$ is given in \eqref{eq_expression_Xlambda3}.
    \item $\Xb_{\lambda}^{(4)}$ is given in \eqref{eq_expression_Xlambda4} and 
    \begin{align}
    \Xb_{\lambda,rl}^{(4)}=\be_r\big(\be_l^{(K-1)}\big)^\T\big(\bLambda_{[\cA_r,-r]}^{(4)}\big)^{-\T}\left(
\begin{array}{cc}
\Ib_{r-1} &  \\
\multicolumn{2}{c}{-\big(\Mb_{ff}^{(4)}\big)_{[r,-r]}} \\
 & \Ib_{K-r}
\end{array}
\right)^\T
\blambda_{\cA_r(l)}^{(4)}\be_r^\T.
\end{align}
\item $\Xb_{\lambda}^{(5)} $ is given as.
\begin{equation}
     \begin{pmatrix}
         \bPhi_{[N],1}^{(5)}\\&\bPhi_{[N],2}^{(5)}\\&&\ddots\\&&&\bPhi_{[N],K}^{(5)}
     \end{pmatrix}
\end{equation}
\item $\bSigma_{[J]}^{(3)}$, $\bSigma_{[J]}^{(4)}$ and $\bSigma_{[J]}^{(5)}$ are given in \eqref{eq_sigma_mff3}, \eqref{eq_sigma_mff4} and \eqref{eq_sigma_mff5}, respectively.
\end{itemize}

\subsection[Limiting Distributions of $\hat\blambda_j^{(0)}$s and $\hat\bbf_i^{(0)}$s]{Limiting Distributions of $\hat\blambda_j^{(0)}$s and $\hat\bbf_i^{(0)}$s}\label{sec:use_clt}
Starting from here we omit the non-asymptotic bound and use $O_p$ as the probabilistic version of the big-$O$ to denote convergence in probability. First, we define some notations.
    Write $\bdelta_{f,i}=\hat\bbf_i^{(0)}-\bbf_i^{(0)}$, $\bdelta_{\lambda,j}=\hat\blambda_j^{(0)}-\blambda_j^{(0)}$,
$\bdelta_{f}=\big(\bdelta_{f,1}^\T,\cdots,\bdelta_{f,N}^\T)^\T$, $\bdelta_{\lambda}=\big(\bdelta_{\lambda,1}^\T,\cdots,\bdelta_{\lambda,J}^\T)^\T$ and $\bdelta_{m}=\big(\bdelta_{\lambda}^\T,\bdelta_{f}^\T)^\T$. We already know from \eqref{eq_derive_accurate_consis} and \eqref{eq_derive_accurate_consis_f} that for any $i\in[N]$ and $j\in[J]$,
\begin{equation}
    \big\|\bdelta_{f,i}\big\| = O_p\big(J^{-1/2}\wedge N^{-1}\big),\quad \big\|\bdelta_{i,\lambda}\big\| = O_p\big(N^{-1/2}\wedge J^{-1}\big).\label{eq_derive_accurate_consis_all}
\end{equation}
Note here $\epsilon_N$ and $\epsilon_J$ are omitted because we are deriving asymptotic bounds instead.

Apply the integral form of the mean value theorem on $[\bS_L(\btheta^{(0)})]_{[(j-1)K+1:jK]} = -\sum_{i=1}^N$ $l_{ij}^{\prime}\big((\blambda_j^{(0)})^\T\bbf_i^{(0)}\big)\bbf_i^{(0)}$ to obtain
\begin{align}
    &-\sum_{i=1}^Nl_{ij}^{\prime}\big((\hat\blambda_j^{(0)})^\T\bbf_i^{(0)}\big)\bbf_i^{(0)}+\sum_{i=1}^Nl_{ij}^{\prime}\big((\blambda_j^{(0)})^\T\bbf_i^{(0)}\big)\bbf_i^{(0)}\label{expand_first_order1}\\
    &=\left(\int_0^1\sum_{i=1}^N-l_{ij}^{\prime\prime}\Big[\big\{\blambda_j^{(0)}+s(\hat\blambda_j^{(0)}-\blambda_j^{(0)})\big\}^\T\bbf_i^{(0)}\Big]\bbf_i^{(0)}\big(\bbf_i^{(0)}\big)^\T ds\right)(\hat\blambda_j^{(0)}-\blambda_j^{(0)}).\label{expand_first_order2}
\end{align}
We expand the integrand in \eqref{expand_first_order2} as
    \begin{align*}
        &\,-l_{ij}^{\prime\prime}\Big\{\big(\blambda_j^{(0)}+s(\hat\blambda_j^{(0)}-\blambda_j^{(0)})\big)^\T\bbf_i^{(0)}\Big\}\bbf_i^{(0)}\big(\bbf_i^{(0)}\big)^\T \\&\,=-l_{ij}^{\prime\prime}\big(\eta_{ij}^*\big)\bbf_i^{(0)}\big(\bbf_i^{(0)}\big)^\T \\&\quad- l_{ij}^{\prime\prime\prime}\Big\{\big(\blambda_j^{(0)}+\tilde s(\hat\blambda_j^{(0)}-\blambda_j^{(0)})\big)^\T\bbf_i^{(0)}\Big\}\bbf_i^{(0)}\big(\bbf_i^{(0)}\big)^\T\tilde s(\hat\blambda_j^{(0)}-\blambda_j^{(0)})\big(\bbf_i^{(0)}\big)^\T.
    \end{align*}
    Here $\tilde s\in[0,1]$. 
    With Assumptions~\ref{assumption: psd covariance} and ~\ref{assumption:smoothness} and \eqref{eq_derive_accurate_consis_all}, the second term on the right side can be bounded by $O_p\big(1/(\sqrt{N}\wedge J)\big)$. Then we can write
    \begin{equation*}
        \eqref{expand_first_order2}=-\sum_{i=1}^N\big\{l_{ij}^{\prime\prime}\big(\eta_{ij}^*\big)\bbf_i^{(0)}\big(\bbf_i^{(0)}\big)^\T\big\}\bdelta_{\lambda,j}+O_p\Big(\frac{N}{N\wedge J^2}\Big).
    \end{equation*}
    Next, we expand the first term in \eqref{expand_first_order1} as 
    \begin{align}
    \sum_{i=1}^Nl_{ij}^{\prime}\big((\hat\blambda_j^{(0)})^\T\bbf_i^{(0)}\big)\bbf_i^{(0)}=&\sum_{i=1}^Nl_{ij}^{\prime}\big(\hat\eta_{ij}\big)\hat\bbf_i^{(0)}-\sum_{i=1}^Nl_{ij}^{\prime}\big(\hat\eta_{ij}\big)(\hat\bbf_i^{(0)}-\bbf_i^{(0)})\label{expand_first_order3}\\&+\sum_{i=1}^N\Big\{l_{ij}^{\prime}\big((\hat\blambda_j^{(0)})^\T\bbf_i^{(0)}\big)-l_{ij}^{\prime}\big((\hat\blambda_j^{(0)})^\T\hat\bbf_i^{(0)}\big)\Big\}\bbf_i^{(0)}\label{expand_first_order4}.
\end{align}
The first term in \eqref{expand_first_order3} is zero with probability approaching 1 by the first order optimality condition. The second term in \eqref{expand_first_order3} can be expressed as
    \begin{equation*}
        -\sum_{i=1}^Nl_{ij}^{\prime}(\eta_{ij}^*)\bdelta_{f,i}-\sum_{i=1}^Nl_{ij}^{\prime\prime}(\tilde\eta_{ij})(\hat\eta_{ij}-\eta_{ij}^*)\bdelta_{f,i}.
    \end{equation*}
    The second term by Assumption~\ref{assumption:smoothness} can be bounded as $b_U\sum_{i=1}^N\big(\|\bdelta_{f,i}\|\|\hat\blambda^{(0)}_j\|+\|\bdelta_{\lambda,j}\|\|\bbf_i^{(0)}\|\big)$ $\|\bdelta_{f,i}\|\le O_p\big(N(N\wedge J)^{-1}\big)$. 
    
    Next for \eqref{expand_first_order4}, since
    \begin{equation*}
        l_{ij}^{\prime}\big((\hat\blambda_j^{(0)})^\T\bbf_i^{(0)}\big)\bbf_i^{(0)}=l_{ij}^{\prime}\big(\hat\eta_{ij}\big)\bbf_i^{(0)}-l_{ij}^{\prime\prime}(\tilde\eta_{ij}^*)\bbf_i^{(0)}\big(\hat\blambda_{j}^{(0)}\big)^\T\bdelta_{f,i}.
    \end{equation*}
    Again $\tilde\eta_{ij}^*=(\hat\blambda_j^{(0)})^\T\tilde\bbf_i^{(0)}$ with $\tilde\bbf_i^{(0)} $ between $\hat\bbf_i^{(0)}$ and $\bbf_i^{(0)}$. Then we have 
    \begin{align*}
        &\,\Big\|l_{ij}^{\prime\prime}(\eta_{ij}^*)(\blambda_{j}^{(0)})^\T\bdelta_{f,i} + \big\{l_{ij}^{\prime}\big((\hat\blambda_j^{(0)})^\T\bbf_i^{(0)}\big)-l_{ij}^{\prime}\big((\hat\blambda_j^{(0)})^\T\hat\bbf_i^{(0)}\big)\big\}\Big\|\\\le &\,\Big\|\big\{l_{ij}^{\prime\prime}(\eta_{ij}^*)-l_{ij}^{\prime\prime}(\tilde\eta_{ij}^*)\big\}(\blambda_{j}^{(0)})^\T\bdelta_{f,i}\Big\|\\\le & \,b_U\big\|\hat\blambda_j^{(0)}\big\|\big(\|\bdelta_{f,i}\|\|\hat\blambda^{(0)}_j\|+\|\bdelta_{\lambda,j}\|\|\bbf_i^{(0)}\|\big)\big\|\bdelta_{f,i}\big\|\\=&\,O_p\Big(\frac{1}{\sqrt{N\wedge J}}\Big).
    \end{align*}
    Then
    \begin{equation*}
        \eqref{expand_first_order4} = -\sum_{i=1}^Nl_{ij}^{\prime\prime}(\eta_{ij}^*)\bbf_i^{(0)}(\blambda_{j}^{(0)})^\T\bdelta_{f,i} + O_p\Big(\frac{N}{N\wedge J}\Big).
    \end{equation*}
    Therefore we conclude
    \begin{equation*}
        \eqref{expand_first_order1}=\sum_{i=1}^Nl_{ij}^{\prime}(\eta_{ij}^*)\bbf_i^{(0)}-\sum_{i=1}^Nl_{ij}^{\prime\prime}(\eta_{ij}^*)\bbf_i^{(0)}\big(\blambda_{j}^{(0)}\big)^\T\bdelta_{f,i}-\sum_{i=1}^Nl_{ij}^{\prime}(\eta_{ij}^*)\bdelta_{f,i}+O_p\Big(\frac{N}{N\wedge J}\Big)
    \end{equation*}
    Equating \eqref{expand_first_order1} and \eqref{expand_first_order2} we have for any $j\in[J]$
    \begin{align}
        \frac{1}{N}\sum_{i=1}^Nl_{ij}^{\prime}(\eta_{ij}^*)\bbf_i^{(0)}=&\,-\frac{1}{N}\sum_{i=1}^N\big\{l_{ij}^{\prime\prime}\big(\eta_{ij}^*\big)\bbf_i^{(0)}\big(\bbf_i^{(0)}\big)^\T\big\}\bdelta_{\lambda,j}\nonumber\\&\,-\frac{1}{N}\sum_{i=1}^Nl_{ij}^{\prime}(\eta_{ij}^*)\bdelta_{f,i}-\frac{1}{N}\sum_{i=1}^Nl_{ij}^{\prime\prime}(\eta_{ij}^*)\bbf_i^{(0)}\big(\blambda_{j}^{(0)}\big)^\T\bdelta_{f,i}+O_p\Big(\frac{1}{N\wedge J}\Big).\label{eq_linear_eq_1}
    \end{align}
    We note that the above bounds can be derived when taking the squared summation of \eqref{expand_first_order1} and \eqref{expand_first_order2} over $j$, which yields
    \begin{align}
        \sum_{j=1}^J\Big\|\sum_{i=1}^N\Big[l_{ij}^{\prime}(\eta_{ij}^*)\bbf_i^{(0)}&\,+\big\{l_{ij}^{\prime\prime}\big(\eta_{ij}^*\big)\bbf_i^{(0)}\big(\bbf_i^{(0)}\big)^\T\big\}\bdelta_{\lambda,j}\nonumber\\&\,+l_{ij}^{\prime}(\eta_{ij}^*)\bdelta_{f,i}+l_{ij}^{\prime\prime}(\eta_{ij}^*)\bbf_i^{(0)}\big(\blambda_{j}^{(0)}\big)^\T\bdelta_{f,i}\Big]\Big\|^2=O_p\Big(\frac{N^2J}{(N\wedge J)^2}\Big).\label{eq_linear_eq_1_sum}
    \end{align}
    Similarly for any $i\in[N]$, we have 
    \begin{align}
        \frac{1}{J}\sum_{j=1}^Jl_{ij}^{\prime}(\eta_{ij}^*)\blambda_j^{(0)}=&\,-\frac{1}{J}\sum_{j=1}^J\big\{l_{ij}^{\prime\prime}\big(\eta_{ij}^*\big)\blambda_j^{(0)}\big(\blambda_j^{(0)}\big)^\T\big\}\bdelta_{f,i}\nonumber\\&\,-\frac{1}{J}\sum_{j=1}^Jl_{ij}^{\prime}(\eta_{ij}^*)\bdelta_{\lambda,j}-\frac{1}{J}\sum_{j=1}^Jl_{ij}^{\prime\prime}(\eta_{ij}^*)\blambda_j^{(0)}\big(\bbf_{i}^{(0)}\big)^\T\bdelta_{\lambda,j}+O_p\Big(\frac{1}{N\wedge J}\Big),\label{eq_linear_eq_2}
    \end{align}
    and
    \begin{align}
        \sum_{i=1}^N\Big\|\sum_{j=1}^J\Big[l_{ij}^{\prime}(\eta_{ij}^*)\blambda_j^{(0)}&\,+\big\{l_{ij}^{\prime\prime}\big(\eta_{ij}^*\big)\blambda_j^{(0)}\big(\blambda_j^{(0)}\big)^\T\big\}\bdelta_{f,i}\nonumber\\&\,+l_{ij}^{\prime}(\eta_{ij}^*)\bdelta_{\lambda,j}+l_{ij}^{\prime\prime}(\eta_{ij}^*)\blambda_j^{(0)}\big(\bbf_{i}^{(0)}\big)^\T\bdelta_{\lambda,j}\Big]\Big\|^2=O_p\Big(\frac{NJ^2}{(N\wedge J)^2}\Big),\label{eq_linear_eq_2_sum}
    \end{align}
    Next, we define $\Hb_P^0$ as follows
    \begin{equation*}
        \Hb_P^0 = NJ\begin{pmatrix}
        \sum_{l=1}^K\sum_{h>l}^K\bv_{\lambda,lh}(\btheta^{(0)})\big\{\bv_{\lambda,lh}(\btheta^{(0)})\big\}^\T&\zero\\\zero&\sum_{l=1}^K\sum_{h\le l}\bv_{f,lh}(\btheta^{(0)})\big\{\bv_{f,lh}(\btheta^{(0)})\big\}^\T
    \end{pmatrix},
    \end{equation*}
    where $$\bv_{f,rr}(\btheta^{(0)})=2N^{-1}\left(f_{1r}^{(0)}\be_r^\T,\cdots,f_{Nr}^{(0)}\be_r^\T\right)^\T\text{ for }r\in[K];$$
$$\bv_{f,lh}(\btheta^{(0)})=\left(N^{-1}\big(f_{1l}^{(0)}\be_h+f_{1h}^{(0)}\be_l\big)^\T,\cdots,N^{-1}\big(f_{Nl}^{(0)}\be_h+f_{Nh}^{(0)}\be_l\big)^\T\right)^\T\text{ for }1\le h\le l\le K;$$
$$\bv_{\lambda,lh}(\btheta^{(0)})=\left(J^{-1}\big(\lambda_{1l}^{(0)}\be_h+\lambda_{1h}^{(0)}\be_l\big)^\T,\cdots,J^{-1}\big(\lambda_{Jl}^{(0)}\be_h+\lambda_{Jh}^{(0)}\be_l\big)^\T\right)^\T\text{ for }1\le l<h\le K.$$ One can verify from the proof of Lemma~\ref{lemma_hessian_property} that $c\Hb_P^0=\Hb_P(\btheta^{(0)})$.
Suppose $\cE_c(C)$ holds. Then
\begin{itemize}
    \item since $N^{-1}\sum_{i=1}^N(\hat f_{ir}^{(0)})^2=N^{-1}\sum_{i=1}^N(f_{ir}^{(0)})^2=1$, we have \begin{align}
        \big\{\bv_{f,rr}(\btheta^{(0)})\big\}^\T\bdelta_f=&\,\frac{1}{N}\sum_{i=1}^Nf_{ir}^{(0)}\big(\hat f_{ir}^{(0)}-f_{ir}^{(0)}\big)\nonumber\\=&\,\frac{1}{2N}\sum_{i=1}^N\big(2f_{ir}^{(0)}\hat f_{ir}^{(0)}-2\big)\nonumber\\=&\,\frac{1}{2N}\sum_{i=1}^N\Big(2f_{ir}^{(0)}\hat f_{ir}^{(0)}-\big(f_{ir}^{(0)}\big)^2-\big(\hat f_{ir}^{(0)}\big)^2\Big)\nonumber\\=&\,\frac{1}{2N}\sum_{i=1}^N\big(\hat f_{ir}^{(0)}-f_{ir}^{(0)}\big)^2\nonumber\\= &\,O\Big(\frac{1}{N\wedge J}\Big);\label{eq_linear_eq_3}
    \end{align}
    \item since $N^{-1}\sum_{i=1}^N\hat f_{il}^{(0)}\hat f_{ih}^{(0)}=N^{-1}\sum_{i=1}^Nf_{il}^{(0)}f_{ih}^{(0)}=0$, we have
     \begin{align}
        \big\{\bv_{f,lh}(\btheta^{(0)})\big\}^\T\bdelta_f=&\,\frac{1}{N}\sum_{i=1}^Nf_{il}^{(0)}\big(\hat f_{ih}^{(0)}-f_{ih}^{(0)}\big)+f_{ih}^{(0)}\big(\hat f_{il}^{(0)}-f_{il}^{(0)}\big)\nonumber\\=&\,-\frac{1}{N}\sum_{i=1}^N\big(\hat f_{il}^{(0)}-f_{il}^{(0)}\big)\big(\hat f_{ih}^{(0)}-f_{ih}^{(0)}\big)\nonumber\\= &\,O\Big(\frac{1}{N\wedge J}\Big);\label{eq_linear_eq_4}
    \end{align}
    \item since $J^{-1}\sum_{j=1}^J\hat\lambda_{jl}^{(0)}\hat\lambda_{jh}^{(0)}=J^{-1}\sum_{j=1}^J\lambda_{jl}^{(0)}\lambda_{jh}^{(0)}=0$, we have
     \begin{align}
        \big\{\bv_{\lambda,lh}(\btheta^{(0)})\big\}^\T\bdelta_{\lambda}=&\,\frac{1}{J}\sum_{j=1}^J\lambda_{jl}^{(0)}\big(\hat \lambda_{jh}^{(0)}-\lambda_{jh}^{(0)}\big)+\lambda_{jh}^{(0)}\big(\hat \lambda_{jl}^{(0)}-\lambda_{jl}^{(0)}\big)\nonumber\\=&\,-\frac{1}{J}\sum_{j=1}^J\big(\hat \lambda_{jl}^{(0)}-\lambda_{jl}^{(0)}\big)\big(\hat \lambda_{jh}^{(0)}-\lambda_{jh}^{(0)}\big)\nonumber\\= &\,O\Big(\frac{1}{N\wedge J}\Big).\label{eq_linear_eq_5}
        \end{align}
\end{itemize}
Then from \eqref{eq_linear_eq_5} and $\|\btheta^{(0)}\|_{\infty}\le D$ we know
\begin{equation}
    \Big\|\big[N^{-1}\Hb_P^0\big(\tilde\btheta-\btheta^{(0)}\big)\big]_{J_K}\Big\|_{\infty} \le \sum_{l=1}^K\sum_{h>l}^KJ\big|\big\{\bv_{\lambda,lh}(\btheta^{(0)})\big\}^\T\bdelta_{\lambda}\big|\big\|\bv_{\lambda,lh}\big\|_{\infty} \lesssim\frac{1}{N\wedge
     J}.\label{eq_bound_aug_term1}
\end{equation}
Similarly from \eqref{eq_linear_eq_3}--\eqref{eq_linear_eq_4}, we have
\begin{equation}
    \Big\|\big[J^{-1}\Hb_P^0\big(\tilde\btheta-\btheta^{(0)}\big)\big]_{JK+N_K}\Big\|_{\infty} =O_p\Big(\frac{1}{N\wedge
     J}\Big).\label{eq_bound_aug_term2}
\end{equation}By the construct of $\Hb_P$, we summarize \eqref{eq_bound_aug_term1} and \eqref{eq_bound_aug_term2} as \begin{align}
\Big\|\Db^{-1}_m\Hb_P^0\bdelta_m\Big\|_{\infty}=\Big\|\Db^{-1/2}_m\Hb_P\Db_m^{-1/2}\big(\tilde\btheta-\btheta^{(0)}\big)\Big\|_{\infty}=O_p\Big(\frac{1}{N\wedge
     J}\Big),\label{eq_bound_penalty_error_term}
\end{align}
Add $c[\Db_m^{-1}\Hb_P^0\bdelta_m]_{[(j-1)K+1:jK]}$ on \eqref{eq_linear_eq_1} and $c[\Db_m^{-1}\Hb_P^0\bdelta_m]_{[JK+(i-1)K+1:JK+iK]}$ on \eqref{eq_linear_eq_2} to get
    \begin{equation}
        \Big\|\Big[\Db_m^{-1}\Hb\big(\btheta^{(0)}\big)\bdelta_m+\Db_m^{-1}\bS_L(\btheta^{(0)})\Big]_{[(i-1)K+1:iK]}\Big\|=O_p\Big(\frac{1}{N\wedge J}\Big),\label{eq_linear_solving_precise_construct}
    \end{equation} and similarly with \eqref{eq_linear_eq_1_sum} and \eqref{eq_linear_eq_2_sum}, we obtain
    \begin{equation}
        \Big\|\Big[\Db_m^{-1/2}\Hb\big(\btheta^{(0)}\big)\bdelta_m+\Db_m^{-1/2}\bS_L(\btheta^{(0)})\Big]\Big\|=O_p\Big(\frac{\sqrt{NJ}}{N\wedge J}\Big),\label{eq_linear_solving_precise_construct_sum}
    \end{equation}
 Here $c$ is a constant as specified in Lemma~\ref{lemma_hessian_property} and $\Hb(\btheta^{(0)})$ is as defined in Lemma~\ref{lemma_hessian_property}. Define 
        \begin{align*}
\bdelta_{\lambda,j}^\ddagger=\bdelta_{\lambda,j}+\big(\sum_{i=1}^Nl_{ij}^{\prime\prime}(\eta_{ij}^*)\bbf_i^{(0)}(\bbf_i^{(0)})^\T\big)^{-1}\sum_{i=1}^Nl_{ij}^{\prime}(\eta_{ij}^*)\bbf_i^{(0)},\\
        \bdelta_{f,i}^\ddagger=\bdelta_{f,i}+\big(\sum_{j=1}^Jl_{ij}^{\prime\prime}(\eta_{ij}^*)\blambda_j^{(0)}(\blambda_j^{(0)})^\T\big)^{-1}\sum_{j=1}^Jl_{ij}^{\prime}(\eta_{ij}^*)\blambda_j^{(0)}.
        \end{align*}
        Similarly, we write $\bdelta_m^\ddagger = \big((\bdelta_{\lambda,1}^\ddagger)^\T,\cdots,(\bdelta_{\lambda,J}^\ddagger)^\T,(\bdelta_{f,1}^\ddagger)^\T,\cdots,(\bdelta_{f,N}^\ddagger)^\T\big)^\T$.

        The following lemma derives bound for $\Db_m^{-1}\Hb\big(\btheta^{(0)}\big)\big(\bdelta^\ddagger_m-\bdelta_m\big)+\Db_m^{-1}\bS_L(\btheta^{(0)})$. The proof is left to Section~\ref{supp_sec_prove_lemma_cor_linear_equation_hold}
\begin{lemma}\it \label{lemma_cor_linear_equation_hold}
    Suppose $\bdelta_{m}$ and $\bdelta_{m}$ are defined as above. Under Assumptions~\ref{assumption: psd covariance} and~\ref{assumption:smoothness}, we have for each $i\in[N+J]$,
    \begin{equation*}
            \Big\|\Big[\Db_m^{-1}\Hb\big(\btheta^{(0)}\big)\big(\bdelta^\ddagger_m-\bdelta_m\big)-\Db_m^{-1}\bS_L(\btheta^{(0)})\Big]_{[(i-1)K+1:iK]}\Big\|=O_p\Big(\frac{1}{N\wedge J}\Big),
        \end{equation*}
        and
        \begin{equation*}
            \Big\|\Db_m^{-1/2}\Hb\big(\btheta^{(0)}\big)\big(\bdelta^\ddagger_m-\bdelta_m\big)-\Db_m^{-1/2}\bS_L(\btheta^{(0)})\Big\|=O_p\Big(\frac{\sqrt{NJ}}{N\wedge J}\Big),
        \end{equation*}
\end{lemma}
Combine Lemma~\ref{lemma_cor_linear_equation_hold} and \eqref{eq_linear_solving_precise_construct} and we have for any $i\in[N+J]$,
        \begin{equation}
        \Big\|\Big[\Db_m^{-1}\Hb\big(\btheta^{(0)}\big)\bdelta_m^\ddagger\Big]_{[(i-1)K+1:iK]}\Big\|=O_p\Big(\frac{1}{N\wedge J}\Big),\label{eq_linear_solving_precise_construct2}
    \end{equation} 
        and combined with \eqref{eq_linear_solving_precise_construct_sum}, we have
            \begin{equation}
        \Big\|\Db_m^{-1/2}\Hb\big(\btheta^{(0)}\big)\bdelta_m^\ddagger\Big\|=O_p\Big(\frac{\sqrt{NJ}}{N\wedge J}\Big).\label{eq_linear_solving_precise_construct3}
    \end{equation}

    Note that $\Hb_{L\lambda\lambda}$ is a block diagonal matrix with the $j$th $K$-by-$K$ block given as $\Hb_{L\lambda_j\lambda_j}:=[\Hb_{L\lambda\lambda}]_{[(j-1)K +1:jK,(j-1)K +1:jK]} = \sum_{i=1}^Nl_{ij}^{\prime\prime}(\eta_{ij})\bbf_i\bbf_i^\T$. Observe that $[\Hb_{L\lambda_j\lambda_j}^{-1}]_{[(j-1)K+1:jK,]} = $ \\$\big(\zero_{K\times (j-1)K},\Hb_{L\lambda_j\lambda_j}^{-1},\zero_{K\times JK-jK}\big)$.
    Then by \eqref{eq_linear_solving_precise_construct2}, \eqref{eq_linear_solving_precise_construct3} and (viii) of Corollary~\ref{cor_lemma_hessian_property2}, we know that when $j\in[J]$,
    \begin{align*}
        \big\|\bdelta_{\lambda,j}^\ddagger\big\|=&\,\frac{1}{\sqrt{N}}\big\|[\Db_m^{1/2}\bdelta^\ddagger_{m}]_{[(j-1)K+1:jK]}\big\|\\=&\,\frac{1}{\sqrt{N}}\big\|[\Db_m^{1/2}\Hb^{-1}\big(\btheta^{(0)}\big)\Db_m^{1/2}\Db_m^{-1/2}\Hb\big(\btheta^{(0)}\big)\bdelta^\ddagger_{m}]_{[(j-1)K+1:jK]}\big\|\\\le&\,\frac{1}{\sqrt{N}}\big\|[N\Hb_{L\lambda\lambda}^{-1}\Db_m^{-1/2}\Hb\big(\btheta^{(0)}\big)\bdelta^\ddagger_{m}]_{[(j-1)K+1:jK]}\big\|\\&\,+\frac{1}{\sqrt{N}}\big\|[\Db_m^{1/2}\Hb^{-1}\big(\btheta^{(0)}\big)\Db_m^{1/2}-N^{-1}\Hb_{L\lambda\lambda}^{-1}\big]_{[(j-1)K+1:jK]}\Db_m^{-1/2}\Hb\big(\btheta^{(0)}\big)\bdelta^\ddagger_{m}\big\|\\\le &\,\frac{1}{\sqrt{N}}\big\|N\Hb_{L\lambda_j\lambda_j}^{-1}\big\|\big\|\big[\Db_m^{-1/2}\Hb\big(\btheta^{(0)}\big)\bdelta^\ddagger_{m}\big]_{[(j-1)K + 1:jK]}\big\|\\&\, + \frac{1}{\sqrt{N}}\big\|[\Db_m^{1/2}\Hb^{-1}\big(\btheta^{(0)}\big)\Db_m^{1/2}-N^{-1}\Hb_{L\lambda\lambda}^{-1}\big]_{[(j-1)K+1:jK]}\big\|\big\|\Db_m^{-1/2}\Hb\big(\btheta^{(0)}\big)\bdelta^\ddagger_{m}\big\| \\= &\, O_p\Big(\frac{1}{N\wedge J}\Big),
    \end{align*}
which implies
\begin{equation}
\bdelta_{\lambda,j}=-\big(\sum_{i=1}^Nl_{ij}^{\prime\prime}(\eta_{ij}^*)\bbf_i^{(0)}(\bbf_i^{(0)})^\T\big)^{-1}\sum_{i=1}^Nl_{ij}^{\prime}(\eta_{ij}^*)\bbf_i^{(0)}+O_p\Big(\frac{1}{N\wedge J}\Big).\label{eq_precise_blambda}\end{equation}By symmetry, we know
\begin{equation}
        \bdelta_{f,i}=-\big(\sum_{j=1}^Jl_{ij}^{\prime\prime}(\eta_{ij}^*)\blambda_j^{(0)}(\blambda_j^{(0)})^\T\big)^{-1}\sum_{j=1}^Jl_{ij}^{\prime}(\eta_{ij}^*)\blambda_j^{(0)}+O_p\Big(\frac{1}{N\wedge J}\Big).\label{eq_precise_f}
        \end{equation}
For the asymptotic normality of $\hat\blambda_j^{(0)}$, it suffices to prove that the $\sqrt{N}$ times the first term in the right-hand side of \eqref{eq_precise_blambda} converges to some normal distribution, and asymptotic normality for $\hat\bbf_i^{(0)}$ can be proved by symmetry. Recall that we have assumed $\EE\big[|l_{ij}^{\prime}(\eta_{ij}^*)|^4\big]\le M$.

Define 
\begin{equation*}
    \bZ_{N,i,j} = -\Big(\sum_{i=1}^N-\EE [l_{ij}^{\prime\prime}(\eta_{ij}^*)]\bbf_i^{(0)}(\bbf_i^{(0)})^\T\Big)^{-1/2}l_{ij}^{\prime}(\eta_{ij}^*)\bbf_i^{(0)}.
\end{equation*}
Then for each $j$, $\{\bZ_{N,i,j}\}_{i\le N}$ is a triangular array. Next, we verify the Lindeberg-Feller conditions \citep{billingsley2017probability}. By $\EE [l_{ij}^{\prime}(\eta_{ij}^*)] =0 $, we know $\EE \bZ_{N,i,j} = 0$. Next note that 
\begin{align*}
    &\,\EE [\bZ_{N,i,j}\bZ_{N,i,j}^\T ] =\\&\, \Big(-\sum_{i=1}^N\EE [l_{ij}^{\prime\prime}(\eta_{ij}^*)]\bbf_i^{(0)}(\bbf_i^{(0)})^\T\Big)^{-1/2}\EE\big[l_{ij}^{\prime}(\eta_{ij}^*)^2\bbf_i^{(0)}(\bbf_i^{(0)})^\T\big]\Big(-\sum_{i=1}^N\EE [l_{ij}^{\prime\prime}(\eta_{ij}^*)]\bbf_i^{(0)}(\bbf_i^{(0)})^\T\Big)^{-1/2}.
\end{align*}
Consequently, $\sum_{i=1}^NCov(\bZ_{N,i,j})=\Ib_K$. Next for any $\epsilon > 0$,
\begin{align*}
    &\, \sum_{i=1}^N\EE\big[\|\bZ_{N,i,j}\|^21_{(\|\bZ_{N,i,j}\|\ge \epsilon)}\big]\\\le &\,  \sum_{i=1}^N\Big[\EE\big[\|\bZ_{N,i,j}\|^4\big]\Big]^{1/2}\Big[\PP\big\{\|\bZ_{N,i,j}\|\ge \epsilon\big\}\Big]^{1/2}\\\le &\, \lambda_{\min}\Big\{\sum_{i=1}^N-\EE [l_{ij}^{\prime\prime}(\eta_{ij}^*)]\bbf_i^{(0)}(\bbf_i^{(0)})^\T\Big\}^{-1}\sum_{i=1}^N\Big[\EE\big[l_{ij}^{\prime}(\eta_{ij}^*)^4\big\|\bbf_i^{(0)}\big\|^4\big]\Big]^{1/2}\\&\,\left[\PP\left\{\lambda_{\min}\Big\{\sum_{i=1}^N-\EE [l_{ij}^{\prime\prime}(\eta_{ij}^*)]\bbf_i^{(0)}(\bbf_i^{(0)})^\T\Big\}^{-1/2}\big\|l_{ij}^{\prime}(\eta_{ij}^*)\bbf_i^{(0)}\big\|\ge \epsilon\right\}\right]^{1/2}\\\le &\, M^{2/\xi}b_L\lambda_{\min}\Big\{N^{-1}\sum_{i=1}^N\bbf_i^{(0)}(\bbf_i^{(0)})^\T\Big\}^{-1}N^{-1}\sum_{i=1}^N\big\|\bbf_i^{(0)}\big\|^4\\&\,\left[\PP\left\{\big\|l_{ij}^{\prime}(\eta_{ij}^*)\bbf_i^{(0)}\big\|\ge b_L\sqrt{N}\epsilon\lambda_{\min}\Big\{N^{-1}\sum_{i=1}^N\bbf_i^{(0)}(\bbf_i^{(0)})^\T\Big\}^{1/2}\right\}\right]^{1/2}\\\le &\, 2^{-1}M^{2/\xi}b_L\lambda_{\min}\big\{\bSigma_f^*\big\}K^2D^4\\&\,\left[\PP\left\{\big\|l_{ij}^{\prime}(\eta_{ij}^*)\bbf_i^{(0)}\big\|\ge \big[2^{-1}b_L\epsilon\lambda_{\min}\big\{\bSigma_f^*\big\}^{1/2}\big]\sqrt{N}\right\}\right]^{1/2}\\\overset{p}{\to} &\, 0
\end{align*}
Here the third and fourth inequality is because $\EE[l_{ij}^{\prime}(\eta_{ij}^*)^4]\le\big(\EE[|l_{ij}^{\prime}(\eta_{ij}^*)|^\xi]\big)^{4/\xi}\le M^{4/\xi} $, $\max_{1\le i\le N}\big\|\bbf_i^{(0)}\big\|_{\max}\le D$ 
and $\lambda_{\min}(\Mb_{ff}^*)>\lambda_{\min}(\bSigma_f^*)/2$ when $N,J$ are large. Finally the term converges to 0 in probability because $l_{ij}^{\prime}(\eta_{ij}^*)$ is sub-exponential. Therefore by the multivariate version of the Lindeberg-Feller theorem, we know
\begin{equation*}
    \sum_{i=1}^N\bZ_{N,i,j} \overset{d}{\to}\cN(0,\Ib_K),
\end{equation*}
Finally since $Var\big(l_{ij}^{\prime}(\eta_{ij}^*)^2\big)=\EE\big[l_{ij}^{\prime}(\eta_{ij}^*)^4\big] + \EE\big[l_{ij}^{\prime}(\eta_{ij}^*)^2\big]^2\le 2M^{4/\xi}$ and $Var\big(l_{ij}^{\prime\prime}(\eta_{ij}^*)\big)\le b_U^2$, we know 
\begin{equation*}
    \frac{1}{N}\sum_{i=1}^Nl_{ij}^{\prime}(\eta_{ij}^*)^2\bbf_i^{(0)}(\bbf_i^{(0)})^\T\overset{p}{\to}\frac{1}{N}\sum_{i=1}^N\EE\big[l_{ij}^{\prime}(\eta_{ij}^*)^2\bbf_i^{(0)}(\bbf_i^{(0)})^\T\big] =-\frac{1}{N}\sum_{i=1}^N\EE\big[l_{ij}^{\prime\prime}(\eta_{ij}^*)\bbf_i^{(0)}(\bbf_i^{(0)})^\T\big] .
\end{equation*}
Then by \eqref{eq_precise_blambda} and $\lambda_{\min}\big(\sum_{i=1}^N-l_{ij}^{\prime\prime}(\eta_{ij}^*)\bbf_i^{(0)}(\bbf_i^{(0)})^\T\big)\ge b_L\lambda_{\min}\big(\sum_{i=1}^N\bbf_i^{(0)}(\bbf_i^{(0)})^\T\big)=\Theta_p(N)$, we have
\begin{align*}
    &\,\hat\blambda_{j}^{(0)}-\blambda_j^{(0)}\\=&\,\Big(\sum_{i=1}^N-l_{ij}^{\prime\prime}(\eta_{ij}^*)\bbf_i^{(0)}(\bbf_i^{(0)})^\T\Big)^{-1}\Big(\sum_{i=1}^N-\EE [l_{ij}^{\prime\prime}(\eta_{ij}^*)](\eta_{ij}^*)\bbf_i^{(0)}(\bbf_i^{(0)})^\T\Big)^{1/2}\sum_{i=1}^N\bZ_{N,i,j} +O_p\Big(\frac{1}{N\wedge J}\Big)\\=&\,\Big(\sum_{i=1}^N-l_{ij}^{\prime\prime}(\eta_{ij}^*)\bbf_i^{(0)}(\bbf_i^{(0)})^\T\Big)^{-1}\Big(\sum_{i=1}^N l_{ij}^{\prime}(\eta_{ij}^*)^2\bbf_i^{(0)}(\bbf_i^{(0)})^\T\Big)^{1/2}\sum_{i=1}^N\bZ_{N,i,j} +o_p(N^{-1/2})+O_p\Big(\frac{1}{N\wedge J}\Big),
\end{align*}
which implies with $\big\|\bPhi_{[N],j}^{(0)}\big\| = O(N^{-1})$ that
\begin{equation}
    \big(\bPhi_{[N],j}^{(0)}\big)^{-1/2}\big(\hat\blambda_{j}^{(0)}-\blambda_j^{(0)}\big) \overset{d}{\to}\cN(\zero_K,\Ib_K),\label{eq_asymptotic_normal}
\end{equation}
as $\sqrt{N}/J\to0$.  The results for $\hat\bbf_i^{(0)}$ follow similarly by symmetry.

To show $\hat\bPhi^{(0)} \to \bPhi^{(0)}$, note that $\hat\blambda_j^{(0)} \overset{p}{\to} \blambda_j^{(0)}$ and $\hat\bbf_i^{(0)} \overset{p}{\to} \bbf_i^{(0)}$ by Section~\ref{sec:supp_prove_consistency_ic0}, and that $l_{ij}^{\prime}(\hat\eta_{ij}) \overset{p}{\to} l_{ij}^{\prime}(\hat\eta_{ij}^*)$ and $l_{ij}^{\prime\prime}(\hat\eta_{ij}) \overset{p}{\to} l_{ij}^{\prime\prime}(\hat\eta_{ij}^*)$ by the individual consistency in Section~\ref{sec:supp_prove_consistency_ic0} and the continuity of $l_{ij}^{\prime}$ and $l_{ij}^{\prime\prime}$. 

Next, we derive the limiting distributions of the MLEs under IC$^{(1)}$–IC$^{(5)}$. While $\bS_L(\hat\btheta^{(S)}) = \zero$ still holds with probability approaching one, the augmentation $\Hb_P^0$ is unavailable under these identifiability conditions. Instead, we adopt a similar approach to that used in establishing the consistency results.
\subsection[Limiting Distributions of $\hat\blambda_j^{(1)}$s and $\hat\bbf_i^{(1)}$s]{Limiting Distributions of $\hat\blambda_j^{(1)}$s and $\hat\bbf_i^{(1)}$s}\label{sec_limit_ic1}
Under IC$^{(1)}$, we have constructed orthogonal matrices $\Gb^{(1)}$ and $\hat\Gb^{(1)}$ in Section~\ref{sec:supp_prove_consistency_ic1} such that $\bLambda^{(1)}=\bLambda^{(0)}\Gb^{(1)}$, $\Fb^{(1)}=\Fb^{(0)}\Gb^{(1)}$, $\hat\bLambda^{(1)}=\hat\bLambda^{(0)}\hat\Gb^{(1)}$ and $\hat\Fb^{(1)}=\hat\Fb^{(0)}\hat\Gb^{(1)}$ with $\|\hat\Gb^{(1)}-\Gb^{(1)}\|=O_p(N^{-1/2}+J^{-1})$. The consistency rate alone is not enough for limiting distributions of the estimation. By the identifiability conditions we know (i) $\bLambda^{(1)}_{[\cA_r,r]}=\hat\bLambda^{(1)}_{[\cA_r,r]}=\zero_{r-1}$; (ii) $\Ib_K=N^{-1}(\Fb^{(1)})^\T\Fb^{(1)}=N^{-1}(\hat\Fb^{(1)})^\T\hat\Fb^{(1)}$. Then
\begin{align*}
    \bLambda_{[\cA_r,]}^{(0)}\Gb_{[,r]}^{(1)} - \hat\bLambda_{[\cA_r,]}^{(0)}\hat\Gb_{[,r]}^{(1)}&\,=\zero_{r-1};\\
    \big(\hat\Gb^{(1)}\big)^{\T}\hat\Gb^{(1)}-\big(\Gb^{(1)}\big)^{\T}\Gb^{(1)}&\,=\zero.
\end{align*}
By $\big\|\Gb^{(1)}-\hat\Gb^{(1)}\big\|=O_p(N^{-1/2} + J^{-1})$ and $\|\hat\blambda_j^{(0)}-\blambda_j^{(0)}\|=O_p\big(N^{-1/2} + J^{-1}\big)$, we can write the first part as
\begin{align*}
    &\,\big[\bLambda^{(0)}\Gb^{(1)} - \hat\bLambda^{(0)}\hat\Gb^{(1)}\big]_{[\cA_r,r]} \\= &\, \big[\bLambda^{(0)}\big(\Gb^{(1)} - \hat\Gb^{(1)}\big)\big]_{[\cA_r,r]} + \big[(\bLambda^{(0)} -\hat\bLambda^{(0)})\Gb^{(1)}\big]_{[\cA_r,r]} \\&\,- \big[(\bLambda^{(0)} - \hat\bLambda^{(0)})(\Gb^{(1)}-\hat\Gb^{(1)})\big]_{[\cA_r,r]} \\&= \, \big[\bLambda^{(0)}\big(\Gb^{(1)} - \hat\Gb^{(1)}\big)\big]_{[\cA_r,r]} + \big[(\bLambda^{(0)} -\hat\bLambda^{(0)})\Gb^{(1)}\big]_{[\cA_r,r]}+O_p\Big(\frac{1}{N\wedge J^{2}}\Big).
\end{align*}
The second part can be written as 
\begin{align*}
    &\,\big(\hat\Gb^{(1)}\big)^{\T}\hat\Gb^{(1)}-\big(\Gb^{(1)}\big)^{\T}\Gb^{(1)}\\ =&\,-\big(\Gb^{(1)}-\hat\Gb^{(1)}\big)^{\T}\Gb^{(1)}-\big(\Gb^{(1)}\big)^\T\big(\Gb^{(1)}-\hat\Gb^{(1)}\big)\\&\,+\big(\Gb^{(1)}-\hat\Gb^{(1)}\big)^\T\big(\Gb^{(1)}-\hat\Gb^{(1)}\big)\\=&\,-\big(\Gb^{(1)}-\hat\Gb^{(1)}\big)^{\T}\Gb^{(1)}-\big(\Gb^{(1)}\big)^\T\big(\Gb^{(1)}-\hat\Gb^{(1)}\big)+O_p\Big(\frac{1}{N\wedge J^{2}}\Big).
\end{align*}
Define $\hat\Xb^{(1)} = \big(\Gb^{(1)} - \hat\Gb^{(1)}\big)^\T\Gb^{(1)}$ and for $r=2,\cdots,K$, define
\begin{equation*}\bdelta_{r}^{(1)}=\big\{\be_r^\T\big(\sum_{t=1}^Nl_{tj}^{\prime\prime}(\eta_{tj}^*)\bbf_t^{(1)}(\bbf_t^{(1)})^\T\big)^{-1}\sum_{t=1}^Nl_{tj}^{\prime}(\eta_{tj}^*)\bbf_t^{(1)}\big\}_{j\in \cA_r}\in\RR^{r-1}.\end{equation*} 
From Section~\ref{sec:use_clt}, we know that 
\begin{align}
    &\,\big(\bPhi_{[N],j}^{(1)}\big)^{-1/2}\big(\sum_{t=1}^Nl_{tj}^{\prime\prime}(\eta_{tj}^*)\bbf_t^{(1)}(\bbf_t^{(1)})^\T\big)^{-1}\sum_{t=1}^Nl_{tj}^{\prime}(\eta_{tj}^*)\bbf_t^{(1)} \nonumber\\= &\, \big(\bPhi_{[N],j}^{(1)}\big)^{-1/2}\Gb^{(1)}\big(\sum_{t=1}^Nl_{tj}^{\prime\prime}(\eta_{tj}^*)\bbf_t^{(0)}(\bbf_t^{(0)})^\T\big)^{-1}\sum_{t=1}^Nl_{tj}^{\prime}(\eta_{tj}^*)\bbf_t^{(0)}\overset{d}{\to}\cN(\zero,\Ib_K),\label{eq_variance_1to21}
\end{align}
as $\bPhi_{[N],j}^{(1)} = \Gb^{(1)}\bPhi_{[N],j}^{(0)}(\Gb^{(1)})^\T$. 
Next we define matrix $\bDelta_{N}^{(1)}$ as 
\begin{equation*}
    \big[\bDelta_{N}^{(1)}\big]_{[K_l,K_h]} = \left\{1_{(j_1=j_2)}\be_l^\T\bPhi_{[N],j_1}^{(1)}\be_h\right\}_{j_1\in \cA_l,j_2\in \cA_h}\in\RR^{(l-1)\times (h-1)},
\end{equation*}
where $K_l = (l-1)(l-2)/2+1,\cdots,(l-1)l/2$ for $l=2,\cdots,K$. 
Then we have $$\big(\bDelta_{N}^{(1)}\big)^{-1/2}\big((\bdelta_{2}^{(1)})^\T\cdots, (\bdelta_{K}^{(1)})^\T\big)^\T\overset{d}{\to}\cN(\zero,\Ib_{(K-1)K/2}).$$
Plug in $\blambda_{j}^{(1)}=(\Gb^{(1)})^\T\blambda_j^{(0)}$, $\bbf_{i}^{(1)}=(\Gb^{(1)})^\T\bbf_i^{(0)}$, \eqref{eq_precise_blambda} and 
we rewrite the equations as
\begin{equation*}
\left\{
\begin{aligned}
    O_p\Big(\frac{1}{N\wedge J}\Big)&=[\bLambda^{(1)}]_{[\cA_r,]}\big(\hat\Xb^{(1)}_{[r,]}\big)^\T - \bdelta_{r}^{(1)}\\
    O_p\Big(\frac{1}{N\wedge J^{2}}\Big)&= \hat\Xb^{(1)}+(\hat\Xb^{(1)})^\T
\end{aligned}
\right.
\end{equation*}
Write the coefficient matrix denoted by $\bDelta_{\lambda}^{(1)}$ being a upper-triangular matrix, with the $[K_l,K_h]$ given as
\begin{equation*}
    \big[\bDelta_{\lambda}^{(1)}\big]_{[K_l,K_h]} = 1_{(l=h)} [\bLambda^{(1)}]_{[\cA_l,1:{(l-1)}]} - 1_{(l < h)}[\bLambda^{(1)}]_{[\cA_h,h]}\big(\be_l^{(h-1)}\big)^\T
\end{equation*}
Then the equations can be solved by
\begin{equation*}
    \hat X_{lh}^{(1)}=\left\{
\begin{aligned}
    &\big(\be_h^{(l-1)}\big)^\T\sum_{k=l}^K\Big[\big(\bDelta_{\lambda}^{(1)}\big)^{-1}\Big]_{[K_l,K_k]}\bdelta_{k}^{(1)}+O_p\big(\frac{1}{N\wedge J}\big)\text{ when }l>h,
    \\&O_p\big(\frac{1}{N\wedge J}\big)\text{ when }l=h,\\&
-\hat X_{hl}^{(1)}+O_p\big(\frac{1}{N\wedge J}\big)\text{ when }l<h.
\end{aligned}
\right.
\end{equation*}
Therefore when $\sqrt{N}/J\to 0$, we have for some semi-positive definite matrix $\Xb_{\lambda}^{(2)}$ that $\big(\Xb_{\lambda}^{(1)}\big)^{-1/2}\mathrm{vec}(\hat\Xb^{(1)}) \overset{d}{\to} \cN(\zero,\Ib_{K^2})$, where the $(l_1+(h_1-1)K, l_2+(h_2-1)K)$th entry of $\Xb_{\lambda}^{(1)}$ is given as follows:
\begin{align}
      \big[\Xb_{\lambda}^{(1)} &\big]_{[l_1+(h_1-1)K, l_2+(h_2-1)K]} \nonumber\\&\,=\text{sgn}(l_1-h_1)\text{sgn}(l_2-h_2)\be_{h_1\wedge l_1}^\T\big(\bDelta_{\lambda}^{(1)}\big)^{-1}\bDelta_{N}^{(1)}\big(\bDelta_{\lambda}^{(1)}\big)^{-\T}\be_{l_2\wedge h_2}.\label{eq_expression_Xlambda1}
\end{align}
Here the sgn$(x)$ function equals 1 if $x>0$, $0$ if $x=0$ and $-1$ if $x<0$, indicating that when either $l_1=h_1$ or $l_2=h_2$ the entry will be zero.

Next we write $\hat\blambda_{j}^{(1)}-\blambda_{j}^{(1)}$ as
\begin{align*}
    \hat\blambda_{j}^{(1)}-\blambda_{j}^{(1)}
    =&\,(\hat\Gb^{(1)})^\T\hat\blambda_{j}^{(0)}-(\Gb^{(1)})^\T\blambda_j^{(0)}\\=&\,\big(\Gb^{(1)}\big)^\T\big(\hat\blambda_{j}^{(0)}-\blambda_j^{(0)}\big)-\hat\Xb^{(1)}\big(\Gb^{(1)}\big)^\T\blambda_j^{(0)}\\&\,+\big(\hat\Gb^{(1)}-\Gb^{(1)}\big)\big(\hat\blambda_j^{(0)}-\blambda_j^{(0)}\big)\\=&\,\big(\Gb^{(1)}\big)^\T\big(\hat\blambda_{j}^{(0)}-\blambda_j^{(0)}\big)-\hat\Xb^{(1)}\blambda_j^{(1)}+O_p\big(N^{-1}\big).
\end{align*}
We already know that $\big(\bPhi_{[N],j}^{(1)}\big)^{-1/2}\big(\Gb^{(1)}\big)^\T\big(\hat\blambda_{j}^{(0)}-\blambda_j^{(0)}\big)\overset{d}{\to} \cN(\zero,\Ib_K)$ by \eqref{eq_variance_1to21} and 
\begin{align*}
    &\,\Big\{\big((\blambda_j^{(1)})^\T\otimes \Ib_K\big)\Xb_{\lambda}^{(1)} \big(\blambda_{j}^{(1)}\otimes \Ib_K\big)\Big\}^{-1/2}\mathrm{vec}\big(\hat\Xb^{(1)}\blambda_j^{(1)}\big)\\=&\, \Big\{\big((\blambda_j^{(1)})^\T\otimes \Ib_K\big)\Xb_{\lambda}^{(1)} \big(\blambda_{j}^{(1)}\otimes \Ib_K\big)\Big\}^{-1/2}\big\{(\blambda_j^{(1)})^\T\otimes \Ib_K\big\}\mathrm{vec}(\hat\Xb^{(1)})\\\overset{d}{\to}&\,\cN(\zero,\Ib_K).
\end{align*}
Then when $j\notin \cA_r$ for any $r$, since $Cov\big((\hat\blambda_{j}^{(0)}-\blambda_j^{(0)}),(\hat\blambda_{r}^{(0)}-\blambda_r^{(0)})\big) = O_p\big((N\wedge J)^{-1}\big)$, by $\|\bPhi_{[N],j}^{(1)}\|=O_p(N^{-1})$, $\|\Xb_{\lambda}^{(1)}\|=O_p(N^{-1})$ and $\sqrt{N}/J\to 0$, we conclude that
\begin{equation*}
    \big(\bSigma_{[N],j}^{(1)}\big)^{-1/2}\big(\hat\blambda_{j}^{(1)}-\blambda_{j}^{(1)}\big)\overset{d}{\to}\cN(\zero,\Ib_K),
\end{equation*}
when $j\notin \cA_r$ for any $r$. Here when $j\in\cA_{r}$, we have $\Db_{(j)} = \Ib_K$.

For the other case, we need to account for $Cov\big(\big(\Gb^{(1)}\big)^\T\big(\hat\blambda_{j}^{(0)}-\blambda_j^{(0)}\big),\hat\Xb^{(1)}\blambda_j^{(1)}\big)$, with its $l$th column expressed as
\begin{align*}
    &\,Cov\Big(\big(\Gb^{(1)}\big)^\T\big(\hat\blambda_{j}^{(0)}-\blambda_j^{(0)}\big),\sum_{h\neq l}\hat X_{lh}^{(1)}\lambda_{jh}^{(1)}\Big) \\= &\,\sum_{h< l}Cov\Big(\big(\Gb^{(1)}\big)^\T\big(\hat\blambda_{j}^{(0)}-\blambda_j^{(0)}\big),\hat X_{lh}^{(1)}\Big)\lambda_{jh}^{(1)}\\&\, - \sum_{h> l}Cov\Big(\big(\Gb^{(1)}\big)^\T\big(\hat\blambda_{j}^{(0)}-\blambda_j^{(0)}\big),\hat X_{hl}^{(1)}\Big)\lambda_{jh}^{(1)}\\=&\, \sum_{h< l}\sum_{k=l}^KCov\Big(\big(\Gb^{(1)}\big)^\T\big(\hat\blambda_{j}^{(0)}-\blambda_j^{(0)}\big),\bdelta_k^{(1)}\Big)\Big[\big(\bDelta_{\lambda}^{(1)}\big)^{-1}\Big]_{[K_k,K_l]}\be_h^{(l-1)}\lambda_{jh}^{(1)}\\ &\, - \sum_{h> l}\sum_{k=h}^KCov\Big(\big(\Gb^{(1)}\big)^\T\big(\hat\blambda_{j}^{(0)}-\blambda_j^{(0)}\big),\bdelta_k^{(1)}\Big)\Big[\big(\bDelta_{\lambda}^{(1)}\big)^{-1}\Big]_{[K_k,K_h]}\be_l^{(h-1)}\lambda_{jh}^{(1)}\\=&\, \sum_{h< l}\sum_{k=l}^K\sum_{r=1}^{k-1}1_{(j= \cA_k(r))}\bPhi_{[N],j}^{(1)}\be_r\Big[\big(\bDelta_{\lambda}^{(1)}\big)^{-1}\Big]_{[K_{k,r},K_{l,h}]}\lambda_{jh}^{(1)}\\ &\, - \sum_{h> l}\sum_{k=h}^K\sum_{r=1}^{k-1}1_{(j= \cA_k(r))}\bPhi_{[N],j}^{(1)}\be_r\Big[\big(\bDelta_{\lambda}^{(1)}\big)^{-1}\Big]_{[K_{k,r},K_{h,l}]}\lambda_{jh}^{(1)}
\end{align*}
Here $K_{l,h} = (l-1)(l-2)/2+h$ for $l=2,\cdots,K$ and $h<l$. Note that if $j\in\cA_{r}$, then the $r$th row in the above form is deterministically $0$. Let
\begin{align}
    \Xb_{\lambda,kr}^{(1)} = 1_{(r<k)}\sum_{l\ge k}\left\{\sum_{h<l}\be_r\Big[\big(\bDelta_{\lambda}^{(1)}\big)^{-1}\Big]_{[K_{k,r},K_{l,h}]}\lambda_{jh}^{(1)}\be_l^\T + \sum_{h>l}\be_r\Big[\big(\bDelta_{\lambda}^{(1)}\big)^{-1}\Big]_{[K_{k,r},K_{h,l}]}\lambda_{jh}^{(1)}\be_l^\T\right\},\label{eq_x_lambda_kr_1}
\end{align}
and we have with the expression of $\bSigma_{[N],j}^{(1)}$ that for any $j\in[J]$, 
\begin{equation*}
    \big(\Db_{(j)}\bSigma_{[N],j}^{(1)}\Db_{(j)}^\T\big)^{-1/2}\Db_{(j)}\big(\hat\blambda_j^{(1)} - \blambda_j^{(1)}\big)\overset{d}{\to}\cN(\zero,\Ib_{|\cS_j|}),
\end{equation*}
with
\begin{equation*}
    \bSigma_{[N],j}^{(1)} = \bPhi_{[N],j}^{(1)} + \big((\blambda_j^{(1)})^\T\otimes \Ib_K\big)\Xb_{\lambda}^{(1)} \big(\blambda_j^{(1)}\otimes \Ib_K\big) + \sum_{k=1}^K\sum_{r=1}^{K}1_{(j= \cA_k(r))} \big\{\bPhi_{[N],j}^{(1)}\Xb_{\lambda,kr}^{(1)} + \big(\Xb_{\lambda,kr}^{(1)}\big)^\T\bPhi_{[N],j}^{(1)}\big\}.
\end{equation*}

Then, we write $(\hat\bbf_i^{(1)}-\bbf_i^{(1)})$ as
\begin{align*}
    \hat\bbf_i^{(1)}-\bbf_i^{(1)}=&\,\big(\hat\Gb^{(1)}-\Gb^{(1)}\big)^\T\bbf_i^{(0)}+(\Gb^{(1)})^\T\big(\hat\bbf_i^{(0)}-\bbf_i^{(0)}\big)\\&\,+\big(\hat\Gb^{(1)}-\Gb^{(1)}\big)^\T\big(\hat\bbf_i^{(0)}-\bbf_i^{(0)}\big)\\=&\,(\hat\Xb^{(1)})^\T\bbf_i^{(1)}+(\Gb^{(1)})^\T\big(\hat\bbf_i^{(0)}-\bbf_i^{(0)}\big)+O_p\big(N^{-1/2}  J^{-1/2}\big).
\end{align*}
Let
\begin{equation*}
     \bSigma_{i,[J]}^{(1)} = \bPhi_{i,[J]}^{(1)}+\big( \Ib_K\otimes (\bbf_i^{(1)})^\T \big)\Xb_{\lambda}^{(1)} \big(\Ib_K\otimes \bbf_i^{(1)}\big).
\end{equation*}
Since 
\begin{equation*}
    Cov\big(\hat\bbf_i^{(0)}-\bbf_i^{(0)}, \hat\blambda_j^{(0)}-\blambda_j^{(0)}\big) = O_p\big((N\wedge J)^{-2}\big)\text{, for any }i\in[N],j\in[J],
\end{equation*}
and $\|\bSigma_{i,[J]}^{(1)}\| = O_p\big(J^{-1}+ N^{-1}\big)$, similar to the derivation of limiting distribution for $\hat\blambda_j^{(1)} - \blambda_{j}^{(1)}$, we have
\begin{equation*}
    \big(\bSigma_{i,[J]}^{(1)}\big)^{-1/2}\big(\hat\bbf_i^{(1)}-\bbf_i^{(1)}\big)\overset{d}{\to} \cN(\zero,\Ib_K).
\end{equation*}
The consistency of the plug-in covariance estimator is implied by $\hat\bPhi^{(0)}\overset{p}{\to} \bPhi^{(0)}$, $\hat\Gb^{(1)}\overset{p}{\to}\Gb^{(1)}$, $\hat\blambda_j^{(1)}\overset{p}{\to}\blambda_j^{(1)}$ and $\hat\bbf_i^{(1)}\overset{p}{\to}\bbf_i^{(1)}$.
\subsection[Limiting Distributions of $\hat\blambda_j^{(2)}$s and $\hat\bbf_i^{(2)}$s]{Limiting Distributions of $\hat\blambda_j^{(2)}$s and $\hat\bbf_i^{(2)}$s}\label{sec_limit_ic2}
Under IC$^{(2)}$, as in the proof of consistency, we find orthogonal matrices $\Gb^{(2)}$ and $\hat\Gb^{(2)}$ such that $\bLambda^{(2)}=\bLambda^{(0)}\Gb^{(2)}$, $\Fb^{(2)}=\Fb^{(0)}\Gb^{(2)}$, $\hat\bLambda^{(2)}=\hat\bLambda^{(0)}\hat\Gb^{(2)}$ and $\hat\Fb^{(2)}=\hat\Fb^{(0)}\hat\Gb^{(2)}$ with $\|\hat\Gb^{(2)}-\Gb^{(2)}\|=O_p(N^{-1/2} + J^{-1})$. The consistency rate alone is not enough for limiting distributions of the estimation. 
From the identifiability condition, we know (1) $\bLambda^{(2)}_{[1:K,]}$ and $\hat\bLambda^{(2)}_{[1:K,]}$ are lower-triangular; (2) $\Ib_K=N^{-1}(\Fb^{(2)})^\T\Fb^{(2)}=N^{-1}(\hat\Fb^{(2)})^\T\hat\Fb^{(2)}$, from which we have the following equations:
\begin{align*}
    \text{lower}\Big\{\big(\hat\Gb^{(2)}\big)^\T\big(\hat\blambda_1^{(0)},\cdots,\hat\blambda_K^{(0)}\big)-\big(\Gb^{(2)}\big)^\T\big(\blambda_1^{(0)},\cdots,\blambda_K^{(0)}\big)\Big\}&\,=\zero;\\
    \big(\hat\Gb^{(2)}\big)^{\T}\hat\Gb^{(2)}-\big(\Gb^{(2)}\big)^{\T}\Gb^{(2)}&\,=\zero.
\end{align*}
By $\big\|\Gb^{(2)}-\hat\Gb^{(2)}\big\|=O_p(N^{-1/2} + J^{-1})$ and $\|\hat\blambda_j^{(0)}-\blambda_j^{(0)}\|=O_p\big(N^{-1/2} + J^{-1}\big)$, we can write the first part as
\begin{align*}
    \text{lower}&\,\Big\{\big(\hat\Gb^{(2)}\big)^\T\big(\hat\blambda_1^{(0)},\cdots,\hat\blambda_K^{(0)}\big)-\big(\Gb^{(2)}\big)^\T\big(\blambda_1^{(0)},\cdots,\blambda_K^{(0)}\big)\Big\}\\=&\,\text{lower}\Big\{\big(\hat\Gb^{(2)}-\Gb^{(2)}\big)^\T\big(\blambda_1^{(0)},\cdots,\blambda_K^{(0)}\big)+\big(\Gb^{(2)}\big)^\T\big(\hat\blambda_1^{(0)}-\blambda_1^{(0)},\cdots,\hat\blambda_K^{(0)}-\blambda_K^{(0)}\big)\Big\}\\&\,+\text{lower}\Big\{\big(\hat\Gb^{(2)}-\Gb^{(2)}\big)^\T\big(\hat\blambda_1^{(0)}-\blambda_1^{(0)},\cdots,\hat\blambda_K^{(0)}-\blambda_K^{(0)}\big)\Big\}\\=&\,\text{lower}\Big\{\big(\hat\Gb^{(2)}-\Gb^{(2)}\big)^\T\big(\blambda_1^{(0)},\cdots,\blambda_K^{(0)}\big)+\big(\Gb^{(2)}\big)^\T\big(\hat\blambda_1^{(0)}-\blambda_1^{(0)},\cdots,\hat\blambda_K^{(0)}-\blambda_K^{(0)}\big)\Big\}\\&\,+O_p\Big(\frac{1}{N\wedge J^{2}}\Big).
\end{align*}
Similarly the second part can be written as 
\begin{align*}
    \big(\hat\Gb^{(2)}\big)^{\T}\hat\Gb^{(2)}-\big(\Gb^{(2)}\big)^{\T}\Gb^{(2)}=&\,-\big(\Gb^{(2)}-\hat\Gb^{(2)}\big)^{\T}\Gb^{(2)}-\big(\Gb^{(2)}\big)^\T\big(\Gb^{(2)}-\hat\Gb^{(2)}\big)\\&\,+\big(\Gb^{(2)}-\hat\Gb^{(2)}\big)^\T\big(\Gb^{(2)}-\hat\Gb^{(2)}\big)\\=&\,-\big(\Gb^{(2)}-\hat\Gb^{(2)}\big)^{\T}\Gb^{(2)}-\big(\Gb^{(2)}\big)^\T\big(\Gb^{(2)}-\hat\Gb^{(2)}\big)+O_p\Big(\frac{1}{N\wedge J^{2}}\Big).
\end{align*}

Define $\hat\Xb^{(2)}=(\Gb^{(2)}-\hat\Gb^{(2)})^{\T}\Gb^{(2)}$ and \begin{equation*}\bdelta_{r}^{(2)}=\big\{\be_r^\T\big(\sum_{t=1}^Nl_{tj}^{\prime\prime}(\eta_{tj}^*)\bbf_t^{(2)}(\bbf_t^{(2)})^\T\big)^{-1}\sum_{t=1}^Nl_{tj}^{\prime}(\eta_{tj}^*)\bbf_t^{(2)}\big\}_{j\in [1:K]}\in\RR^{K}.\end{equation*} 
From Section~\ref{sec:use_clt}, we know that 
\begin{align}
    &\,\big(\bPhi_{[N],j}^{(2)}\big)^{-1/2}\big(\sum_{t=1}^Nl_{tj}^{\prime\prime}(\eta_{tj}^*)\bbf_t^{(2)}(\bbf_t^{(2)})^\T\big)^{-1}\sum_{t=1}^Nl_{tj}^{\prime}(\eta_{tj}^*)\bbf_t^{(2)} \nonumber\\= &\, \big(\bPhi_{[N],j}^{(2)}\big)^{-1/2}\Gb^{(2)}\big(\sum_{t=1}^Nl_{tj}^{\prime\prime}(\eta_{tj}^*)\bbf_t^{(0)}(\bbf_t^{(0)})^\T\big)^{-1}\sum_{t=1}^Nl_{tj}^{\prime}(\eta_{tj}^*)\bbf_t^{(0)}\overset{d}{\to}\cN(\zero,\Ib_K),\label{eq_variance_1to23}
\end{align}
as $\bPhi_{[N],j}^{(2)} = \Gb^{(2)}\bPhi_{[N],j}^{(0)}(\Gb^{(2)})^\T$. 
Therefore for $\bdelta_r^{(2)}$ if we define matrix $\bDelta_{N}^{(2)}$ as 
\begin{equation*}
    \bDelta_{N}^{(2)} = \sum_{j=1}^K \bPhi_{[N],j}^{(2)}\otimes \big(\be_j\be_j^\T\big),
\end{equation*}
we have $\big(\bDelta_{N}^{(2)}\big)^{-1/2}\big((\bdelta_{1}^{(2)})^\T,\cdots, (\bdelta_{K}^{(2)})^\T\big)^\T\overset{d}{\to}\cN(\zero,\Ib_{K^2})$.
Plug in $\blambda_{j}^{(2)}=(\Gb^{(2)})^\T\blambda_j^{(0)}$, $\bbf_{i}^{(2)}=(\Gb^{(2)})^\T\bbf_i^{(0)}$, \eqref{eq_precise_blambda} and 
we rewrite the equations as
\begin{equation*}
\left\{
\begin{aligned}
    O_p\Big(\frac{1}{N\wedge J}\Big)&=\text{lower}\Big\{\hat\Xb^{(2)}\big(\blambda_1^{(2)},\cdots,\blambda_K^{(2)}\big)\\&\quad-\Big(\bdelta_{1}^{(2)},\cdots,\bdelta_{K}^{(2)}\Big)^\T\Big\}, \\
    O_p\Big(\frac{1}{N\wedge J^{2}}\Big)&= \hat\Xb^{(2)}+(\hat\Xb^{(2)})^\T.
\end{aligned}
\right.
\end{equation*}
Since $(\bLambda_{[1:K,]}^{(2)})^\T$ is upper-triangular, $\text{lower}\big(\hat\Xb^{(2)}(\bLambda_{[1:K,]}^{(2)})^\T\big)=\text{lower}\big\{\text{lower}\big(\hat\Xb^{(2)}\big)(\bLambda_{[1:K,]}^{(2)})^\T\big\}$. Then the equations can be solved by
\begin{equation*}
    \hat X_{lh}^{(2)}=\left\{
\begin{aligned}
    &\be_h^\T(\bLambda_{[1:K,]}^{(2)})^{-1}\bdelta^{(2)}_l+O_p\big(\frac{1}{N\wedge J}\big)\text{ when }l>h,
    \\&O_p\big(\frac{1}{N\wedge J}\big)\text{ when }l=h,\\&
-\hat X_{hl}^{(2)}+O_p\big(\frac{1}{N\wedge J}\big)\text{ when }l<h.
\end{aligned}
\right.
\end{equation*}

Therefore when $\sqrt{N}/J\to 0$, we have for some semi-positive definite matrix $\Xb_{\lambda}^{(2)}$ that $\big(\Xb_{\lambda}^{(2)}\big)^{-1/2}\mathrm{vec}(\hat\Xb^{(2)}) \overset{d}{\to} \cN(\zero,\Ib_{K^2})$, where the $(l_1+(h_1-1)K, l_2+(h_2-1)K)$th entry of $\Xb_{\lambda}^{(2)}$ is given as follows:
\begin{align}
       \big[\Xb_{\lambda}^{(2)}& \big]_{l_1+(h_1-1)K, l_2+(h_2-1)K}\nonumber\\&\,= \text{sgn}(l_1-h_1)\text{sgn}(l_2-h_2)\be_{h_1\wedge l_1}^\T(\bLambda_{[1:K,]}^{(2)})^{-1}\bDelta_{N,l_1\vee h_1,l_2\vee h_2}^{(2)}(\bLambda_{[1:K,]}^{(2)})^{-\T}\be_{l_2\wedge h_2}\label{eq_expression_Xlambda2}
\end{align}
Here $\bDelta_{N,lh}^{(2)} := \big[\bDelta_{N}^{(2)}\big]_{[(l-1)K + 1:lK,(h-1)K + 1:hK]}$ denotes the $(l,h)$th $K$-by-$K$ block of $\bDelta_{N}^{(2)}$. 


Next we write $\hat\blambda_{j}^{(2)}-\blambda_{j}^{(2)}$ as
\begin{align*}
    \hat\blambda_{j}^{(2)}-\blambda_{j}^{(2)}
    =&\,(\hat\Gb^{(2)})^\T\hat\blambda_{j}^{(0)}-(\Gb^{(2)})^\T\blambda_j^{(0)}\\=&\,\big(\Gb^{(2)}\big)^\T\big(\hat\blambda_{j}^{(0)}-\blambda_j^{(0)}\big)-\hat\Xb^{(2)}\big(\Gb^{(2)}\big)^\T\blambda_j^{(0)}\\&\,+\big(\hat\Gb^{(2)}-\Gb^{(2)}\big)\big(\hat\blambda_j^{(0)}-\blambda_j^{(0)}\big)\\=&\,\big(\Gb^{(2)}\big)^\T\big(\hat\blambda_{j}^{(0)}-\blambda_j^{(0)}\big)-\hat\Xb^{(2)}\blambda_j^{(2)}+O_p\big(N^{-1}\big).
\end{align*}
We already know that $\big(\bPhi_{[N],j}^{(2)}\big)^{-1/2}\big(\Gb^{(2)}\big)^\T\big(\hat\blambda_{j}^{(0)}-\blambda_j^{(0)}\big)\overset{d}{\to} \cN(\zero,\Ib_K)$ by \eqref{eq_variance_1to23} and 
\begin{align*}
    &\,\Big\{\big((\blambda_j^{(2)})^\T\otimes \Ib_K\big)\Xb_{\lambda}^{(2)} \big(\blambda_{j}^{(2)}\otimes \Ib_K\big)\Big\}^{-1/2}\mathrm{vec}\big(\hat\Xb^{(2)}\blambda_j^{(2)}\big)\\=&\, \Big\{\big((\blambda_j^{(2)})^\T\otimes \Ib_K\big)\Xb_{\lambda}^{(2)} \big(\blambda_{j}^{(2)}\otimes \Ib_K\big)\Big\}^{-1/2}(\blambda_j^{(2)})^\T\otimes \Ib_K)\mathrm{vec}(\hat\Xb^{(2)})\\\overset{d}{\to}&\,\cN(\zero,\Ib_K).
\end{align*}
Then when $j> K$, we know $\Db_{(j)} = \Ib_K$ where no component in $\blambda_j$ is fixed. Since $Cov\big((\hat\blambda_{j}^{(0)}-\blambda_j^{(0)}),(\hat\blambda_{r}^{(0)}-\blambda_r^{(0)})\big) = O_p\big((N\wedge J)^{-1}\big)$, by $\|\bPhi_{[N],j}^{(2)}\|=O_p(N^{-1})$, $\|\Xb_{\lambda}^{(2)}\|=O_p(N^{-1})$ and $\sqrt{N}/J\to 0$, we conclude that
\begin{equation*}
    \big(\bSigma_{[N],j}^{(2)}\big)^{-1/2}\big(\hat\blambda_{j}^{(2)}-\blambda_{j}^{(2)}\big)\overset{d}{\to}\cN(\zero,\Ib_K)
\end{equation*}
hold for $j>K$.
For the case of $j\le K$, we select the part in $\hat\Xb^{(2)}$ that relates to $\hat\blambda_j^{(0)} - \blambda_j^{(0)}$. First since $\hat\Xb^{(2)}$ is solved by $\text{lower}(\hat\Xb^{(2)})=\text{lower}\big[(\bdelta_{1}^{(2)},\cdots,\bdelta_K^{(2)})^\T(\bLambda_{[1:K,]}^{(2)})^{-\T}\big]+O_p\big((N\wedge J)^{-1}\big)$, we know that the last $K-j$ elements of $\sqrt{N}\hat\Xb^{(2)}\blambda_j^{(2)}$ can be expressed as $(\Ib_K-\Ib_K^{j})(\Gb^{(2)})^\T\big(\hat\blambda_j^{(0)}-\blambda_j^{(0)}\big)+O_p\big((N\wedge J)^{-1}\big)$. Then 
\begin{align*}
    Cov\big((\Gb^{(2)})^\T(\hat\blambda_j^{(0)}-\blambda_j^{(0)}),(\Ib_K-\Ib_K^j)\hat\Xb^{(2)}\blambda_j^{(2)}\big)= &\,(\Gb^{(2)})^\T Var\big(\hat\blambda_j^{(0)}-\blambda_j^{(0)}\big)\Gb^{(2)}(\Ib_K-\Ib_K^{j})  \\&\,+ O_p\big((N\wedge J)^{-2}\big).
\end{align*}
For the first $j$ elements in $\sqrt{N}\hat\Xb^{(2)}\blambda_j^{(2)}$, note that it is only related to $\text{lower}\big(\hat\Xb^{(2)}_{[1:j,1:j]}\big)=\text{lower}\big[(\bdelta_{1}^{(2)},\cdots,\bdelta_K^{(2)})^\T(\bLambda_{[1:K,]}^{(2)})^{-\T}\big]_{[1:K,1:K]}$. Note that $(\bLambda_{[1:K,]}^{(2)})^{-\T}$ is upper-triangular, then this expression will not involve any term in $(\Gb^{(2)})^\T\big(\hat\blambda_j^{(0)}-\blambda_j^{(0)}\big)$. Therefore 
\begin{equation*}
    Cov\big((\Gb^{(2)})^\T(\hat\blambda_j^{(0)}-\blambda_j^{(0)}),\Ib_K^j\hat\Xb^{(2)}\blambda_j^{(2)}\big) = O_p\big((N\wedge J)^{-2}\big).
\end{equation*}
the $(j+1)$--$K$ rows in the above form are deterministically $0$. Therefore, with the expression of $\bSigma_{[N],j}^{(2)}$,  we conclude that
\begin{equation*}
    \big(\Db_{(j)}\bSigma_{[N],j}^{(2)}\Db_{(j)}^\T\big)^{-1/2}\Db_{(j)}\big(\hat\blambda_{j}^{(2)}-\blambda_{j}^{(2)}\big)\overset{d}{\to}\cN(\zero,\Ib_{|\cS_j|})
\end{equation*}
hold for $j\le K$.

And we write $(\hat\bbf_i^{(2)}-\bbf_i^{(2)})$ as
\begin{align*}
    \hat\bbf_i^{(2)}-\bbf_i^{(2)}=&\,\big(\hat\Gb^{(2)}-\Gb^{(2)}\big)^\T\bbf_i^{(0)}+(\Gb^{(2)})^\T\big(\hat\bbf_i^{(0)}-\bbf_i^{(0)}\big)\\&\,+\big(\hat\Gb^{(2)}-\Gb^{(2)}\big)^\T\big(\hat\bbf_i^{(0)}-\bbf_i^{(0)}\big)\\=&\,(\hat\Xb^{(2)})^\T\bbf_i^{(2)}+(\Gb^{(2)})^\T\big(\hat\bbf_i^{(0)}-\bbf_i^{(0)}\big)+O_p\big(N^{-1/2}J^{-1/2}\big).
\end{align*}
Since 
\begin{equation*}
    Cov\big(\hat\bbf_i^{(0)}-\bbf_i^{(0)}, \hat\blambda_j^{(0)}-\blambda_j^{(0)}\big) = O_p\big((N\wedge J)^{-2}\big)\text{, for any }i\in[N],j\in[J],
\end{equation*}
we know by $\|\bSigma_{i,[J]}^{(2)}\| = O_p\big(J^{-1}+ N^{-1}\big)$ that
\begin{equation*}
    \big(\bSigma_{i,[J]}^{(2)}\big)^{-1/2}\big(\hat\bbf_i^{(2)}-\bbf_i^{(2)}\big)\overset{d}{\to} \cN(\zero,\Ib_K).
\end{equation*}
The consistency of the plug-in covariance estimator is implied by $\hat\bPhi^{(0)}\overset{p}{\to} \bPhi^{(0)}$, $\hat\Gb^{(2)}\overset{p}{\to}\Gb^{(2)}$, $\hat\blambda_j^{(2)}\overset{p}{\to}\blambda_j^{(2)}$ and $\hat\bbf_i^{(2)}\overset{p}{\to}\bbf_i^{(2)}$.

\subsection[Limiting Distributions of $\hat\blambda_j^{(3)}$s and $\hat\bbf_i^{(3)}$s]{Limiting Distributions of $\hat\blambda_j^{(3)}$s and $\hat\bbf_i^{(3)}$s}\label{sec_limit_ic3}
Under IC$^{(3)}$, similarly we find in the proof of Theorem~\ref{thm_consis} $\Gb^{(3)}$ and $\hat\Gb^{(3)}$ such that $\bLambda^{(3)}=\bLambda^{(0)}\Gb^{(3)}$, $\Fb^{(3)}=\Fb^{(0)}(\Gb^{(3)})^{-\T}$, $\hat\bLambda^{(3)}=\hat\bLambda^{(0)}\hat\Gb^{(3)}$ and $\hat\Fb^{(3)}=\hat\Fb^{(0)}(\hat\Gb^{(3)})^{-\T}$ with $\|\hat\Gb^{(3)}-\Gb^{(3)}\|=O_p(N^{-1/2} + J^{-1})$. From IC$^{(3)}$, (1) $\bLambda^{(3)}$ and $\hat\bLambda^{(3)}$ are lower-triangular with unit diagonal elements; (2) $N^{-1}(\Fb^{(3)})^\T\Fb^{(3)}$ and $N^{-1}(\hat\Fb^{(3)})^\T\hat\Fb^{(3)}$ are diagonal. Similar to what we did previously, we can construct equations for $\hat\Gb^{(3)}$ as follows:
\begin{align*}
    \text{nupper}\Big\{\big(\hat\Gb^{(3)}\big)^\T\big(\hat\blambda_1^{(0)},\cdots,\hat\blambda_K^{(0)}\big)-\big(\Gb^{(3)}\big)^\T\big(\blambda_1^{(0)},\cdots,\blambda_K^{(0)}\big)\Big\}&\,=\zero;\\
    \text{ndiag}\big\{(\hat\Gb^{(3)})^{-1}\big(\hat\Gb^{(3)}\big)^{-\T}-(\Gb^{(3)})^{-1}\big(\Gb^{(3)}\big)^{-\T}\big\}&\,=\zero.
\end{align*}
By $\big\|\Gb^{(3)}-\hat\Gb^{(3)}\big\|=O_p(N^{-1/2} + J^{-1})$ and $\|\hat\blambda_j^{(0)}-\blambda_j^{(0)}\|=O_p\big(N^{-1/2} + J^{-1}\big)$, we can write the first part as
\begin{align*}
    \text{nupper}&\,\Big\{\big(\hat\Gb^{(3)}\big)^\T\big(\hat\blambda_1^{(0)},\cdots,\hat\blambda_K^{(0)}\big)-\big(\Gb^{(3)}\big)^\T\big(\blambda_1^{(0)},\cdots,\blambda_K^{(0)}\big)\Big\}\\=&\,\text{nupper}\Big\{\big(\hat\Gb^{(3)}-\Gb^{(3)}\big)^\T\big(\blambda_1^{(0)},\cdots,\blambda_K^{(0)}\big)+\big(\Gb^{(3)}\big)^\T\big(\hat\blambda_1^{(0)}-\blambda_1^{(0)},\cdots,\hat\blambda_K^{(0)}-\blambda_K^{(0)}\big)\Big\}\\&\,+\text{nupper}\Big\{\big(\hat\Gb^{(3)}-\Gb^{(3)}\big)^\T\big(\hat\blambda_1^{(0)}-\blambda_1^{(0)},\cdots,\hat\blambda_K^{(0)}-\blambda_K^{(0)}\big)\Big\}\\=&\,\text{nupper}\Big\{\big(\hat\Gb^{(3)}-\Gb^{(3)}\big)^\T\big(\blambda_1^{(0)},\cdots,\blambda_K^{(0)}\big)+\big(\Gb^{(3)}\big)^\T\big(\hat\blambda_1^{(0)}-\blambda_1^{(0)},\cdots,\hat\blambda_K^{(0)}-\blambda_K^{(0)}\big)\Big\}\\&\,+O_p\Big(\frac{1}{N\wedge J^{2}}\Big).
\end{align*}
Similarly with $(\Gb^{(3)})^{-1}\big(\Gb^{(3)}\big)^{-\T}=\Mb_{ff}^{(3)}$ and 
\begin{equation*}
    (\Gb^{(3)})^{-1}-(\hat\Gb^{(3)})^{-1}=(\Gb^{(3)})^{-1}\big(\hat\Gb^{(3)}-\Gb^{(3)}\big)(\Gb^{(3)})^{-1}+O_p\Big(\frac{1}{N\wedge J^{2}}\Big),
\end{equation*}
we can write the second part as 
\begin{align*}
    \text{ndiag}\big\{(\hat\Gb^{(3)})^{-1}&\,\big(\hat\Gb^{(3)}\big)^{-\T}-(\Gb^{(3)})^{-1}\big(\Gb^{(3)}\big)^{-\T}\big\}\\=&\,\text{ndiag}\big\{-\big[(\Gb^{(3)})^{-1}-(\hat\Gb^{(3)})^{-1}\big](\Gb^{(3)})^{-\T}-\big(\Gb^{(3)}\big)^{-1}\big[(\Gb^{(3)})^{-1}-(\hat\Gb^{(3)})^{-1}\big]^{\T}\\&\,+\big[(\Gb^{(3)})^{-1}-(\hat\Gb^{(3)})^{-1}\big]\big[(\Gb^{(3)})^{-1}-(\hat\Gb^{(3)})^{-1}\big]^{\T}\big\}\\=&\,\text{ndiag}\big\{-(\Gb^{(3)})^{-1}\big(\Gb^{(3)}-\hat\Gb^{(3)}\big)\Mb_{ff}^{(3)}-\Mb_{ff}^{(3)}\big(\Gb^{(3)}-\hat\Gb^{(3)}\big)^{\T}\big(\Gb^{(3)}\big)^{-\T}\big\}+O_p\Big(\frac{1}{N\wedge J^{2}}\Big).
\end{align*}
Define $\hat\Xb^{(3)}=\Mb_{ff}^{(3)}(\Gb^{(3)}-\hat\Gb^{(3)})^{\T}(\Gb^{(3)})^{-\T}$ and \begin{equation*}\bdelta_{r}^{(3)}=\big\{\be_r^\T\big(\sum_{t=1}^Nl_{tj}^{\prime\prime}(\eta_{tj}^*)\bbf_t^{(3)}(\bbf_t^{(3)})^\T\big)^{-1}\sum_{t=1}^Nl_{tj}^{\prime}(\eta_{tj}^*)\bbf_t^{(3)}\big\}_{j\in [1:K]}\in\RR^{K}.\end{equation*} 
Similar to the derivation in Section~\ref{sec_limit_ic2}, we know $\big(\bDelta_{N}^{(3)}\big)^{-1/2}\big((\bdelta_{1}^{(3)})^\T,\cdots, (\bdelta_{K}^{(3)})^\T\big)^\T\overset{d}{\to}\cN(\zero,\Ib_{K^2})$ where $\bDelta_{N}^{(3)}$
 is defined as
\begin{equation*}
    \bDelta_{N}^{(3)} = \sum_{j=1}^K \bPhi_{[N],j}^{(3)}\otimes \big(\be_j\be_j^\T\big).
\end{equation*}

Plug in $\blambda_{j}^{(3)}=(\Gb^{(3)})^\T\blambda_j^{(0)}$, \eqref{eq_precise_blambda} and 
we rewrite the equations as
\begin{equation*}
\left\{
\begin{aligned}
    O_p\Big(\frac{1}{N\wedge J}\Big)&=\text{nupper}\Big\{\big(\Mb_{ff}^{(3)}\big)^{-1}\hat\Xb^{(3)}\big(\blambda_1^{(3)},\cdots,\blambda_K^{(3)}\big)\\&-\Big[\big(\sum_{i=1}^Nl_{1j}^{\prime\prime}(\eta_{1j}^*)\bbf_i^{(3)}(\bbf_i^{(3)})^\T\big)^{-1}\sum_{i=1}^Nl_{1j}^{\prime}(\eta_{1j}^*)\bbf_i^{(3)},\cdots,\\&\qquad \big(\sum_{i=1}^Nl_{Kj}^{\prime\prime}(\eta_{Kj}^*)\bbf_i^{(3)}(\bbf_i^{(3)})^\T\big)^{-1}\sum_{i=1}^Nl_{Kj}^{\prime}(\eta_{Kj}^*)\bbf_i^{(3)}\Big]\Big\}, \\
    O_p\Big(\frac{1}{N\wedge J^{2}}\Big)&=\text{ndiag}\big( \hat\Xb^{(3)}+(\hat\Xb^{(3)})^\T\big).
\end{aligned}
\right.
\end{equation*}The equations can be solved by
\begin{equation*}
    \hat X_{lh}^{(3)}=\left\{
\begin{aligned}&
     m_{f,l}^{(3)}\be_h^\T\big[(\bLambda_{[1:K,]}^{(3)})^{-1}\big]\bdelta_l^{(3)}+O_p\big(\frac{1}{N\wedge J}\big)\text{ when }l\ge h\\
&-\hat X_{hl}^{(3)}+O_p\big(\frac{1}{N\wedge J}\big)\text{ when }l<h
\end{aligned}
\right.
\end{equation*}
Here $m_{f,l}^{(3)}$ is the $l$th diagonal elements in $\Mb_{ff}^{(3)}$.
Then when $\sqrt{N}/J\to 0$, we have for some semi-positive definite matrix $\Xb_{\lambda}^{(3)}$ that $\big(\Xb_{\lambda}^{(3)}\big)^{-1/2}\mathrm{vec}(\hat\Xb^{(3)})\overset{d}{\to}\cN(\zero,\Ib_{K^2})$, where the $(l_1+(h_1-1)K,l_2+(h_2-1)K)$th entry of $\Xb_{\lambda}^{(3)}$ is given as follows:
\begin{align}
        \big[&\Xb_{\lambda}^{(3)}\big]_{[l_1+(h_1-1)K,l_2+(h_2-1)K]}\nonumber\\&=\text{sgn}^+(l_1-h_1)\text{sign}^+(l_2-h_2)m_{f,l_1\vee h_1}^{(3)}m_{f,l_2\vee h_2}^{(3)}\be_{h_1\wedge l_1}^\T(\bLambda_{[1:K,]}^{(3)})^{-1}\bDelta_{N,l_1\vee h_1,l_2\vee h_2}^{(3)}(\bLambda_{[1:K,]}^{(3)})^{-\T}\be_{l_2\wedge h_2}.\label{eq_expression_Xlambda3}
\end{align}
Here $\bDelta_{N,lh}^{(3)} := \big[\bDelta_{N}^{(3)}\big]_{[(l-1)K + 1:lK,(h-1)K + 1:hK]}$ denotes the $(l,h)$th $K$-by-$K$ block of $\bDelta_{N}^{(3)}$ and function sgn$^+$ differs from sgn in that ${\rm sgn}^+(0)=0$.

Next we compute $\hat\blambda_{j}^{(3)}-\blambda_{j}^{(3)}$ as
\begin{align*}
    \hat\blambda_{j}^{(3)}-\blambda_{j}^{(3)}
    &\,=(\hat\Gb^{(3)})^\T\hat\blambda_{j}^{(0)}-(\Gb^{(3)})^\T\blambda_j^{(0)}\\&\,=\big(\Gb^{(3)}\big)^\T\big(\hat\blambda_{j}^{(0)}-\blambda_j^{(0)}\big)-(\Mb_{ff}^{(3)})^{-1}\hat\Xb^{(3)}\big(\Gb^{(3)}\big)^\T\blambda_j^{(0)}\\&\,+\big(\hat\Gb^{(3)}-\Gb^{(3)}\big)^\T\big(\hat\blambda_j^{(0)}-\blambda_j^{(0)}\big)\\&\,=\big(\Gb^{(3)}\big)^\T\big(\hat\blambda_{j}^{(0)}-\blambda_j^{(0)}\big)-(\Mb_{ff}^{(3)})^{-1}\hat\Xb^{(3)}\blambda_j^{(3)}+O_p\Big(\frac{1}{N\wedge J^{2}}\Big).
\end{align*}
Similar to the analysis for the two cases of $j\le K$ and $j>K$, we can conclude with the expression of $\bSigma_{[N],j}^{(3)}$ that when $\sqrt{N}/J\to 0$ and $j>1$,
\begin{equation*}
    \big(\Db_{(j)}\bSigma_{[N],j}^{(3)}\Db_{(j)}^\T\big)^{-1/2}\Db_{(j)}\big(\hat\blambda_j^{(3)} - \blambda_j^{(3)}\big)\overset{d}{\to}\cN(\zero,\Ib_{|\cS_j|}).
\end{equation*}
Next for $\hat\bbf_i^{(3)}-\bbf_i^{(3)}$, similarly we have
\begin{align*}
    \hat\bbf_i^{(3)}-\bbf_i^{(3)}=&\,\big((\hat\Gb^{(3)})^{-1}-(\Gb^{(3)})^{-1}\big)\bbf_i^{(0)}+(\Gb^{(3)})^{-1}\big(\hat\bbf_i^{(0)}-\bbf_i^{(0)}\big)\\&\,+\big((\hat\Gb^{(3)})^{-1}-(\Gb^{(3)})^{-1}\big)\big(\hat\bbf_i^{(0)}-\bbf_i^{(0)}\big)\\=&\,(\hat\Xb^{(3)})^\T(\Mb_{ff}^{(3)})^{-1}\bbf_i^{(3)}+(\Gb^{(3)})^{-1}\big(\hat\bbf_i^{(0)}-\bbf_i^{(0)}\big)+O_p\big(N^{-1/2}J^{-1/2}\big),
\end{align*}
and therefore $\big(\bSigma_{i,[J]}^{(3)}\big)^{-1/2}\big(\hat\bbf_i^{(3)} - \bbf_i^{(3)}\big)\overset{d}{\to}\cN(\zero,\Ib_K)$. The consistency of the plug-in covariance estimator is implied by $\hat\bPhi^{(0)}\overset{p}{\to} \bPhi^{(0)}$, $\hat\Gb^{(3)}\overset{p}{\to}\Gb^{(3)}$, $\hat\blambda_j^{(3)}\overset{p}{\to}\blambda_j^{(3)}$ and $\hat\bbf_i^{(3)}\overset{p}{\to}\bbf_i^{(3)}$.

\subsection[Limiting Distributions of $\hat\blambda_j^{(4)}$s and $\hat\bbf_i^{(4)}$s]{Limiting Distributions of $\hat\blambda_j^{(4)}$s and $\hat\bbf_i^{(4)}$s}\label{sec_limit_ic4}
Under IC$^{(4)}$, similarly we find in the proof of Theorem~\ref{thm_consis} $\Gb^{(4)}$ and $\hat\Gb^{(4)}$ such that $\bLambda^{(4)}=\bLambda^{(0)}\Gb^{(4)}$, $\Fb^{(4)}=\Fb^{(0)}(\Gb^{(4)})^{-\T}$, $\hat\bLambda^{(4)}=\hat\bLambda^{(0)}\hat\Gb^{(4)}$ and $\hat\Fb^{(4)}=\hat\Fb^{(0)}(\hat\Gb^{(4)})^{-\T}$ with $\|\hat\Gb^{(4)}-\Gb^{(4)}\|=O_p(N^{-1/2} + J^{-1})$. From IC$^{(4)}$, (1) $\mathrm{diag}\big(N^{-1}(\Fb^{(4)})^\T\Fb^{(4)}\big)$ and $\mathrm{diag}\big(N^{-1}(\hat\Fb^{(4)})^\T\hat\Fb^{(4)}\big)$ are $\Ib_K$; (2) $\bLambda^{(4)}_{[\cA_r,r]}=\hat\bLambda^{(4)}_{[\cA_r,r]}=\zero_{K-1}$. Then $\Gb^{(4)}$ can be solved from the following equations:
\begin{align*}\bLambda^{(0)}_{[\cA_r,]}\Gb^{(4)}_{[,r]}-\hat\bLambda^{(4)}_{[\cA_r,]}\hat\Gb^{(4)}_{[,r]}=&\,\zero,\;r\in[K];
    \\\mathrm{diag}\big\{(\Gb^{(4)})^{-1}(\Gb^{(4)})^{-\T}-(\hat\Gb^{(0)})^{-1}(\hat\Gb^{(4)})^{-\T}\big\}=&\,\zero.
\end{align*}
By $\big\|\Gb^{(4)}-\hat\Gb^{(4)}\big\|=O_p(N^{-1/2})$ and $\|\hat\blambda_j^{(0)}-\blambda_j^{(0)}\|=O_p\big(N^{-1/2}\big)$, we can write the first part as
\begin{align*}
    &\,\big[\bLambda^{(0)}\Gb^{(4)}-\hat\bLambda^{(0)}\hat\Gb^{(4)}\big]_{[\cA_r,r]}\\=&\,\big[\big(\bLambda^{(0)}-\hat\bLambda^{(0)}\big)\Gb^{(4)}\big]_{[\cA_r,r]}+\big[\bLambda^{(0)}\big(\Gb^{(4)}-\hat\Gb^{(4)}\big)\big]_{[\cA_r,r]}\\&\,-\big[\big(\bLambda^{(0)}-\hat\bLambda^{(0)}\big)\big(\Gb^{(4)}-\hat\Gb^{(4)}\big)\big]_{[\cA_r,r]}\\=&\,\big[\big(\bLambda^{(0)}-\hat\bLambda^{(0)}\big)\Gb^{(4)}\big]_{[\cA_r,r]}+\big[\bLambda^{(0)}\big(\Gb^{(4)}-\hat\Gb^{(4)}\big)\big]_{[\cA_r,r]}+O_p\Big(\frac{1}{N\wedge J^2}\Big).
\end{align*}
Similarly with $(\Gb^{(4)})^{-1}\big(\Gb^{(4)}\big)^{-\T}=\Mb_{ff}^{(4)}$ and 
\begin{equation*}
    (\Gb^{(4)})^{-1}-(\hat\Gb^{(4)})^{-1}=(\Gb^{(4)})^{-1}\big(\hat\Gb^{(4)}-\Gb^{(4)}\big)(\Gb^{(4)})^{-1}+O_p\Big(\frac{1}{N\wedge J^2}\Big),
\end{equation*}
we can write the second part as 
\begin{align*}
    \mathrm{diag}\big\{(\hat\Gb^{(4)})^{-1}&\,\big(\hat\Gb^{(4)}\big)^{-\T}-(\Gb^{(4)})^{-1}\big(\Gb^{(4)}\big)^{-\T}\big\}\\=&\,\mathrm{diag}\big\{-\big[(\Gb^{(4)})^{-1}-(\hat\Gb^{(4)})^{-1}\big](\Gb^{(4)})^{-\T}-\big(\Gb^{(4)}\big)^{-1}\big[(\Gb^{(4)})^{-1}-(\hat\Gb^{(4)})^{-1}\big]^{\T}\\&\,+\big[(\Gb^{(4)})^{-1}-(\hat\Gb^{(4)})^{-1}\big]\big[(\Gb^{(4)})^{-1}-(\hat\Gb^{(4)})^{-1}\big]^{\T}\big\}\\=&\,\mathrm{diag}\big\{-\big(\Gb^{(4)}\big)^{-1}\big(\Gb^{(4)}-\hat\Gb^{(4)}\big)\Mb_{ff}^{(4)}-\Mb_{ff}^{(4)}\big(\Gb^{(4)}-\hat\Gb^{(4)}\big)^{\T}(\Gb^{(4)})^{-\T}\big\}\\&\,+O_p\Big(\frac{1}{N\wedge J^2}\Big).
\end{align*}
Define $\hat\Xb^{(4)}=(\Gb^{(4)}-\hat\Gb^{(4)})^{\T}(\Gb^{(4)})^{-\T}$. Define \begin{equation*}\bdelta_{r}^{(4)}=\big\{\be_r^\T\big(\sum_{t=1}^Nl_{tj}^{\prime\prime}(\eta_{tj}^*)\bbf_t^{(4)}(\bbf_t^{(4)})^\T\big)^{-1}\sum_{t=1}^Nl_{tj}^{\prime}(\eta_{tj}^*)\bbf_t^{(4)}\big\}_{j\in \cA_r}\in\RR^{K-1}.\end{equation*}  
Similar to Section~\ref{sec_limit_ic2}, we know $\big(\bDelta_N^{(4)}\big)^{-1/2}\big((\bdelta_1^{(4)})^\T,\cdots,(\bdelta_K^{(4)})^\T\big)^\T\overset{d}{\to} \cN(\zero,\Ib_{(K-1)K}\big)$ where $\bDelta_{N}^{(4)}$ is defined as
\begin{equation*}
    \big[\bDelta_{N}^{(4)}\big]_{[(l-1)(K-1)+1:l(K-1),(h-1)(K-1)+1:h(K-1)]} = \Big[1_{(j_1=j_2)}\be_l^\T\bPhi_{[N],j}^{(4)}\be_h
    \Big]_{j_1\in \cA_l,j_2\in \cA_h}\in\RR^{(K-1)\times(K-1)},
\end{equation*}for $k,h\in[K]$.
Plug in $\blambda_{j}^{(4)}=(\Gb^{(4)})^\T\blambda_j^{(0)}$, \eqref{eq_precise_blambda} and 
we rewrite the equations as\begin{equation*}
\left\{
\begin{aligned}
    O_p\Big(\frac{1}{N\wedge J}\Big)&=[\bLambda^{(4)}]_{[\cA_r,r]}(\hat\Xb^{(4)}_{[r,]})^\T - \bdelta_{r}^{(4)}, \\
    O_p\Big(\frac{1}{N\wedge J^2}\Big)&=\mathrm{diag}\big( (\hat\Xb^{(4)})^\T\Mb_{ff}^{(4)}+\Mb_{ff}^{(4)}\hat\Xb^{(4)}\big).
\end{aligned}
\right.
\end{equation*}
The solution can be given as follows:
\begin{equation*}
    \left(\hat\Xb^{(4)}_{[l,]}\right)^\T=\left(\begin{array}{cc}\Ib_{l-1} &  \\\multicolumn{2}{c}{(-\Mb_{ff}^{(4)})_{[l,-l]}} \\ & \Ib_{K-l}\end{array}\right)
\big(\bLambda^{(4)}_{[\cA_l,-l]}\big)^{-1}\bdelta_{l}^{(4)} + O_p\Big(\frac{1}{N\wedge J}\Big).
\end{equation*}
Then when $\sqrt{N}/J\to 0$, we have for some semi-positive definite matrix $\Xb_{\lambda}^{(4)}$ that $\big(\Xb_{\lambda}^{(4)}\big)^{-1/2}$ $\mathrm{vec}(\hat\Xb^{(4)})\overset{d}{\to} \cN(\zero,\Ib_{K^2})$, where
the $(l_1+(h_1-1)K,l_2+(h_2-1)K)$th entry of $\Xb_{\lambda}^{(4)}$ is given as follows:
\begin{align}\big[&\Xb_{\lambda}^{(4)}\big]_{[l_1+(h_1-1)K,l_2+(h_2-1)K]}=\nonumber\\&
    \be_{h_1}^\T\left(\begin{array}{cc}\Ib_{l_1-1} &  \\\multicolumn{2}{c}{(-\Mb_{ff}^{(4)})_{[l_1,-l_1]}} \\ & \Ib_{K-l_1}\end{array}\right)\big(\bLambda^{(4)}_{[\cA_{l_1},-l_1]}\big)^{-1}\bDelta^{(4)}_{N,l_1l_2}\big(\bLambda^{(4)}_{[\cA_{l_2},-l_2]}\big)^{-\T}
    \left(\begin{array}{cc}\Ib_{l_2-1} &  \\\multicolumn{2}{c}{(-\Mb_{ff}^{(4)})_{[l_2,-l_2]}} \\ & \Ib_{K-l_2}\end{array}\right)^\T\be_{h_2}.\label{eq_expression_Xlambda4}
\end{align}
Next we compute $\hat\blambda_{j}^{(4)}-\blambda_{j}^{(4)}$ given as
\begin{align*}
    \hat\blambda_{j}^{(4)}-\blambda_{j}^{(4)}
    &\,=(\hat\Gb^{(4)})^\T\hat\blambda_{j}^{(0)}-(\Gb^{(4)})^\T\blambda_j^{(0)}\\&\,=\big(\Gb^{(4)}\big)^\T\big(\hat\blambda_{j}^{(0)}-\blambda_j^{(0)}\big)-\hat\Xb^{(4)}\big(\Gb^{(4)}\big)^\T\blambda_j^{(0)}\\&\,+\big(\hat\Gb^{(4)}-\Gb^{(4)}\big)\big(\hat\blambda_j^{(0)}-\blambda_j^{(0)}\big)\\&\,=\big(\Gb^{(4)}\big)^\T\big(\hat\blambda_{j}^{(0)}-\blambda_j^{(0)}\big)-\hat\Xb^{(4)}\blambda_j^{(4)}+O_p\Big(\frac{1}{N\wedge J^2}\Big).
\end{align*}
The covariance matrix between the first term and second term is given as
\begin{align*}
    &\,Cov\big(\big(\Gb^{(4)}\big)^\T\big(\hat\blambda_{j}^{(0)}-\blambda_j^{(0)}\big),\hat\Xb^{(4)}\blambda_j^{(4)}\big)\\=&\,\sum_{r=1}^{K}\sum_{l=1}^{K-1}1_{(j=\cA_r(l))}\bPhi_{[N],j}^{(4)}\be_r\be_l^\T\big(\bLambda_{[\cA_r,-r]}^{(4)}\big)^{-\T}\left(
\begin{array}{cc}
\Ib_{r-1} &  \\
\multicolumn{2}{c}{\big(-\Mb_{ff}^{(4)}\big)_{[r,-r]}} \\
 & \Ib_{K-r}
\end{array}
\right)^\T\blambda_j^{(4)}\be_r^\T +O_p\big((N\wedge J)^{-2}\big).
\end{align*}
Define
\begin{align}
    \Xb_{\lambda,rl}^{(4)}=\be_r\big(\be_l^{(K-1)}\big)^\T\big(\bLambda_{[\cA_r,-r]}^{(4)}\big)^{-\T}\left(
\begin{array}{cc}
\Ib_{r-1} &  \\
\multicolumn{2}{c}{-\big(\Mb_{ff}^{(4)}\big)_{[r,-r]}} \\
 & \Ib_{K-r}
\end{array}
\right)^\T
\blambda_{\cA_r(l)}^{(4)}\be_r^\T.\label{eq_x_lambda_kr_4}
\end{align}
Still, when $j\in\cA_{r}$, the $r$th row in $\Xb_{\lambda,rl}^{(4)}$ is deterministically $\zero$.
Then when $\sqrt{N}/J\to 0$, we know with the expression of $\bSigma_{[N],j}^{(4)}$ that
\begin{equation*}
    \big(\Db_{(j)}\bSigma_{[N],j}^{(4)}\Db_{(j)}^\T\big)^{-1/2}\Db_{(j)}\big(\hat\blambda_{j}^{(4)}-\blambda_{j}^{(4)}\big)\overset{d}{\to}\cN(\zero,\Ib_{|\cS_j|}).
\end{equation*}

And next for $\hat\bbf_i^{(4)}-\bbf_i^{(4)}$, we have
\begin{align*}
    \hat\bbf_i^{(4)}-\bbf_i^{(4)}=&\,\big((\hat\Gb^{(4)})^{-1}-(\Gb^{(4)})^{-1}\big)\bbf_i^{(0)}+(\Gb^{(4)})^{-1}\big(\hat\bbf_i^{(0)}-\bbf_i^{(0)}\big)\\&\,+\big((\hat\Gb^{(4)})^{-1}-(\Gb^{(4)})^{-1}\big)\big(\hat\bbf_i^{(0)}-\bbf_i^{(0)}\big)\\=&\,(\hat\Xb^{(4)})^\T\bbf_i^{(4)}+(\Gb^{(4)})^{-1}\big(\hat\bbf_i^{(0)}-\bbf_i^{(0)}\big)+O_p\big(N^{-1/2}J^{-1/2}\big).
\end{align*}
Similar to the derivation in Section~\ref{sec_limit_ic2}, we have
\begin{equation*}
    \big(\bSigma_{i,[J]}^{(4)}\big)^{-1/2}\big(\hat\bbf_i^{(4)}-\bbf_i^{(4)}\big)\overset{d}{\to}\cN(\zero,\Ib_K).
\end{equation*}
The consistency of the plug-in covariance estimator is implied by $\hat\bPhi^{(0)}\overset{p}{\to} \bPhi^{(0)}$, $\hat\Gb^{(4)}\overset{p}{\to}\Gb^{(4)}$, $\hat\blambda_j^{(4)}\overset{p}{\to}\blambda_j^{(4)}$ and $\hat\bbf_i^{(4)}\overset{p}{\to}\bbf_i^{(4)}$.
\subsection[Limiting Distributions of $\hat\blambda_j^{(5)}$s and $\hat\bbf_i^{(5)}$s]{Limiting Distributions of $\hat\blambda_j^{(5)}$s and $\hat\bbf_i^{(5)}$s}\label{sec_limit_ic5}
Under IC$^{(5)}$, we find $\Gb^{(5)}=\big(\bLambda_{[1:K,]}^{(0)}\big)^{-1}$ and $\hat\Gb^{(5)}=\big(\hat\bLambda_{[1:K,]}^{(0)}\big)^{-1}$ such that $\bLambda^{(5)}=\bLambda^{(0)}\Gb^{(5)}$, $\Fb^{(5)}=\Fb^{(0)}(\Gb^{(5)})^{-\T}$, $\hat\bLambda^{(5)}=\hat\bLambda^{(0)}\hat\Gb^{(5)}$ and $\hat\Fb^{(5)}=\hat\Fb^{(0)}(\hat\Gb^{(5)})^{-\T}$ with $\|\hat\Gb^{(5)}-\Gb^{(5)}\|=O_p(N^{-1/2} + J^{-1})$. Next for asymptotic distribution of $\hat\blambda^{(5)}_j$ when $j> K$, we have
\begin{align*}
    \hat\blambda_j^{(5)}-\blambda_j^{(5)}=&\,(\Gb^{(5)})^\T\big(\hat\blambda_j^{(0)}-\blambda_j^{(0)}\big)\\&\,+\big(\hat\Gb^{(5)}-\Gb^{(5)}\big)^\T\blambda_j^{(0)}+\big(\hat\Gb^{(5)}-\Gb^{(5)}\big)^\T\big(\hat\blambda_j^{(0)}-\blambda_j^{(0)}\big)\\=&\,(\Gb^{(5)})^\T\big(\sum_{i=1}^Nl_{ij}^{\prime\prime}(\eta_{ij}^*)\bbf_i^{(5)}(\bbf_i^{(0)})^\T\big)^{-1}\sum_{i=1}^Nl_{ij}^{\prime}(\eta_{ij}^*)\bbf_i^{(5)}+O_p\Big(\frac{1}{N\wedge J}\Big)\\&\,+(\Gb^{(5)})^\T\big((\Gb^{(5)})^{-1}-(\hat\Gb^{(5)})^{-1}\big)^\T\blambda_j^{(5)}+O_p\Big(\frac{1}{N\wedge J^2}\Big).
\end{align*}
Define $\hat\Xb^{(5)} = (\Gb^{(5)})^\T\big((\Gb^{(5)})^{-1}-(\hat\Gb^{(5)})^{-1}\big)^\T$ and $\Xb_{\lambda}^{(5)}$ as
\begin{equation}
    [\Xb_{\lambda}^{(5)}]_{[(r-1)K+1:rK,(r-1)K+1:rK]} = \bPhi_{[N],r}^{(5)}.\label{eq_expression_Xlambda5}
\end{equation}
Then we know by Section~\ref{sec:use_clt} that $\big(\Xb_{\lambda}^{(5)}\big)^{-1/2}\mathrm{vec}\big(\hat\Xb^{(5)}\big) \overset{d}\to \cN(\zero,\Ib_{K^2})$. Note that 
\begin{equation*}
    \big((\blambda_j^{(5)})^\T\otimes \Ib_K\big)\Xb_{\lambda}^{(5)}\big(\blambda_j^{(5)}\otimes \Ib_K\big) = \sum_{r=1}^K\big(\lambda_{jr}^{(5)}\big)^2\bPhi_{[N],r}^{(5)}
\end{equation*}
Therefore, for the second term, we have
\begin{align*}
    \Big\{\sum_{r=1}^K\big(\lambda_{jr}^{(5)}\big)^2\bPhi_{[N],r}^{(5)}\Big\}^{-1/2}(\Gb^{(5)})^\T\big((\Gb^{(5)})^{-1}-(\hat\Gb^{(5)})^{-1}\big)^\T\blambda_j^{(5)}\overset{d}{\to}\cN(\zero,\Ib_K).
\end{align*}
Then we know that when $\sqrt N/J\to0$, $j> K$, 
\begin{equation*}
    \bSigma_{[N],j}^{(5)} \big(\hat\blambda_j^{(5)}-\blambda_j^{(5)}\big)\overset{d}{\to}\cN(\zero,\Ib_K).
\end{equation*}
When $j\le K$, $|\cS_j|=0$.
For $\hat\bbf_i^{(5)}$, note that
\begin{align*}
    \hat\bbf_i^{(5)}-\bbf_i^{(5)}=&\,(\hat\Gb^{(5)})^{-\T}\hat\bbf_i^{(0)}-(\Gb^{(5)})^{-\T}\bbf_i^{(0)}\\=&\,-\big((\hat\Gb^{(5)})^{-\T}-(\Gb^{(5)})^{-\T}\big)\bbf_i^{(0)}+(\Gb^{(5)})^{-\T}(\hat\bbf_i^{(0)}-\bbf_i^{(0)})\\&\,+\big((\hat\Gb^{(5)})^{-\T}-(\Gb^{(5)})^{-\T}\big)\big(\hat\bbf_i^{(0)}-\bbf_i^{(0)}).
\end{align*}
Similarly we have
\begin{equation*}
    \big(\bSigma_{i,[J]}^{(5)}\big)^{-1/2}\big(\hat\bbf_i^{(5)}-\bbf_i^{(5)}\big)\overset{d}{\to}\cN(\zero,\Ib_K).
\end{equation*}
The consistency of the plug-in covariance estimator is implied by $\hat\bPhi^{(0)}\overset{p}{\to} \bPhi^{(0)}$, $\hat\Gb^{(5)}\overset{p}{\to}\Gb^{(5)}$, $\hat\blambda_j^{(5)}\overset{p}{\to}\blambda_j^{(5)}$ and $\hat\bbf_i^{(5)}\overset{p}{\to}\bbf_i^{(5)}$.


\subsection{Proof of Theorem~\ref{thm_mff_limiting_distri}}\label{prove_thm_mff_limiting_dist}
\noindent\textbf{Limiting Distribution of $\hat\Mb_{ff}^{(3)}$. }
In Section~\ref{sec_limit_ic3}, we know
\begin{align*}
    \hat\bbf_i^{(3)}-\bbf_i^{(3)}=(\hat\Xb^{(3)})^\T(\Mb_{ff}^{(3)})^{-1}\bbf_i^{(3)}+(\Gb^{(3)})^{-1}\big(\hat\bbf_i^{(0)}-\bbf_i^{(0)}\big)+O_p\big(N^{-1/2}J^{-1/2}\big).
\end{align*}
Then by $\|\hat\bbf_i^{(3)} - \bbf_i^{(3)}\| = O_p(N^{-1/2} + J^{-1/2})$, we know
\begin{align}
    \hat\Mb_{ff}^{(3)} - \Mb_{ff}^{(3)} = &\, N^{-1}\sum_{i=1}^N \hat\bbf_i^{(3)} \big(\hat\bbf_i^{(3)}\big)^\T - N^{-1}\sum_{i=1}^N \bbf_i^{(3)} \big(\bbf_i^{(3)}\big)^\T\nonumber\\=&\, N^{-1}\sum_{i=1}^N \big(\hat\bbf_i^{(3)} - \bbf_i^{(3)}\big) \big(\bbf_i^{(3)}\big)^\T + N^{-1}\sum_{i=1}^N \bbf_i^{(3)} \big(\hat\bbf_i^{(3)} - \bbf_i^{(3)}\big)^\T\nonumber\\&\, +N^{-1}\sum_{i=1}^N\big(\hat\bbf_i^{(3)} - \bbf_i^{(3)}\big)\big(\hat\bbf_i^{(3)} - \bbf_i^{(3)}\big)^\T\nonumber\\ = &\,(\hat\Xb^{(3)})^\T(\Mb_{ff}^{(3)})^{-1}N^{-1}\sum_{i=1}^N\bbf_i^{(3)}\big(\bbf_i^{(3)}\big)^\T \label{eq_diff_mff_eq1}\\ &\,+ (\Gb^{(3)})^{-1}N^{-1}\sum_{i=1}^N \big(\hat\bbf_i^{(0)}-\bbf_i^{(0)}\big)\big(\bbf_i^{(0)}\big)^\T(\Gb^{(3)})^{-\T}\label{eq_diff_mff_eq2}\\&\, +N^{-1}\sum_{i=1}^N\bbf_i^{(3)}\big(\bbf_i^{(3)}\big)^\T(\Mb_{ff}^{(3)})^{-1} \hat\Xb^{(3)}\label{eq_diff_mff_eq3}\\ &\,+ (\Gb^{(3)})^{-1}N^{-1}\sum_{i=1}^N \big(\bbf_i^{(0)}\big)\big(\hat\bbf_i^{(0)}-\bbf_i^{(0)}\big)^\T(\Gb^{(3)})^{-\T} + O_p\Big(\frac{1}{N\wedge J}\Big).\label{eq_diff_mff_eq4}
\end{align}
By $\text{ndiag}\big(\hat\Xb^{(3)} + (\hat\Xb^{(3)})^\T\big) = O_p\big(N^{-1} + J^{-2}\big)$ we know
\begin{equation*}
    \eqref{eq_diff_mff_eq1} + \eqref{eq_diff_mff_eq3} = \mathrm{diag}\Big(\hat\Xb^{(3)} + \big(\hat\Xb^{(3)}\big)^\T\Big) + O_p\Big(\frac{1}{N\wedge J^2}\Big).
\end{equation*}
By \eqref{eq_precise_f} and Lemma~\ref{lemma_concentration} we know \eqref{eq_diff_mff_eq2} can be bounded as 
\begin{align}
    \|\eqref{eq_diff_mff_eq2}\|\le &\,N^{-1}\big\|(\Gb^{(3)})^{-1}\big\|^2\Big\|\sum_{i=1}^N\big(\sum_{j=1}^J-l_{ij}^{\prime\prime}(\eta_{ij}^*)\blambda_j^{(0)}(\blambda_j^{(0)})^\T\big)^{-1}\sum_{j=1}^Jl_{ij}^{\prime}(\eta_{ij}^*)\blambda_j^{(0)}\bbf_i^{(0)}\Big\|\nonumber\\&\,+O_p\Big(\frac{1}{N\wedge J}\Big)\nonumber\\ \le &\, J^{-1}\big\|(\Gb^{(3)})^{-1}\big\|^2\Big(\lambda_{\min}\big\{J^{-1}\sum_{j=1}^Jb_L\blambda_j^{(0)}(\blambda_j^{(0)})^\T\big\}\Big)^{-1}\Big\|\sum_{i=1}^N\sum_{j=1}^Jl_{ij}^{\prime}(\eta_{ij}^*)\blambda_j^{(0)}\bbf_i^{(0)}\Big\|\nonumber\\&\,+O_p\Big(\frac{1}{N\wedge J}\Big)\nonumber\\\le &\, O_p\Big(\frac{1}{N\wedge J}\Big).\label{eq_bound_sum_diff_ff}
\end{align}For the same reason $\|\eqref{eq_diff_mff_eq4}\|\le O_p(N^{-1}+J^{-1})$.
Therefore we conclude 
\begin{equation*}
    \hat\Mb_{ff}^{(3)} - \Mb_{ff}^{(3)} = 2\mathrm{diag}\big(\hat\Xb^{(3)}\big) + O_p\Big(\frac{1}{N\wedge J}\Big).
\end{equation*}
Let $S_{d} = \{(r-1)K + r: r\in[K]\}$. With $\bSigma_{[J]}^{(3)}$ given as 
\begin{equation}
    \bSigma_{[J]}^{(3)} = 4[\Xb_{\lambda}^{(3)}]_{[S_{d},S_{d}]},\label{eq_sigma_mff3}
\end{equation}
we have by $\big(\Xb_{\lambda}^{(3)}\big)^{-1/2}\mbox{vec}\big(\hat\Xb^{(3)}\big)\overset{d}{\to} \cN(\zero,\Ib_{K^2})$ and $\|\Xb_{\lambda}^{(3)}\| = O_p(N^{-1})$ that
\begin{equation*}
    \big(\bSigma_{[J]}^{(3)}\big)^{-1/2}\big(\hat\Mb_{ff}^{(3)} - \Mb_{ff}^{(3)}\big)\overset{d}{\to}\cN(\zero_K,\Ib_K),
\end{equation*}when $\sqrt{N}/J\to0$.

\noindent\textbf{Limiting Distribution of $\hat\Mb_{ff}^{(4)}$. }
In Section~\ref{sec_limit_ic4}, we know
\begin{align*}
    \hat\bbf_i^{(4)}-\bbf_i^{(4)}=(\hat\Xb^{(4)})^\T\bbf_i^{(4)}+(\Gb^{(4)})^{-1}\big(\hat\bbf_i^{(0)}-\bbf_i^{(0)}\big)+O_p\big(N^{-1/2}J^{-1/2}\big).
\end{align*}
Then by $\|\hat\bbf_i^{(4)} - \bbf_i^{(4)}\| = O_p(N^{-1/2} + J^{-1/2})$ and similar to deriving $\hat\Mb_{ff}^{(3)} - \Mb_{ff}^{(3)}$, we know
\begin{align}
    \hat\Mb_{ff}^{(4)} - \Mb_{ff}^{(4)} = &\,(\hat\Xb^{(4)})^\T N^{-1}\sum_{i=1}^N\bbf_i^{(4)}\big(\bbf_i^{(4)}\big)^\T +N^{-1}\sum_{i=1}^N\bbf_i^{(4)}\big(\bbf_i^{(4)}\big)^\T \hat\Xb^{(4)}\label{eq_diff_mff_ic4_eq1}\\ &\,+ (\Gb^{(4)})^{-1}N^{-1}\sum_{i=1}^N \big(\hat\bbf_i^{(0)}-\bbf_i^{(0)}\big)\big(\bbf_i^{(0)}\big)^\T(\Gb^{(4)})^{-\T}\label{eq_diff_mff_ic4_eq2}\\ &\,+ (\Gb^{(4)})^{-1}N^{-1}\sum_{i=1}^N \big(\bbf_i^{(0)}\big)\big(\hat\bbf_i^{(0)}-\bbf_i^{(0)}\big)^\T(\Gb^{(4)})^{-\T} + O_p\Big(\frac{1}{N\wedge J}\Big).\label{eq_diff_mff_ic4_eq4}
\end{align}
By $\mathrm{diag}\big((\hat\Xb^{(4)})^\T\Mb_{ff}^{(4)} + \Mb_{ff}^{(4)}\hat\Xb^{(4)}\big) = O_p\big(N^{-1} + J^{-2}\big)$ and \eqref{eq_bound_sum_diff_ff}, we have
\begin{equation*}
     \hat\Mb_{ff}^{(4)} - \Mb_{ff}^{(4)}= \text{ndiag}\big((\hat\Xb^{(4)})^\T\Mb_{ff}^{(4)} + \Mb_{ff}^{(4)}\hat\Xb^{(4)}\big) + O_p\Big(\frac{1}{N\wedge J}\Big).
\end{equation*}
Let $\Db^{(4)}_K$ be a matrix of zeros and ones such that $\mathrm{veck}(\Ab) = \Db_{K}^{(4)}\mathrm{vec}(\Ab)$ for any symmetric matrix $\Ab$ with zero diagonal entries. Then
\begin{equation*}
    \mathrm{veck}\big((\hat\Xb^{(4)})^\T\Mb_{ff}^{(4)} + \Mb_{ff}^{(4)}\hat\Xb^{(4)}\big) = 2\Db^{(4)}_K \big(\Ib_K\otimes \Mb_{ff}^{(4)}\big)\mathrm{vec}\big(\hat\Xb^{(4)}\big).
\end{equation*}
Therefore with 
\begin{equation}
    \bSigma_{[J]}^{(4)} = 4\Db^{(4)}_K \big(\Ib_K\otimes \Mb_{ff}^{(4)}\big)\Xb_{\lambda}^{(4)} \big(\Ib_K\otimes \Mb_{ff}^{(4)}\big)(\Db^{(4)}_K)^\T,\label{eq_sigma_mff4}
\end{equation}
$\big(\Xb_{\lambda}^{(4)}\big)^{-1/2}\mathrm{vec}\big(\hat\Xb^{(4)}\big)\overset{d}{\to} \cN(\zero,\Ib_{K^2})$ and $\|\Xb_{\lambda}^{(4)}\| = O_p(N^{-1})$, we have
\begin{equation*}
    \big(\bSigma_{[J]}^{(4)}\big)^{-1/2}\mathrm{veck}\big(\hat\Mb_{ff}^{(4)} - \Mb_{ff}^{(4)}\big) \overset{d}{\to}\cN(\zero_{(K-1)K/2},\Ib_{{(K-1)K/2}}),
\end{equation*}when $\sqrt{N}/J\to0$.

\noindent\textbf{Limting Distribution of $\hat\Mb_{ff}^{(5)}$. }
In Section~\ref{sec_limit_ic5}, we know
\begin{align*}
    \hat\bbf_i^{(5)}-\bbf_i^{(5)}=(\hat\Xb^{(5)})^\T\bbf_i^{(5)}+(\Gb^{(5)})^{-1}\big(\hat\bbf_i^{(0)}-\bbf_i^{(0)}\big)+O_p\big(N^{-1/2}J^{-1/2}\big).
\end{align*}
Then by $\|\hat\bbf_i^{(5)} - \bbf_i^{(5)}\| = O_p(N^{-1/2} + J^{-1/2})$ and similar to deriving $\hat\Mb_{ff}^{(3)} - \Mb_{ff}^{(3)}$, we know
\begin{align}
    \hat\Mb_{ff}^{(5)} - \Mb_{ff}^{(5)} = &\,(\hat\Xb^{(5)})^\T N^{-1}\sum_{i=1}^N\bbf_i^{(5)}\big(\bbf_i^{(5)}\big)^\T +N^{-1}\sum_{i=1}^N\bbf_i^{(5)}\big(\bbf_i^{(5)}\big)^\T \hat\Xb^{(5)}\label{eq_diff_mff_ic5_eq1}\\ &\,+ (\Gb^{(5)})^{-1}N^{-1}\sum_{i=1}^N \big(\hat\bbf_i^{(0)}-\bbf_i^{(0)}\big)\big(\bbf_i^{(0)}\big)^\T(\Gb^{(5)})^{-\T}\label{eq_diff_mff_ic5_eq2}\\ &\,+ (\Gb^{(5)})^{-1}N^{-1}\sum_{i=1}^N \big(\bbf_i^{(0)}\big)\big(\hat\bbf_i^{(0)}-\bbf_i^{(0)}\big)^\T(\Gb^{(5)})^{-\T} + O_p\Big(\frac{1}{N\wedge J}\Big).\label{eq_diff_mff_ic5_eq4}
\end{align}
Still, we know \eqref{eq_diff_mff_ic5_eq2} and \eqref{eq_diff_mff_ic5_eq4} can be bounded by $O_p\big(N^{-1}+J^{-1}\big)$, therefore we have
\begin{equation*}
    \hat\Mb_{ff}^{(5)} - \Mb_{ff}^{(5)} = (\hat\Xb^{(5)})^\T\Mb_{ff}^{(5)} + \Mb_{ff}^{(5)}\hat\Xb^{(5)}.
\end{equation*}
Let $\Db^{(5)}_K$ be a matrix of zeros and ones such that $\mathrm{vech}(\Ab) = \Db_{K}^{(5)}\mathrm{vec}(\Ab)$ for any symmetric matrix $\Ab$. Then
\begin{equation*}
    \mathrm{vech}\big((\hat\Xb^{(5)})^\T\Mb_{ff}^{(5)} + \Mb_{ff}^{(5)}\hat\Xb^{(5)}\big) = 2\Db^{(5)}_K \big(\Ib_K\otimes \Mb_{ff}^{(5)}\big)\mathrm{vec}\big(\hat\Xb^{(5)}\big).
\end{equation*}
Therefore with 
\begin{equation}
    \bSigma_{[J]}^{(5)} = 4\Db^{(5)}_K \big(\Ib_K\otimes \Mb_{ff}^{(5)}\big)\Xb_{\lambda}^{(5)} \big(\Ib_K\otimes \Mb_{ff}^{(5)}\big)(\Db^{(5)}_K)^\T,\label{eq_sigma_mff5}
\end{equation}
$\big(\Xb^{(5)}_{\lambda}\big)^{-1/2}\mathrm{vec}\big(\hat\Xb^{(5)}\big)\overset{d}{\to} \cN(\zero,\Ib_{K^2})$ and $\|\Xb_{\lambda}^{(5)}\|=O_p(N^{-1})$, we have
\begin{equation*}
    \big(\bSigma_{[J]}^{(5)}\big)^{-1/2}\mathrm{vech}\big(\hat\Mb_{ff}^{(5)} - \Mb_{ff}^{(5)}\big) \overset{d}{\to}\cN(\zero_{(K+1)K/2},\Ib_{{(K+1)K/2}}),
\end{equation*}when $\sqrt{N}/J\to0$.
\section{Proofs of Technical Lemmas}\label{sec:supp_techni}

\subsection{Proof of Lemma~\ref{lemma_concentration_l_mat}}\label{prove_lemma_concentration_l_mat}

Let $Z_{ij}:=l_{ij}'(\eta_{ij}^*)$ and then $\Lb_\eta=(Z_{ij})_{N\times J}$. Under Assumption~\ref{assumption:smoothness}, the random variables $Z_{ij}$ are independent and satisfy $\|Z_{ij}\|_{\psi_1}\le C$ uniformly over $(i,j)$, where $C>0$ is a universal constant. In particular, $\EE Z_{ij}^2\le C$ and $\|Z_{ij}^2-\EE Z_{ij}^2\|_{\psi_{1/2}}\le C$ uniformly over $(i,j)$. Moreover, in the score setting, we have $\EE Z_{ij}=0$.

We first prove (i) and (ii), which are standard consequences of concentration for sums of independent sub-Weibull variables. For each fixed $i\in[N]$, let $S_i:=\sum_{j=1}^J Z_{ij}^2=\|[\Lb_\eta]_{[i,]}\|^2$.
Then $S_i-\EE S_i=\sum_{j=1}^J (Z_{ij}^2-\EE Z_{ij}^2)$ is a sum of independent centered sub-Weibull$(1/2)$ random variables. Hence a Bernstein-type inequality for such sums yields $\PP(S_i-\EE S_i\ge t) \le 2\exp\big\{-c\min(\frac{t^2}{J},\,\sqrt{t})\big\}$,
for some universal constant $c>0$. Since $\EE S_i\le CJ$, by taking $t=C_\delta J$ we know $\PP(S_i\ge C_\delta J)\le J^{-\delta}$, after enlarging $C_\delta$ if necessary. Therefore,
\begin{equation*}
\PP\Big(\|[\Lb_\eta]_{[i,]}\|\le C_\delta \sqrt{J}\Big)\ge 1-J^{-\delta}.
\end{equation*}
This proves the first statement in part (i).

Next, again by the same inequality, taking $t=C_\delta J\log N$ gives $\PP\big(S_i\ge C_\delta J\log N\big)\le N^{-1}(NJ)^{-\delta}$,
provided $C_\delta$ is chosen large enough. Hence, by the union bound,
\begin{equation*}
\PP\Big(\max_{1\le i\le N}\|[\Lb_\eta]_{[i,]}\|^2\ge C_\delta J\log N\Big)
\le \sum_{i=1}^N \PP\Big(S_i\ge C_\delta J\log N\Big)
\le (NJ)^{-\delta},
\end{equation*}
that is,
\begin{equation*}
\PP\Big(\max_{1\le i\le N}\|[\Lb_\eta]_{[i,]}\|\le C_\delta \sqrt{J\log N}\Big)
\ge 1-(NJ)^{-\delta}.
\end{equation*}
This proves part (i). Part (ii) can be similarly obtained by symmetry.
It remains to prove part (iii).

Let $\widetilde{\Lb}_\eta=(\widetilde Z_{ij})_{N\times J}$ be an independent copy of $\Lb_\eta$, and define $W_{ij}:=Z_{ij}-\widetilde Z_{ij}$.
Then the $W_{ij}$ are independent, symmetric, satisfying  $\|W_{ij}\|_{\psi_1}\le C$ and $\EE W_{ij}^2\le C$ uniformly over $(i,j)$.
Fix a truncation level $\nu>0$, and let $\Wb^{(\nu)}:=\big(W_{ij}^{(\nu)}\big)_{N\times J}$ with  $W_{ij}^{(\nu)}:={\rm trun}_\nu(W_{ij})$. Consider the self-adjoint dilation
\begin{equation*}
\Xb := \begin{pmatrix}
\zero & (\Wb^{(\nu)})^\T\\ \Wb^{(\nu)} & \zero
\end{pmatrix}.
\end{equation*}
Then $\Xb$ is an $(N+J)\times(N+J)$ symmetric random matrix with independent symmetric entries with $\|\Xb\|=\|\Wb^{(\nu)}\|$. We now apply Corollary 3.12 in \cite{bandeira2016sharp}. For the variance parameter,
\begin{equation*}
\tilde\sigma := \max_{1\le a\le N+J} \Big(\sum_{b=1}^{N+J}\EE X_{ab}^2\Big)^{1/2} \le C\sqrt{N\vee J},
\end{equation*}
since $\EE (W_{ij}^{(\nu)})^2\le \EE W_{ij}^2\le C$. Moreover, there is $\tilde\sigma_*:=\max_{a,b}\|X_{ab}\|_\infty\le \nu$.
Therefore, for any $t>0$, Corollary 3.12 in \cite{bandeira2016sharp} gives
\begin{equation*}
\PP\Big(\|\Wb^{(\nu)}\|\ge C\sqrt{N\vee J}+t\Big) = \PP\Big(\|\Xb\|\ge C\sqrt{N\vee J}+t\Big) \le (N+J)\exp\Big(-\frac{c t^2}{\nu^2}\Big)
\end{equation*}
for some universal constant $c>0$. 
Next, if $\Wb\neq \Wb^{(\nu)}$, then at least one entry satisfies $|W_{ij}|>\nu$. Hence, we know $\PP(\Wb\neq \Wb^{(\nu)})\le \sum_{i=1}^N\sum_{j=1}^J \PP(|W_{ij}|>\nu)\le cNJe^{-c\nu}$,
because the $W_{ij}$ are uniformly sub-exponential. Combining the last two estimates, we obtain
\begin{equation*}
\PP\Big(\|\Lb_\eta-\widetilde{\Lb}_\eta\| \ge C\sqrt{N\vee J}+t\Big) \le (N+J)\exp\Big(-\frac{c t^2}{\nu^2}\Big) + CNJ e^{-c\nu}.
\end{equation*}
Take $\nu=C_\delta \log(NJ)$ and $t=C_\delta \nu \sqrt{\log(N+J)}$
with $C_\delta$ sufficiently large, and we arrive at
\begin{equation*}
(N+J)\exp\Big(-\frac{c t^2}{\nu^2}\Big) \le (N+J)^{-\delta}, \qquad CNJ e^{-c\nu}\le (NJ)^{-\delta},
\end{equation*}
and therefore
\begin{equation}\label{eq:symm_bound_Leta}
\PP\Big(\|\Lb_\eta-\widetilde{\Lb}_\eta\| \ge C\sqrt{N\vee J}+C_\delta \log(NJ)\sqrt{\log(N+J)}\Big) \le
(N+J)^{-\delta}+(NJ)^{-\delta}.
\end{equation}

Finally, we transfer this bound from the symmetrized matrix to $\Lb_\eta$. Since $\EE Z_{ij}=0$, a standard symmetrization inequality for Banach valued sums gives
\begin{equation*}
\PP\big(\|\Lb_\eta\|\ge 2u\big)
\le
2\PP\big(\|\Lb_\eta-\widetilde{\Lb}_\eta\|\ge u\big),
\qquad u>0.
\end{equation*}
Applying this with $u = C\sqrt{N\vee J}+C_\delta \log(NJ)\sqrt{\log(N+J)}$ and \eqref{eq:symm_bound_Leta}, we know
\begin{equation*}
\PP\Big(\|\Lb_\eta\| \ge C\sqrt{N\vee J}+C_\delta \log(NJ)\sqrt{\log(N+J)}\Big) \le C\Big((N+J)^{-\delta}+(NJ)^{-\delta}\Big).
\end{equation*}
After enlarging $C_\delta$, the constant $C$ in front of the probability bound can be absorbed, so\begin{equation*}
\PP\Big(\|\Lb_\eta\| \ge C_\delta\sqrt{N\vee J} + C_\delta \log(NJ)\sqrt{\log(N+J)}\Big) \le (N+J)^{-\delta}+(NJ)^{-\delta}.
\end{equation*}

Under the standing scaling condition $(N\vee J)\gg \log^3(N+J)$, the second term on the left-hand side is of smaller order than $\sqrt{N\vee J}$, and therefore can be absorbed into the first one. 
This proves part (iii), and hence the lemma.

    \subsection{Proof of Lemma~\ref{lemma_hessian_property}}\label{supp_sec_proof_hessian_property_1}


    We write for short $\tilde\btheta^s = \btheta^{(0)} + s(\tilde\btheta-\btheta^{(0)})$ and $\tilde\Hb_L=-\int_0^1\big\{\partial^2_{\btheta\btheta}L(\tilde\btheta^s)|\Yb)\big\}ds$ with each part expressed as follows:
    \begin{equation*}
        \tilde\Hb_L=\begin{pmatrix}
            \tilde\Hb_{L\lambda\lambda}&\tilde\Hb_{L\lambda f}\\\tilde\Hb_{Lf\lambda}&\tilde\Hb_{Lf f}
        \end{pmatrix},
    \end{equation*}
    where 
    \begin{align*}
        [\tilde\Hb_{L\lambda\lambda}]_{[K_j,K_j]}&\,=-\int_0^1\sum_{i=1}^Nl_{ij}^{\prime\prime}(\tilde\eta_{ij}^s)\tilde\bbf_i^s(\tilde\bbf_i^s)^\T ds,\;[\tilde\Hb_{Lff}]_{[K_i,K_i]}=-\int_0^1\sum_{j=1}^Jl_{ij}^{\prime\prime}(\tilde\eta_{ij}^s)\tilde\blambda_j^s(\tilde\blambda_j^s)^\T ds,\\ [\tilde\Hb_{L\lambda f}]_{[K_j,K_i]}&\,=-\int_0^1\Big\{l_{ij}^{\prime\prime}(\tilde\eta_{ij}^s)\tilde\bbf_i^s(\tilde\blambda_j^s)^\T+l_{ij}^{\prime}(\tilde\eta_{ij}^s)\Ib_{K}\Big\}ds,\;\tilde\Hb_{Lf\lambda }=\tilde\Hb_{L\lambda f}^\T.
    \end{align*}
Here $\tilde\blambda_j^s = \blambda_j^{(0)} + s(\tilde\blambda_j-\blambda_j^{(0)})$, $\tilde\bbf_i^s = \bbf_i^{(0)}+s(\tilde\bbf_i-\bbf_i^{(0)})$, and $\tilde\eta_{ij}^s$ is defined as $(\tilde\blambda_j^s)^\T\tilde\bbf_i^s$ for any $i\in[N]$ and $j\in[J]$. We write for short $K_i=\{(i-1)K+1,\cdots,iK\}$. 
For $\Hb_P(\tilde\btheta^s)$, by trapezoid rule, we have $\int_0^1\big\{\bx+s(\by-\bx)\big\}\big\{\bx+s(\by-\bx)\big\}^\T ds = (\by+\bx)(\by+\bx)^\T/2 - (\by-\bx)(\by-\bx)^\T/12$, and therefore, 
\begin{equation*}
\int_0^1\Hb_P(\btheta)ds = NJ\sum_{s\in\{+,-\}}\sum_{l=1}^K\sum_{h=1}^K\tilde\bv_{lh}^s(\tilde\bv_{lh}^s)^\T,
\end{equation*}where $\tilde\bv_{lh}^+ =\Big[ 1_{(h>l)}\big\{\big(\bv_{\lambda,lh}(\tilde\btheta)+\bv_{\lambda,lh}(\btheta^{(0)})\big)^\T,\zero_{NK}^\T\big\}^\T + 1_{(h\le l)}\big\{\zero_{JK}^\T,\big(\bv_{f,lh}(\tilde\btheta)+\bv_{f,lh}(\btheta^{(0)})\big)^\T\big\}^\T\Big]/\sqrt{2}$ and $\tilde\bv_{lh}^- = \Big[1_{(h>l)}\big\{\big(\bv_{\lambda,lh}(\tilde\btheta)-\bv_{\lambda,lh}(\btheta^{(0)})\big)^\T,\zero_{NK}^\T\big\}^\T + 1_{(h\le l)}\big\{\zero_{JK}^\T,\big(\bv_{f,lh}(\tilde\btheta)-\bv_{f,lh}(\btheta^{(0)})\big)^\T\big\}^\T\Big]/(2\sqrt 3)$.
Since all entries in $\bv_{lh}(\tilde\btheta)$ and $\bv_{lh}(\btheta^{(0)})$ are bounded by $2D/J$ by construct, we know that $\big\|\tilde\bv_{lh}^+\big\|_{\infty}\le \sqrt2D/J$ and $\big\|\tilde\bv_{lh}^-\big\|_{\infty}\le D/(\sqrt3J)$ uniformly over $l,h\in[K]$.
We first estimate the $l_2$ and $l_{\infty}$ norm of these blocks. By Assumption~\ref{assumption: psd covariance}, we know that when $N$ and $J$ are large enough, we have $\lambda_{\min}(\bSigma_f)/2\le \lambda_{\min}(\Mb_{ff})\le \lambda_{\max}(\Mb_{ff})\le 2\lambda_{\min}(\bSigma_f)$ and $\lambda_{\min}(\bSigma_\lambda)/2\le \lambda_{\min}(\Mb_{\lambda\lambda})\le \lambda_{\max}(\Mb_{\lambda\lambda})\le 2\lambda_{\min}(\bSigma_\lambda)$. For $[\tilde\Hb_{L\lambda\lambda}]_{[K_j,K_j]}$, since $N^{-1}\sum_{i=1}^N\|\tilde\bbf_i-\bbf_i^{(0)}\|^2 + J^{-1}\sum_{j=1}^J\|\tilde\blambda_j-\blambda_j^{(0)}\|^2\le c_{NJ}$, we have by Assumption~\ref{assumption:smoothness} and continuity of eigenvalues that
\begin{align}
    \lambda_{\max}\Big\{[\tilde\Hb_{L\lambda\lambda}]_{[K_j,K_j]}\Big\}\le&\, \max_{s\in[0,1]}\lambda_{\max}\Big\{\sum_{i=1}^Nl_{ij}^{\prime\prime}(\tilde\eta_{ij}^s)\tilde\bbf_i^s(\tilde\bbf_i^s)^\T\Big\}\nonumber\\\le&\, b_U\max_{s\in[0,1]}\lambda_{\max}\Big\{\sum_{i=1}^N\tilde\bbf_i^s(\tilde\bbf_i^s)^\T\Big\}\nonumber\\\le &\, b_U \lambda_{\max}\Big\{\sum_{i=1}^N\bbf_i^{(0)}(\bbf_i^{(0)})^\T\Big\}+2b_U\max_{s\in[0,1]}\Big\|\sum_{i=1}^N\big(\tilde\bbf_i^s-\bbf_i^{(0)}\big)\big(\bbf_i^{(0)}\big)^\T\Big\|\nonumber\\&\,+b_U\max_{s\in[0,1]}\Big\|\sum_{i=1}^N\big(\tilde\bbf_i^s-\bbf_i^{(0)}\big)\big(\tilde\bbf_i^s-\bbf_i^{(0)}\big)^\T\Big\|\nonumber\\\lesssim &\,b_U \lambda_{\max}\Big\{\sum_{i=1}^N\bbf_i^{(0)}(\bbf_i^{(0)})^\T\Big\}+N\sqrt{c_{NJ}}\nonumber\\\lesssim&\,N.\label{eq_estimate_lambda_max_ff}
\end{align}
Here the second inequality follows from the fact that for any $s\in[0,1]$, $\tilde\btheta^s\in\cB(D)$ and therefore $\big|\tilde\eta_{ij}^s\big|\le KD^2$ and therefore $\big|l_{ij}^{\prime\prime}(\tilde\eta_{ij}^s)\big|\le b_U$ by Assumption~\ref{assumption:smoothness}. Similarly,
\begin{align}
    \lambda_{\min}\Big\{[\tilde\Hb_{L\lambda\lambda}]_{[K_j,K_j]}\Big\}\ge&\, b_L \min_{s\in[0,1]}\lambda_{\min}\Big\{\sum_{i=1}^N\tilde\bbf_i^s(\tilde\bbf_i^s)^\T\Big\}\nonumber\\\ge &\, b_L \lambda_{\min}\Big\{\sum_{i=1}^N\bbf_i^{(0)}(\bbf_i^{(0)})^\T\Big\}-2b_L\min_{s\in[0,1]}\Big\|\sum_{i=1}^N\big(\tilde\bbf_i^s-\bbf_i^{(0)}\big)\big(\bbf_i^{(0)}\big)^\T\Big\|\nonumber\\&\,-b_L\min_{s\in[0,1]}\Big\|\sum_{i=1}^N\big(\tilde\bbf_i^s-\bbf_i^{(0)}\big)\big(\tilde\bbf_i^s-\bbf_i^{(0)}\big)^\T\Big\|\nonumber\\\gtrsim &\,b_L \lambda_{\min}\Big\{\sum_{i=1}^N\bbf_i^{(0)}(\bbf_i^{(0)})^\T\Big\}-N\sqrt{c_{NJ}}\nonumber\\ \gtrsim&\,N.\label{eq_estimate_lambda_min_ff}
\end{align}
By the block diagonal structure we know that $\lambda_{\min}(\tilde\Hb_{L\lambda\lambda})\gtrsim N$, $\lambda_{\max}(\tilde\Hb_{L\lambda\lambda})\lesssim N$ and $\big\|\tilde\Hb_{L\lambda\lambda}\big\|_{\infty}\lesssim N$ when $N$ is sufficiently large. 
For $\tilde\Hb_{Lff}$, similarly we have $\lambda_{\min}([\tilde\Hb_{Lff}]_{[K_i,K_i]})\gtrsim J$, $\lambda_{\max}([\tilde\Hb_{Lff}]_{[K_i,K_i]})\lesssim J$, $\lambda_{\min}(\tilde\Hb_{Lff})\gtrsim J$, $\lambda_{\max}(\tilde\Hb_{Lff})\lesssim J$ and $\big\|\tilde\Hb_{Lff}\big\|_{\infty}\lesssim J$ when $J$ is sufficiently large. For $\tilde\Hb_{L\lambda f}$, note $\|\tilde\bbf_i^s\|\le D$ and $\|\tilde\blambda_j^s\|\le D$. Then by Assumption~\ref{assumption:smoothness} and $N^{-1}\sum_{i=1}^N\|\tilde\bbf_i-\bbf_i^{(0)}\|^2 + J^{-1}\sum_{j=1}^J\|\tilde\blambda_j-\blambda_j^{(0)}\|^2\le c_{NJ}$,
\begin{align*}
    \big\|\tilde\Hb_{L\lambda f}\big\|_{\max}\le&\,\max_{s\in[0,1]}\max_{i,j}\big|l_{ij}^{\prime\prime}(\tilde\eta_{ij}^s)\big|\big\|\tilde\bbf_i^s\big\|\big\|\tilde\blambda_j^s\big\|+\max_{i,j}\big|l_{ij}^{\prime}(\tilde\eta_{ij})\big|\\\le &\,b_UKD^2+\max_{i,j}\big|l_{ij}^{\prime}(\eta_{ij}^*)\big|+\max_{s\in[0,1]}\max_{i,j}\big|l_{ij}^{\prime\prime}(\tilde\eta_{ij}^{s*})(\tilde\eta_{ij}^s-\eta_{ij}^*)\big|\\\lesssim &\,\max_{i,j}\big|l_{ij}^{\prime}(\eta_{ij}^*)\big|+ 3b_UKD^2.\end{align*}
Here $\tilde\eta_{ij}^{s*}$ is between $\tilde\eta_{ij}^{s}$ and $\eta_{ij}^{*}$. The last inequality follows from the fact that $\big|\tilde\eta_{ij}^{s*}\big| \le KD^2$ and $\big|\tilde\eta_{ij}^s-\eta_{ij}^*\big|\le 2KD^2$ uniformly over $i\in[N]$ and $j\in[J]$.
By (iii) of Lemma~\ref{lemma_concentration}, $\big\|\tilde\Hb_{L\lambda f}\big\|_{\max}\lesssim \sqrt{\epsilon_{NJ}\log (NJ)}$ with probability at least $1 - \exp\big\{-C\epsilon_{NJ}\big\}$ for some constant $C>0$ and $\epsilon_{NJ}\to\infty$.
Recall that $\Lb_{\eta} = \{l_{ij}^{\prime}(\eta_{ij}^*)\}_{N\times J}$. Similarly, we have
\begin{align*}
    \max_{1\le j\le J}\Big\|\big[\tilde\Hb_{L\lambda f}\big]_{[K_j,]}\Big\|= &\,\max_{s\in[0,1]}\max_{1\le j\le J}\Big\|\big\{\cdots,l_{ij}^{\prime\prime}(\tilde \eta_{ij}^s)\tilde\blambda_j^s(\tilde\bbf_i^s)^\T + l_{ij}^{\prime}(\tilde\eta_{ij}^s)\Ib_K,\cdots\big\}_i\Big\|\\\le&\, \max_{s\in[0,1]}\max_{1\le j\le J}b_U\big\|\tilde\blambda_j^s\big\|\big\|\tilde\Fb^s\big\|+\max_{1\le j\le J}\big\|[\Lb_{\eta}]_{[,j]}\big\|\\&\,+\max_{s\in[0,1]}\max_{1\le j\le J}\Big(\sum_{i=1}^Nl_{ij}^{\prime\prime}\big(\tilde\eta_{ij}^{s*}\big)^2(\tilde\eta_{ij}^s-\eta_{ij}^{*})\Big)^{1/2}\\ \le &\, b_UD^2 C_{\delta}\sqrt{N\log J} + b_U\max_{s\in[0,1]}\max_{1\le j\le J}\Big\{\|\tilde\lambda_j^s\|^2\sum_{i=1}^N\big(\|\tilde\bbf_i^s-\bbf_i^{(0)}\|^2\big)\Big\}^{1/2}\\\le &\,C_{\delta}\sqrt{N\log J},\end{align*}
with probability at least $1-(NJ)^{-\delta}$ by the second part of (ii) of Lemma~\ref{lemma_concentration_l_mat}. Similarly, apply the first part of (ii) of Lemma~\ref{lemma_concentration_l_mat} to get, with probability at least $1-N^{-\delta}$,
\begin{align*}
    \Big\|\big[\tilde\Hb_{L\lambda f}\big]_{[K_j,]}\Big\|\le &\,b_U\max_{s\in[0,1]}\big\|\tilde\blambda_j^s\big\|\big\|\tilde\Fb^s\big\|+\Big(\sum_{i=1}^Nl_{ij}^{\prime}\big(\eta_{ij}^*\big)^2\Big)^{1/2}\\&\,+\max_{s\in[0,1]}\Big(\sum_{i=1}^Nl_{ij}^{\prime\prime}\big(\tilde\eta_{ij}^{s*}\big)^2(\tilde\eta_{ij}^s-\eta_{ij}^{*})\Big)^{1/2}\\\le &\,C_{\delta} \sqrt{N}.
\end{align*} Similarly, use (iii) of Lemma~\ref{lemma_concentration_l_mat} to obtain 
\begin{align*}
    \big\|\tilde\Hb_{L\lambda f}\big\| &\,\le \max_{s\in[0,1]}b_U\big\|\tilde\bLambda^s\big\|\big\|\tilde\Fb^s\big\| + \big\|\Lb_{\eta}\big\| + \max_{s\in[0,1]}\Big\|\big\{l_{ij}^{\prime}(\tilde \eta_{ij}^s) - l_{ij}^{\prime}(\eta_{ij}^*)\big\}_{N\times J}\otimes \Ib_K\Big\|.
\end{align*}
The first and second terms can be bounded, with probability at least $1-(NJ)^{-\delta} - (N+J)^{-\delta}$, by $C\sqrt{NJ}$ by the boundedness of $\tilde\btheta$ and (iii) of Lemma~\ref{lemma_concentration_l_mat}. For the third term, we have
\begin{align}
    \max_{s\in[0,1]}\Big\|\big\{l_{ij}^{\prime}(\tilde \eta_{ij}^s) - l_{ij}^{\prime}(\eta_{ij}^*)\big\}_{N\times J}\otimes \Ib_K\Big\| &\,\le \max_{s\in[0,1]}\Big\|\big\{l_{ij}^{\prime\prime}(\tilde \eta_{ij}^{s*})(\tilde\eta_{ij}^s-\eta_{ij}^*)\big\}_{N\times J}\otimes \Ib_K\Big\|_F^2\nonumber\\\lesssim &\, \Big\{\sum_{i=1}^N\sum_{j=1}^J\big(\tilde\eta_{ij}-\eta_{ij})^2\Big\}^{1/2}\nonumber\\\lesssim &\, \sqrt{c_{NJ}}.\label{eq_noise_bound}
\end{align}
as $N^{-1}\sum_{i=1}^N\|\tilde\bbf_i-\bbf_i^{(0)}\|^2 + J^{-1}\sum_{j=1}^J\|\tilde\blambda_j-\blambda_j^{(0)}\|^2\le c_{NJ}$. To conclude, we have
$\big\|\tilde\Hb_{L\lambda f}\big\|\le C_{\delta}\sqrt{NJ}$ with probability at least $1-(NJ)^{-\delta} - (N+J)^{-\delta}$. In the following, we omit this constant $C_{\delta}$ and write `$\lesssim$' instead for simplicity.
In summary, we have,
\begin{enumerate}[label = (\arabic*)]
    \item $\big\|[\tilde\Hb_{L\lambda\lambda}^{-1}]_{[K_j,K_j]}\big\|\lesssim N^{-1}$, $\big\|\tilde\Hb_{L\lambda\lambda}^{-1}\big\|\lesssim N^{-1}$, $\blambda_{\min}\big(\tilde\Hb_{L\lambda\lambda}\big) \gtrsim N$, $\big\|\tilde\Hb_{L\lambda\lambda}^{-1}\big\|_{\infty}\lesssim N^{-1}$;
    \item $\big\|[\tilde\Hb_{Lff}^{-1}]_{[K_i,K_i]}\big\|\lesssim J^{-1}$, $\big\|\tilde\Hb_{Lff}^{-1}\big\|\lesssim J^{-1}$,  $\blambda_{\min}\big(\tilde\Hb_{Lff}\big) \gtrsim J$, $\big\|\tilde\Hb_{Lff}^{-1}\big\|_{\infty}\lesssim J^{-1}$;
    \item with probability at least $1-\exp\big\{-C\epsilon_{NJ}\big\}$, $\max_{i,j}\big\|[\tilde\Hb_{L\lambda f}]_{[K_j,K_i]}\big\|\lesssim \sqrt{\log(NJ)}$
    \item with probability at least $1-(NJ)^{-\delta}$, $\max_{j}\big\|[\tilde\Hb_{L\lambda f}]_{[K_j,]}\big\|\lesssim (N\log J)^{1/2}$;
    \item with probability at least $1-(NJ)^{-\delta}$, $\max_{i}\big\|[\tilde\Hb_{L\lambda f}]_{[,K_i]}\big\|\lesssim (J\log N)^{1/2}$;
    \item with probability at least $1-N^{-\delta}$, $\big\|[\tilde\Hb_{L\lambda f}]_{[K_j,]}\big\|\lesssim N^{1/2}$;
    \item with probability at least $1-J^{-\delta}$, $\big\|[\tilde\Hb_{L\lambda f}]_{[,K_i]}\big\|\lesssim J^{1/2}$;
    \item with probability at least $1-(NJ)^{-\delta} - (N+J)^{-\delta}$, $\big\|\tilde\Hb_{L\lambda f}\big\|\lesssim (NJ)^{1/2}$.
\end{enumerate}
    \subsubsection*{Part (i).}  
   Note that we can write $\Hb(\tilde\btheta)$ as 
    \begin{equation*}
        \Hb(\tilde\btheta)=\Hb_L(\tilde\btheta)+cNJ\sum_{l=1}^K\sum_{h=1}^K\tilde\bv_{lh}\tilde\bv_{lh}^\T,
    \end{equation*}
    where $\tilde\bv_{lh}=(\zero_{JK}^\T,\{\bv_{f,lh}(\tilde\btheta)\}^\T)^\T$ for $h\le l$ and $\tilde\bv_{lh}=(\{\bv_{\lambda,lh}(\tilde\btheta)\}^\T,\zero_{NK}^\T)^\T$ when $l<h$.
    Then by the continuity of eigenvalues, it suffices to show that for any $\delta>0$, \begin{align*}
        &P\Bigg\{\lambda_{\min}\Big(\Db_{m}^{-1/2}\Hb_L(\tilde\btheta)\Db_{m}^{-1/2}+cNJ\Db_{m}^{-1/2}\sum_{l=1}^K\sum_{h=1}^K\tilde\bv_{lh}\tilde\bv_{lh}^\T\Db_{m}^{-1/2}\Big)\\&\qquad\qquad<Cb_L\min\{\lambda_{\min}(\bSigma_f^*),\lambda_{\min}(\bSigma_\lambda^*)\}\Bigg\}\le (NJ)^{-\delta} + (N+J)^{-\delta}.
    \end{align*}
 Similarly to Lemma~\ref{lemma_concentration_l_mat}, let $\tilde\Lb_{\eta} = \big\{l_{ij}^{\prime}(\tilde\eta_{ij})\big\}_{N\times J}$ be an $N\times J$ matrix with the $(i,j)$th entry being $l_{ij}^{\prime}(\tilde\eta_{ij})$. Here we slightly abuse the notation where $\tilde\Lb_{\eta}$ defined in this proof is different from that in the proof of Lemma~\ref{lemma_concentration_l_mat}.  Then with $\big|l_{ij}^{\prime}(\tilde\eta_{ij}) - l_{ij}^{\prime}(\eta_{ij}^*)\big|^2 \le b_U^2\big|\tilde\eta_{ij} - \eta_{ij}^*\big|^2 \le b_U^2\big| (\tilde\blambda_j-\blambda_j^*)^\T\bbf_i^*\big|^2  + b_U^2\big|(\blambda_j^*)^\T(\tilde\bbf_i - \bbf_i^*) \big|^2 + b_U^2\big|(\tilde\blambda_j-\blambda_j^*)^\T(\tilde\bbf_i - \bbf_i^*)\big|^2$ and $N^{-1}\sum_{i=1}^N\|\tilde\bbf_i-\bbf_i^*\|^2 + J^{-1}\sum_{j=1}^J\|\tilde\blambda_j - \blambda_j^*\|^2\le c_{NJ}$, we know 
    \begin{align*}
        \|\tilde\Lb_{\eta} - \Lb_{\eta}\|\le\|\tilde\Lb_{\eta} - \Lb_{\eta}\|_F\le \Big\{\sum_{i=1}^N\sum_{j=1}^Jb_U^2\big(\tilde\eta_{ij}-\eta_{ij}^*\big)^2\Big\}^{1/2}\lesssim \sqrt{NJc_{NJ}}.
    \end{align*}
    Define 
    \begin{equation*}
        \Db_{m}^{-1/2}\Hb_L^0(\tilde\btheta)\Db_{m}^{-1/2} = \Db_{m}^{-1/2}\Hb_L(\tilde\btheta)\Db_{m}^{-1/2} + \Db_{m}^{-1/2}\begin{pmatrix}
            \zero & \tilde\Lb_{\eta}^\T\otimes \Ib_K\\\tilde\Lb_{\eta}\otimes \Ib_K&\zero
        \end{pmatrix}\Db_{m}^{-1/2}.
    \end{equation*}
    Together Lemma~\ref{lemma_concentration_l_mat}, we know that, with probability at least $1-(NJ)^{-\delta} - (N+J)^{-\delta}$, 
    \begin{equation}
        \left\|\begin{pmatrix}
            \zero & (NJ)^{-1/2}\tilde\Lb_{\eta}^\T\otimes \Ib_K\\(NJ)^{-1/2}\tilde\Lb_{\eta}\otimes \Ib_K&\zero
        \end{pmatrix}\right\|\lesssim(N\wedge J)^{-1/2} + \sqrt{c_{NJ}}.\label{eq_bound_jocob_hessian}
    \end{equation}Then we focus on $\Db_{m}^{-1/2}\Hb_L^0(\tilde\btheta)\Db_{m}^{-1/2}+cNJ\Db_{m}^{-1/2}\sum_{l=1}^K\sum_{h=1}^K\tilde\bv_{lh}\tilde\bv_{lh}^\T\Db_{m}^{-1/2}$.
    
    We define 
    \begin{equation*}
        \bv_{lh}^{*}=\left(J^{-1/2}\lambda_{1h}^{(0)}\be_l^\T,\cdots,J^{-1/2}\lambda_{Jh}^{(0)}\be_l^\T,-N^{-1/2} f_{1l}^{(0)}\be_h^\T,\cdots,-N^{-1/2} f_{Nl}^{(0)}\be_h^\T\right)^\T,
    \end{equation*} 
    and $\Vb^*=\left(\bv_{11}^*,\bv_{12}^*,\cdots,\bv_{1K}^*,\bv_{21}^*,\cdots,\bv_{KK}^*\right)$. Write $\Vb_\lambda^*=J^{-1/2}\big(\bLambda^{(0)}\otimes \be_1,\cdots,\bLambda^{(0)}\otimes \be_K\big)$ and $\Vb_{f}^*=-N^{-1/2}\Fb^{(0)}\otimes \Ib_K$. Then $\Vb^*=\big((\Vb_\lambda^*)^\T,(\Vb_f^*)^\T\big)^\T$. Similarly we denote $\Vb_{\tilde \lambda}^*=J^{-1/2}\big(\tilde\bLambda\otimes \be_1,\cdots,\tilde\bLambda\otimes \be_K\big)$ and $\Vb_{\tilde f}^*=[\Vb^*]_{[JK+N_K,]}=-N^{-1/2}\tilde\Fb\otimes \Ib_K$.
    Next, we find that
    \begin{align}
     &\,\Db_{m}^{-1/2}\Hb^0_L(\tilde\btheta)\Db_{m}^{-1/2}+c\sum_{l=1}^K\sum_{h=1}^K\bv_{lh}^*(\bv_{lh}^*)^\T\nonumber\\=&\,  -\Db_{m}^{-{1}/{2}}\sum_{i=1}^N \sum_{j=1}^J  l_{i j}^{\prime\prime}\big(\tilde\eta_{ij})\left[\begin{array}{c}
\be_j^{(J)} \otimes \tilde\bbf_i \\
\be_i^{(N)} \otimes \tilde\blambda_j 
\end{array}\right]\left[\begin{array}{c}
\be_j^{(J)} \otimes \tilde\bbf_i \\
\be_i^{(N)} \otimes \tilde\blambda_j 
\end{array}\right]^{\intercal} \Db_{m}^{-{1}/{2}} + c\Vb^*(\Vb^*)^\T   \nonumber \\
     =&\, -\Db_{m}^{-{1}/{2}}\sum_{i=1}^N \sum_{j=1}^J \big( l_{i j}^{\prime\prime}(\tilde\eta_{ij})+c\big)\left[\begin{array}{c}
\be_j^{(J)} \otimes \tilde\bbf_i \\
\be_i^{(N)} \otimes \tilde\blambda_j 
\end{array}\right]\left[\begin{array}{c}
\be_j^{(J)} \otimes \tilde\bbf_i \\
\be_i^{(N)} \otimes \tilde\blambda_j 
\end{array}\right]^{\intercal} \Db_{m}^{-{1}/{2}}  \nonumber \\
 \label{eq:breve H expand term 1} \\
&\,+\left[\begin{array}{ccc}
I_J \otimes \frac{c}{N} \sum_{i=1}^N\tilde\bbf_i\tilde\bbf_i^\T& \zero   \\
\zero & I_N \otimes \frac{c}{J} \sum_{j=1}^J \tilde\blambda_j \tilde\blambda_j^\T
\end{array}\right] \nonumber \\
 \label{eq:breve H expand term 2}\\
&\,+c\left[\begin{array}{ccc}
 \Vb_\lambda^*\big(\Vb_\lambda^*\big)^\T&\zero\\\zero&\Vb_f^*\big(\Vb_f^*\big)^\T\end{array}\right]\label{eq:breve H expand term 3}\\&\,+c\left[\begin{array}{ccc}
 \zero&\Vb_\lambda^*\big(\Vb_f^*\big)^\T-\Vb_{\tilde\lambda}^*\big(\Vb_{\tilde f}^*\big)^\T\\\Vb_f^*\big(\Vb_\lambda^*\big)^\T-\Vb_{\tilde f}^*\big(\Vb_{\tilde\lambda}^*\big)^\T&\zero\end{array}\right]
 \label{eq:breve H expand term 4}
\end{align}
Take $0<c<b_L/2$ and we know that \eqref{eq:breve H expand term 1} and \eqref{eq:breve H expand term 3} are positive semi-definite. For \eqref{eq:breve H expand term 4}, its off-diagonal term can be bounded as
\begin{align*}
    \Big\|\Vb_\lambda^*\big(\Vb_f^*\big)^\T-\Vb_{\tilde\lambda}^*\big(\Vb_{\tilde f}^*\big)^\T\Big\|\le&\,\Big\|\big(\Vb_\lambda^*-\Vb_{\tilde\lambda}^*\big)\big(\Vb_{ f}^*\big)^\T\Big\| + \Big\|\Vb_\lambda^*\big(\Vb_f^*-\Vb_{\tilde f}^*\big)^\T\Big\|\\&\,+\Big\|\big(\Vb_\lambda^*-\Vb_{\tilde\lambda}^*\big)\big(\Vb_{ f}^*-\Vb_{\tilde f}^*\big)^\T\Big\|\\\lesssim &\,\sqrt{c_{NJ}},
\end{align*}
as $J^{-1}\sum_{j=1}^J\big(\tilde\lambda_{jl}-\lambda_{jl}^{(0)}\big)^2+N^{-1}\sum_{i=1}^N\big(\tilde f_{il}-f_{il}^{(0)}\big)^2\le c_{NJ}$ for any $l\in[K]$.
Next similar to the derivation of \eqref{eq_estimate_lambda_min_ff}, the smallest eigenvalue of \eqref{eq:breve H expand term 2} can be bounded by $\min\{\lambda_{\min}(\bSigma_{f}),$ $\lambda_{\min}(\bSigma_{\lambda})\}/2$ when $N,J$ are large enough. 
 Then we conclude that \begin{equation*}
\lambda_{\min}\big(\Db_{m}^{-1/2}\Hb^0_L(\tilde\btheta)\Db_{m}^{-1/2}+c\sum_{l=1}^K\sum_{h=1}^K\bv_{lh}^*(\bv_{lh}^*)^\T\big)>C b_L\min\{\lambda_{\min}(\bSigma_{f}^*),\lambda_{\min}(\bSigma_{\lambda}^*)\},\end{equation*}  for some constant $C$.

Next, define $\Vb(\tilde\btheta) = \sqrt{NJ}\Db_m^{-1/2}(\tilde\bv_{11},\tilde\bv_{12},\cdots,\tilde\bv_{1K},\tilde\bv_{21},\cdots,\tilde\bv_{KK})$ and $\Vb(\btheta^{(0)}) $ \\$=\sqrt{NJ} \Db_m^{-1/2}(\bv_{11},\bv_{12},\cdots,\bv_{1K},\bv_{21},\cdots,\bv_{KK})$ with $\bv_{lh} = \tilde\bv_{lh}$ when $\tilde\btheta = \btheta^{(0)}$. Note 
\begin{align*}
    \big\|\Vb(\tilde\btheta)-\Vb(\btheta^{(0)})\big\| \le \Big[\sum_{l=1}^K\sum_{h=1}^K\big\{&\,J^{-1}\sum_{j=1}^J(\tilde\lambda_{jl}+\tilde\lambda_{jh} - \lambda_{jl}^{(0)} - \lambda_{jh}^{(0)})^2\\&\,+N^{-1}\sum_{i=1}^N\big(\tilde f_{il}+\tilde f_{ih}-f_{il}^{(0)}-f_{ih}^{(0)}\big)^2\big\}\Big]^{1/2}\lesssim \sqrt{c_{NJ}},
\end{align*}as $J^{-1}\sum_{j=1}^J\big(\tilde\lambda_{jl}-\lambda_{jl}^{(0)}\big)^2+N^{-1}\sum_{i=1}^N\big(\tilde f_{il}-f_{il}^{(0)}\big)^2\le c_{NJ}$ for any $l\in[K]$.
Write $\sigma_r^{(0)}=J^{-1}\sum_{j=1}^J(\lambda_{jr}^{(0)})^2$. Since $\Mb_{ff}^{(0)}$ and $\Mb_{\lambda\lambda}^{(0)}$ are diagonal, we know 
\begin{equation*}
    \big((\Vb^*)^\T\Vb^*\big)^{-1}=\left\{\Ib_K\otimes\begin{pmatrix}
        1+\sigma_1^{(0)}\\&\ddots\\&&1+\sigma_K^{(0)}
    \end{pmatrix}\right\}^{-1}.
\end{equation*}
    Then it is easy to check that 
    \begin{align*}
        \sigma_{V,\min}:=&\,\sigma_{\min}\big\{\big((\Vb^*)^\T\Vb^*\big)^{-1}\big(\Vb^*)^\T\Vb(\tilde\btheta)\big\}\\\ge &\,\sigma_{\min}\big\{\big((\Vb^*)^\T\Vb^*\big)^{-1}\big(\Vb^*)^\T\Vb(\btheta^{(0)})\big\}\\&\,-\sigma_{\max}\Big\{\big((\Vb^*)^\T\Vb^*\big)^{-1}\big(\Vb^*)^\T\big(\Vb(\tilde\btheta)-\Vb(\btheta^{(0)})\big)\Big\}\\\ge &\,\min_{l\neq h}\sigma_{\min}\begin{pmatrix}
            \sigma_l^{(0)}/(\sigma_l^{(0)}+1)&\sigma_h^{(0)}/(\sigma_l^{(0)}+1)\\1/(\sigma_l^{(0)}+1)&1/(\sigma_l^{(0)}+1)
        \end{pmatrix}\\&\,-\big\|\big((\Vb^*)^\T\Vb^*\big)^{-1}\big(\Vb^*)^\T\big\|\big\|\Vb(\tilde\btheta)-\Vb(\btheta^{(0)})\big\|.
    \end{align*}
    The first term depends only on $\sigma_{l}^{(0)}$ and has some positive lower bound as $\big\{\sigma_{l}^{(0)}\big\}_l$ are mutually distinct. The second term is bounded by $C\sqrt{c_{NJ}}$ for some constant C. Therefore  $\sigma_{V,\min}$ is bounded from below by some positive constant.    
   Then we apply the following lemma.
   \begin{lemma}\label{lemma_projection_non_diminish}\it
    Suppose $\Ab\in\RR^{n\times n}$ and $\Vb\in\RR^{n\times K}$ such that $\lambda_{\min}\big(\Ab+\Vb\Vb^\T\big)>\gamma>0$. Suppose $\sigma_{K}(\Vb)\ge \sigma_0$. For any $\Ub\in\RR^{n\times K}$ such that $\sigma_{\max}(\Ub)\le \sigma_1$ and $\sigma_{\min}\big((\Vb^\T\Vb)^{-1}\Vb^\T\Ub\big)\ge \sigma_2$, we have
    \begin{equation*}
        \lambda_{\min}\big(\Ab + c\Ub\Ub^\T\big)\ge\min\{\gamma - c\sigma_1^2, c\sigma_0^2\sigma_2^2\}.
    \end{equation*}
\end{lemma} 
The proof is in Section~\ref{prove_lemma_projection_non_diminish}. Set $\Ab = \Db^{-1/2}_m\tilde\Hb_L^0\Db^{-1/2}_m$, $\Vb = \Vb^*$ and $\Ub = \tilde\Vb$ and $0<c<Cb_L\min\{\lambda_{\min}(\bSigma_{f}^*),\lambda_{\min}(\bSigma_{\lambda}^*)\}/(2\sigma_{\max}(\Vb^*))<Cb_L\min\{\lambda_{\min}(\bSigma_{f}^*),\lambda_{\min}(\bSigma_{\lambda}^*)\}/(2D)$ to obtain, with probability at least $1-(NJ)^{-\delta}-(N+J)^{-\delta}$, 
\begin{equation*}
\lambda_{\min}\big(\Db_{m}^{-1/2}\Hb^0_L(\tilde\btheta)\Db_{m}^{-1/2}+c\sum_{l=1}^K\sum_{h=1}^K\tilde\bv_{lh}\tilde\bv_{lh}^\T\big)>C b_L\min\{\lambda_{\min}(\bSigma_{f}^*),\lambda_{\min}(\bSigma_{\lambda}^*)\},\end{equation*} for some constant $C$.  Combine with \eqref{eq_bound_jocob_hessian} and we finish the proof.

\subsubsection*{Part (ii) and (iii). }
By union bound, we know, with probability at least $1-N^{-\delta}-3(NJ)^{-\delta} - (N+J)^{-\delta}$  that (a) $\lambda_{\min}\big(\Db^{-1/2}_{m}\Hb(\tilde\btheta^s)\Db_m^{-1/2}\big)\ge \gamma$ for any $s\in[0,1]$ with $c\le 1$ selected as in (i); (b) $\big\|[\tilde\Hb_{\lambda f}]_{[K_j,]}\big\|\lesssim N^{1/2}$; (c) $\max_{1\le j\le J}\big\|[\tilde\Hb_{\lambda f}]_{[K_j,]}\big\|\lesssim \sqrt{N\log J}$; (d) $\big\|\tilde\Hb_{L\lambda f}\big\|\lesssim (NJ)^{1/2}$.
Suppose for the rest of the proof that events (a)--(d) hold.

Write 
\begin{equation*}
    \tilde\Hb=\begin{pmatrix}
        \tilde\Hb_{\lambda\lambda}&\tilde\Hb_{\lambda f}\\\tilde\Hb_{f\lambda}&\tilde\Hb_{ff}
    \end{pmatrix}=\begin{pmatrix}
        \tilde\Hb_{L\lambda\lambda}&\tilde\Hb_{L\lambda f}\\\tilde\Hb_{Lf\lambda}&\tilde\Hb_{Lff}
    \end{pmatrix}+cNJ\tilde\Vb^{p+}(\tilde\Vb^{p-})^\T.
\end{equation*}
where $\tilde\Vb^{p+} = (\tilde\bv_{11}^+,\tilde\bv_{12}^+,\cdots,\tilde\bv_{1K}^+,\tilde\bv_{21}^+,\cdots,\tilde\bv_{KK}^+,\tilde\bv_{11}^-,\tilde\bv_{12}^-,\cdots,\tilde\bv_{1K}^-,\tilde\bv_{21}^-,\cdots,\tilde\bv_{KK}^-)$ and $\tilde\Vb^{p-} = (\tilde\bv_{11}^+,\tilde\bv_{12}^+,\cdots,\tilde\bv_{1K}^+,\tilde\bv_{21}^+,\cdots,\tilde\bv_{KK}^+,-\tilde\bv_{11}^-,-\tilde\bv_{12}^-,\cdots,-\tilde\bv_{1K}^-,-\tilde\bv_{21}^-,\cdots,-\tilde\bv_{KK}^-)$.
It is easy to observe that $\tilde\Hb_{\lambda f} = \tilde\Hb_{L\lambda f}$ and $\tilde\Hb_{f\lambda} = \tilde\Hb_{Lf\lambda}$.
Next by Sherman–Morrison–Woodbury formula, the inverse of $\tilde\Hb$ can be expressed as 
\begin{equation*}
    \tilde\Hb^{-1}=\begin{pmatrix}
        \tilde\Hb_{\lambda\lambda}^{-1}+\tilde\Hb_{\lambda\lambda}^{-1}\tilde\Hb_{\lambda f}\tilde\Hb_{-f}\tilde\Hb_{ f\lambda}\tilde\Hb_{\lambda\lambda}^{-1}&\big[\tilde\Hb_{\lambda\lambda}^{-1}+\tilde\Hb_{\lambda\lambda}^{-1}\tilde\Hb_{\lambda f}\tilde\Hb_{-f}\tilde\Hb_{ f\lambda}\tilde\Hb_{\lambda\lambda}^{-1}\big]\tilde\Hb_{\lambda f}\tilde\Hb_{ff}^{-1}\\\big[\tilde\Hb_{ff}^{-1}+\tilde\Hb_{ff}^{-1}\tilde\Hb_{ f\lambda}\tilde\Hb_{-\lambda}\tilde\Hb_{\lambda f}\tilde\Hb_{ff}^{-1}\big]\tilde\Hb_{ f\lambda}\tilde\Hb_{\lambda\lambda}^{-1}&\tilde\Hb_{ff}^{-1}+\tilde\Hb_{ff}^{-1}\tilde\Hb_{ f\lambda}\tilde\Hb_{-\lambda}\tilde\Hb_{\lambda f}\tilde\Hb_{ff}^{-1}
        \end{pmatrix}
\end{equation*}
Here $\tilde\Hb_{-f}$ is the Schur complement of $\tilde\Hb_{ff}$ and $\tilde\Hb_{-\lambda}$ is the Schur complement of $\tilde\Hb_{\lambda\lambda}$. By $\lambda_{\min}\big(\Db^{-1/2}_{m}\Hb(\tilde\btheta)\Db_m^{-1/2}\big)\ge \gamma$, we know that $\big\|\tilde\Hb_{-f}\big\|\lesssim J^{-1}$ and $\big\|\tilde\Hb_{-\lambda}\big\|\lesssim N^{-1}$.
 Let 
\begin{align*}
     \tilde\Vb^{p+}_{\lambda} = \sqrt{cJ}\big[\tilde\Vb^{p+}\big]_{[1:JK,]}\text{ and }\tilde\Vb^{p+}_{f} = \sqrt{cN}\big[\tilde\Vb^{p+}\big]_{[JK+1:NK+JK,]};\\
     \tilde\Vb^{p-}_{\lambda} = \sqrt{cJ}\big[\tilde\Vb^{p-}\big]_{[1:JK,]}\text{ and }\tilde\Vb^{p-}_{f} = \sqrt{cN}\big[\tilde\Vb^{p-}\big]_{[JK+1:NK+JK,]}
 \end{align*}
Then we can rewrite $cNJ\tilde\Vb^{p+}\Ib_{2K^2}(\tilde\Vb^{p-})^\T$ as
 \begin{equation*}
     cNJ\tilde\Vb^{p+}(\tilde\Vb^{p-})^\T = \begin{pmatrix}
         N\tilde\Vb^{p+}_{\lambda}\Ib_{2K^2}(\tilde\Vb^{p-}_{\lambda})^\T&\zero\\\zero&J\tilde\Vb^{p+}_{f}\Ib_{2K^2}(\tilde\Vb^{p-}_{f})^\T.
     \end{pmatrix}
 \end{equation*}
Since $\big\|\tilde\bv_{lh}^+\big\|_{\infty}\le \sqrt2D/J$ and $\big\|\tilde\bv_{lh}^-\big\|_{\infty}\le D/(\sqrt3J)$, we have $\|\tilde\Vb^{p+}_\lambda\|\le  {4K}D$, $\|[\tilde\Vb^{p+}_\lambda]_{[K_j,]}\| \le 4KDJ^{-1/2}$, $\|\tilde\Vb^{p-}_\lambda\|\le  {4K}D$ and $\|[\tilde\Vb^{p-}_\lambda]_{[K_j,]}\| \le 4KDJ^{-1/2}$. Next we compute $\tilde\Hb_{\lambda\lambda}^{-1}$. By Woodbury identity, 
\begin{align*}
    \tilde\Hb_{\lambda\lambda}^{-1}=&\,\Big(\tilde\Hb_{L\lambda\lambda}+N\tilde\Vb^{p+}_{\lambda}(\tilde\Vb^{p-}_{\lambda})^\T\Big)^{-1}\\=&\,\tilde\Hb_{L\lambda\lambda}^{-1}-\tilde\Hb_{L\lambda\lambda}^{-1}\tilde\Vb^{p+}_{\lambda}\big\{N^{-1}\Ib_{2K^2}+(\tilde\Vb^{p-}_{\lambda})^\T\tilde\Hb_{L\lambda\lambda}^{-1}\tilde\Vb^{p+}_{\lambda}\big\}^{-1}(\tilde\Vb^{p-}_{\lambda})^\T\tilde\Hb_{L\lambda\lambda}^{-1}.
\end{align*}
Note $(\tilde\Vb^{p-}_{\lambda})^\T\tilde\Hb_{L\lambda\lambda}^{-1}\tilde\Vb^{p+}_{\lambda} = \sqrt{NJ}\{\Db_m^{-1/2}(\tilde\bv_{lh}^+)^\T\tilde\Hb_{L\lambda\lambda}^{-1}\tilde\bv_{lh}^+\Db_m^{-1/2} - \Db_m^{-1/2}(\tilde\bv_{lh}^-)^\T\tilde\Hb_{L\lambda\lambda}^{-1}\tilde\bv_{lh}^-\Db_m^{-1/2}\}_{l,h=1}^{K^2}$. Here the first term is semi-positive definite, and the second term can be bounded as
\begin{align*}
    \big\|\sqrt{NJ}\Db_m^{-1/2}(\tilde\bv_{lh}^-)^\T\tilde\Hb_{L\lambda\lambda}^{-1}\tilde\bv_{lh}^-\Db_m^{-1/2}\big\|&\,\le \max_{l,h}\sqrt{NJ}\big\|\tilde\Hb_{L\lambda\lambda}^{-1}\big\|\big\|\Db_m^{-1/2}\tilde\bv_{lh}^-\big\|^2\\&\,\lesssim \sqrt{NJ} \times N^{-1}\times c_{NJ}/\sqrt{NJ}\asymp N^{-1}c_{NJ}.
\end{align*}
Therefore we know 
\begin{equation*}
    \Big\|\tilde\Vb^{p+}_{\lambda}\big\{N^{-1}\Ib_{2K^2}+(\tilde\Vb^{p-}_{\lambda})^\T\tilde\Hb_{L\lambda\lambda}^{-1}\tilde\Vb^{p+}_{\lambda}\big\}(\tilde\Vb^{p+}_{\lambda})^\T\Big\|\lesssim N\big\|\tilde\Vb^{p+}_{\lambda}\big\|\big\|\tilde\Vb^{p-}_{\lambda}\big\|\asymp N.
\end{equation*}
By $\lambda_{\min}\big([\tilde\Hb_{L\lambda\lambda}^{-1}]_{[K_j,K_j]}\big)\lesssim N^{-1}$, $\big\|[\tilde\Hb_{\lambda f}]_{[K_j,]}\big\|\lesssim N^{1/2}$ and $\big\|\tilde\Hb_{\lambda f}\big\|\lesssim \sqrt{NJ}$, we have \begin{align}
    &\,\big\|[\tilde\Hb_{\lambda\lambda}^{-1}\tilde\Hb_{\lambda f}]_{[K_j,]}\big\|\le\big\|[\tilde\Hb_{L\lambda\lambda}^{-1}]_{[K_j,K_j]}\big\|\big\|[\tilde\Hb_{\lambda f}]_{[K_j,]}\big\|\nonumber\\&\,+\big\|[\tilde\Hb_{L\lambda\lambda}^{-1}]_{[K_j,K_j]}\big\|\big\|[\tilde\Vb^{p+}_{\lambda}]_{[K_j,]}\big\|\big\{N^{-1}\Ib_{2K^2}+(\tilde\Vb^{p-}_{\lambda})^\T\tilde\Hb_{L\lambda\lambda}^{-1}\tilde\Vb^{p+}_{\lambda}\big\}^{-1}(\tilde\Vb^{p+}_{\lambda})^\T\big\|\big\|\tilde\Hb_{L\lambda\lambda}^{-1}\big\|\big\|\tilde\Hb_{\lambda f}\big\|\nonumber\\&\, \lesssim N^{-1/2},\label{eq_bound_row_hessian_1}
\end{align} and \begin{align}
    \big\|\tilde\Hb_{\lambda\lambda}^{-1}\big\|\le&\,\big\|\tilde\Hb_{L\lambda\lambda}^{-1}\big\|+\big\|\tilde\Hb_{L\lambda\lambda}^{-1}\big\|\big\|\tilde\Vb^{p+}_{\lambda}\big\{N^{-1}\Ib_{2K^2}+(\tilde\Vb^{p-}_{\lambda})^\T\tilde\Hb_{L\lambda\lambda}^{-1}\tilde\Vb^{p+}_{\lambda}\big\}^{-1}(\tilde\Vb^{p-}_{\lambda})^\T\big\|\big\|[\tilde\Hb_{L\lambda\lambda}^{-1}\big\|\nonumber\\\lesssim &\, N^{-1}.\label{eq_bound_row_hessian_2}
\end{align}
Similarly, we can derive $\big\|[\tilde\Hb_{ff}^{-1}]_{[K_i,K_i]}\big\|\lesssim J^{-1}$ and $\big\|\tilde\Hb_{ff}^{-1}\big\|\lesssim J^{-1}$.

Then we compute the $l_\infty$ bound for $\tilde\Hb$. Note that $\big\|\tilde\Hb_{L\lambda\lambda}^{-1}\big\|_{\infty}=\max_{1\le j\le J}\big\|[\tilde\Hb_{L\lambda\lambda}]_{[K_j,K_j]}^{-1}\big\|_{\infty} \lesssim N^{-1} $. Then
\begin{align*}
    &\, \big\|\tilde\Hb_{\lambda\lambda}^{-1}\big\|_{\infty} \le \big\|\tilde\Hb_{L\lambda\lambda}^{-1}\big\|_{\infty} \\&\,+\max_{1\le j\le J} \big\|[\tilde\Hb_{L\lambda\lambda}]_{[K_j,K_j]}^{-1}\big\|_{\infty}\big\|[\Vb_{\lambda}^{p+}]_{[K_j,]}\big\|_{\infty}\big\|\big(N^{-1}\Ib_{2K^2}+(\Vb_{\lambda}^{p-})^\T\tilde\Hb_{L\lambda\lambda}^{-1}\Vb_{\lambda}^{p+}\big)^{-1}\big\|_{\infty}\\&\quad \big\|(\Vb_{\lambda}^{p-})^\T\big\|_{\infty}\big\|\tilde\Hb_{L\lambda\lambda}^{-1}\big\|_{\infty}
\end{align*}
Note that $\big\|[\tilde\Vb^{p+}_{\lambda}]_{[K_j,]}\big\|_{\infty}\le  2J^{-1/2}KD$, $\big\|[\tilde\Vb^{p-}_{\lambda}]_{[K_j,]}\big\|_{\infty}\le  2J^{-1/2}KD$, $\big\|(\tilde\Vb_{\lambda}^{p+})^\T\big\|_{\infty}\le\sqrt{JK}$ $ \big\|(\tilde\Vb^{p+}_{\lambda})^\T\big\|\le 4KD\sqrt{JK}$, $\big\|(\tilde\Vb_{\lambda}^{p-})^\T\big\|_{\infty}\le 4KD\sqrt{JK}$, and $\big\|\big(N^{-1}\Ib_{2K^2}+(\Vb_{\lambda}^{p-})^\T\tilde\Hb_{L\lambda\lambda}^{-1}\Vb_{\lambda}^{p+}\big)^{-1}\big\|_{\infty} $ $ \le \sqrt{K}\big\|\big(N^{-1}\Ib_{2K^2} + \Vb_{\lambda}^\T\tilde\Hb_{L\lambda\lambda}^{-1}\Vb_{\lambda}\big)^{-1}\big\|\lesssim N$. Therefore we conclude $\big\|\tilde\Hb_{\lambda\lambda}^{-1}\big\|_{\infty}\lesssim N^{-1}$. Similarly we have $\big\|\tilde\Hb_{ff}^{-1}\big\|_{\infty}\lesssim J^{-1}$.
Since $\max_{1\le j\le J}\big\|[\tilde\Hb_{\lambda f}]_{[K_j,]}\big\|\le C_{\delta}\sqrt{N\log J}$, and $\big\|\tilde\Hb_{\lambda f}\big\| = \big\|\tilde\Hb_{\lambda f}\big\|\le C_{\delta}\sqrt{NJ}$, we have
\begin{align*}
    &\,\Big\|\big[\tilde\Hb^{-1}\Db_m\big]_{[J_K,]}\Big\|_{\infty}\\\le &\,N\left\{\big\|\tilde\Hb_{\lambda\lambda}^{-1}\big\|_{\infty}+\big\|\tilde\Hb_{\lambda\lambda}^{-1}\tilde\Hb_{\lambda f}\tilde\Hb_{-f}\tilde\Hb_{ f\lambda}\tilde\Hb_{\lambda\lambda}^{-1}\big\|_{\infty}\right\}\\&\,+J\left\{\big\|\tilde\Hb_{\lambda\lambda}^{-1}\tilde\Hb_{\lambda f}\tilde\Hb_{ff}^{-1}\big\|_{\infty}+\big\|\tilde\Hb_{\lambda\lambda}^{-1}\tilde\Hb_{\lambda f}\tilde\Hb_{-f}\tilde\Hb_{ f\lambda}\tilde\Hb_{\lambda\lambda}^{-1}\tilde\Hb_{\lambda f}\tilde\Hb_{ff}^{-1}\big\|_{\infty}\right\}\\\le &\, N\left\{\big\|\tilde\Hb_{\lambda\lambda}^{-1}\big\|_{\infty}+\max_{1\le j\le J}\big\|\big[\tilde\Hb_{\lambda\lambda}^{-1}\tilde\Hb_{\lambda f}\tilde\Hb_{-f}\tilde\Hb_{ f\lambda}\tilde\Hb_{\lambda\lambda}^{-1}\big]_{[K_j,]}\big\|_{\infty}\right\}\\&\,+J\max_{1\le j\le J}\left\{\big\|\big[\tilde\Hb_{\lambda\lambda}^{-1}\tilde\Hb_{\lambda f}\tilde\Hb_{ff}^{-1}\big]_{[K_j,]}\big\|_{\infty}+\big\|\big[\tilde\Hb_{\lambda\lambda}^{-1}\tilde\Hb_{\lambda f}\tilde\Hb_{-f}\tilde\Hb_{ f\lambda}\tilde\Hb_{\lambda\lambda}^{-1}\tilde\Hb_{\lambda f}\tilde\Hb_{ff}^{-1}\big]_{[K_j,]}\big\|_{\infty}\right\}\\\le&\,N\left\{\big\|\tilde\Hb_{\lambda\lambda}^{-1}\big\|_{\infty}+\max_{1\le j\le J}\sqrt{JK}\big\|\big[\tilde\Hb_{\lambda\lambda}^{-1}\big]_{[K_j,K_j]}\big\|\big\|[\tilde\Hb_{\lambda f}]_{[K_j,]}\big\|\big\|\tilde\Hb_{-f}\big\|\big\|\tilde\Hb_{ f\lambda}\big\|\big\|\tilde\Hb_{\lambda\lambda}^{-1}\big\|\right\}\\&\,+J\max_{1\le j\le J}\sqrt{NK}\left\{\big\|\big[\tilde\Hb_{\lambda\lambda}^{-1}]_{[K_j,K_j]}\big\|\big\|[\tilde\Hb_{\lambda f}]_{[K_j,]}\big\|\big\|\tilde\Hb_{ff}^{-1}\big\|\right.\\&\,\left.+\big\|\big[\tilde\Hb_{\lambda\lambda}^{-1}]_{[K_j,K_j]}\big\|\big\|[\tilde\Hb_{\lambda f}]_{[K_j,]}\big\|\big\|\tilde\Hb_{-f}\big\|\big\|\tilde\Hb_{ f\lambda}\big\|\big\|\tilde\Hb_{\lambda\lambda}^{-1}\big\|\big\|\tilde\Hb_{\lambda f}\big\|\big\|\tilde\Hb_{ff}^{-1}\big\|\right\}\\\le &\, C_{\delta}\log J,
\end{align*}
where $C_{\delta}$ is some constant depending only on $\delta$. The probability is at least $1-N^{-\delta} - 2(NJ)^{-\delta} - (N+J)^{-\delta}$. 
Similarly, we have $$\Big\|\big[\tilde\Hb^{-1}\Db_m\big]_{[JK+N_K,]}\Big\|_{\infty}\le C_{\delta}\log N,$$
 with probability at least $1-J^{-\delta}-2(NJ)^{-\delta} - (N+J)^{-\delta}$.
\subsubsection*{Part (iv) and (v).}
By union bound, we know, with probability at least $1-N^{-\delta}-3(NJ)^{-\delta} - (N+J)^{-\delta}$  that (a) $\lambda_{\min}\big(\Db^{-1/2}_{m}\Hb(\tilde\btheta^s)\Db_m^{-1/2}\big)\ge \gamma$ for any $s\in[0,1]$ with $c\le 1$ selected as in (i); (b) $\big\|[\tilde\Hb_{\lambda f}]_{[K_j,]}\big\|\lesssim N^{1/2}$; (c) $\max_{1\le j\le J}\big\|[\tilde\Hb_{\lambda f}]_{[K_j,]}\big\|\lesssim \sqrt{N\log J}$; (d) $\big\|\tilde\Hb_{L\lambda f}\big\|\lesssim (NJ)^{1/2}$
Suppose for the rest of the proof that events (a)--(d) hold.

Note that
\begin{equation*}
    \big[\tilde\Hb^{-1}\big]_{[J_K,J_K]}  = \tilde\Hb_{\lambda\lambda}^{-1}+\tilde\Hb_{\lambda\lambda}^{-1}\tilde\Hb_{\lambda f}\tilde\Hb_{-f}\tilde\Hb_{ f\lambda}\tilde\Hb_{\lambda\lambda}^{-1}.
\end{equation*}
and
\begin{equation*}
    \big[\tilde\Hb^{-1}\big]_{[J_K,NK+J_K]} =\big[\tilde\Hb_{\lambda\lambda}^{-1}+\tilde\Hb_{\lambda\lambda}^{-1}\tilde\Hb_{\lambda f}\tilde\Hb_{-f}\tilde\Hb_{ f\lambda}\tilde\Hb_{\lambda\lambda}^{-1}\big]\tilde\Hb_{\lambda f}\tilde\Hb_{ff}^{-1}.
\end{equation*}
Note $\big\|\tilde\Hb_{\lambda\lambda}^{-1}\big\|\lesssim N^{-1}$, $\big\|\tilde\Hb_{ff}^{-1}\big\|\lesssim J^{-1}$ when $N,J$ are large enough, and $\big\|\tilde\Hb_{\lambda f}\big\|\lesssim \sqrt{NJ}$, and $\big\|\tilde\Hb_{-f}\big\|\lesssim J^{-1}$. Then there is
\begin{align*}
    \Big\|\big[\tilde\Hb^{-1}\big]_{[J_K,J_K]}\Big\|\le C_{\delta} N^{-1}\text{ and }\Big\|\big[\tilde\Hb^{-1}\big]_{[J_K,NK+J_K]}\Big\|\le C_{\delta}(NJ)^{-1/2},
\end{align*}which proves the first half of (iv).
For the second half of (iv), use the expansion again and we have
\begin{align*}
    \big[\tilde\Hb^{-1}\bS_{L}^*\big]_{[J_K,]}  = &\,\tilde\Hb_{\lambda\lambda}^{-1}\bS_{L\lambda}^*+\tilde\Hb_{\lambda\lambda}^{-1}\tilde\Hb_{\lambda f}\tilde\Hb_{-f}\tilde\Hb_{ f\lambda}\tilde\Hb_{\lambda\lambda}^{-1}\bS_{L\lambda}^*\\&\,+\big[\tilde\Hb_{\lambda\lambda}^{-1}+\tilde\Hb_{\lambda\lambda}^{-1}\tilde\Hb_{\lambda f}\tilde\Hb_{-f}\tilde\Hb_{ f\lambda}\tilde\Hb_{\lambda\lambda}^{-1}\big]\tilde\Hb_{\lambda f}\tilde\Hb_{ff}^{-1}\bS_{Lf}^*.
\end{align*}
For the first term, we know by $\big\|\tilde\Hb_{\lambda\lambda}^{-1}\big\|\lesssim N^{-1}$ and  (i) of Corollary~\ref{coro_first_order_concern} that for some $C_{\delta}$,
\begin{equation*}
    \big\|\tilde\Hb_{\lambda\lambda}^{-1}\bS_{L\lambda}^*\big\|\le \big\|\tilde\Hb_{\lambda\lambda}^{-1} \big\|\big\|\bS_{L\lambda}^*\big\|\le C_{\delta} N^{-1}\times \sqrt{NJ} = C_{\delta}\sqrt{J/N},
\end{equation*}
with probability at least $1-J^{-\delta}$. Similarly we can bound the second term by $C_{\delta}\sqrt{J/N}$ with probability at least $1-J^{-\delta}$. For the third term, we first note that, by $\big\|\tilde\Hb_{\lambda\lambda}^{-1}\big\|\lesssim N^{-1}$, $\big\|\tilde\Hb_{-f}\big\|\lesssim J^{-1}$ and $\big\|\tilde\Hb_{\lambda f}\big\|\lesssim\sqrt{NJ}$, we have
\begin{equation*}
    \Big\|\tilde\Hb_{\lambda\lambda}^{-1}+\tilde\Hb_{\lambda\lambda}^{-1}\tilde\Hb_{\lambda f}\tilde\Hb_{-f}\tilde\Hb_{ f\lambda}\tilde\Hb_{\lambda\lambda}^{-1}\Big\|\le C_{\delta}N^{-1}.
\end{equation*}
For $\tilde\Hb_{\lambda f}\tilde\Hb_{ff}^{-1}\bS_{Lf}^*$, use the Woodbury identity on $\tilde\Hb_{ff}^{-1} = \big\{\tilde\Hb_{ff} +J\tilde\Vb_f^{p+}(\tilde\Vb_f^{p-})^\T\big\}^{-1}$ to get
\begin{align}
    \Big\|\tilde\Hb_{\lambda f}\tilde\Hb_{ff}^{-1}\bS_{Lf}^*\Big\|\le &\,\Big\|\Hb_{\lambda f}(\btheta^{(0)})\tilde\Hb_{ff}^{-1}\bS_{Lf}^*\Big\|+\Big\|\big\{\tilde\Hb_{\lambda f}-\Hb_{\lambda f}(\btheta^{(0)})\big\}\tilde\Hb_{ff}^{-1}\bS_{Lf}^*\Big\|\nonumber\\\le &\,\Big\|\Hb_{\lambda f}(\btheta^{(0)})\tilde\Hb_{Lff}^{-1}\bS_{Lf}^*\Big\|\label{eq_bound_resi_term_1} \\&\,+\Big\|\Hb_{\lambda f}(\btheta^{(0)})\tilde\Hb_{Lff}^{-1}\tilde\Vb^{p-}_f\big\{J^{-1}\Ib_{2K^2}+(\tilde\Vb_{f}^{p-})^\T\tilde\Hb_{Lff}^{-1}\tilde\Vb_f^{p+}\big\}^{-1}(\tilde\Vb_{f}^{p+})^\T\tilde\Hb_{Lff}^{-1}\bS_{Lf}^*\Big\|\label{eq_bound_resi_term_2}\\&\, +\Big\|\big\{\tilde\Hb_{\lambda f}-\Hb_{\lambda f}(\btheta^{(0)})\big\}\tilde\Hb_{ff}^{-1}\bS_{Lf}^*\Big\| .\label{eq_bound_resi_term_3}
\end{align}
For \eqref{eq_bound_resi_term_1}, we expand $[\Hb_{\lambda f}(\btheta^{(0)})\tilde\Hb_{Lff}^{-1}\bS_{Lf}^*]_{[K_j]}$ as
\begin{align*}
    &\,[\Hb_{\lambda f}(\btheta^{(0)})\tilde\Hb_{Lff}^{-1}\bS_{Lf}^*]_{[K_j]} \\ &\,= \sum_{i=1}^N\big\{l_{ij}^{\prime\prime}(\eta_{ij}^*)\bbf_i^{(0)}(\blambda_j^{(0)})^\T + l_{ij}^{\prime}(\eta_{ij}^*)\Ib_K\big\}\big\{\sum_{t=1}^Jl_{it}^{\prime\prime}(\tilde\eta_{it})\tilde\blambda_t\tilde\blambda_t^\T\big\}^{-1}\big\{\sum_{t=1}^Jl_{it}^{\prime}(\eta_{it}^*)\blambda_t^*\big\}.
\end{align*}
Since $\max_{1\le i\le N}\lambda_{\min}\big\{\sum_{t=1}^Jl_{it}^{\prime\prime}(\tilde\eta_{it})\tilde\blambda_t\tilde\blambda_t^\T\big\}\gtrsim J^{-1}$, similarly to deriving (iii) and (v) of Lemma~\ref{lemma_linear_equation_hold}, 
with probability at least $1-N^{-\delta}-J^{-\delta}-\exp(-C\epsilon_{NJ})$,
\begin{equation}
    \max_{1\le j\le J}\big\|[\Hb_{\lambda f}(\btheta^{(0)})\tilde\Hb_{Lff}^{-1}\bS_{Lf}^*]_{[K_j]}\big\|\le C_{\delta}\left(\sqrt{N\log (NJ)/J} + N/J+\sqrt{N\epsilon_{NJ}/J}\right).\label{eq_bound_row_hessian_4}
\end{equation}
Also, similar to deriving (i) of Lemma~\ref{lemma_linear_equation_hold}, we have $\max_{1\le j\le 2K^2}\big\|\big[N(\tilde\Vb_{f}^p)^\T\tilde\Hb_{Lff}^{-1}\bS_{Lf}^*\big]_{[j]}\big\|\le C_{\delta}\sqrt{N\epsilon_{NJ}/J}$ with probability at least $1-\exp(-C\epsilon_{NJ})$, as each entry in $\tilde\Vb^{p+}_f$ and $\tilde\Vb^{p-}_f$ is bounded by $N^{-1/2}$. Therefore, taking $\epsilon_{NJ}=C^{-1}\delta\log (NJ)$, by the results in proving (ii) and (iii)
\begin{align*}
    \eqref{eq_bound_resi_term_1} + \eqref{eq_bound_resi_term_2}\le &\,\sqrt{J}\max_{1\le j\le J}\big\|[\Hb_{\lambda f}(\btheta^{(0)})\tilde\Hb_{Lff}^{-1}\bS_{Lf}^*]_{[K_j]}\big\| \\&\, +\Big\|\Hb_{\lambda f}(\btheta^{(0)})\Big\|\Big\|\tilde\Hb_{Lff}^{-1}\Big\|\Big\|\tilde\Vb^{p-}_f\Big\|\Big\|\big\{J^{-1}\Ib_{2K^2}+(\tilde\Vb_{f}^{p-})^\T\tilde\Hb_{Lff}^{-1}\tilde\Vb_f^{p+}\big\}^{-1}\Big\|\\&\,\qquad\Big\|(\tilde\Vb_{f}^{p+})^\T\tilde\Hb_{Lff}^{-1}\bS_{Lf}^*\Big\|\\\le &\,C_{\delta}\left(\sqrt{N\log (NJ)} + N/\sqrt{J}\right)
\end{align*}
with probability at least $1-N^{-\delta}-J^{-\delta} - 2(NJ)^{-\delta}$. For \eqref{eq_bound_resi_term_3}, note 
\begin{equation*}
    \Big[\Hb_{\lambda f}(\btheta^{(0)}) -\tilde\Hb_{\lambda f}\Big]_{[K_j,K_j]} = \int_0^1\Big\{l_{ij}^{\prime\prime}(\tilde\eta_{ij}^s)\tilde\bbf_i^s(\tilde\blambda_j^s)^\T-l_{ij}^{\prime}(\tilde\eta_{ij}^s)\Ib_{K} - l_{ij}^{\prime\prime}(\tilde\eta_{ij}^*)\tilde\bbf_i^{(0)}(\tilde\blambda_j^{(0)})^\T-l_{ij}^{\prime}(\eta_{ij}^{*})\Ib_{K}\Big\}ds.
\end{equation*}
Since $N^{-1}\sum_{i=1}^N\big\|\tilde\bbf_i-\bbf_i^{(0)}\big\|^2 + J^{-1}\sum_{j=1}^J\big\|\tilde\blambda_j-\blambda_j^{(0)}\big\|^2\le c_{NJ}$, we know
\begin{align*}
    \max_{s\in[0,1]}\big\|\Hb_{\lambda f}(\btheta^{(0)}) -\Hb_{\lambda f}(\btheta^s)\big\|\le &\, b_U\big\|\tilde\bLambda^s\big\|\big\|\tilde\Fb^s-\Fb^{(0)}\big\|+b_U\big\|\tilde\bLambda^s-\bLambda^{(0)}\big\|\big\|\Fb^{(0)}\big\| \\&\,+ \big\|\big(\Lb_{\eta}-\tilde\Lb_{\eta}^s\big)\otimes \Ib_K\big\|\\\le &\,\sqrt{NJc_{NJ}},
\end{align*}
which gives $\eqref{eq_bound_resi_term_3}\lesssim N\sqrt{c_{NJ}}$,
with probability at least $1-(NJ)^{-\delta}-(N+J)^{-\delta}$. Here $\tilde\Lb_{\eta}^s = \big\{l_{ij}^{\prime}(\tilde\eta_{ij}^s)\big\}_{N\times J}$ and the bound for $\big\|(\Lb_{\eta}-\tilde\Lb_{\eta}^s)\otimes \Ib_K\big\|$ follows \eqref{eq_noise_bound}. Finally we have that, with probability at least $1-N^{-\delta}-J^{-\delta}-3(NJ)^{-\delta} - (N+J)^{-\delta}$, 
\begin{equation}
    \big\|\tilde\Hb_{\lambda f}\tilde\Hb_{ff}^{-1}\bS_{Lf}^*\big\|\le C_{\delta}\left(\sqrt{N\log (NJ)} + N/\sqrt{J}+N\sqrt{c_{NJ}}\right).\label{eq_bound_row_hessian_5}
\end{equation}
To conclude with probability at least $1-N^{-\delta}-J^{-\delta}-3(NJ)^{-\delta} - (N+J)^{-\delta}$, 
\begin{equation*}
    \Big\|\big[\tilde\Hb^{-1}\bS_{L}^*\big]_{[J_K,]}\Big\|\le C_{\delta}\left(\sqrt{J/N}+\sqrt{c_{NJ}}+\sqrt{\log (NJ)/N} + 1/\sqrt{J}\right).
\end{equation*}
The probability bound is obtained by union bounding the events (a)--(c).

For (v), note \begin{align*}
    \Big\|\big[\tilde\Hb^{-1}\bS_{L}^*\big]_{[K_j]}\Big\|  = &\,\underbrace{\Big\|\big[\tilde\Hb_{\lambda\lambda}^{-1}\bS_{L\lambda}^*\big]_{[K_j]}\Big\|}_{(H1)}+\underbrace{\Big\|\big[\tilde\Hb_{\lambda\lambda}^{-1}\tilde\Hb_{\lambda f}\big]_{[K_j]}\tilde\Hb_{-f}\tilde\Hb_{ f\lambda}\tilde\Hb_{\lambda\lambda}^{-1}\bS_{L\lambda}^*\Big\|}_{(H2)}\\&\,+\underbrace{\Big\|\big[\tilde\Hb_{\lambda\lambda}^{-1}\tilde\Hb_{\lambda f}\big]_{[K_j]}\tilde\Hb_{ff}^{-1}\bS_{Lf}^*\Big\|}_{(H3)}+\underbrace{\Big\|\big[\tilde\Hb_{\lambda\lambda}^{-1}\tilde\Hb_{\lambda f}\big]_{[K_j]}\tilde\Hb_{-f}\tilde\Hb_{ f\lambda}\tilde\Hb_{\lambda\lambda}^{-1}\tilde\Hb_{\lambda f}\tilde\Hb_{ff}^{-1}\bS_{Lf}^*\Big\|}_{(H4)}.
\end{align*}
 For $(H1)$, by $\big\|[\tilde\Hb_{L\lambda\lambda}^{-1}]_{[K_j,K_j]}\big\|\lesssim N^{-1}$, $\big\|\tilde\Hb_{L\lambda\lambda}^{-1}\big\|\lesssim N^{-1}$, $\big\|[\tilde\Vb_{\lambda}^{p-}]_{[K_j,]}\big\|\lesssim J^{-1/2}$, $\big\|\tilde\Vb_{\lambda}^{p-}\big\|\lesssim 1$, $\big\|\big(N^{-1}\Ib_{2K^2}+(\tilde\Vb_{\lambda}^{p-})^\T\tilde\Hb_{L\lambda\lambda}^{-1}\tilde\Vb_{\lambda}^{p+}\big)^{-1}\big\|\lesssim N^{-1}$, (i) of Lemma~\ref{lemma_concentration} and (i) of Corollary~\ref{coro_first_order_concern}, we have
\begin{align}
    &\,\big\|[\tilde\Hb_{\lambda\lambda}^{-1}\bS_{L}^*]_{[K_j,]}\big\|\le\big\|[\tilde\Hb_{L\lambda\lambda}^{-1}]_{[K_j,K_j]}\big\|\big\|[\bS_{L}^*]_{[K_j,]}\big\|\nonumber\\&\,+\big\|[\tilde\Hb_{L\lambda\lambda}^{-1}]_{[K_j,K_j]}\big\|\big\|[\tilde\Vb_{\lambda}^{p-}]_{[K_j,]}\big\|\big(N^{-1}\Ib_{2K^2}+(\tilde\Vb_{\lambda}^{p-})^\T\tilde\Hb_{L\lambda\lambda}^{-1}\tilde\Vb_{\lambda}^{p+}\big)^{-1}\big\|\big\|(\tilde\Vb_{\lambda}^{p+})^\T\big\|\big\|\tilde\Hb_{L\lambda\lambda}^{-1}\big\|\big\|\bS_{L\lambda}^*\big\|\nonumber\\&\, \le C_{\delta}\sqrt{\epsilon_N/N},\nonumber
\end{align}
with probability at least $1-\exp(-C\epsilon_N)-J^{-\delta}$. For $(H2)$, similar to bound the second term in $\big[\tilde\Hb^{-1}\bS_L^*\big]_{[J_K,]}$, we can bound it by $CN^{-1/2}$ using \eqref{eq_bound_row_hessian_1} for some constant $C$.
For $(H3)$, use \eqref{eq_bound_row_hessian_2} and \eqref{eq_bound_row_hessian_5} to get
\begin{align}
    &\,\big\|[\tilde\Hb_{\lambda\lambda}^{-1}\tilde\Hb_{\lambda f}\big]_{[K_j]}\tilde\Hb_{ff}^{-1}\bS_{Lf}^*\big\|\le\big\|[\tilde\Hb_{L\lambda\lambda}^{-1}]_{[K_j,K_j]}\big\|\big\|[\tilde\Hb_{\lambda f}\tilde\Hb_{ff}^{-1}\bS_{Lf}^*]_{[K_j,]}\big\|\nonumber\\&\,+\big\|[\tilde\Hb_{L\lambda\lambda}^{-1}]_{[K_j,K_j]}\big\|\big\|[\tilde\Vb^{p-}_{\lambda}]_{[K_j,]}\big\|\big(N^{-1}\Ib_{2K^2}+(\tilde\Vb^{p-}_{\lambda})^\T\tilde\Hb_{L\lambda\lambda}^{-1}\tilde\Vb^{p+}_{\lambda}\big)^{-1}(\tilde\Vb^{p+}_{\lambda})^\T\big\|\\&\,\qquad\big\|\tilde\Hb_{L\lambda\lambda}^{-1}\big\|\big\|\tilde\Hb_{\lambda f}\tilde\Hb_{ff}^{-1}\bS_{Lf}^*\big\|\nonumber\\&\, \le C_{\delta}\left(\sqrt{\frac{c_{NJ}}{N}}+\sqrt{\frac{\log(NJ)}{NJ}} + \frac{1}{J}\right),\nonumber
\end{align}with probability at least $1-N^{-\delta}-J^{-\delta} - 3(NJ)^{-\delta} - (N+J)^{-\delta}$.
Similarly, using \eqref{eq_bound_row_hessian_1} and \eqref{eq_bound_row_hessian_5}, we can bound $(H4)$ by $C_{\delta}(\sqrt{c_{NJ}/N}+\sqrt{\log(NJ)/(NJ)} + 1/J)$ with the same probability. Combining the bounds for $(H1)$--$(H4)$ and the probability for events to hold, we finish the proof.

\subsubsection*{Part (vi) and (vii).}Parts (vi) and (vii) follow similarly from Parts (iv) and (v).
\subsubsection*{Part (viii) and (ix).}
By union bound, we know, with probability at least $1-N^{-\delta}-3(NJ)^{-\delta} - (N+J)^{-\delta}$, that (a) $\lambda_{\min}\big(\Db^{-1/2}_{m}\Hb(\tilde\btheta^s)\Db_m^{-1/2}\big)\ge \gamma$ for any $s\in[0,1]$ with $c\le 1$ selected as in (i); (b) $\big\|[\tilde\Hb_{\lambda f}]_{[K_j,]}\big\|\lesssim N^{1/2}$; (c) $\max_{1\le j\le J}\big\|[\tilde\Hb_{\lambda f}]_{[K_j,]}\big\|\lesssim \sqrt{N\log J}$; and (d) $\big\|\tilde\Hb_{L\lambda f}\big\|\lesssim (NJ)^{1/2}$.
Suppose for the rest of the proof that events (a)--(d) hold.

Since $\big\|[\tilde\Hb_{L\lambda\lambda}^{-1}]_{[K_j,K_j]}\big\|\lesssim N^{-1}$, $\big\|\tilde\Hb_{L\lambda\lambda}^{-1}\big\|\lesssim N^{-1}$, $\big\|[\tilde\Vb_{\lambda}^{p+}]_{[K_j,]}\big\|\lesssim J^{-1/2}$, $\big\|\tilde\Vb_{\lambda}^{p+}\big\|\lesssim 1$, $\big\|\big(N^{-1}\Ib_{2K^2}+(\tilde\Vb_{\lambda}^{p-})^\T\tilde\Hb_{L\lambda\lambda}^{-1}\tilde\Vb_{\lambda}^{p+}\big)^{-1}\big\|\lesssim N^{-1}$, we have
\begin{align}
    \Big\|\big[\tilde\Hb_{\lambda\lambda}^{-1} - \tilde\Hb_{L\lambda\lambda}^{-1}\big]_{[K_j,]}\Big\|\le &\, \big\|\big[\tilde\Hb_{L\lambda\lambda}^{-1}]_{[K_j,K_j]}\big\|\big\|[\tilde\Vb^{p+}_{\lambda}]_{[K_j,]}\big\|\nonumber\\&\,\big\|\big(N^{-1}\Ib_{2K^2}+(\tilde\Vb^{p-}_{\lambda})^\T\tilde\Hb_{L\lambda\lambda}^{-1}\tilde\Vb^{p-}_{\lambda}\big)^{-1}\big\|\big\|(\tilde\Vb^{p+}_{\lambda})^\T\big\|\big\|\tilde\Hb_{L\lambda\lambda}^{-1}\big\|\nonumber\\\lesssim &\, N^{-1}J^{-1/2}.\label{eq_bound_row_hessian_3}
\end{align}
Combine \eqref{eq_bound_row_hessian_1}, \eqref{eq_bound_row_hessian_2} and \eqref{eq_bound_row_hessian_3} all together and we have, with probability at least $1-N^{-\delta} - 2(NJ)^{-\delta} - (N + J)^{-\delta}$, that
\begin{align*}
    &\,\Big\|\big[\Db_m^{1/2}\tilde\Hb^{-1}\Db_m^{1/2} - N\tilde\Hb_{L\lambda\lambda}^{-1}\big]_{[K_j,]}\Big\|\\\le &\,N\left\{\big\|[\tilde\Hb_{\lambda\lambda}^{-1} - \tilde\Hb_{L\lambda\lambda}]_{[K_j,]}\big\|+\big\|\big[\tilde\Hb_{\lambda\lambda}^{-1}\tilde\Hb_{\lambda f}\tilde\Hb_{-f}\tilde\Hb_{ f\lambda}\tilde\Hb_{\lambda\lambda}^{-1}\big]_{[K_j,]}\big\|\right\}\\&\,+\sqrt{NJ}\left\{\big\|\big[\tilde\Hb_{\lambda\lambda}^{-1}\tilde\Hb_{\lambda f}\tilde\Hb_{ff}^{-1}\big]_{[K_j,]}\big\|+\big\|\big[\tilde\Hb_{\lambda\lambda}^{-1}\tilde\Hb_{\lambda f}\tilde\Hb_{-f}\tilde\Hb_{ f\lambda}\tilde\Hb_{\lambda\lambda}^{-1}\tilde\Hb_{\lambda f}\tilde\Hb_{ff}^{-1}\big]_{[K_j,]}\big\|\right\}\\\le &\, N\left\{\big\|[\tilde\Hb_{\lambda\lambda}^{-1} - \tilde\Hb_{L\lambda\lambda}]_{[K_j,]}\big\|+\big\|\big[\tilde\Hb_{\lambda\lambda}^{-1}\tilde\Hb_{\lambda f}]_{[K_j,]}\big\|\big\|\tilde\Hb_{-f}\big\|\big\|\tilde\Hb_{ f\lambda}\big\|\big\|\tilde\Hb_{\lambda\lambda}^{-1}\big\|\right\}\\&\,+\sqrt{NJ}\left\{\big\|\big[\tilde\Hb_{\lambda\lambda}^{-1}\tilde\Hb_{\lambda f}]_{[K_j,]}\big\|\big\|\tilde\Hb_{ff}^{-1}\big\|\right.\\&\,\left.+\big\|\big[\tilde\Hb_{\lambda\lambda}^{-1}\tilde\Hb_{\lambda f}]_{[K_j,]}\big\|\big\|\tilde\Hb_{-f}\big\|\big\|\tilde\Hb_{ f\lambda}\big\|\big\|\tilde\Hb_{\lambda\lambda}^{-1}\big\|\big\|\tilde\Hb_{\lambda f}\big\|\big\|\tilde\Hb_{ff}^{-1}\big\|\right\} \\\le &\,C_{\delta} J^{-1/2},
\end{align*}
where $C_{\delta}$ is some constant depending only on $\delta$. This dependence arises from the omission when bounding the terms relating to $\tilde\Hb_{\lambda f}$.
Similarly we can prove that when $i=N+1,\cdots N+J$, $\big\|[\Db_m^{1/2}\tilde\Hb^{-1}\Db_m^{1/2}]_{[(i-1)K+1:iK,]}\big\|\le C_{\delta} N^{-1/2}$ with probability at least $1-J^{-\delta} - 3(NJ)^{-\delta} - (N + J)^{-\delta}$.


\subsubsection{Proof of Corollary~\ref{cor_lemma_hessian_property}}\label{supp_sec_prove_cor_lemma_hessian_property}

   We compute $\partial_{\btheta\btheta}^2\bS_P(\btheta)$ as follows.
\begin{align}
    \partial_{\btheta\btheta}^2\bS_P(\btheta) =&\, cNJ\begin{pmatrix}
        \sum_{l=1}^K\sum_{h>l}^K\bv_{\lambda,lh}(\btheta)\bv_{\lambda,lh}(\btheta)^\T&\zero\\\zero&\sum_{l=1}^K\sum_{h\le l}\bv_{f,lh}(\btheta)\bv_{f,lh}(\btheta)^\T
    \end{pmatrix} \label{eq_hp_term1} \\&\,+cN\begin{pmatrix}
        \sum_{l=1}^K\sum_{h>l}\big(J^{-1}\lambda_{jl}\lambda_{jh}\big)\Ib_J\otimes(\be_l\be_h^\T+\be_h\be_l^\T)&\zero\\\zero&\zero
    \end{pmatrix}\label{eq_hp_term2}\\&\,+cJ\begin{pmatrix}
        \zero&\zero\\\zero&\sum_{l=1}^K\sum_{h\le l}\big(N^{-1}\sum_{i=1}^Nf_{il}f_{ih}-\delta_{hl}\big)\Ib_N\otimes \big(\be_l\be_h^\T+\be_h\be_l^\T\big)
    \end{pmatrix}\label{eq_hp_term3}
\end{align}
Note $\eqref{eq_hp_term1} = \Hb_P(\btheta)$.
When $\max_{j,l}|\lambda_{jl}|\le D$ and $J^{-1}\sum_{j=1}^J\big\|\blambda_j-\blambda_j^{(0)}\big\|^2\le c_{NJ}$, with $\max_{j,l}|\lambda_{jl}^{(0)}|\le D$ and $J^{-1}\sum_{j=1}^J\lambda_{jl}^{(0)}\lambda_{jh}^{(0)} = 0$ for any $l\neq h$ we have
\begin{align*}
    J^{-1}\lambda_{jl}\lambda_{jh} = &\,J^{-1}\lambda_{jl}\lambda_{jh} - J^{-1}\sum_{j=1}^J\lambda_{jl}^{(0)}\lambda_{jh}^{(0)}\\= &\,J^{-1}\lambda_{jl}(\lambda_{jh} - \lambda_{jh}^{(0)}) + J^{-1}(\lambda_{jl}-\lambda_{jl}^{(0)})\lambda_{jh}^{(0)}\\\le &\,\sqrt{Dc_{NJ}}.
\end{align*}
Then for \eqref{eq_hp_term2} and \eqref{eq_hp_term3}, we have
\begin{equation}
    \big\|[\eqref{eq_hp_term2}]_{[K_{j_1},K_{j_2}]}\big\| \lesssim 1_{(1\le j_1=j_2\le J)}\sqrt{c_{NJ}}N\label{eq_bound_hp_res2}
\end{equation}
holds for any $\tilde\btheta$ satisfying the conditions in Lemma~\ref{lemma_hessian_property}. Similarly
\begin{equation}
    \big\|[\eqref{eq_hp_term3}]_{[K_{i_1},K_{i_2}]}\big\| \lesssim 1_{(J<i_1=i_2\le N+J)}\sqrt{c_{NJ}}J\label{eq_bound_hp_res3}
\end{equation}
holds for any $\tilde\btheta$ satisfying the conditions in Lemma~\ref{lemma_hessian_property}. 
 Note $cN\le\lambda\big([\tilde\Hb_{\lambda\lambda}]_{[K_j,K_j]}\big)\le CN$ for some constant $c$ and $C$. Therefore \eqref{eq_bound_hp_res2} can be added to $\tilde\Hb_{\lambda\lambda}$ without changing the upper and lower bound. Similarly \eqref{eq_bound_hp_res3} can be added to $\tilde\Hb_{ff}$. We conclude that the results in Lemma~\ref{lemma_hessian_property} still hold as \eqref{eq_bound_hp_res2} and \eqref{eq_bound_hp_res3} can be absorbed into $\tilde\Hb_{\lambda\lambda}$ and $\tilde\Hb_{ff}$, respectively.

\subsubsection{Proof of Corollary~\ref{cor_lemma_hessian_property2}}\label{supp_sec_prove_cor_lemma_hessian_property2}

The proof is omitted as all the bounds in the proof of Lemma~\ref{lemma_hessian_property} still hold without the integral.

\subsubsection{Proof of Corollary~\ref{cor_lemma_hessian_property_bound_l_prime}}\label{supp_sec_prove_cor_lemma_hessian_property_bound_l_prime}

    Note that $\|\Lb_{\eta}\|$ can be bounded by some constant. Therefore the log-terms in the bounds for $\max_{i,j}\|[\tilde\Hb_{L\lambda f}]_{[K_j,K_i]}\|$, $\max_{j}\|[\tilde\Hb_{L\lambda f}]_{[K_j,]}\|$ and $\max_{i}\|[\tilde\Hb_{L\lambda f}]_{[,K_i]}\| $ can be removed. The other part follows similarly.

\subsection{Proof of Lemma~\ref{lemma_concentration}}\label{prove_lemma_concentation}

Follow the Bernstein-type probability bound as in Proposition 5.16 in \cite{vershynin2010introduction}:
\[
\mathbb{P} \left\{ \left| \sum_{i=1}^N a_i l_{ij}^{\prime}(\eta_{ij}^*) \right| \geq t \right\} \leq 2 \exp \left[ -c \min \left( \frac{t^2}{(MA)^2N}, \frac{t}{MAN} \right) \right],\;\forall j\in[J],
\]where $c > 0$ is an absolute constant and $M$ is the defined in Assumption~\ref{assumption:smoothness} with $\|l_{ij}^{\prime}(\eta_{ij}^*)\|_{\varphi_1}\le M$. 
Take $t = (CN\epsilon_N/c)^{1/2}MA$, and we obtain the bound for $\big| \sum_{i=1}^N a_i l_{ij}^{\prime}(\eta_{ij}^*) \big| $ in (i). Next by union bound, 
\[
\mathbb{P} \left\{\max_{1\le j\le J} \left| \sum_{i=1}^N a_i l_{ij}^{\prime}(\eta_{ij}^*) \right| \geq t \right\} \leq 2 J\exp \left[ -c \min \left( \frac{t^2}{(MA)^2N}, \frac{t}{MAN} \right) \right].
\]Take $t = (CN\log J\epsilon_N/c)^{1/2}MA$ and the second part in (i) is proved.
Similarly, we can derive (ii) and (iii)

\subsubsection{Proof of Corollary~\ref{coro_first_order_concern}}\label{sup_sec_prove_coro_first_order_concern}

   This corollary follows from the same argument as in the proof of Lemma~\ref{lemma_concentration_l_mat}, using the fact that each entry of $\Db_m^{-1/2}\bS_L(\btheta^*)$ is sub-exponential by Lemma~\ref{lemma_concentration}.

\subsection{Proof of Lemma~\ref{lemma_for_convergenceG}}\label{prove_lemma_for_convergenceG}

Since $\hat\Ab$ and $\hat \Bb $ are of finite dimension, we have $\|\hat\Ab-\Ab\|_{\max}=O_p(\nu)$, $\|\hat\Bb -\Bb \|_{\max}=O_p(\nu)$, with \begin{align*}\|\hat\Ab\hat\Bb -\Ab\Bb \|&\le \|\hat\Ab-\Ab\|\|\hat\Bb-\Bb\|+\|\Ab\|\|\hat\Bb-\Bb\|+\|\hat\Ab-\Ab\|\|\Bb\|\\&=O_p(\nu).\end{align*} Let $\gamma$ be the eigen-gap of $\Ab\Bb $. By Weyl's theorem, we know that when $\nu$ is small enough, the eigen-gap of $\hat\Ab\hat\Bb $ is larger than $\gamma/2$. Similarly if we denote $\gamma_L=\min\big\{\lambda_{\min}(\Ab),\lambda_{\min}(\Bb )\big\}>0$ and $\gamma_U=\max\big\{\lambda_{\max}(\Ab),\lambda_{\max}(\Bb )\big\}>0$, then we have $$\min\big\{\lambda_{\min}(\hat\Ab),\lambda_{\min}(\hat\Bb )\big\}>\gamma_L/2\text{ and }\max\big\{\lambda_{\max}(\hat\Ab),\lambda_{\max}(\hat\Bb )\big\}<2\gamma_U$$ when $\nu$ is below some threshold. Next, we discuss only this regime.

    Let the singular value decomposition of $\Ab$ be $\cU_1\bUpsilon_1\cU_1^\T$, and the singular value decomposition of $\Ab^{1/2}\Bb \Ab^{1/2}$ be $\cU_2\bUpsilon_2\cU_2^\T$ with $\Ab^{1/2}$ defined as $\cU_1\bUpsilon_1^{1/2}\cU_1^\T$. Then $\Gb=\Ab^{-1/2}\cU_2$ is the unique solution up to a permutation of $\bUpsilon_2$. The uniqueness is a result of the eigenvalues of $\Ab\Bb $ being different, which implies that there exists a unique order for the diagonal entries of $\bUpsilon_2$, and thus $\cU_2$ is uniquely determined.
    Similarly we write the SVD of $\hat \Ab$ and $\hat \Ab^{1/2}\hat \Bb \hat \Ab^{1/2}$ as $\hat \cU_1\hat \bUpsilon_1\hat \cU_1^\T$ and $\hat \cU_2\hat \bUpsilon_2\hat \cU_2^\T$. Therefore $\hat\Gb$ can be uniquely expressed as $\hat \Ab^{-1/2}\hat \cU_2$. The existence of $\hat\Ab^{-1/2}$ is implied by $\lambda_{\min}(\hat\Ab)>\gamma_L/2$, $\lambda_{\max}(\hat\Ab)<\gamma_U/2$ and uniqueness is implied by the fact that the eigen-gap of $\hat\Ab\hat\Bb $ is larger than $\gamma/2$.

    Implied by $\|\hat\Ab-\Ab\|\lesssim \nu$ and that $\lambda_{\max}(\hat\Ab),\lambda_{\max}(\Ab), \lambda_{\min}(\hat\Ab),\lambda_{\max}(\Ab)$ are bounded in $(2\gamma_U,\gamma_L/2)$, we have
    \begin{align}
        \big\|\hat\Ab^{-1/2}-\Ab^{-1/2}\big\|&\le \big\|\big(\hat\Ab^{-1/2}+\Ab^{-1/2}\big)\big\|\big\|\hat\Ab^{-1}-\Ab^{-1}\big\|\nonumber\\&\le \big\|\big(\hat\Ab^{-1/2}+\Ab^{-1/2}\big)\big\|\big\|\hat\Ab^{-1}\big\|\big\|\Ab^{-1}\big\|\big\|\hat\Ab-\Ab\big\|\nonumber\\&\lesssim \nu.\label{lemma_eq_dkthm}
    \end{align}
     Next by the variant of Davis-Kahan theorem \citep{yu2015useful},
    \begin{equation*}
        \big\|\hat \cU_2-\cU_2\big\|\lesssim\nu,
    \end{equation*}
    as we have fixed the order of the distinctive diagonal entries in $\hat\bUpsilon_2$ and $\bUpsilon_2$. Combining all these together, we have 
    \begin{equation*}
        \|\hat \Gb- \Gb\| \lesssim\nu.
    \end{equation*}

\subsection{Proof of Lemma~\ref{lemma_linear_equation_perturb}}\label{prove_lemma_linear_equation_perturb}

        When $c$ is small enough, we must have $\mathrm{rank}(\tilde\Hb) = K-1$ as well. Without loss of generality we find $\bx_h,\tilde\bx_h$ such that $\|\bx_h\| = \|\tilde \bx_h\| = 1$. Define projection matrix $\Pb_{\Hb^\T} = \Hb^\T(\Hb\Hb^\T)^{-1}\Hb$. By definition we know $(\Ib_{K} - \Pb_{\Hb^\T})\bx \in \{\bx_h\nu:\nu\in\RR\}$ for any $\bx\in\RR^K$. Next
        \begin{align*}
            \Big|\{\bx_h^\T\tilde\bx_h\}^2 - \|\bx_h\|^2\|\tilde\bx_h\|^2\Big| = &\,\Big|\big\{\bx_h^\T\big(\Ib_{K} - \Pb_{\Hb^\T}\big)\tilde\bx_h\big\}^2 - 1\Big|\\=&\,\Big|\|\bx_h\|^2\big\|\big(\Ib_{K} - \Pb_{\Hb^\T}\big)\tilde\bx_h\big\|^2 - 1\Big| \\ \le &\, \Big|\|\bx_h\|^2\big\|\big(\Ib_{K} - \Pb_{\tilde\Hb^\T}\big)\tilde\bx_h\big\|^2 - 1\Big| + \Big|\big(\Pb_{\Hb^\T}-\Pb_{\tilde\Hb^\T}\big)\tilde \bx_h\Big|\\\le &\,\big\|\Pb_{\Hb^\T}-\Pb_{\tilde\Hb^\T}\big\|^2.
        \end{align*}
Suppose for some $C_1$ we have $\|(\Hb\Hb^\T)^{-1}\| , \|\Hb\Hb^\T\|\le  C_1$. Then
\begin{align*}
    \big\|(\tilde\Hb\tilde\Hb^\T)^{-1}\big\| \le&\, \big\|(\tilde\Hb\tilde\Hb^\T)^{-1}-(\Hb\Hb^\T)^{-1}\big\| +C_1\\\le &\, C_1\big\|(\tilde\Hb\tilde\Hb^\T)^{-1}\big\|\big\|\tilde\Hb\tilde\Hb^\T-\Hb\Hb^\T\big\|+C_1\\&\, C_1 + C_1\big\|(\tilde\Hb\tilde\Hb^\T)^{-1}\big\|\big(2\|\Hb\| + \epsilon\big)\epsilon.
\end{align*}
When $\epsilon<\min\{\|\Hb\|/2, (5C_1\|\Hb\|)^{-1}\}$, we have $\|\tilde\Hb\|\le 3\|\Hb\|/2$ and $\big\|(\tilde\Hb\tilde\Hb^\T)^{-1}\big\|\le 2C_1$, $\big\|(\tilde\Hb\tilde\Hb^\T)^{-1}-(\Hb\Hb^\T)^{-1}\big\| \le 6C_1^2\|\Hb\|\epsilon$.
    By simple telescoping we know that 
    \begin{align*}
        \big\|\Pb_{\Hb^\T}-\Pb_{\tilde\Hb^\T}\big\| \le &\,\big\|\Hb\big\|\big\|(\Hb\Hb^\T)^{-1}\big\|\big\|\Hb-\tilde\Hb\big\| + \big\|\tilde\Hb\big\|\big\|(\tilde\Hb\tilde\Hb^\T)^{-1}\big\|\big\|\Hb-\tilde\Hb\big\| \\&\, + \big\|\Hb\big\|\big\|(\Hb\Hb^\T)^{-1}-(\tilde\Hb\tilde\Hb^\T)^{-1}\big\|\big\|\tilde\Hb\big\|\\\le &\, 4C_1\big\|\Hb\big\|\epsilon+\frac{27}{2}C_1\|\Hb\|^4\epsilon.
    \end{align*}
    Therefore
    \begin{equation*}
        \sin\big\langle\bx_h,\tilde\bx_h\rangle = \Big|\{\bx_h^\T\tilde\bx_h\}^2 - \|\bx_h\|^2\|\tilde\bx_h\|^2\Big|^{1/2}\|\bx_h\|^{-1}\|\tilde\bx_h\|^{-1} \le 4C_1\big\|\Hb\big\|\epsilon+\frac{27}{2}C_1\|\Hb\|^4\epsilon,
    \end{equation*}which proved the result.

\subsection{Proof of Lemma~\ref{lemma_cor_linear_equation_hold}}\label{supp_sec_prove_lemma_cor_linear_equation_hold}
Before we prove Lemma~\ref{lemma_cor_linear_equation_hold}, we introduce the following lemma.
\begin{lemma}\it\label{lemma_linear_equation_hold}
    Suppose Assumptions~\ref{assumption: psd covariance} and \ref{assumption:smoothness} hold. For any $\delta>0$ and some $\epsilon_{NJ}\to\infty$ as $N$ and $J\to\infty$, there exists some constant $C$ such that:
    \begin{enumerate}[label = (\roman*)]
        \item For any $j\in[J]$, for any $r,l\in[K]$, with probability at least $1-\exp(-C\epsilon_{NJ})$,
    \begin{equation*}
        \left|\sum_{i=1}^Nf_{il}^{(0)}\be_r^\T\Big\{\sum_{t=1}^Jl_{it}^{\prime\prime}\big(\eta_{it}^*\big)\blambda_t^{(0)}\big(\blambda_t^{(0)}\big)^\T\Big\}^{-1}\big\{\sum_{t=1}^J\blambda_t^{(0)}l_{it}^{\prime}(\eta_{it}^*)\big\}\right|\lesssim\sqrt{N\epsilon_{NJ}/J};
    \end{equation*}
    \item For any $j\in[J]$, for any $r\in[K]$, with probability at least $1 - N^{-\delta}-\exp(-C\epsilon_{NJ})$,
    \begin{align*}
       &\, \left|\sum_{i=1}^Nl_{ij}^{\prime}(\eta_{ij}^*)\be_r^\T\Big\{\sum_{t=1}^Jl_{it}^{\prime\prime}\big(\eta_{it}^*\big)\blambda_t^{(0)}\big(\blambda_t^{(0)}\big)^\T\Big\}^{-1}\big\{\sum_{t=1}^J\blambda_t^{(0)}l_{it}^{\prime}(\eta_{it}^*)\big\}\right|\\&\,\le C_{\delta}\left( \sqrt{N\log N\epsilon_{NJ}/J} + N/J\right),
    \end{align*}
    where $C_{\delta}$ is some constant depending only on $\delta$;
    \item For any $r\in[K]$, with probability at least $1-J^{-\delta} - N^{-\delta}$,
    \begin{align*}
        &\,\max_{1\le j\le J}\left|\sum_{i=1}^Nl_{ij}^{\prime}(\eta_{ij}^*)\be_r^\T\big\{\sum_{t=1}^Jl_{it}^{\prime\prime}\big(\eta_{it}^*\big)\blambda_t^{(0)}\big(\blambda_t^{(0)}\big)^\T\big\}^{-1}\big\{\sum_{t=1}^J\blambda_t^{(0)}l_{it}^{\prime}(\eta_{it}^*)\big\}\right|\\&\,\le  C_{\delta}\left(\sqrt{{N\log {NJ}}/{J}} + {N}/{J}\right),
    \end{align*}where $C_{\delta}$ is some constant depending only on $\delta$;
    \item For any $j\in[J]$, for any $r\in[K]$, with probability at least $1-\exp(-C\epsilon_{NJ})$,
    \begin{equation*}
        \left|\sum_{i=1}^Nl_{ij}^{\prime\prime}(\eta_{ij}^*)f_{ir}^{(0)}\big(\blambda_{j}^{(0)}\big)^\T\Big\{\sum_{t=1}^Jl_{it}^{\prime\prime}\big(\eta_{it}^*\big)\blambda_t^{(0)}\big(\blambda_t^{(0)}\big)^\T\Big\}^{-1}\big\{\sum_{t=1}^J\blambda_t^{(0)}l_{it}^{\prime}(\eta_{it}^*)\big\}\right|\lesssim\sqrt{N\epsilon_{NJ}/J};
    \end{equation*}
    \item For any $r\in[K]$, with probability at least $1-\exp(-C\epsilon_{NJ})$,
    \begin{equation*}
        \max_{1\le j\le J}\left|\sum_{i=1}^Nl_{ij}^{\prime\prime}(\eta_{ij}^*)f_{ir}^{(0)}\big(\blambda_{j}^{(0)}\big)^\T\Big\{\sum_{t=1}^Jl_{it}^{\prime\prime}\big(\eta_{it}^*\big)\blambda_t^{(0)}\big(\blambda_t^{(0)}\big)^\T\Big\}^{-1}\big\{\sum_{t=1}^J\blambda_t^{(0)}l_{it}^{\prime}(\eta_{it}^*)\big\}\right|\lesssim\sqrt{N\epsilon_{NJ}/J};
    \end{equation*}
    
    \item For any $i\in[N]$, for any $r,l\in[K]$, with probability at least $1-\exp(-C\epsilon_{NJ})$,
    \begin{equation*}
        \left|\sum_{j=1}^J\lambda_{jl}^{(0)}\be_r^\T\Big\{\sum_{t=1}^Nl_{tj}^{\prime\prime}\big(\eta_{tj}^*\big)\bbf_t^{(0)}\big(\bbf_t^{(0)}\big)^\T\Big\}^{-1}\big\{\sum_{t=1}^N\bbf_t^{(0)}l_{tj}^{\prime}(\eta_{tj}^*)\big\}\right|\lesssim\sqrt{J\epsilon_{NJ}/N};
    \end{equation*}
    \item For any $i\in[N]$, for any $r\in[K]$, with probability at least $1-J^{-\delta}-\exp(-C\epsilon_{NJ})$,
    \begin{align*}
        &\,\left|\sum_{j=1}^Jl_{ij}^{\prime}(\eta_{ij}^*)\be_r^\T\Big\{\sum_{t=1}^Nl_{tj}^{\prime\prime}\big(\eta_{tj}^*\big)\bbf_t^{(0)}\big(\bbf_t^{(0)}\big)^\T\Big\}^{-1}\big\{\sum_{t=1}^N\bbf_t^{(0)}l_{tj}^{\prime}(\eta_{tj}^*)\big\}\right|\\&\,\le C_{\delta}\left(\sqrt{J\log N\epsilon_{NJ}/N} + J/N\right),
    \end{align*}
    where $C_{\delta}$ is some constant depending only on $\delta$;
    \item For any $r\in[K]$, with probability at least $1-J^{-\delta}-N^{-\delta}$,
    \begin{align*}
        &\,\max_{1\le i\le N}\left|\sum_{j=1}^Jl_{ij}^{\prime}(\eta_{ij}^*)\be_r^\T\big\{\sum_{t=1}^Nl_{tj}^{\prime\prime}\big(\eta_{tj}^*\big)\bbf_t^{(0)}\big(\bbf_t^{(0)}\big)^\T\big\}^{-1}\big\{\sum_{t=1}^N\bbf_t^{(0)}l_{tj}^{\prime}(\eta_{tj}^*)\big\}\right|\\&\,\le C_{\delta}\left(\sqrt{\frac{J\log (NJ)}{N}} + \frac{J}{N}\right),
    \end{align*}where $C_{\delta}$ is some constant depending only on $\delta$;
    \item For any $i\in[N]$, for any $r\in[K]$, with probability at least $1-\exp(-C\epsilon_{NJ})$,
    \begin{equation*}
        \left|\sum_{j=1}^Jl_{ij}^{\prime\prime}(\eta_{ij}^*)\lambda_{jr}^{(0)}\big(\bbf_{i}^{(0)}\big)^\T\Big\{\sum_{t=1}^Nl_{tj}^{\prime\prime}\big(\eta_{tj}^*\big)\bbf_t^{(0)}\big(\bbf_t^{(0)}\big)^\T\Big\}^{-1} \big\{\sum_{t=1}^N\bbf_t^{(0)}l_{tj}^{\prime}(\eta_{tj}^*)\big\}\right|\lesssim\sqrt{J\epsilon_{NJ}/N}.
    \end{equation*}
    \item For any $r\in[K]$, with probability at least $1-\exp(-C\epsilon_{NJ})$,
    \begin{equation*}
        \max_{1\le i\le N}\left|\sum_{j=1}^Jl_{ij}^{\prime\prime}(\eta_{ij}^*)\lambda_{jr}^{(0)}\big(\bbf_{i}^{(0)}\big)^\T\Big\{\sum_{t=1}^Nl_{tj}^{\prime\prime}\big(\eta_{tj}^*\big)\bbf_t^{(0)}\big(\bbf_t^{(0)}\big)^\T\Big\}^{-1} \big\{\sum_{t=1}^N\bbf_t^{(0)}l_{tj}^{\prime}(\eta_{tj}^*)\big\}\right|\lesssim\sqrt{J\epsilon_{NJ}/N}.
    \end{equation*}
    \end{enumerate}
\end{lemma}
\begin{proof}
    It suffices to prove (i)--(v), and (vi)--(x) follow by symmetry. First by $-l_{ij}^{\prime\prime}(\eta_{ij}^*)\le b_U$, we have
    \begin{equation*}
        \lambda_{\max}\big\{-J^{-1}\sum_{t=1}^Jl_{it}^{\prime\prime}\big(\eta_{it}^*\big)\blambda_t^{(0)}\big(\blambda_t^{(0)}\big)^\T\big\}\le b_U\lambda_{\max}\big\{J^{-1}\sum_{t=1}^J\blambda_t^{(0)}\big(\blambda_t^{(0)}\big)^\T\big\}\le b_U/2 \lambda_{\max}\big\{\bSigma_{\lambda}\big\},
    \end{equation*}
    for large $N,J$, which implies that, for any $t\in[J]$,
    \begin{equation*}
        \left|f_{il}^{(0)}\be_r^\T\Big\{\sum_{t=1}^Jl_{it}^{\prime\prime}\big(\eta_{it}^*\big)\blambda_t^{(0)}\big(\blambda_t^{(0)}\big)^\T\Big\}^{-1}\blambda_t^{(0)}\right|\lesssim J^{-1}.
    \end{equation*}
    Then by (iii) of Lemma~\ref{lemma_concentration}, with probability at least $1-\exp(-C\epsilon_{NJ})$,
    \begin{equation*}
    \left|\sum_{i=1}^Nf_{il}^{(0)}\be_r^\T\Big\{\sum_{t=1}^Jl_{it}^{\prime\prime}\big(\eta_{it}^*\big)\blambda_t^{(0)}\big(\blambda_t^{(0)}\big)^\T\Big\}^{-1}\big\{\sum_{t=1}^J\blambda_t^{(0)}l_{it}^{\prime}(\eta_{it}^*)\big\}\right|\lesssim\sqrt{N\epsilon_{NJ}/J}.
    \end{equation*}
    This proves (i).
    For (ii), we have
    \begin{align}
        &\,\left|\sum_{i=1}^Nl_{ij}^{\prime}(\eta_{ij}^*)\be_r^\T\Big\{\sum_{t=1}^Jl_{it}^{\prime\prime}\big(\eta_{it}^*\big)\blambda_t^{(0)}\big(\blambda_t^{(0)}\big)^\T\Big\}^{-1}\big\{\sum_{t=1}^J\blambda_t^{(0)}l_{it}^{\prime}(\eta_{it}^*)\big\}\right|\\\le &\,\left|\sum_{i=1}^N\be_r^\T\Big\{\sum_{t=1}^Jl_{it}^{\prime\prime}\big(\eta_{it}^*\big)\sum_{t=1,t\neq j}^J\blambda_t^{(0)}\big(\blambda_t^{(0)}\big)^\T\Big\}^{-1}\Big\{\sum_{t=1,t\neq j}^J\blambda_t^{(0)}l_{it}^{\prime}(\eta_{it}^*)\Big\}l_{ij}^{\prime}(\eta_{ij}^*)\right|\label{eq_bound_residual1}\\&\,+\left|\sum_{i=1}^N\big[l_{ij}^{\prime}(\eta_{ij}^*)\big]^2\be_r^\T\Big\{\sum_{t=1}^Jl_{it}^{\prime\prime}\big(\eta_{it}^*\big)\blambda_t^{(0)}\big(\blambda_t^{(0)}\big)^\T\Big\}^{-1}\blambda_j^{(0)}\right|.\label{eq_bound_residual2}
    \end{align}
    For the first term \eqref{eq_bound_residual1}, denote $a_{il} = \be_r^\T\Big\{J^{-1}\sum_{t=1}^Jl_{it}^{\prime\prime}\big(\eta_{it}^*\big)\sum_{t=1,t\neq j}^J\blambda_t^{(0)}\big(\blambda_t^{(0)}\big)^\T\Big\}^{-1}\be_l$ with $\max_{i,l}|a_{il}|\le A$. By union bound and the Bernstein type bound as in the proof of Lemma~\ref{lemma_concentration}
    \begin{align*}
        &\,P\left\{\left|\sum_{i=1}^Na_{il}\Big\{\sum_{t\neq j}\lambda_{tl}^{(0)}l_{it}^{\prime}(\eta_{it}^*)\Big\}l_{ij}^{\prime}(\eta_{ij}^*)\right|\ge t\right\}\\\le&\, P\left\{\left|\sum_{i=1}^Na_{il}\Big\{\sum_{t\neq j}\lambda_{tl}^{(0)}l_{it}^{\prime}(\eta_{it}^*)\Big\}l_{ij}^{\prime}(\eta_{ij}^*)\right|\ge t\text{ and }\Big|\sum_{t\neq j}\lambda_{tl}^{(0)}l_{it}^{\prime}(\eta_{it}^*)\Big|< \nu\right\} \\&\,+NP\left\{\Big|\sum_{t\neq j}\lambda_{tl}^{(0)}l_{it}^{\prime}(\eta_{it}^*)\Big|\ge \nu\right\}\\\le &\, 2 \exp \left[ -c \min \left( \frac{t^2}{(MA\nu)^2N}, \frac{t}{MA\nu N} \right) \right]\\&\,+2N\exp \left[ -c \min \left( \frac{\nu^2}{(MD)^2(J-1)}, \frac{\nu}{MD(J-1)} \right) \right],
    \end{align*}
    for some constant $c$. Take $\nu = \sqrt{J\log N}$ and $t = MA\nu \sqrt{N\epsilon_{NJ}}$ and we know that the first term can be bounded by $C_{\delta}\sqrt{NJ\epsilon_{NJ}\log N}/J$ with probability at least $1-\exp(-C\epsilon_{NJ}) - N^{-\delta}$ for any $\delta>0$ and some constant $C_{\delta}$ only depending on $\delta$.
    The second term \eqref{eq_bound_residual2} can be bounded by $C_{\delta}N/J$ with probability at least $1-N^{-\delta}$ analogous to (ii) of Lemma~\ref{lemma_concentration_l_mat}, regardless of $j$. Finally (ii) is proved by union bound. Similarly, with probability at least $1-J^{-\delta} - N^{-\delta}$,
    \begin{align*}
        \max_{1\le j\le J}\left|\sum_{i=1}^Nl_{ij}^{\prime}(\eta_{ij}^*)\be_r^\T\Big\{\sum_{t=1}^Jl_{it}^{\prime\prime}\big(\eta_{it}^*\big)\blambda_t^{(0)}\big(\blambda_t^{(0)}\big)^\T\Big\}^{-1}\big\{\sum_{t=1}^J\blambda_t^{(0)}l_{it}^{\prime}(\eta_{it}^*)\big\}\right|\le C_{\delta}\big( \sqrt{N\log {(NJ)}/J} + N/J\big),
    \end{align*} where $C_{\delta}$ is some constant only depending $\delta$. This proves (iii).
    
    For (vi), note that for any $t\in[J]$,
    \begin{equation}
        \left|l_{ij}^{\prime\prime}(\eta_{ij}^*)f_{ir}^{(0)}\big(\blambda_{j}^{(0)}\big)^\T\Big\{\sum_{t=1}^Jl_{it}^{\prime\prime}\big(\eta_{it}^*\big)\blambda_t^{(0)}\big(\blambda_t^{(0)}\big)^\T\Big\}^{-1}\blambda_t^{(0)}\right|\lesssim J^{-1}.\label{eq_bound_residual3}
    \end{equation}
    Then by applying Lemma~\ref{lemma_concentration} we have
    \begin{equation*}
        \left|\sum_{i=1}^Nl_{ij}^{\prime\prime}(\eta_{ij}^*)\bbf_i^{(0)}\big(\blambda_{j}^{(0)}\big)^\T\Big\{\sum_{t=1}^Jl_{it}^{\prime\prime}\big(\eta_{it}^*\big)\blambda_t^{(0)}\big(\blambda_t^{(0)}\big)^\T\Big\}^{-1}\big\{\sum_{t=1}^Jl_{it}^{\prime}(\eta_{it}^*)\blambda_t^{(0)}\big\}\right|\lesssim \sqrt{N\epsilon_{NJ}/J},
    \end{equation*}
    with probability at least $1-\exp(-C\epsilon_{NJ})$. Since \eqref{eq_bound_residual3} holds regardless of $j$, with probability at least $1-\exp(-C\epsilon_{NJ})$
    \begin{equation*}
        \max_{1\le j\le J}\left|\sum_{i=1}^Nl_{ij}^{\prime\prime}(\eta_{ij}^*)\bbf_i^{(0)}\big(\blambda_{j}^{(0)}\big)^\T\Big\{\sum_{t=1}^Jl_{it}^{\prime\prime}\big(\eta_{it}^*\big)\blambda_t^{(0)}\big(\blambda_t^{(0)}\big)^\T\Big\}\big\{\sum_{t=1}^Jl_{it}^{\prime}(\eta_{it}^*)\blambda_t^{(0)}\big\}\right|\lesssim \sqrt{N\epsilon_{NJ}/J},
    \end{equation*}which proves (v).\end{proof}

Now we prove Lemma~\ref{lemma_cor_linear_equation_hold}.

    When $j\in[J]$, by the expression for $\Hb(\btheta^{(0)})$ as in Corollary~\ref{cor_lemma_hessian_property_bound_l_prime}, we have
\begin{align}
    &\,\Big[\Db_m^{-1}\Hb\big(\btheta^{(0)}\big)\big(\bdelta^\ddagger_m-\bdelta_m\big)-\Db_m^{-1}\bS_L(\btheta^{(0)})\Big]_{[(j-1)K+1:jK]}\nonumber \\ =&\, N^{-1}\big[\Hb\big(\btheta^{(0)}\big)\big]_{[(j-1)K+1:jK,]}\big(\bdelta^\ddagger_m-\bdelta_m\big) + N^{-1}\sum_{i=1}^Nl_{ij}^{\prime}(\eta_{ij}^*)\bbf_i^{(0)}\nonumber\\=&\,-N^{-1}\big\{\sum_{i=1}^Nl_{ij}^{\prime\prime}(\eta_{ij}^*)\bbf_i^{(0)}(\bbf_i^{(0)})^\T\big\}\big\{\sum_{i=1}^Nl_{ij}^{\prime\prime}(\eta_{ij}^*)\bbf_i^{(0)}(\bbf_i^{(0)})^\T\big\}^{-1} \sum_{i=1}^Nl_{ij}^{\prime}(\eta_{ij}^*)\bbf_i^{(0)}\nonumber\\&\,-N^{-1}\Big[\sum_{t=1}^N\big\{l_{tj}^{\prime\prime}(\eta_{tj}^*)\bbf_t^{(0)}(\blambda_j^{(0)})^\T - l_{tj}^{\prime}(\eta_{tj}^*)\Ib_K\big\}\big\{\sum_{s=1}^Jl_{ts}^{\prime\prime}(\eta_{ts}^*)\blambda_s^{(0)}(\blambda_s^{(0)})^\T\big\}^{-1} \sum_{s=1}^Jl_{ts}^{\prime}(\eta_{ts}^*)\blambda_s^{(0)}\nonumber\\&\,+J\sum_{l=1}^K\sum_{h>l}^K\big[\bv_{\lambda,lh}(\btheta^{(0)})\big]_{[(j-1)K+1:jK]}\big\{\bv_{\lambda,lh}(\btheta^{(0)})\big\}^\T\big\{\bdelta_{m}^{\ddagger}-\bdelta_{m}\big\}_{[1:JK]}
    \nonumber\\  &\,+ N^{-1}\sum_{i=1}^Nl_{ij}^{\prime}(\eta_{ij}^*)\bbf_i^{(0)}\nonumber\\=&\,-N^{-1}\Big[\sum_{t=1}^Nl_{tj}^{\prime\prime}(\eta_{tj}^*)\bbf_t^{(0)}(\blambda_j^{(0)})^\T \big\{\sum_{s=1}^Jl_{ts}^{\prime\prime}(\eta_{ts}^*)\blambda_s^{(0)}(\blambda_s^{(0)})^\T\big\}^{-1} \sum_{s=1}^Jl_{ts}^{\prime}(\eta_{ts}^*)\blambda_s^{(0)}\label{eq_bound_res_term_1}\\&\,+N^{-1}\Big[\sum_{t=1}^N l_{tj}^{\prime}(\eta_{tj}^*)\Ib_K\big\{\sum_{s=1}^Jl_{ts}^{\prime\prime}(\eta_{ts}^*)\blambda_s^{(0)}(\blambda_s^{(0)})^\T\big\}^{-1} \sum_{s=1}^Jl_{ts}^{\prime}(\eta_{ts}^*)\blambda_s^{(0)}\label{eq_bound_res_term_2}\\&\, +\sum_{l=1}^K\sum_{h>l}^K(\lambda_{jl}^{(0)}\be_h + \lambda_{jh}^{(0)}\be_l)^\T\nonumber\\&\quad\times J^{-1}\sum_{s=1}^J(\lambda_{sl}^{(0)}\be_h + \lambda_{sh}^{(0)}\be_l)^\T \big\{\sum_{i=1}^Nl_{is}^{\prime\prime}(\eta_{is}^*)\bbf_i^{(0)}(\bbf_i^{(0)})^\T\big\}^{-1}\sum_{i=1}^Nl_{is}^{\prime}(\eta_{is}^*)\bbf_i^{(0)}\label{eq_bound_res_term_3}.
\end{align}
By (iv) of Lemma~\ref{lemma_linear_equation_hold}, $|\eqref{eq_bound_res_term_1}|$ can be bounded by $O_p\big((NJ)^{-1/2}\big)$; by (ii) of Lemma~\ref{lemma_linear_equation_hold}, $|\eqref{eq_bound_res_term_2}|$ can be bounded by $O_p\big((NJ)^{-1/2} + J^{-1}\big)$; by (vi) of Lemma~\ref{lemma_linear_equation_hold}, $|\eqref{eq_bound_res_term_3}|$ can be bounded by $O_p\big((NJ)^{-1/2}\big)$. Therefore the summation can be bounded by $O_p\big(1/\sqrt{J(N\wedge J)})\le O_p\big(1/(N\wedge J))$. Similarly we can derive the bound for $i\in\{J+1,\cdots,N+J\}$ using (i), (vii) and (ix) of Lemma~\ref{lemma_linear_equation_hold}. The bound for $\Db_m^{-1/2}\Hb(\btheta^{(0)})(\delta_m^{\ddagger} - \bdelta_m) - \Db_m^{-1/2}\bS_L(\btheta^{(0)})$ can be analogously derived by noting that the row-wise bounds hold uniformly.

\subsection{Proof of Lemma~\ref{lemma_projection_non_diminish}}\label{prove_lemma_projection_non_diminish}

By $\lambda_{\min}\big(\Ab+\Vb\Vb^\T\big)>\gamma>0$, we know that for any $\bx\in \mathrm{Col}(\Vb)$,
    \begin{equation*}
        \bx^\T\Ab\bx \ge \gamma \|\bx\|^2.
    \end{equation*}
    For general $\bx\in\RR^n$, decompose it into $\bx = \by + \bz$ where $\by\in\mathrm{Col}(\Vb)$ and $\bz=\bx - \by$. Such decomposition is unique as $\by = \Vb(\Vb^\T\Vb)^{-1}\Vb^\T\bx$. Then
    \begin{equation*}
        \bx^\T\big(\Ab + c\Ub\Ub^\T\big)\bx\ge \gamma\big\|\bz\big\|^2 + c\bx^\T\Ub\Ub^\T\bx
    \end{equation*}
 Write $\Ub = \Ub_y + \Ub_z$ such that $\Ub_y = \Vb(\Vb^\T\Vb)^{-1}\Vb^\T\Ub$ and $\Ub_z = \Ub - \Ub_y$. Then $\Ub^\T\bx = \Ub^\T\by + \Ub^\T\bz = \Ub^\T\Vb(\Vb^\T\Vb)^{-1}\Vb^\T\by + \Ub^\T(\Ib_{n} - \Vb(\Vb^\T\Vb)^{-1}\Vb^\T)\bz= \Ub_y^\T\by + \Ub_z^\T\bz$. By $\sigma_{\min}\big((\Vb^\T\Vb)^{-1}\Vb^\T\Ub\big)\ge \sigma_2$, we know \begin{equation*}\by^\T\Ub_y\Ub_y^\T\by =(\by^\T\Vb)(\Vb^\T\Vb)^{-1}\Vb^\T\Ub\Ub^\T\Vb(\Vb^\T\Vb)^{-1}(\Vb^\T\by)\ge\sigma_{2}^2\by^\T\Vb\Vb^\T\by\ge \sigma_0^2\sigma_2^2\|\by\|^2.\end{equation*} Next for $\Ub_z^\T\bz = \Ub (\Ib_n - \Vb(\Vb^\T\Vb)^{-1}\Vb^\T)\bz$, we have
 \begin{equation*}
     \bz^\T\Ub_z\Ub_z^\T\bz\le  \sigma_1^2 \big\|\bz\big\|^2
 \end{equation*}
 Then we further have
 \begin{equation*}
        \bx^\T\big(\Ab + c\Ub\Ub^\T\big)\bx\ge \gamma\big\|\bz\big\|^2 + c\by^\T\Ub_y\Ub_y^\T\by - c\bz^\T\Ub_z\Ub_z^\T\bz\ge ( \gamma-c\sigma_1^2)\big\|\bz\big\|^2  + c\sigma_0^2\sigma_2^2\|\by\|^2,
    \end{equation*}
    which finishes the proof.

\section{Proofs of Theoretical Results of the Generalized Latent Factor Model with Intercept}\label{se:integfm proof}

This section is organized as follows. Section~\ref{sec_zemo} details the expressions for $\bSigma_{[N],-j}^{(S)}$s and $\bSigma_{-i,[J]}^{(S)}$s omitted in the main text and Section~\ref{sec_sketch_proof} outlines the proof sketches for the theoretical properties.

We slightly abuse notation in this section for convenience. These notations are used only within this section to sketch the theoretical results for the generalized latent factor model with intercepts. We write for $i\in[N]$ and $j\in[J]$ $\blambda_j^c = (\lambda_{j0},1,\lambda_{j1},\cdots,\lambda_{jK})^\T$ and $\bbf_i^c = (1,f_{i0},f_{i1},\cdots,f_{iK})^\T$. We define $\blambda_{-j} = (\lambda_{j0},\lambda_{j1},\cdots,\lambda_{jK})^\T$ and $\bbf_{-i} = (f_{i0},f_{i1},\cdots,f_{iK})^\T$, and the model parameter as $\btheta = (\blambda_{-1}^\T,\cdots,\blambda_{-J}^{\T}, \bbf_{-1}^\T,\bbf_{-N}^\T)^\T$. The true parameters are expressed as $\blambda_j^{c*} = (\lambda_{j0}^*,1,\lambda_{j1}^*,\cdots,\lambda_{jK}^*)^\T$, $\bbf_i^{c*} = (1,f_{i0}^*,f_{i1}^*,\cdots,f_{iK}^*)^\T$, $\blambda_{-j}^* = (\lambda_{j0}^*,\lambda_{j1}^*,\cdots,\lambda_{jK}^*)^\T$ and $\bbf_{-i}^* = (f_{i0}^*,f_{i1}^*,\cdots,f_{iK}^*)^\T$. The product $\eta_{ij}^*$ is then expressed as $f_{i0}^* + \lambda_{j0}^* + (\blambda_j^*)^\T\bbf_i^* = (\blambda_j^{c*})^\T\bbf_i^{c*}$ for any $i\in[N]$ and $j\in[J]$. Define the compact set $\cB(D) = \{\btheta: \|\btheta\|_{\infty}\le D\}$ similarly. Write $\eta_{-ij} = \sum_{k=1}^K\lambda_{jk}f_{ik} = \eta_{ij} - \lambda_{j0}-f_{i0}$.

\subsection[Detailed Expressions for $\bSigma_{[N],-j}^{(S)}$ and $\bSigma_{-i,[J]}^{(S)}$]{Detailed Expressions for $\bSigma_{[N],-j}^{(S)}$ and $\bSigma_{-i,[J]}^{(S)}$}\label{sec_zemo}
For each $j\in[J]$, $\bSigma_{[N],-j}^{(S)}$ is given as 
\begin{equation}
    \bSigma_{[N],-j}^{(S)} =\begin{pmatrix}
        1&\zero_K^\T\\\zero_K&\Db_{(j)}
    \end{pmatrix}\left\{ \bPsi_{[N],j}^{*} + \bOmega_{[N],-j}^{(S)} + \big((\blambda_{-j}^{(S)})^\T\otimes \Ib_{K+1}\big)\Xb_{-\lambda}^{(S)} \big(\blambda_{-j}^{(S)}\otimes \Ib_{K+1}\big)\right\}\begin{pmatrix}
        1&\zero_K^\T\\\zero_K&\Db_{(j)}^\T
    \end{pmatrix},  \label{eq_expression_int_gfm_cov_lambda}
\end{equation}
where $\bPsi_{[N],j}^{*}$ is given as
\begin{equation*}
    \bPsi_{[N],j}^{*} = \left\{-\sum_{i=1}^Nl_{ij}^{\prime\prime}(\eta_{ij}^*)\bbf_i^{c*}(\bbf_i^{c*})^\T\right\}^{-1}\left\{\sum_{i=1}^N[l_{ij}^{\prime}(\eta_{ij}^*)]^2\bbf_i^{c*}(\bbf_i^{c*})^\T\right\}\left\{-\sum_{i=1}^Nl_{ij}^{\prime\prime}(\eta_{ij}^*)\bbf_i^{c*}(\bbf_i^{c*})^\T\right\}^{-1}, 
\end{equation*} 
$\Xb_{-\lambda}^{(S)}$ is given as
\begin{equation*}
    \Xb_{-\lambda}^{(S)} = \begin{pmatrix}
        0&\zero_K^\T\\\zero_K&\Xb_{\lambda}^{(S)}
    \end{pmatrix},
\end{equation*}
and $ \bOmega_{[N],j}^{(S)}$ is given as
\begin{itemize}
\item when $S= 0 $, $\bOmega_{[N],-j}^{(0)} = \zero$;
    \item when $S=1$, $\bOmega_{[N],-j}^{(1)} = \sum_{r=1}^K\sum_{l=1}^{K}1_{(j= \cA_r(l))} \big\{\bPsi_{[N],j}^*\Xb_{-\lambda,rl}^{(1)} + \big(\Xb_{-\lambda,rl}^{(1)}\big)^\T\bPsi_{[N],j}^*\big\}$;
    \item when $S=2$, $\bOmega_{[N],-j}^{(2)} = 1_{(1\le j\le K)}\big(\Ib_{K+1}^{j+1}-\Ib_{K+1}\big)\bPsi_{[N],j}^* + 1_{(1\le j\le K)}\bPsi_{[N],j}^*\big(\Ib_{K+1}^{j+1}-\Ib_{K+1}\big)$.
    \item when $S=3$, $\bOmega_{[N],-j}^{(3)} = 1_{(1\le j\le K)}\big(\Ib_{K+1}^{j}-\Ib_{K+1}\big)\bPsi_{[N],j}^* + 1_{(1\le j\le K)}\bPsi_{[N],j}^*\big(\Ib_{K+1}^{j}-\Ib_{K+1}\big)$;
    \item when $S=4$, $\bOmega_{[N],-j}^{(4)} = \sum_{r=1}^K\sum_{l=1}^{K-1}1_{(j\in \cA_r(l))}\bPsi_{[N],j}^{*}\Xb_{-\lambda,rl}^{(4)} + (\Xb_{-\lambda,rl}^{(4)})^\T\bPsi_{[N],j}^{*}$;
    \item when $S=5$, $\bOmega_{[N],-j}^{(5)} = \zero$.
\end{itemize}
Here $\Xb_{-\lambda,rl}^{(1)}$ can be obtained by replacing $\bPhi_{[N],r}^{(1)}$ and $\bPhi_{[N],l}^{(1)}$ in $\Xb_{\lambda,rl}^{(1)}$ with $\bPsi_{[N],r}^{(1)}$ and $\bPsi_{[N],l}^{(1)}$; $\Xb_{-\lambda,rl}^{(4)}$ can be obtained by replacing $\bPhi_{[N],r}^{(4)}$ and $\bPhi_{[N],l}^{(4)}$ in $\Xb_{\lambda,rl}^{(4)}$ with $\bPsi_{[N],r}^{(4)}$ and $\bPsi_{[N],l}^{(4)}$.

For $i\in[N]$, $\bSigma_{-i,[J]}^{(S)}$ is given as 
\begin{equation}
    \bSigma_{-i,[J]}^{(S)} = \bPsi_{-i,[J]}^{(S)} + \big( \Ib_{K+1}\otimes (\bbf_{-i}^{(S)})^\T\big)\Xb_{-\lambda}^{(S)} \big(\Ib_{K+1}\otimes \bbf_{-i}^{(S)}\big),  \label{eq_expression_int_gfm_cov_f}
\end{equation}
where \begin{equation*}
    \bPsi_{-i,[J]}^{*}:=\left\{-\sum_{j=1}^Jl_{ij}^{\prime\prime}(\eta_{ij}^{*})\blambda_j^{{c*}}(\blambda_j^{{c*}})^\T\right\}^{-1}\left\{\sum_{j=1}^Jl_{ij}^{\prime}(\eta_{ij}^*)^2\blambda_j^{{c*}}(\blambda_j^{{c*}})^\T\right\}\left\{-\sum_{j=1}^Jl_{ij}^{\prime\prime}(\eta_{ij}^{*})\blambda_j^{{c*}}(\blambda_j^{{c*}})^\T\right\}^{-1}.
\end{equation*}

\subsection{Proof Sketch}\label{sec_sketch_proof}
The proof framework for results in Section~\ref{sec:mainres} is readily adaptable to configurations that incorporate intercepts, albeit with certain modifications. Here, we focus solely on models with both intercepts. The proofs for models with a single intercept can be similarly derived by applying the same approach. Details of these proofs are omitted for brevity.

We start with proving the consistency results under IC$^{(0)}$ and ICC. Step 1 in Section~\ref{sec:supp_prove_consistency_ic1} can be extended as follows.
\begin{proposition}\label{lemma_consistency_int}\it
 Under Assumptions~\ref{assump:psd covariance int}--\ref{assumption:smoothness int}, we have for $S=0$ that for any $\delta>0$, there exists some constant $C$ such that
\begin{equation*}
    P\big(\cE^{\prime}_1(C)\big)\ge 1-(NJ)^{-\delta} - (N+J)^{-\delta},
\end{equation*}
where $\cE^{\prime}_1(C) = \{{J}^{-1}\sum_{j=1}^J\big\|\hat\blambda_{-j}^{(0)}-\blambda_{-j}^{(0)}\big\|^2\le C(N\wedge J)^{-1}\text{ and }{N}^{-1}\sum_{i=1}^N\big\|\hat\bbf_{-i}^{(0)}-\bbf_{-i}^{(0)}\big\|^2\le C(N\wedge J)^{-1}\}$.
   \end{proposition}

   Similar to the proof of average consistency for $\hat\btheta^{(0)}$, we can show that, with probability at least $1-(NJ)^{-\delta} - (N+J)^{-\delta}$, we have
    \begin{equation*}
        \Big(\sum_{i=1}^N\sum_{j=1}^J(\hat\eta_{ij}-\eta_{ij}^*)^2\Big)^{1/2}\le C_{\delta}\sqrt{N\vee J},
    \end{equation*}
    where $C_{\delta}$ is some constant depending only on $\delta$. Still, we omit this dependence on $\delta$ for simplicity.
    Since both the true parameters and the estimation satisfy ICC, we know $[N^{-1}$ $\sum_{i=1}^N\bbf_{-i}^{(0)}]_{[2:K+1]} =[N^{-1}\sum_{i=1}^N\hat\bbf_{-i}^{(0)}]_{[2:K+1]} = \zero$ and $J^{-1}\sum_{j=1}^J\blambda_{-j}^{(0)} = J^{-1}\sum_{j=1}^J\hat\blambda_{-j}^{(0)} = \zero$, which implies that 
    \begin{equation*}
        \sum_{i=1}^N\hat\eta_{-ij} = \sum_{i=1}^N \eta_{-ij}^* =0,\;\sum_{j=1}^J\hat\eta_{-ij} = \sum_{j=1}^J \eta_{-ij}^* =0,\; \sum_{j=1}^J\hat\lambda_{j0}^{(0)}=\sum_{j=1}^J\lambda_{j0}^{(0)} = 0.
    \end{equation*}
    Then
    \begin{align*}
        &\,\sum_{i=1}^N\sum_{j=1}^J\big(\hat\eta_{ij}-\eta_{ij}^*\big)^2 = \sum_{i=1}^N\sum_{j=1}^J\big(\hat\eta_{-ij} - \eta_{-ij}^* + \hat\lambda_{j0}^{(0)} - \lambda_{j0}^{(0)} + \hat f_{i0}^{(0)} - f_{i0}^{(0)}\big)^2\\=&\,\sum_{i=1}^N\sum_{j=1}^J\big(\hat\eta_{-ij} - \eta_{-ij}^*\big)^2 + \sum_{i=1}^N\sum_{j=1}^J\big(\hat\lambda_{j0}^{(0)} - \lambda_{j0}^{(0)}\big)^2 + \sum_{i=1}^N\sum_{j=1}^J\big(\hat f_{i0}^{(0)} - f_{i0}^{(0)}\big)^2 \\&\,+\sum_{j=1}^J\Big\{\sum_{i=1}^N\big(\hat\eta_{-ij} - \eta_{-ij}^*\big)\Big\}\big(\hat\lambda_{j0}^{(0)} - \lambda_{j0}^{(0)}\big)+\sum_{i=1}^N\Big\{\sum_{j=1}^J\big(\hat\eta_{-ij} - \eta_{-ij}^*\big)\Big\}\big(\hat f_{i0}^{(0)} - f_{i0}^{(0)}\big)\\&\,+\Big\{\sum_{j=1}^J\big(\hat\lambda_{j0}^{(0)} - \lambda_{j0}^{(0)}\big)\Big\}\Big\{\sum_{i=1}^N\big(\hat f_{i0}^{(0)} - f_{i0}^{(0)}\big)\Big\}\\=&\,\sum_{i=1}^N\sum_{j=1}^J\big(\hat\eta_{-ij} - \eta_{-ij}^*\big)^2 + \sum_{i=1}^N\sum_{j=1}^J\big(\hat\lambda_{j0}^{(0)} - \lambda_{j0}^{(0)}\big)^2 + \sum_{i=1}^N\sum_{j=1}^J\big(\hat f_{i0}^{(0)} - f_{i0}^{(0)}\big)^2
    \end{align*}
    Therefore we conclude that with probability at least $1-(NJ)^{-\delta} - (N+J)^{-\delta}$, 
    \begin{equation*}
        \frac{1}{N}\sum_{i=1}^N\big(\hat f_{i0}^{(0)} - f_{i0}^{(0)}\big)^2 \lesssim \frac{1}{N\wedge J},\quad \frac{1}{J}\sum_{j=1}^J\big(\hat \lambda_{j0}^{(0)} - \lambda_{j0}^{(0)}\big)^2\lesssim \frac{1}{N\wedge J},
    \end{equation*}
    and $\sum_{i=1}^N\sum_{j=1}^J\big(\hat\eta_{-ij} - \eta_{-ij}^*\big)^2 \lesssim N\vee J$. Following the similar application of the result in~\cite{yu2015useful} in Section~\ref{sec:supp_prove_consistency_ic0} for $\hat\btheta^{(0)}$, we finish the proof.

   Abusing the notation of $\be_r$ to denote the indicator of dimension $K+1$, we define the augmentation to the Hessian matrix as
\begin{equation*}
    \Hb_P(\btheta)=NJ\begin{pmatrix}
        \sum_{l=0}^K\sum_{h> l}^K\bv_{\lambda,lh}(\btheta)\{\bv_{\lambda,lh}(\btheta)\}^\T\\&\sum_{l=0}^K\sum_{h\le l}^K\bv_{f,lh}(\btheta)\{\bv_{f,lh}(\btheta)\}^\T
    \end{pmatrix}
\end{equation*}
with
$$\bv_{f,rr}(\btheta)=N^{-1}\left(f_{1r}^{(0)}\be_r^\T,\cdots,f_{Nr}^{(0)}\be_r^\T\right)^\T,\;r=1,\cdots,K;$$
$$\bv_{f,0r}(\btheta)=N^{-1}\left(\be_r^\T,\cdots,\be_r^\T\right)^\T,\;r=1,\cdots,K;$$
$$\bv_{f,lh}(\btheta)=\left(N^{-1}\big(f_{1l}^{(0)}\be_h+f_{1h}^{(0)}\be_l\big)^\T,\cdots,N^{-1}\big(f_{Nl}^{(0)}\be_h+f_{Nh}^{(0)}\be_l\big)^\T\right)^\T,1\le l <h \le K;$$
$$\bv_{\lambda,r0}(\btheta)=\left(J^{-1}\be_r^\T,\cdots,J^{-1}\be_r^\T\right)^\T,\;r=0,\cdots,K;$$
$$\bv_{\lambda,lh}(\btheta)=J^{-1}\left(\lambda_{1l}^{(0)}\be_h+\lambda_{1h}^{(0)}\be_l\big)^\T,\cdots,J^{-1}\big(\lambda_{Jl}^{(0)}\be_h+\lambda_{Jh}^{(0)}\be_l\big)^\T\right)^\T,1\le h \le l\le K.$$
With the augmentation to the Hessian matrix $\Hb_L(\btheta)=-\partial^2_{\btheta\btheta}L(\btheta)$. Define $J_{K+1} = \{1,2,\cdots,JK+K\}$ and $N_{K+1} = \{1,2,\cdots,NK+K\}$. We have the following result.
\begin{lemma}\label{lemma_hessian_property_int}
\it
Fix a $\tilde\btheta\in\cB(D)$ that satisfy IC$^{(0)}$ and ICC with $N^{-1}\sum_{i=1}^N\|\tilde\bbf_{-i}-\bbf_{-i}^{(0)}\|^2+J^{-1}\sum_{j=1}^J\|\tilde\blambda_{-j}-\blambda_{-j}^{(0)}\|^2\le c_{NJ}$ for some $c_{NJ}\to0$ as $N,J\to\infty$. Let $$\tilde\Hb(\tilde\btheta)=\int_0^1\Big\{\Hb_L\big(\btheta^{(0)} + s(\tilde\btheta-\btheta^{(0)})\big)+\Hb_P\big(\btheta^{(0)} + s(\tilde\btheta-\btheta^{(0)})\big)\Big\}ds,$$
with $\Hb_L=-\partial^2_{\btheta\btheta}L(\btheta)$ and $\Hb_P(\btheta)$ given as above. Here $L(\btheta) = \sum_{i=1}^N\sum_{j=1}^Jl_{ij}(\blambda_j^\T\bbf_i +\lambda_{j0}+ f_{i0})$. Write \begin{align*}\Hb_{L\lambda\lambda}(\tilde\btheta) =[\Hb_L(\tilde\btheta)]_{[J_{K+1}, J_{K+1}]} ,\quad \Hb_{Lff}(\tilde\btheta) =[\Hb_L(\tilde\btheta)]_{[J(K+1)+N_{K+1},J(K+1)+N_{K+1}]} .\end{align*}Define a similar scaling matrix as
\begin{equation*}
    \Db_{m}=\begin{pmatrix}
            N\Ib_{JK+J}\\&J\Ib_{NK+N}
        \end{pmatrix}.
\end{equation*}
    Under Assumptions~\ref{assump:psd covariance int} and \ref{assumption:smoothness int}, we have
    \begin{enumerate}[label = (\roman*)]
   \item for any $\delta>0$, there exists some $\gamma>0$ such that $P\Big\{\lambda_{\min}(\Db_{m}^{-1/2}\tilde\Hb(\tilde\btheta)\Db_{m}^{-1/2})<\gamma\Big\}\le (N+J)^{-\delta} + (NJ)^{-\delta}$;
    \item  With probability at least $1 - 2(NJ)^{-\delta} - (N+J)^{-\delta}$, $\big\|[\tilde\Hb^{-1}\Db_m]_{[J_{K+1},]}\big\|_{\infty}\le C_{\delta}\log J$;
    \item  With probability at least $1 - 2(NJ)^{-\delta} - (N+J)^{-\delta}$, $\big\|[\tilde\Hb^{-1}\Db_m]_{[J(K+1)+N_{K+1},]}\big\|_{\infty}\le C_{\delta}\log N$.
    \item With probability at least $1-N^{-\delta}-3J^{-\delta} - 2(NJ)^{-\delta} - (N+J)^{-\delta}$,\\
    $\big\|\big[\{\tilde\Hb(\tilde\btheta)\}^{-1}\big]_{[J_{K+1},J_{K+1}]}\big\|\le C_{\delta} N^{-1}$; $\big\|\big[\{\tilde\Hb(\tilde\btheta)\}^{-1}\big]_{[J_{K+1},J(K+1)+N_{K+1}]}\big\|\le C_{\delta} (NJ)^{-1/2}$;
    \\ $\big\|\big[\tilde\Hb^{-1}\bS_{L}^*\big]_{[J_{K+1},]}\big\|\le C_{\delta}\left( \sqrt{J/N}+\sqrt{c_{NJ}}+\sqrt{\log (NJ)/N} + 1/\sqrt{J}\right)$;
    \item With probability at least $1-\exp(-C\epsilon_N)-N^{-\delta}-4J^{-\delta} - 2(NJ)^{-\delta} - (N+J)^{-\delta}$\\
    $\big\|\big[\tilde\Hb^{-1}\bS_{L}^*\big]_{[(j-1)(K+1)+1:j(K+1),]}\big\|\le C_{\delta}\big( \sqrt{\epsilon_N/N}+\sqrt{c_{NJ}/J}+\sqrt{\log (NJ)/(NJ)} + 1/{J}\big)$.
    \item With probability at least $1-3N^{-\delta}-J^{-\delta} - 2(NJ)^{-\delta} - (N+J)^{-\delta}$,\\
    $\big\|\big[\{\tilde\Hb(\tilde\btheta)\}^{-1}\big]_{[J(K+1)+N_{K+1},J_{K+1}]}\big\|\le C_{\delta}(NJ)^{-1/2}$; $\big\|\big[\{\tilde\Hb(\tilde\btheta)\}^{-1}\big]_{[J(K+1)+N_{K+1},J(K+1)+N_{K+1}]}\big\|\le C_{\delta}J^{-1}$;\\
    $\big\|\big[\tilde\Hb^{-1}\bS_{L}^*\big]_{[J(K+1)+N_{K+1},]}\Big\|\le C_{\delta}\left( \sqrt{N/J}+\sqrt{c_{NJ}}+\sqrt{\log (NJ)/J} + 1/\sqrt{N}\right)$. 
    \item With probability at least $1-\exp(-C\epsilon_J)-4N^{-\delta}-J^{-\delta} - 2(NJ)^{-\delta} - (N+J)^{-\delta}$
    \begin{align*}\big\|\big[\tilde\Hb^{-1}\bS_{L}^*\big]_{[J(K+1)+(i-1)(K+1)+1:J(K+1)+i(K+1),]}&\Big\|\le \\C_{\delta}\Big(\sqrt{\epsilon_J/J}&+\sqrt{c_{NJ}/N}+\sqrt{\log (NJ)/(NJ)} + 1/{N}\Big);\end{align*}
    \item With probability at least $1 - N^{-\delta} - (NJ)^{-\delta} - (N+J)^{-\delta}$, for any $j\in[J]$, \begin{align*}\Big\|\big[\Db_m^{1/2}\{\tilde\Hb(\tilde\btheta)\}^{-1}&\Db_m^{1/2}\big]_{[(j-1)(K+1)+1:j(K+1),]}\\& - N\big[\{\Hb_{L\lambda\lambda}(\tilde\btheta)\}^{-1}\big]_{[(j-1)(K+1)+1:j(K+1),]}\Big\|\le C_{\delta} J^{-1/2};\end{align*} 
    \item With probability at least $1 - J^{-\delta} - (NJ)^{-\delta} - (N+J)^{-\delta}$, for any $i\in[N]$, we have\begin{align*}
        \Big\|\big[\Db_m^{1/2}\{\tilde\Hb(\tilde\btheta)\}^{-1}&\Db_m^{1/2}\big]_{[J(K+1)+(i-1)(K+1)+1:J(K+1)+i(K+1),]} \\&- J[\{\Hb_{Lff}(\tilde\btheta)\}^{-1}]_{[(i-1)(K+1)+1:i(K+1),]}\Big\|\le  C_{\delta}N^{-1/2}.\end{align*}
    \end{enumerate}
    Define $P(\btheta) = NJ\big\|N^{-1}\sum_{i=1}^N\bbf_i\bbf_i^\T-\Ib_K\big\|_F^2/2+NJ\big\|\text{ndiag}\big(J^{-1}\sum_{j=1}^J\blambda_j\blambda_j^\T\big)\big\|_F^2/2 $\\$+ NJ\big\|N^{-1}\sum_{i=1}^N\bbf_i\big\|^2 + NJ\big\|J^{-1}\sum_{j=1}^J\blambda_j\big\|^2 + NJ\big(N^{-1}\sum_{i=1}^Nf_{i0}\big)^2$ and $\Hb^*(\btheta) = \Hb_L(\btheta) + c\partial^2_{\btheta\btheta}P(\btheta)$. Then (i)--(ix) also hold for $\int_0^1\Hb^*\big(\btheta^{(0)} +s(\tilde\btheta-\btheta^{(0)})\big)ds$.
\end{lemma}
The proof is similar to that Lemma~\ref{lemma_hessian_property}. With Lemma~\ref{lemma_hessian_property_int}, following the step 2 and step 3 in Section~\ref{sec:supp_prove_consistency_ic0}, we have the following result.
\begin{proposition}\label{lemma_first_order_condition_int}\it
    Under Assumptions~\ref{assump:psd covariance int}, \ref{assumption:smoothness int}, and~\ref{assump:scaling}, for any constant $\delta>1$, when $N$ and $J$ are large enough, with probability at least $1-(N\wedge J)^{-\delta}$
     \begin{equation*}
        \frac{1}{J}\sum_{j=1}^J\big\|\hat\blambda_{-j}^{(0)}-\blambda_{-j}^{(0)}\big\|^2\le C_\delta\big(N^{-1}+J^{-2}\big),\quad \frac{1}{N}\sum_{i=1}^N\big\|\hat\bbf_{-i}^{(0)}-\bbf_{-i}^{(0)}\big\|^2\le C_\delta\big(J^{-1} + N^{-2}\big),
    \end{equation*}where $C_\delta$ is a constant only depending on $\delta$. Write $\epsilon_{NJ} := (\epsilon_N\log J + \epsilon_J\log N)^{1/2}$. For any $\delta>1$ and any $\epsilon_N,\epsilon_J\to\infty$, with probability at least $1-(N\wedge J)^{-\delta} - \exp(-\epsilon_N\wedge \epsilon_J)$, we have  \begin{equation*}
        \max_{1\le j\le J}\big\|\hat\blambda_{-j}^{(0)}-\blambda_{-j}^{(0)}\big\|\le C_{\delta} \frac{\epsilon_{NJ}\log J}{N\wedge J},
        \quad \max_{1\le i\le N}\big\|\hat\bbf_{-i}^{(0)}-\bbf_{-i}^{(0)}\big\|\le  C_{\delta}\frac{\epsilon_{NJ}\log N}{N\wedge J},
    \end{equation*}where $C_{\delta}$ is some constant only depending on $\delta$.
\end{proposition}
We have the following result following the proof in Section~\ref{sec:use_clt}.
   \begin{proposition}\label{lemma_precise_construct_int}\it
Under Assumptions~\ref{assump:psd covariance int}, \ref{assumption:smoothness int}, and~\ref{assump:scaling}, for any constant $\delta>1$, when $N$ and $J$ are large enough
\begin{align}
\hat\blambda_{-j}^{(0)}-\blambda_{-j}^{(0)}=-\big(\sum_{i=1}^Nl_{ij}^{\prime\prime}(\eta_{ij}^*)\bbf_i^{(c0)}(\bbf_i^{(c0)})^\T\big)^{-1}\sum_{i=1}^Nl_{ij}^{\prime}(\eta_{ij}^*)\bbf_i^{(c0)}+O_p\Big(\frac{1}{N\wedge J}\Big),\label{eq_precise_blambda_int}\\
        \hat\bbf_{-i}^{(0)}-\bbf_{-i}^{(0)}=-\big(\sum_{j=1}^Jl_{ij}^{\prime\prime}(\eta_{ij}^*)\blambda_j^{(c0)}(\blambda_j^{(c0)})^\T\big)^{-1}\sum_{j=1}^Jl_{ij}^{\prime}(\eta_{ij}^*)\blambda_j^{(c0)}+O_p\Big(\frac{1}{N\wedge J}\Big).\label{eq_precise_f_int}
        \end{align}
        Here $\bbf_i^{(c0)} = \big(1,(\bbf_i^{(0)})^\T\big)^\T$ and $\blambda_j^{(c0)} = \big(1,(\blambda_j^{(0)})^\T\big)^\T$
   \end{proposition}

Theorem~\ref{thm_consis_int} can be obtained from Proposition~\ref{lemma_first_order_condition_int}. Theorem~\ref{main_thm_asymp} can be proved with Proposition~\ref{lemma_precise_construct_int} and a similar verification of Lindeberg-Feller conditions as in Section~\ref{sec:use_clt}.

To derive the results under the other identifiability conditions, we use the same reduction as for factor models without intercepts. Suppose that the true parameters $(\bLambda^{(S)},\Fb^{(S)},\bbeta_{\lambda0}^{(S)},\bbeta_{f0}^{(S)})$ that satisfy IC$^{(S)}$ and ICC. Following a similar strategy introduced at the start of Section~\ref{sec:supp_prove_consistency_ic1}, we know that there exists an invertible matrix $\Gb^{(S)}$ such that $\big(\bLambda^{(S)}(\Gb^{(S)})^{-1},\Fb^{(S)}(\Gb^{(S)})^{\T},\bbeta_{\lambda0}^{(S)},\bbeta_{f0}^{(S)}\big)$ satisfy IC$^{(0)}$ and ICC. Here $\bbeta_{\lambda0}^{(S)}$ and $\bbeta_{f0}^{(S)}$ remain unchanged. Then $\big(\bLambda^{(S)}(\Gb^{(S)})^{-1},$ $\Fb^{(S)}(\Gb^{(S)})^{\T},\bbeta_{\lambda0}^{(S)},\bbeta_{f0}^{(S)}\big)$ can be viewed as a variant of the true parameters. Note that the intercepts are unchanged because ICC is always required. Therefore we can treat $\big(\bLambda^{(S)}(\Gb^{(S)})^{-1},\Fb^{(S)}(\Gb^{(S)})^{\T},\bbeta_{\lambda0}^{(S)},\bbeta_{f0}^{(S)}\big)$ as the true parameters that satisfy IC$^{(0)}$ and ICC, and Propositions~\ref{lemma_consistency_int}--\ref{lemma_precise_construct_int} still hold for the true parameters and $\hat\btheta^{(0)}$ obtained in \eqref{eq_joint_mle_intij} when setting $S=0$. Inspired by the proof in Section~\ref{sec:supp_prove_consistency}, it suffices to show that there exists $\hat\Gb^{(S)}$ such that 
\begin{itemize}
    \item $(\hat\bLambda^{(0)}\hat\Gb^{(S)},\hat\Fb^{(0)}(\hat\Gb^{(S)})^{-\T},\hat\bbeta_{\lambda0}^{(0)},\hat\bbeta_{f0}^{(0)})$ satisfy identifiability condition IC$^{(S)}$ and ICC;
    \item $\|\hat\Gb^{(S)}-\Gb^{(S)}\|=o_p(1)$ for a data-dependent matrix $\hat\Gb^{(S)}$;
\end{itemize}
then $\big(\hat\bLambda^{(0)}\hat\Gb^{(S)},\hat\Fb^{(0)}(\hat\Gb^{(S)})^{-\T},\hat\bbeta_{\lambda0}^{(0)},\hat\bbeta_{f0}^{(0)}\big)$ is inside $\cB(D)$ w.h.p., and therefore by definition $\hat\bLambda^{(S)}=\hat\bLambda^{(0)}\hat\Gb^{(S)},\hat\Fb^{(S)}=\hat\Fb^{(0)}(\hat\Gb^{(S)})^{-\T}$, up to a signed permutation and $\hat\bbeta_{\lambda0}^{(S)}= \hat\bbeta_{\lambda0}^{(0)}$ and $\hat\bbeta_{f0}^{(S)} = \hat\bbeta_{f0}^{(0)}$. For the intercepts we have, with probability at least $1-(N+J)^{-\delta} - (NJ)^{-\delta}$,
\begin{equation*}
    \frac{1}{J}\sum_{j=1}^J\big(\hat\lambda_{j0}^{(S)} - \lambda_{j0}^{(S)}\big)^2 \le C_{\delta}\frac{1}{N\wedge J^2},\; \frac{1}{N}\sum_{i=1}^N\big(\hat f_{i0}^{(S)} - f_{i0}^{(S)}\big)^2 \le C_{\delta}\frac{1}{N^2\wedge J},
\end{equation*}where $C_{\delta}$ is some constant only depending on $\delta$.
Similar to the proofs in Section~\ref{sec:supp_prove_consistency}, $\|\hat\Gb^{(S)} - \Gb^{(S)}\| \lesssim \sqrt{\epsilon_N/N}+1/J$ with probability at least $1-\exp(-C\epsilon_N)-4N^{-\delta}-J^{-\delta}-2(NJ)^{-\delta} - (N+J)^{-\delta}$,  and the consistency results for $\btheta^{(S)}$ in Theorem~\ref{thm_consis_int} can be readily derived.

Next, we sketch the proofs of the limiting distributions of $\hat\btheta^{(S)}$. Define 
\begin{equation*}
    \hat\Gb_{-}^{(S)}=\begingroup
    \renewcommand{\arraystretch}{0.8}
    \begin{pmatrix}
            1&\zero_K^\T\\\zero_k&\hat\Gb^{(S)}
    \end{pmatrix}
    \endgroup\text{ and } \Gb_{-}^{(S)}=\begingroup
    \renewcommand{\arraystretch}{0.8}
    \begin{pmatrix}
            1&\zero_K^\T\\\zero_k&\Gb^{(S)}
    \end{pmatrix}
    \endgroup
\end{equation*}
Then we can write
\begin{align*}
    \hat\blambda_{-j}^{(S)} - \blambda_{-j}^{(S)} = &\,\Gb_{-}^{(S)}\big(\hat\blambda_{-j}^{(0)} - \blambda_{-j}^{(0)}\big) +  \big(\hat\Gb_{-}^{(S)} - \Gb_{-}^{(S)}\big)\blambda_{-j}^{(0)}\\&\, + \big(\hat\Gb_{-}^{(S)} - \Gb_{-}^{(S)}\big)\big(\hat\blambda_{-j}^{(0)} - \blambda_{-j}^{(0)}\big)\\ = &\, \Gb_{-}^{(S)}\big(\hat\blambda_{-j}^{(0)} - \blambda_{-j}^{(0)}\big)  + \Gb_{-}^{(S)}\big(\Xb_{-}^{(S)}\big)^\T\blambda_{-j}^{(0)} + O_p\big(N^{-1}+J^{-1}\big)
\end{align*}
with 
\begin{equation*}
    \Xb_{-}^{(S)} = \begingroup
    \renewcommand{\arraystretch}{0.8}
    \begin{pmatrix}
            0&\zero_K^\T\\\zero_k&(\Gb^{(S)} - \hat\Gb^{(S)})^\T(\Gb^{(S)})^{-\T}
    \end{pmatrix}=
    \begin{pmatrix}
            0&\zero_K^\T\\\zero_K&\Xb^{(S)}
    \end{pmatrix}
    \endgroup
\end{equation*}
Similar to the proofs in Section~\ref{sec:proof limit} we can derive by Lemma~\ref{lemma_precise_construct_int} that for $S\in\{1,\cdots,5\}$ when $\sqrt{N}/J\to 0$,
\begin{equation*}
    \big(\bSigma_{[N],-j}^{(S)}\big)^{-1/2}\big(\hat\blambda_{-j}^{(S)} - \blambda_{-j}^{(S)}\big)\overset{d}{\to}\cN(\zero_{|\cS_j|+1},\Ib_{|\cS_j|+1}).
\end{equation*}
For $\hat\bbf_{-i}^{(S)}$, similarly we can show for $S\in\{1,\cdots,5\}$
\begin{equation*}
    \big(\bSigma_{-i,[J]}^{(S)}\big)^{-1/2}\big(\hat\bbf_{-i}^{(S)} - \bbf_{-i}^{(S)}\big)\overset{d}{\to}\cN(\zero_{K+1},\Ib_{K+1}).
\end{equation*}

\section{Proofs of Theoretical Results of the Generalized Latent Factor Model with Random Factors}\label{sec_sketch_proof_random}
In this Section, we outline the proofs for  Theorems~\ref{thm_consis_random} and~\ref{main_thm_asymp_random}. We establish the results under $\overline{\mathrm{IC}}^{(1)}$ and $\overline{\mathrm{IC}}^{(4)}$, and the results under other identifiability conditions can be similarly derived. 

For $\overline{\mathrm{IC}}^{(1)}$, we first show that for $\Mb_{ff} = N^{-1}\sum_{i=1}^N\bbf_i\bbf_i^\T$, there is
\begin{equation}
    \big\|\Mb_{ff} - \Ib_K\big\| = O_p(N^{-1/2}\big).\label{eq_random_mff1}
\end{equation} Specifically, note that by Assumption~\ref{assump_random_factor},
\begin{align*}
    \Mb_{ff} - \Ib_K = &\, N^{-1}\sum_{i=1}^N\bbf_i\bbf_i^\T - N^{-1}\sum_{i=1}^N\EE[\bbf_i\bbf_i^\T]\\&\, = N^{-1}\sum_{i=1}^N\big\{\bbf_i - \EE[\bbf_i]\big\}\bbf_i^\T + N^{-1}\sum_{i=1}^N\bbf_i\big\{\bbf_i - \EE[\bbf_i]\big\}^\T\\&\, +N^{-1}\sum_{i=1}^N\big\{\bbf_i - \EE[\bbf_i]\big\}\big\{\bbf_i - \EE[\bbf_i]\big\}^\T.
\end{align*}
With $\PP(\max_{i\in[N]}\|\bbf_i\|_{\infty}\le D/2)\to 1$, $\EE[\bbf_i\bbf_i^\T] = \Ib_K$, Cauchy's inequality and Chebyshev’s inequality, we obtain \eqref{eq_random_mff1}. Obviously \eqref{eq_random_mff1} also holds for $\Mb_{ff}^* = N^{-1}\sum_{i=1}^N\bbf_i^*\bbf_i^*{}^\T$. 

Recall that we have defined $\bar\bbf_i^{(1)} = \Mb_{ff}^{-1/2}\bbf_i$ and $\bar\blambda_j^{(1)} = \Mb_{ff}^{1/2}\blambda_j^*$, and the associated matrix forms as $\bar\Fb^{(1)}$ and $\bar\bLambda^{(1)}$, respectively. One can check that $(\bar\bLambda^{(1)},\bar\Fb^{(1)})$ as the true parameters satisfy IC$^{(1)}$, as well as Assumption~\ref{assump_random_factor} with probability approaching $1$ as $N\to\infty$ (with a slightly large $D$). Therefore Theorems~\ref{thm_consis}--\ref{main_thm_factor_asym} hold for $(\hat\bLambda^{(1)} ,\hat\Fb^{(1)})$ as the estimator and $(\bar\bLambda^{(1)},\bar\Fb^{(1)})$ as the true parameters. Consequently, we know
\begin{equation*}
        \frac{1}{J}\sum_{j=1}^J\big\|\hat\blambda_j^{(1)}-\bar\blambda_j^{(1)}\big\|^2 = O_p \big(N^{-1}+J^{-2}\big),\quad \frac{1}{N}\sum_{i=1}^N\big\|\hat\bbf_i^{(1)}-\bar\bbf_i^{(1)}\big\|^2 = O_p\big(J^{-1} + N^{-1}\big),
    \end{equation*}
and
    \begin{equation*}
        \max_{1\le j\le J}\big\|\hat\blambda_j^{(1)}-\bar\blambda_j^{(1)}\big\|_{\infty} = O_p\Big( \frac{\epsilon_{NJ}\log J}{\sqrt{N\wedge J}}\Big),
        \quad \max_{1\le i\le N}\big\|\hat\bbf_i^{(1)}-\bar\bbf_i^{(1)}\big\|_{\infty} = O_p\Big(\frac{\epsilon_{NJ}\log N}{\sqrt{N\wedge J}}\Big),
    \end{equation*}
With \eqref{eq_random_mff1} holding for $\Mb_{ff}$ and the expression for $(\bar\bLambda^{(1)},\bar\Fb^{(1)})$, we establish Theorem~\ref{thm_consis_random}. 

Next, we derive the asymptotic distributions. By standard multivariate Lindeberg–Feller CLT~\citep{billingsley2017probability} and Assumption~\ref{assump_random_factor}, we know
\begin{equation*}
    \bZ_N^{(1)}:=\sqrt{N}\mathrm{vec}(\Mb_{ff} - \Ib_K)\overset{d}{\to } \bZ^{(1)}\sim \cN\big(\zero_{K^2}, Cov\big\{\mathrm{vec}(\bbf_1\bbf_1^\T)\big\}\big) = \cN(\zero_K,\Ib_K + \Sbb_K),
\end{equation*}
where $\Sbb_K$ denotes the commutation matrix. Because $\bZ_N$ is finite, and thus is uniformly stochastically bounded. By Theorems 1 and 2 in \cite{sweeting1980uniform} and that $\{\bbf_i\}$ are i.i.d. normal, we know that
\begin{equation}
    \bZ_N^{(1)}|\cG^{(1)} = \bZ_N^{(1)}|\sigma\big(\bar\bbf_1^{(1)},\dots,\bar\bbf_N^{(1)}\big)\overset{d}{\to} \bZ^{(1)}.\label{eq_random_f_i_dis_base}
\end{equation}

Note that conditioned on $\cG^{(1)}$, we can derive a similar asymptotic distributions in Theorem~\ref{main_thm_loading_asym} as
\begin{equation}
    \big\{\Db_{(j)}\bSigma_{[N],j}^{(1)}\Db_{(j)}\big\}^{-1/2}\big\{\Db_{(j)}\big(\hat\blambda_j^{(1)} - \bar\blambda_j^{(1)}\big)\big\} \overset{d}{\to} \cN\big(\zero_{|\cS_j|},\Ib_{|\cS_j|}\big)\text{ as }N,J\to\infty.\label{eq_random_lambda_j_dis_s1}
\end{equation}
and
\begin{equation}
    \big\{\bSigma_{i,[J]}^{(1)}\big\}^{-1/2}\big(\hat\bbf_i^{(1)} - \bar\bbf_i^{(1)}\big)\overset{d}{\to}\cN(\zero_K,\Ib_K),\text{ as }N,J\to\infty.\label{eq_random_f_i_dis_s1}
\end{equation}
Here, $\bar \blambda_j^{(1)} = (\Mb_{ff})^{1/2}\blambda_j^*$ and $\bar \bbf_i^{(1)} = (\Mb_{ff})^{-1/2}\bbf_i$.
Then we can decompose $\hat\blambda_j^{(1)} - \blambda_j^*$ into
\begin{align*}
    \hat\blambda_{j}^{(1)} - \blambda_j^* = &\,\hat\blambda_{j}^{(1)} -\bar\blambda_j^{(1)} + \bar\blambda_j^{(1)} - \blambda_j^* \\ = &\, \hat\blambda_{j}^{(1)} -\bar\blambda_j^{(1)} + \big(\Mb_{ff} - \Ib_K\big)\big(\Mb_{ff}^{1/2} + \Ib_K\big)^{-1}\blambda_j^*.
\end{align*}
For the first term, from the proof in Section~\ref{sec_limit_ic1} we note that it can be decomposed as 
\begin{equation*}
    \hat\blambda_j^{(1)} - \bar\blambda_j^{(1)} = \Big\{\sum_{i=1}^Nl_{ij}^{\prime\prime}\big(\bar\blambda_j^{(1)}{}^\T\bar\bbf_i^{(1)}\big)\bar\bbf_i^{(1)}\bar\bbf_i^{(1)}{}^\T\Big\}^{-1}\sum_{i=1}^Nl_{ij}^{\prime}\big(\bar\blambda_j^{(1)}{}^\T\bar\bbf_i^{(1)}\big)\bar\bbf_i^{(1)} + \hat\Xb^{(1)}\bar\blambda_j^{(1)} + O_p\big((N\wedge J)^{-1}\big),
\end{equation*}
where $(\bar\bLambda^{(1)},\bar\Fb^{(1)})$ is now viewed as the true parameters, as our analysis is conditioned on $\cG^{(1)} = \sigma(\bar\bbf_1^{(1)},\dots,\bar\bbf_N^{(1)})$. By Theorem~\ref{main_thm_loading_asym}, we know this term is asymptotically normal with \\$\big\{\Db_{(j)}\check{\bSigma}_{[N],j}^{(1)}\Db_{(j)}^\T\big\}^{-1/2}\Db_{(j)}(\hat\blambda_j^{(1)} - \bar\blambda_j^{(1)}) \overset{d}{\to}\cN(\zero_{|\cS_j|},\Ib_{|\cS_j|})$. Here $\check{\bSigma}_{[N],j}^{(1)}$ denotes $\bSigma_{[N],j}^{(1)}$ with $(\bLambda^*,\Fb^*)$ replaced by $(\bar\bLambda^{(1)},\bar\Fb^{(1)})$. Since $(\bar\bLambda^{(1)},\bar\Fb^{(1)})\overset{p}{\to}(\bLambda^*,\Fb^*)$ by \eqref{eq_random_mff1}, the asymptotic distribution also holds when replacing $\check{\bSigma}_{[N],j}^{(1)}$ with $\bSigma_{[N],j}^{(1)}$ by Slutsky's Theorem. For the second term, conditioned on $\cG^{(1)} = \sigma(\bar\bbf_1^{(1)},\dots,\bar\bbf_N^{(1)})$, by \eqref{eq_random_mff1} and \eqref{eq_random_f_i_dis_base}, we know that
\begin{equation*}
    \Big\{(2N)^{-1}(\blambda_j^*{}^\T\otimes \Ib_K)(\Ib_{K} + \Sbb_K)(\blambda_j^*\otimes \Ib_K)\Big\}^{-1/2}\big\{\Mb_{ff} - \Ib_K\big\}\big\{(\Mb_{ff})^{1/2} + \Ib_K\big\}^{-1}\blambda_j^*\overset{d}{\to}\cN(\zero_K,\Ib_K).
\end{equation*}
Finally, with the moment condition assumed in Assumption~\ref{assump_random_factor} under the orthogonal case and Markov inequality, we know that
\begin{equation*}
    \Big\|Cov\Big[\Big\{\sum_{i=1}^Nl_{ij}^{\prime\prime}\big(\bar\blambda_j^{(1)}{}^\T\bar\bbf_i^{(1)}\big)\bar\bbf_i^{(1)}\bar\bbf_i^{(1)}{}^\T\Big\}^{-1}\sum_{i=1}^Nl_{ij}^{\prime}\big(\bar\blambda_j^{(1)}{}^\T\bar\bbf_i^{(1)}\big)\bar\bbf_i^{(1)}, \bar\blambda_j^{(1)} - \blambda_j^*\Big]\Big\| = O_p\big((N\wedge J)^{-1}\big),
\end{equation*}for each $j\in[J]$.
For the covariance between $\hat\Xb^{(1)}\bar\blambda_j^{(1)}$ can be bounded similarly, as the expression $\hat\Xb^{(1)}\bar\blambda_j^{(1)}$ involves \begin{equation*}
    \Big\{\sum_{i=1}^Nl_{ij}^{\prime\prime}\big(\bar\blambda_j^{(1)}{}^\T\bar\bbf_i^{(1)}\big)\bar\bbf_i^{(1)}\bar\bbf_i^{(1)}{}^\T\Big\}^{-1}\sum_{i=1}^Nl_{ij}^{\prime}\big(\bar\blambda_j^{(1)}{}^\T\bar\bbf_i^{(1)}\big)\bar\bbf_i^{(1)},
\end{equation*}
for some finite $j$. Therefore, we obtain \eqref{eq_converge_d_lambda_random}. For the asymptotic distributions for $\hat\bbf_i^{(1)}$, we note that
\begin{align*}
    \hat\bbf_i^{(1)} - \bbf_i &\, = \hat\bbf_i^{(1)} - \bar\bbf_i^{(1)} + \bar\bbf_i^{(1)} - \bbf_i \\&\,= \hat\bbf_i^{(1)} - \bar\bbf_i^{(1)} - (\Mb_{ff} - \Ib_K)(\Mb_{ff}^{1/2} + \Ib_K)^{-1} \bar\bbf_i^{(1)}
\end{align*}
Conditioned on $\cG^{(1)}$, we know that both terms are asymptotically normal, where $\big\{\bSigma_{i,[J]}^{(1)}\big\}^{-1/2}$ $(\hat\bbf_i^{(1)} - \bar\bbf_i^{(1)})\overset{d}{\to}\cN(\zero_K,\Ib_K)$, and for the second term, we have
\begin{equation*}
    \Big\{(2N)^{-1}(\bar\bbf_i^{(1)}{}^\T\otimes \Ib_K)(\Ib_{K} + \Sbb_K)(\bar\bbf_i^{(1)}\otimes \Ib_K)\Big\}^{-1/2}\big(\Mb_{ff} - \Ib_K\big)\big(\Mb_{ff}^{1/2} + \Ib_K\big)^{-1}\bar\bbf_i^{(1)}\overset{d}{\to}\cN(\zero_K,\Ib_K).
\end{equation*}
The covariance between $\hat\bbf_i^{(1)} - \bar\bbf_i^{(1)}$ and $ \bar\bbf_i^{(1)} - \bbf_i $, by the expansion for $\hat\bbf_i^{(1)} - \bar\bbf_i^{(1)}$ derived in Section~\ref{sec_limit_ic1}, moment condition assumed in Assumption~\ref{assump_random_factor} under the orthogonal case, and Markov inequality, it can be bounded by $O_p\big((N\wedge J)^{-1}\big)$. Therefore, \eqref{eq_asym_int_f_random} can be similarly obtained.

For $\overline{\mathrm{IC}}^{(4)}$, similarly, we have $\bar\bbf_i^{(4)} = \mathrm{diag}(\Mb_{ff}^*)^{-1/2}\bbf_i$ and $\bar\blambda_j^{(4)} = \mathrm{diag}(\Mb_{ff}^*)^{1/2}\blambda_j^{*}$, and the associated matrix forms as $\bar\Fb^{(4)}$ and $\bar\bLambda^{(4)}$, respectively. One can check that $(\bar\bLambda^{(4)},\bar\Fb^{(4)})$ as the true parameters satisfy IC$^{(4)}$, as well as Assumption~\ref{assump_random_factor} with probability approaching $1$ as $N\to\infty$ (with a slightly large $D$). Then Theorems~\ref{thm_consis}--\ref{main_thm_factor_asym} hold for $(\hat\bLambda^{(4)} ,\hat\Fb^{(4)})$ as the estimator and $(\bar\bLambda^{(4)},\bar\Fb^{(4)})$ as the true parameters.
We can similarly show that
\begin{equation}
    \big\|\mathrm{diag}(\Mb_{ff}) - \Ib_K\big\| = O_p(N^{-1/2}\big).\label{eq_random_mff4}
\end{equation}
The consistency results can be immediately obtained via a similar procedure. For the asymptotic distribution, define $\bZ_N^{(4)} = \big\{(\Mb_{ff})_{[1,1]}-1,\dots, (\Mb_{ff})_{[K,K]}-1\big\}$ and $\bZ^{(4)}$ be a normal random vector with distribution $\cN(\zero_K,\bSigma\circ\bSigma)$. Then one can verify by a similar argument that  
\begin{equation}
    \bZ_N^{(4)}|\cG^{(4)} = \bZ_N^{(4)}|\sigma\big(\mathrm{diag}(\Mb_{ff})^{-1/2}\bbf_1,\dots, \mathrm{diag}(\Mb_{ff})^{-1/2}\bbf_N\big)\overset{d}{\to} \bZ^{(4)}.\label{eq_random_f_i_dis_base_S4}
\end{equation}
For $\hat\blambda_j^{(4)}$, one has
\begin{align*}
    \hat\blambda_{j}^{(4)} - \blambda_j^* = &\,\hat\blambda_{j}^{(4)} -\bar\blambda_j^{(4)} + \big\{\mathrm{diag}(\Mb_{ff})^{1/2} - \Ib_K\big\}\blambda_j^* \\ = &\, \hat\blambda_{j}^{(4)} -\bar\blambda_j^{(4)} + \big\{\mathrm{diag}(\Mb_{ff}) - \Ib_K\big\}\big\{\mathrm{diag}(\Mb_{ff})^{1/2} + \Ib_K\big\}^{-1}\blambda_j^*.
\end{align*}
Similarly, we can obtain \eqref{eq_converge_d_lambda_random} by the expansion of $\hat\blambda_j^{(4)} - \bar\blambda_j^{(4)}$ derived in Section~\ref{sec_limit_ic4} and the moment condition in Assumption~\ref{assump_random_factor}. And \eqref{eq_asym_int_f_random} follows by symmetry.

\end{document}